\documentclass{aa}  

\usepackage{graphicx}
\usepackage{txfonts}
\usepackage{gensymb}
\usepackage{verbatim}
\usepackage{siunitx}
\usepackage{amsmath}
\usepackage{caption, subcaption}
\usepackage[colorlinks,allcolors=blue]{hyperref}

\usepackage{mathtools}
\usepackage{xcolor}
\usepackage[normalem]{ulem}
\usepackage{natbib}
\bibpunct{(}{)}{;}{a}{}{,} 

\makeatletter
\renewcommand*\aa@pageof{, page \thepage{} of \pageref*{LastPage}}
\makeatother
\begin{document}

   \title{High resolution observations of $^{12}$CO and $^{13}$CO $J = 3 \rightarrow 2$ toward the NGC 6334 extended filament}

   \subtitle{I. Emission morphology and velocity structure}

   \author{S. Neupane \inst{1}
   \and F. Wyrowski \inst{1} 
   \and K. M. Menten\inst{1}
   \and J. Urquhart\inst{2}
   \and D. Colombo\inst{1,3}
   \and L.-H. Lin\inst{1}
   \and G. Garay\inst{4}
   }
   \institute{Max-Planck-Institut f\"{u}r Radioastronomie, Auf dem H\"{u}gel 69, 53121 Bonn, Germany \\
   \email{sneupane@mpifr-bonn.mpg.de}          
    \and 
    Centre for Astrophysics and Planetary Science, University of Kent, Canterbury, CT2 7NH, UK 
    \and 
    Argelander-Institut f\"{u}r Astronomie, Auf dem H\"{u}gel 71, 53121 Bonn
    \and 
    Departamento de Astronomia, Universidad de Chile, Casilla 36-D, Santiago, Chile}

   \date{Received 2024; accepted 2024}

%
  \abstract
    {NGC 6334 is a giant molecular cloud (GMC) complex that exhibits elongated filamentary structure and harbours numerous OB-stars, H II regions and star forming clumps.
    To study the emission morphology and velocity structure of the gas in the extended NGC 6334 region using high-resolution molecular line data, we made observations of the $^{12}$CO and $^{13}$CO $J=3 \rightarrow 2$ lines with the LAsMA instrument at APEX telescope . The LAsMA data provided a spatial resolution of 20" ($\sim$0.16 pc) and sensitivity of ~0.4 K at a spectral resolution of 0.25 km s$^{-1}$.
    Our observations revealed that gas in the extended NGC 6334 region exhibits connected velocity coherent structure over $\sim$80 pc  parallel to the galactic plane. The NGC 6334 complex has its main velocity component at $\sim -3.9$ km s$^{-1}$ with two connected velocity structures at velocities $\sim -9.2$ km s$^{-1}$ (the `bridge' features) and $\sim-20$ km s$^{-1}$ (the  Northern Filament, NGC 6334-NF). We observed local velocity fluctuations at smaller spatial scales along the filament that are likely tracing local density enhancement and infall while the broader V-shaped velocity fluctuations observed toward the NGC 6334 central ridge and G352.1 region located in the eastern filament EF1 indicate globally collapsing gas onto the filament. We investigated the $^{13}$CO emission and velocity structure around 42 WISE H II regions located in the extended NGC 6334 region and found that most H II regions show signs of molecular gas dispersal from the center (36 of 42) and intensity enhancement at their outer radii (34 of 42). Furthermore most H II regions (26 of 42) are associated with least one ATLASGAL clump within or just outside of their radii 
    the formation of which may have been triggered by H II bubble expansion. Typically toward larger size H II regions we found visually clear signatures of bubble shells emanating from the filamentary structure. Overall the NGC 6334 filamentary complex exhibits sequential star formation from west to east. Located in the west, the GM-24 region exhibits bubbles within bubbles and is at a relatively evolved stage of star formation. The NGC 6334 central ridge is undergoing global gas infall and exhibits two gas `bridge' features possibly connected to the cloud-cloud collision scenario of the NGC 6334-NF and the NGC 6334 main gas component. The relatively quiescent eastern filament (EF1 - G352.1) is a hub-filament in formation which shows the kinematic signature of global gas infall onto the filament. Our observations highlight the important role of H II regions in shaping the molecular gas emission and velocity structure as well as the overall evolution of the molecular filaments in the NGC 6334 complex.
    } 
  %
\titlerunning{NGC 6334: I. Gas emission morphology and velocity structure}
\authorrunning{Neupane et al.}
\maketitle
%
\section{Introduction}
NGC 6334 is a giant molecular cloud (GMC) complex, harbouring a central dense filament and a large number of OB-type stars, H II regions/bubbles and star forming clumps in the extended region (e.g., \citealt{Loughran1986, Kraemer+Jackson1999, Munoz2007, persitapia2008, andre2016, Russeil2013, Tige2017, Arzoumanian2022}). This complex is located at a distance of $\sim$1.7 kpc (e.g., \citealt{Russeil2012}) and its large-scale molecular appearance is dominated by a $\sim$70 pc long filamentary structure (within 352.6 $\geq l \geq$ 350.2 (where $l$ is Galactic longitude). The central dense ridge of NGC 6334, extending along the filament,  contains sites of high-mass star formation in a sequence of evolutionary stages (e.g, \citealt{Tige2017}) and has been  reported to be undergoing longitudinal collapse, possibly  triggered by a past high-mass star forming event (e.g., \citealt{Zernickel2013,Russeil2013}).
\cite{Arzoumanian2022} studied the extended NGC 6334 region based on their $^{13}$CO and C$^{18}$O J = 2$\rightarrow$1 observations obtained with the APEX telescope and highlighted cases of multiple gas compression in the region. In addition, they found a large number of velocity coherent filaments (VCFs) and interpreted their formation resulting from large-scale compression by propagating shock fronts.  Presenting a broader picture \cite{FukuiY2018CC} proposed a cloud-cloud collision scenario in which a collision is happening from lower to higher longitudes of the extended NGC 6334 region, giving rise to the evolutionary sequence of star formation. A similar  evolutionary sequence from lower longitude (GUM 1-24) to higher (in the NGC 6334 extended filament) is also reported by \cite{Russeil2013}.

\begin{figure*}[htbp!]
    \centering
    \includegraphics[width=0.99\linewidth]{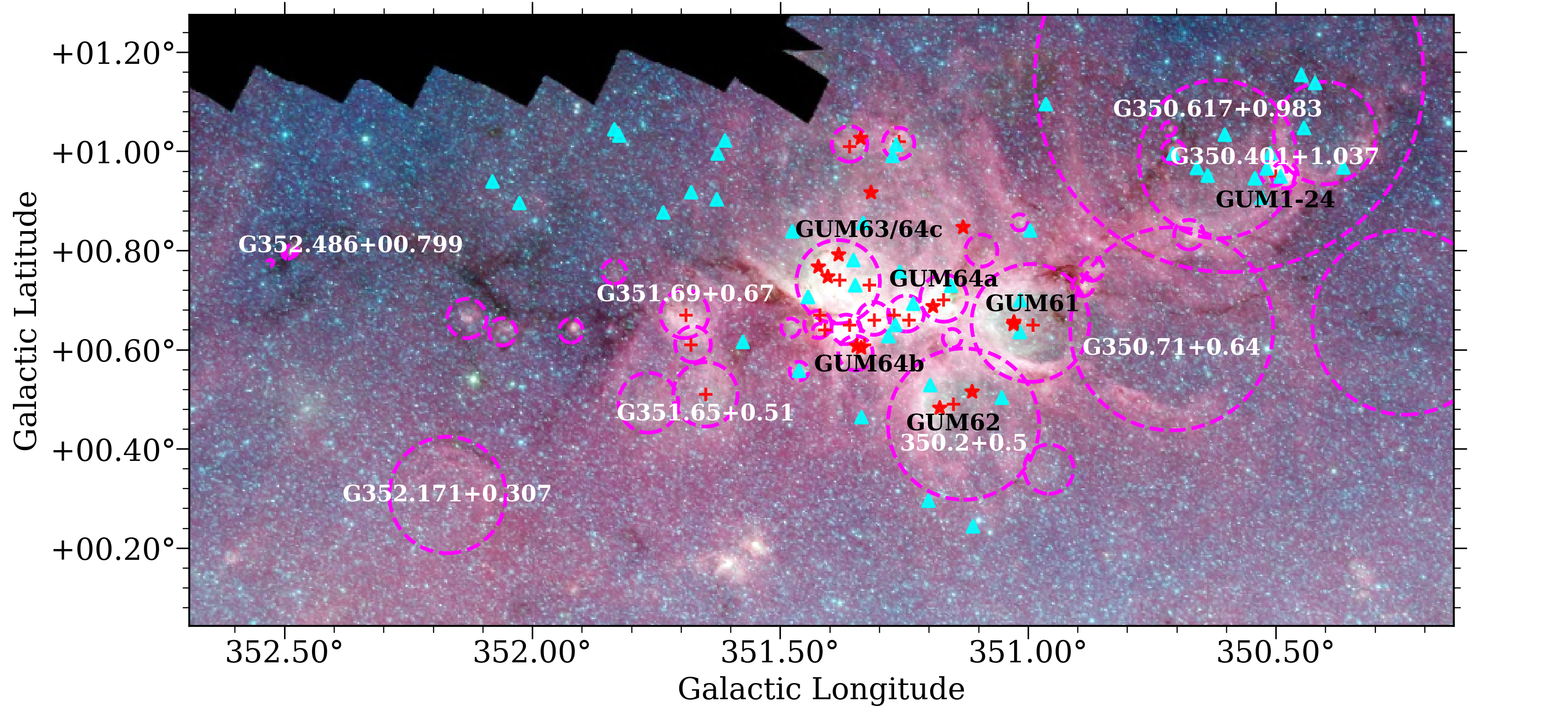}
    \caption{Spitzer three color image (Red: 8 $\mu$m, Green:4.5 $\mu$m, Blue:3.6 $\mu$m) of the NGC 6334 extended region. OB stars from \cite{Russeil2020OBcat} and \cite{persitapia2008} are indicated by cyan triangles and red stars, respectively. H II bubbles from \cite{Anderson2014} are shown by dotted magenta circles, some of which are labelled in white/black text. Plus red markers show positions of radio sources from \cite{Russeil2017}.}
    \label{fig:3color}
\end{figure*}

The formation of the large scale filaments and their fragmentation processes are not understood well (see the review by \citealt{Hacar2023}). High angular resolution continuum and molecular line observations are essential to understand the small to large scale gas kinematics through which gas flows along or onto the filament. Regarding the formation of molecular clouds in the filaments, \cite{Elmegreen1998} have highlighted the important role of gas compression due to H II regions (see also \citealt{Inutsuka2015}). H II regions are signposts of (high-mass) star formation within molecular clouds (e.g., \citealt{Churchwell2002}). New H II regions are formed inside proto-stellar cores forming  high mass stars (OB-type) and their influence is evident on the evolution of the cloud as they emerge. At evolved stages, H II regions can shape the parental cloud material into structures with bubble- or shell-like morphology, in which triggered cases of star formation can proceed (cf. \citealt{Elmegreen1998}). 
The young and evolved H II regions, either found in isolation or in groups, shape the gas emission and velocity structure around them, therefore, their role in natal cloud dispersal and in creating a new generation of star formation requires detailed observational study of the gas kinematics around them.

The extended NGC 6334 region (including GM-24) (see Figure \ref{fig:3color}) contains 42 infrared H II regions/bubbles reported in the catalog\footnote{WISE catalog V2.2: http://astro.phys.wvu.edu/wise/} of \cite{Anderson2014}, of which 14 are classified as `known' (K), 9 as `group' (G), 6 as `candidate' (C) and 13 as `radio quiet' (Q) H II regions. Identification of these regions were primarily based on the mid-infrared characteristics of the WISE all sky survey data (see \citealt{Anderson2014} for more details) and their sizes range between 0.2 pc to 12 pc (see Table \ref{tab1}). 

In this work we aim to study the gas emission morphology and velocity structure of the NGC 6334 extended region using high-resolution observations of $^{12}$CO and $^{13}$CO $J=3 \rightarrow 2$ molecular lines obtained with  the LAsMA instrument on the APEX Telescope. 
Our goal is to disentangle, with this new large-scale, sensitive and high resolution spectral line data, the different origins of the velocity structure in NGC 6334. This is done by  investigating the impact of a large number of H II regions already formed in the giant molecular complex as well as the large scale inflow of gas onto and through the filaments.

The paper is organized as follows. In Section \ref{sec:data} we describe the APEX LAsMA observations toward the region and provide an overview of the data reduction process and resulting sensitivities. In Section \ref{RandA} we present the results. In Section \ref{sec:velocitystructure} we present the velocity structure of the extended filament. In Section \ref{sec:channelmap} gas emission properties around the H II sources will be presented. The Sections \ref{sec:analysis} and \ref{sec:disc} will include analysis and discussion of the results. Finally, in Section \ref{sec:summary} summarizes highlights of this work.

\section{Observations and data reduction}\label{sec:data}
\subsection{LAsMA observations of $^{12}$CO and $^{13}$CO $J=3\rightarrow2$}
We mapped the NGC 6334 star forming complex in the $^{12}$CO $J = 3 \rightarrow 2$ and $^{13}$CO $J = 3 \rightarrow 2$ lines, using the Large APEX sub-Millimetre Array (LAsMA), a 7-pixel heterodyne array receiver installed on the Atacama Pathfinder EXperiment 12 meter submillimeter  telescope  (APEX) located on the Llano de Chajnantor (elevation of $\sim$5100 m) in the Atacama desert, Chile. The LAsMA receiver operates in the 870 $\mu$m (345 GHz) atmospheric window and its 7-pixels are arranged in a hexagonal shape with one central pixel and $40''$ spacing between pixels, which corresponds to $\sim 2$ FWHM beam wdiths. The map  was centered at Galactic coordinates $l$ = 351.415$^{\circ}$ and $b$ = +0.66$^{\circ}$. The area of the extended NGC 6334 region mapped in this project covers $\sim$2.5$^{\circ}$ $\times$ 1.2$^{\circ}$.
   
The observations were made during June to September 2021 under good atmospheric conditions of $\leq$1.5 mm precipitable water vapor (PWV) content. The mapped region was divided into $10^{'} \times 10^{'}$-sized sub-maps (`tiles') that were observed in on-the-fly (OTF) mode in both $l$ and $b$ directions. The  $^{12}$CO $J = 3 \rightarrow 2$ and $^{13}$CO J = $3 \rightarrow 2$ lines were observed simultaneously using a local oscillator frequency of 338.3 GHz. The $^{13}$CO ($\nu_{rest}$ = 330.587 GHz ) line is observed in the lower side band and $^{12}$CO ($\nu_{rest}$ = 345.796 GHz ) line in the upper side band. An advanced  versions of the APEX Fast Fourier Transform Spectrometer (FFTS, \citealt{Klein2012}) was used as a backends, resulting in a spectral resolution of 0.1 km s$^{-1}$. At this observing frequency, the full width half maximum (FWHM) beam width of the telescope is $\sim$19$''$. The OTF observing time for each tile coverage was $\sim$35 minutes and the total time spent to complete this project, including overheads, was approximately $\sim$62 hours.

\subsection{Data reduction process}
Data reduction was performed using the GILDAS{\footnote{https://www.iram.fr/IRAMFR/GILDAS/}} software package. The following steps were taken to obtain the final spectral cubes of the full mapped region: first, we extracted the spectra in the $-$100 km s$^{-1}$ to +100 km s$^{-1}$ LSR velocity range and re-sampled them to a common velocity resolution of 0.25 km s$^{-1}$. The baseline subtracted data sets from different days and sub regions were then combined to produce spectral cubes for the full mapped region. We use the $table$ and $xy\_map$ packages in CLASS-GILDAS to regrid and smooth the data to the desired pixel size and resolution. The pixel size of the final cubes is set to 6$\arcsec$ $\times$ 6$\arcsec$, chosen to affort better than Nyquist sampling as well as to match that of other complementary data sets. In this data reduction procedure, we carefully flagged and removed spectra with bad baseline and those that contain ripples and artifacts. The final spatial resolution for the cubes is 20$\arcsec$, 0.16 pc at the distance of 1.7 kpc. 

\begin{figure}[htbp!]
    \centering
    \includegraphics[width=0.95\linewidth]{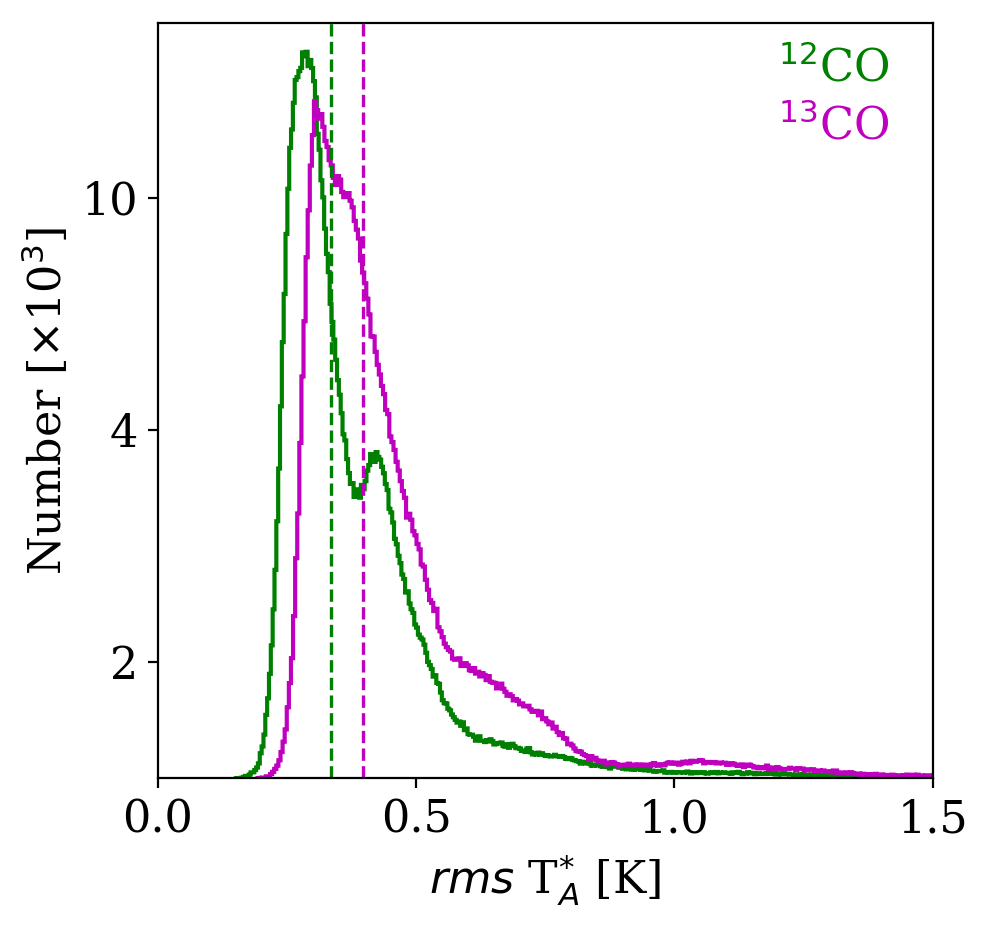}
    \caption{Histogram of the $rms$ T$^{*}_{A}$ [K] in the $^{13}$CO and $^{12}$CO lines. The dotted vertical lines in color indicate the median rms values. }
    \label{fig:rms}
\end{figure}

\subsection{rms sensitivity}
Figure \ref{fig:rms} presents the histogram of the $root~mean~square$ noise ($rms$ in corrected antenna temperature T$^{*}_{A}$ K)\footnote{We used antenna temperature units T$^{*}_{A}$ throughout this work unless otherwise stated in the text.} for both the $^{13}$CO and $^{12}$CO spectra estimated from the first order baseline fit to the spectra. The average $rms$ in T$^{*}_{A}$ (computed per channel per pixel at the velocity resolution of 0.25 km s$^{-1}$) for $^{12}$CO and $^{13}$CO is 0.39 K and 0.46 K, respectively. The rms distribution of the $^{13}$CO emission is presented in Figure \ref{fig:rmsnoisemap}. The median $rms$ in T$^{*}_{A}$ for $^{12}$CO and $^{13}$CO is 0.34 K and 0.40 K, respectively.  A 3$\sigma$ detection in $^{13}$CO roughly correspond to a column density of $N_\mathrm{^{13}CO}\sim \num{3e14}\, \si{\per\square\cm}$  ($N_\mathrm{H_2} \sim 2 \times 10^{20}$ cm$^{-2}$), estimated with RADEX\footnote{RADEX: Non-LTE molecular radiative transfer in an isothermal homogeneous medium by \cite{RADEX2007}, also available online at {https://personal.sron.nl/$\sim$vdtak/radex/index.shtml}} assuming a kinetic temperature of 20 K, density (n$_{H_2}$) of $10^{4}$ cm$^{-3}$ and abundance ratios of $^{12}$CO/$^{13}$CO=77 (\citealt{wilson-rood1994}), $^{12}$CO/H$_2\sim8.5\times 10^{-5}$ (\citealt{Frerking1982}). 

Previous studies of NGC 6334 in the $^{12}$CO $J = 2\rightarrow1$ line line using the NANTEN2 Telescope had a spectral resolution of $\sim$0.1 km s$^{-1}$ and rms noise level of $\sim$1.1 K per channel at an angular resolution of $\sim$90$\arcsec$ (\citealt{FukuiY2018CC}). Furthermore, previous $^{13}$CO and C$^{18}$O $J = 2\rightarrow1$ observations with APEX had a spectral resolution of 0.3 km s$^{-1}$ and sensitivity of about $\sim$0.5 K at an angular resolution of $\sim$30$\arcsec$ covering 2.2$^{\circ}\times0.7^{\circ}$ region (see, \citealt{Arzoumanian2022}). In comparison to both of these previous studies, we have at least 1.5 times higher spatial resolution, better sensitivity and map twice as large an area to cover the extended emission region. In addition, since we observe the higher excitation $^{13}$CO $J = 3\rightarrow2$ lines,
we also probe higher density gas (e.g., $n_{crit}$ $\sim$ \num{1e4} cm$^{-3}$ at 10 K) directly participating in the star forming activity.

\begin{figure}[ht!]
    \centering
    \includegraphics[width=0.95\linewidth]{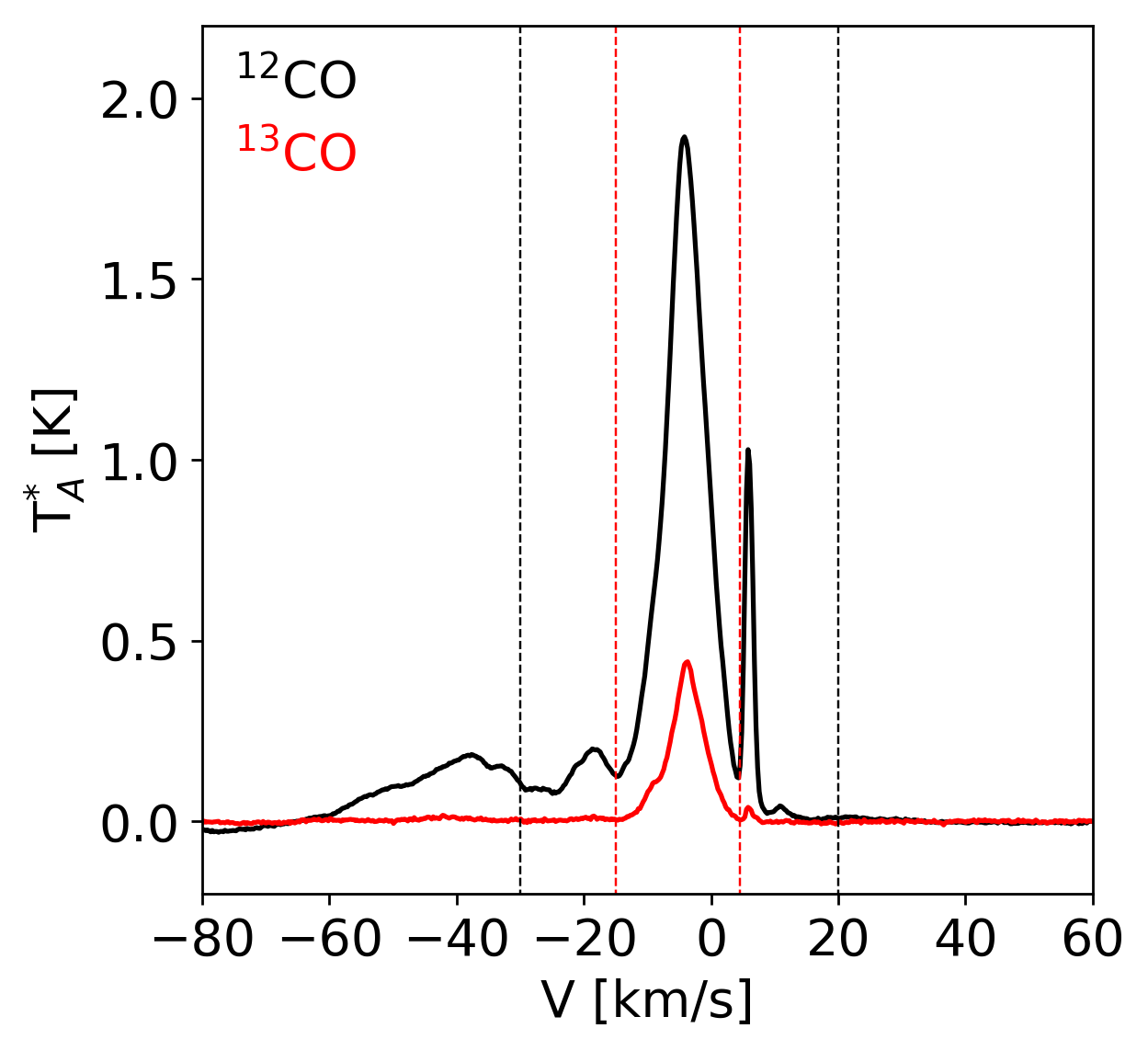}
    \caption{Average of the $^{13}$CO (in red) and $^{12}$CO (in black) spectra toward the NGC 6334 extended filament. The vertical dotted lines indicate velocity ranges of [$-$30,+20] and [$-$15,+5] km s$^{-1}$ in black and red color, respectively.}
    \label{fig:sumspec}
\end{figure}

\section{Results}\label{RandA}

\begin{figure*}[hbtp!]
    \centering
    \includegraphics[width=0.95\linewidth]
    {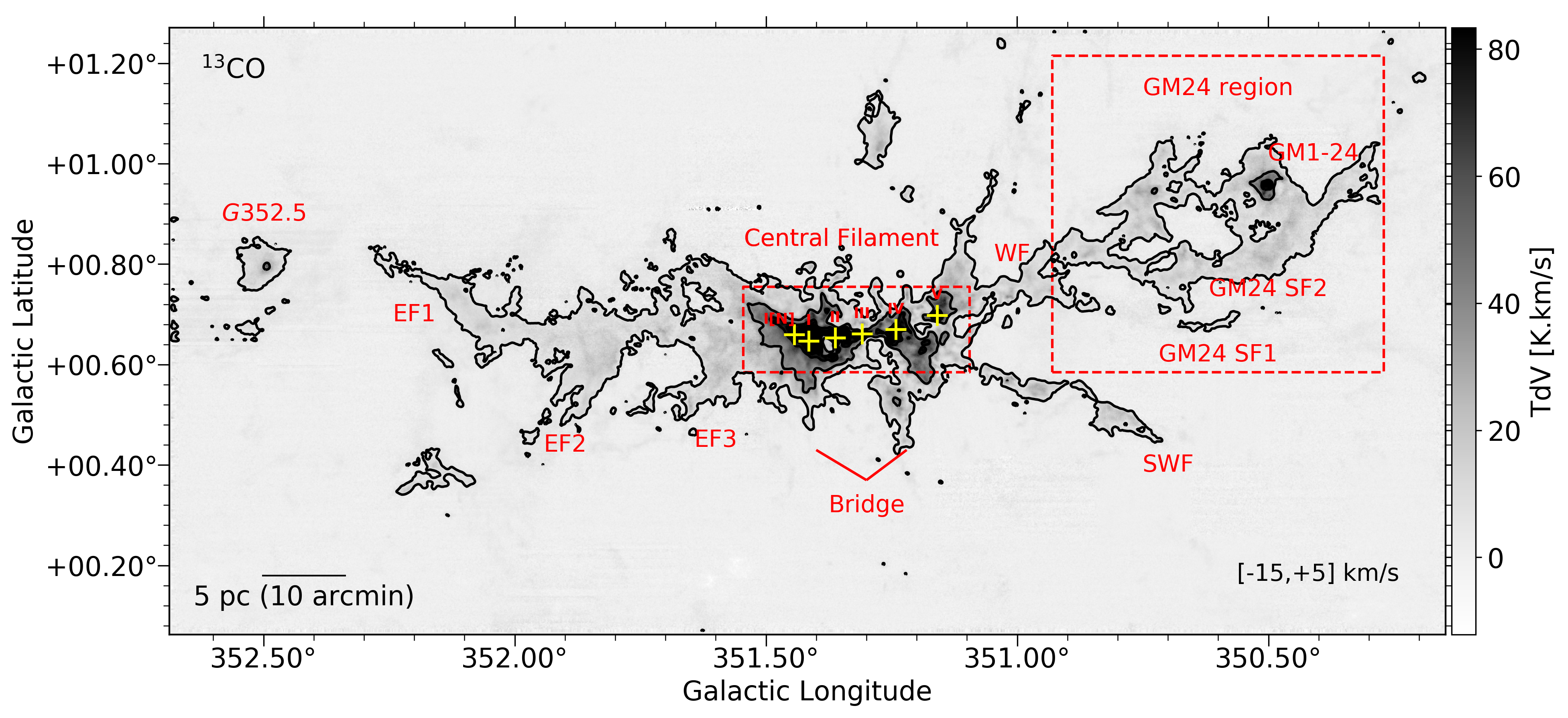}\\
    \includegraphics[width=0.96\linewidth]{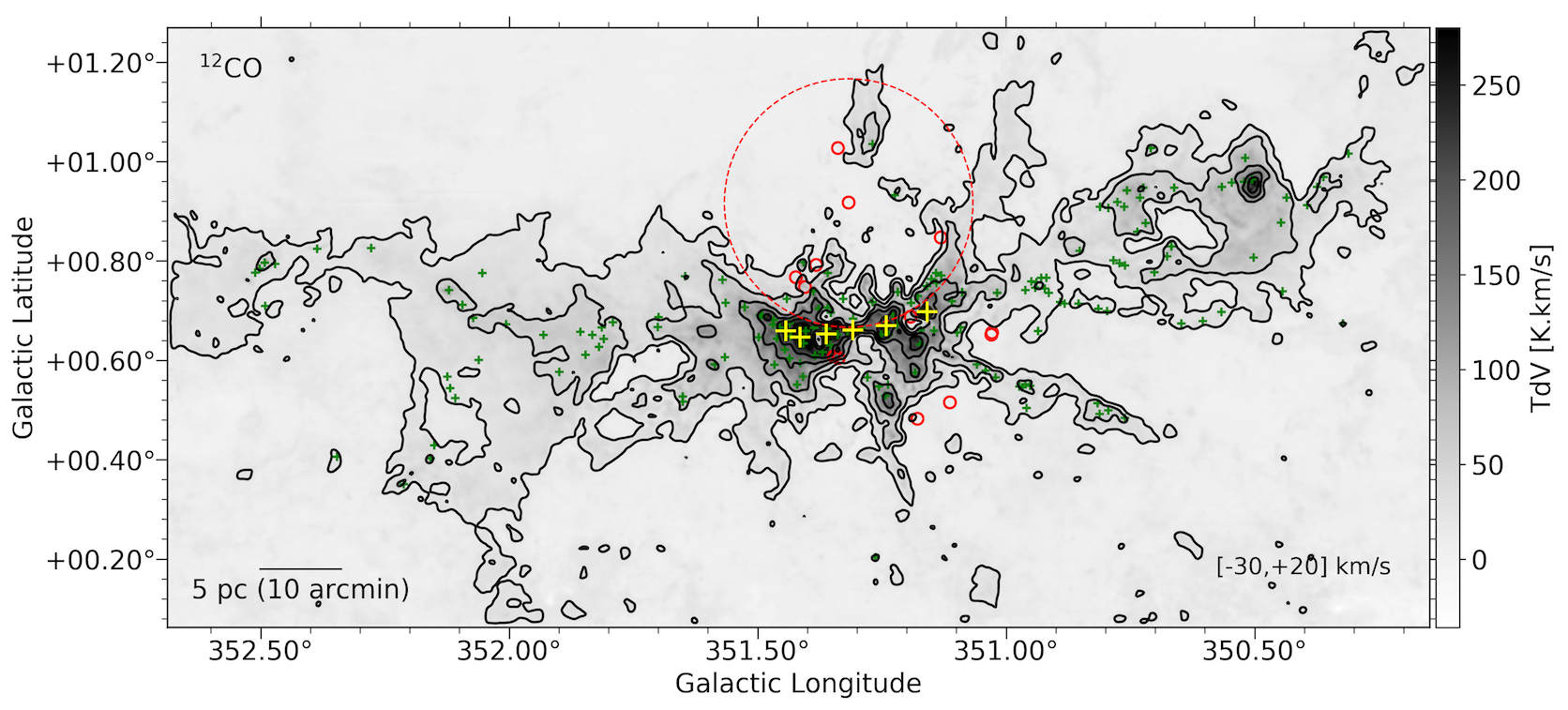}
    \caption{$^{13}$CO (top) and $^{12}$CO (bottom) emission morphologies (moment 0 maps) toward the extended NGC 6334 filament. Contours are drawn at 7, 30 and 70 K km s$^{-1}$ for $^{13}$CO and at 15, 50, 100, 150 and 200 K km s$^{-1}$ for $^{12}$CO. Velocity ranges used for integrating the intensities are given in bottom right corner of the maps. Dashed red boxes in top image outline the central filament and the GM24 region. Various regions discussed in the text are labeled in red. Far-Infrared (FIR) sources I[N], I to V (corresponding to radio sources F to A, see \citealt{Kraemer+Jackson1999}) are marked with yellow plus markers on the images. In the $^{12}$CO moment 0 map  also the OB star positions from \cite{persitapia2008} are shown as red open dots. A circle is drawn centered at $l=351.317, b=0.918$, the position of the dominant O6.5 star, to match the arc-like emission morphology observed toward the NGC 6334 central filament.
    The green plus markers indicate the position of ATLASGAL sources in the NGC 6334 extended region (\citealt{Urquhart2018}).}
    \label{fig:mommaps}
\end{figure*}

\subsection{CO emission morphology}
Figure \ref{fig:sumspec} presents the  $^{13}$CO  and $^{12}$CO  J = 3$\rightarrow2$ spectra as in red and black, respectively, averaged over the extended NGC 6334 filament. 
Most of the $^{13}$CO emission is confined within the velocity range of $-$15 to +5 km s$^{-1}$, while $^{12}$CO emission is observed also at significantly blue-shifted velocities. In both lines, the emission peaks around $-$4 km s$^{-1}$, the systemic velocity of the gas in the main filament. In $^{13}$CO, a less prominent peak is seen at around $-$9 km s$^{-1}$ blended with the main component. Another emission peak is seen at a redder velocity of +7 km s$^{-1}$ in both lines, which is associated with local molecular clouds (e.g., \citealt{Russeil2017}). Blue-shifted $^{12}$CO emission is found down to $-$60 km s$^{-1}$ and has two peaks around $-$20 km s$^{-1}$ and $-$40 km s$^{-1}$ that do not have clear $^{13}$CO counterparts. The $-$15 to +5 km s$^{-1}$ range, corresponding to the bulk of the emission in the $^{13}$CO line, is indicated by the dotted red lines and $-$30 to +20 km s$^{-1}$, corresponding to the bulk of the$^{12}$CO emission, is indicated by the dotted black lines in the Figure \ref{fig:sumspec}. We use these velocity ranges to compute the moment 0 maps. 

Figure \ref{fig:mommaps} presents the $^{13}$CO (top panel) and $^{12}$CO (bottom panel) velocity integrated moment 0 maps of the extended NGC 6334 filament. The prominent features discussed in the text are also labelled in the top panel image. Also indicated in the figure are the OB stars from \cite{persitapia2008} (red circles) and ATLASGAL clumps from \cite{Urquhart2018} (green plus markers).

\begin{figure*}[htbp!]
    \centering
        \includegraphics[width=0.9\linewidth]{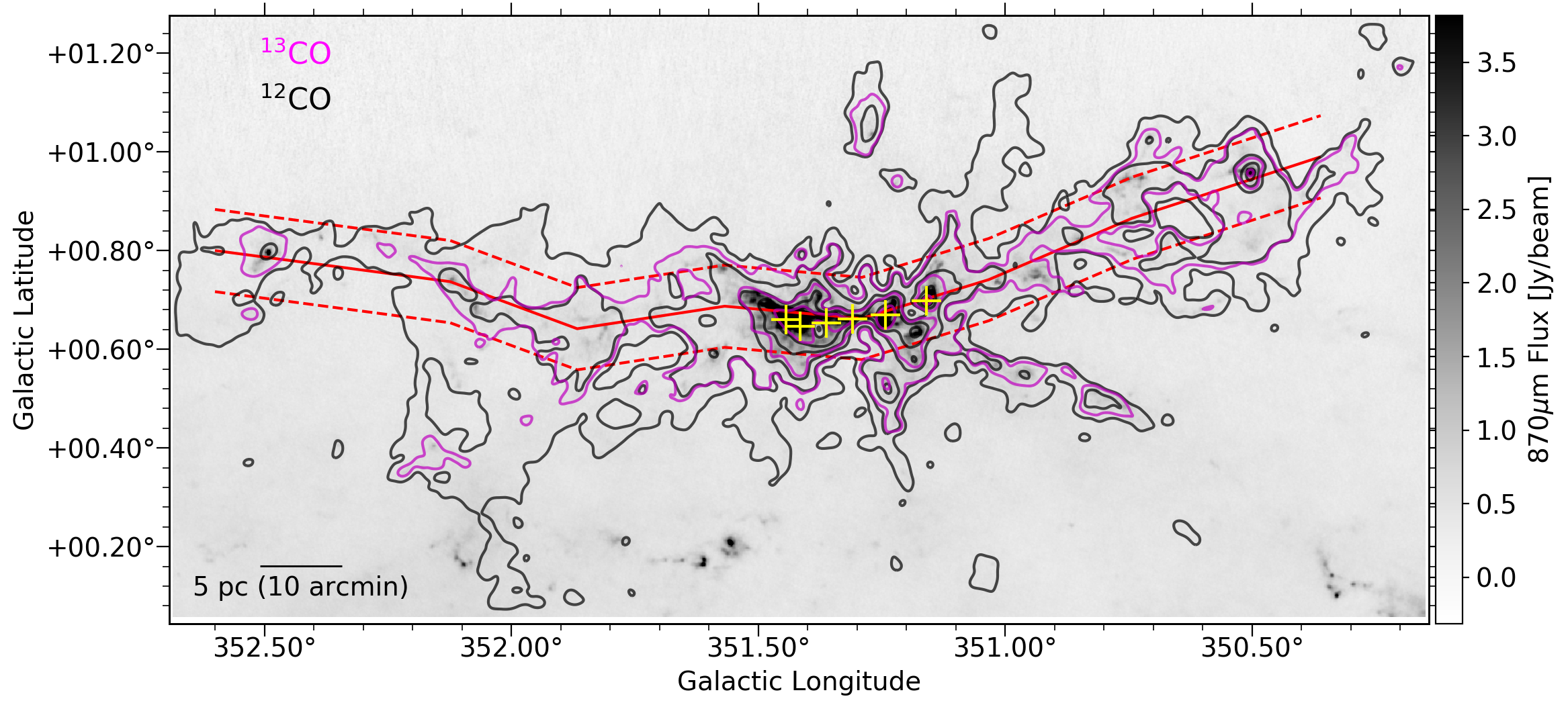}
        \includegraphics[width=0.9\linewidth]{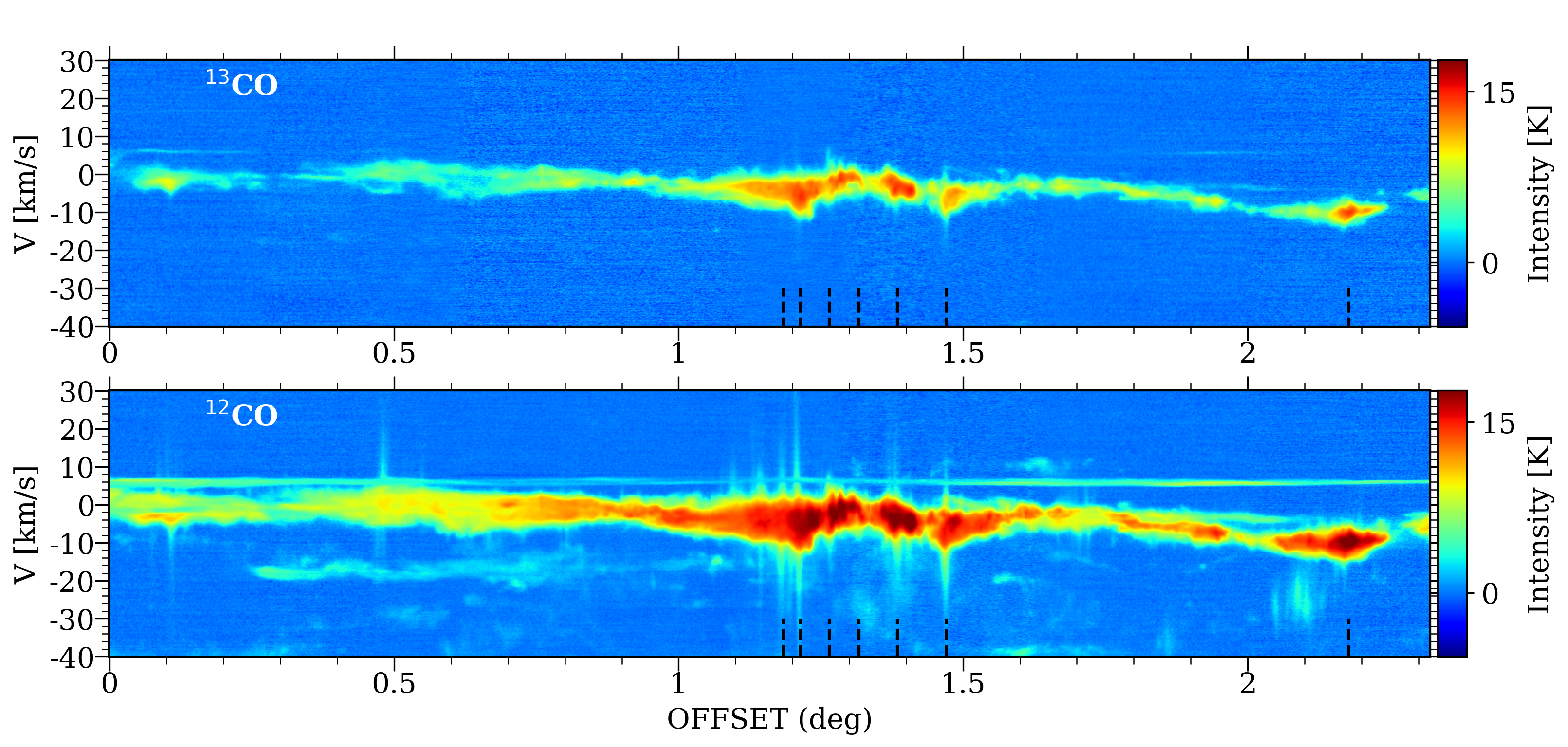}
    \caption{Top: $^{13}$CO (red) and $^{12}$CO (black) emission contours overlaid on the 870$\mu$m ATLASGAL dust emission (gray scale). Contour levels are the same as in Figure \ref{fig:mommaps}. Middle and Bottom: Position-velocity maps of $^{13}$CO and $^{12}$CO, respectively. The dotted red lines on the top image show the path along which the PV maps are constructed. The offsets are the projected path length along the galactic longitude from east to west along the strip. On the PV-maps, projected offset positions of the far-infrared (FIR) sources (I[N], I to V from right to left) in the central filament and GM-24 region are indicated by dotted black markers at the bottom.}
    \label{fig:agal+pvmap}
\end{figure*}

\begin{figure*}[htbp!]
    \centering
    \includegraphics[width=0.99\linewidth]{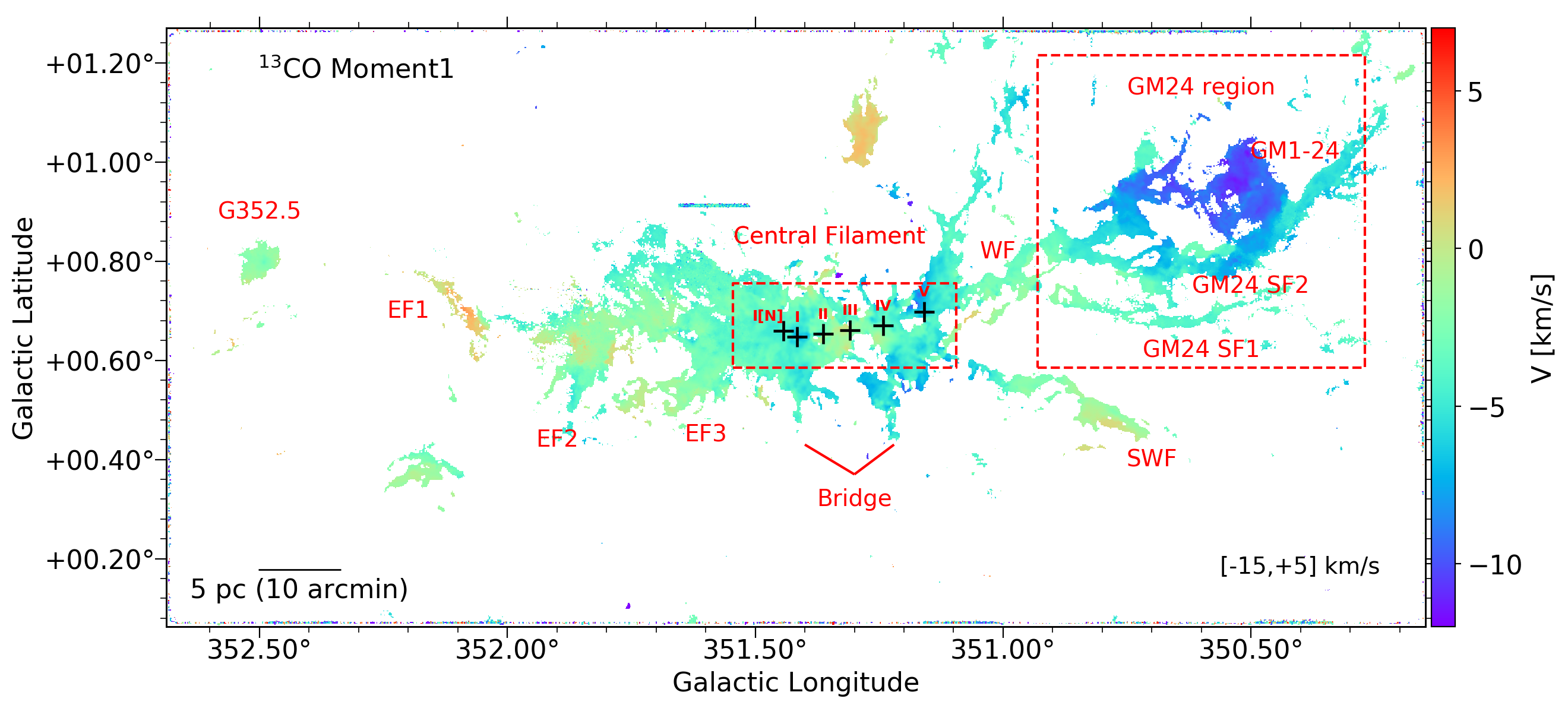}
    \caption{Intensity weighted velocity map (moment 1) of the $^{13}$CO emission toward the NGC 6334 extended filament. Labels are the same as in Figure \ref{fig:mommaps}}
    \label{fig:mom1}
\end{figure*}

The morphology of the CO line emission shows that the NGC 6334 central filament has a spatially concentrated dense gas reservoir. In both CO lines, the central filament is generally very bright and exhibits multiple bright emission spots. In areas that connect the central filament to the GM-24 region, $^{13}$CO emission shows elongated and finger-like filamentary structures originating from the filament WF and spreading towards the south-west and north-west\footnote{This study employs the Galactic coordinate system. We use `east' (`west') to mean directions of higher (lower) Galactic longitude, while `north' (`south') mean higher (lower) latitude.}. These features merge and reveal wide-spread emission in $^{12}$CO (see bottom panel). Some elongated finger-like (pillar-like) emission structures going north-ward and south-ward from the central filament are prominently seen in both lines. In the eastern region, however, $^{13}$CO emission is tracing only the trunk of the extended filament while $^{12}$CO exhibits a more extended emission morphology that connects NGC 6334 with the star forming region NGC 6357 ($l$=353.166$\degree$, $b$=0.89$\degree$, see Fig. 1 in \citealt{Russeil2017}). 

Most of the bright emission spots in the central filament correspond to the far-infrared (FIR) sources shown by the yellow markers in Figure \ref{fig:mommaps} (top panel). Numerous ATLASGAL 870 $\mu$m dust emission clumps are located in this region.  One interesting feature of the emission north of the central filament is that the gas emission seem to exhibit a pinched morphology. We over-plotted the position of the OB type stars from \cite{persitapia2008} on the moment0 map of $^{12}$CO. A circle of radius 8 pc (0.25 deg at 1.7 kpc) centered on the dominant O6.5 star is drawn in the map to match the arc-like emission morphology observed toward the NGC6334 central filament and it seems to perfectly match the morphological shape of the emission toward the north indicating that the high-mass star is likely interacting with the gas in the central ridge. However we also note that there are other OB stars, radio sources and H II bubbles located in the north to the ridge (see Figure \ref{fig:3color}) that could have contributed in shaping the emission morphology of the central ridge.

\begin{figure}[htbp!]
    \centering
    \includegraphics[width=0.9\linewidth]{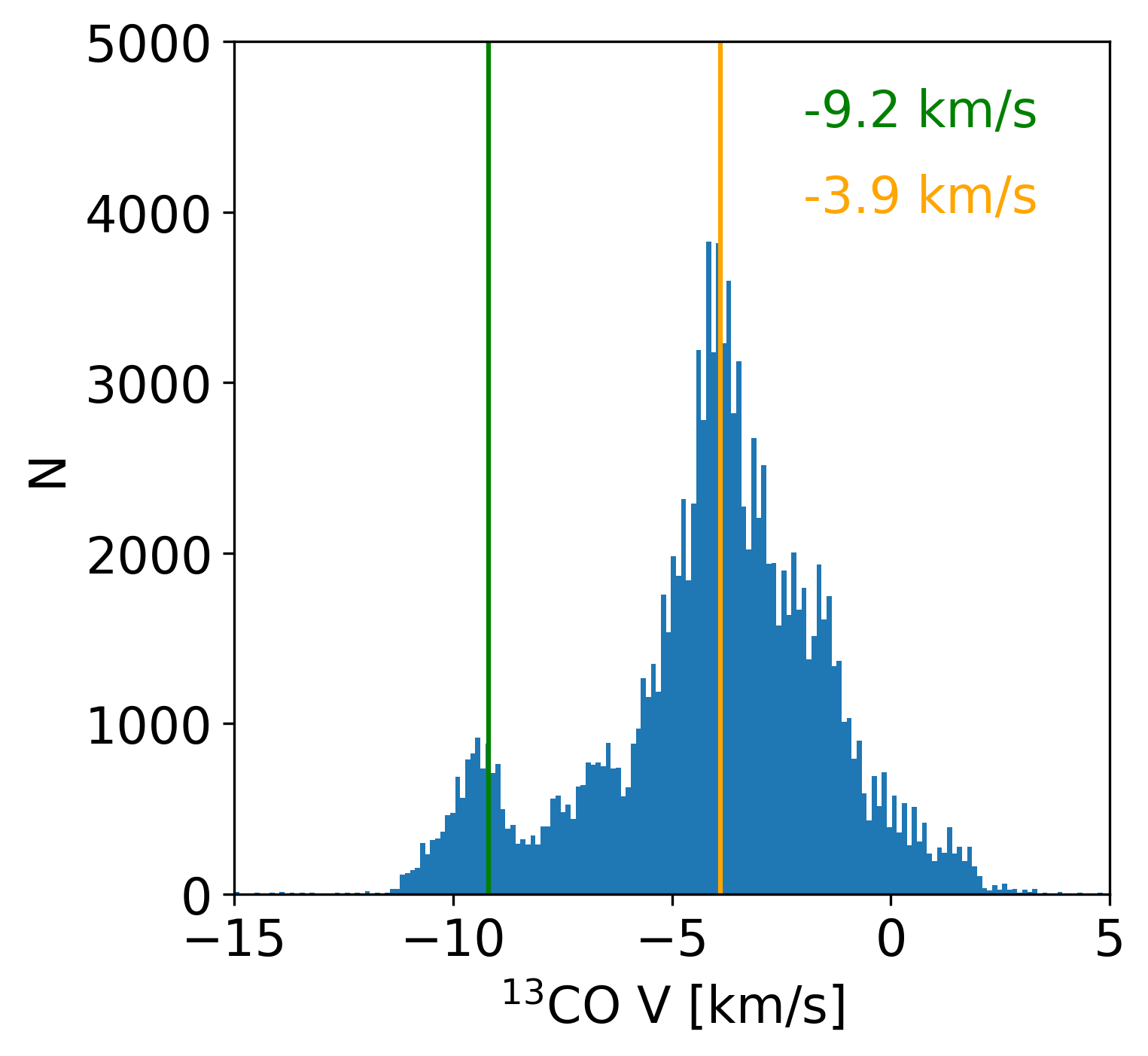}
    \caption{Histogram of the $^{13}$CO velocities obtained from the moment 1 map. }
    \label{fig:mom1_hist}
\end{figure}

\begin{figure*}[htbp!]
    \centering
    \includegraphics[width=0.75\linewidth]{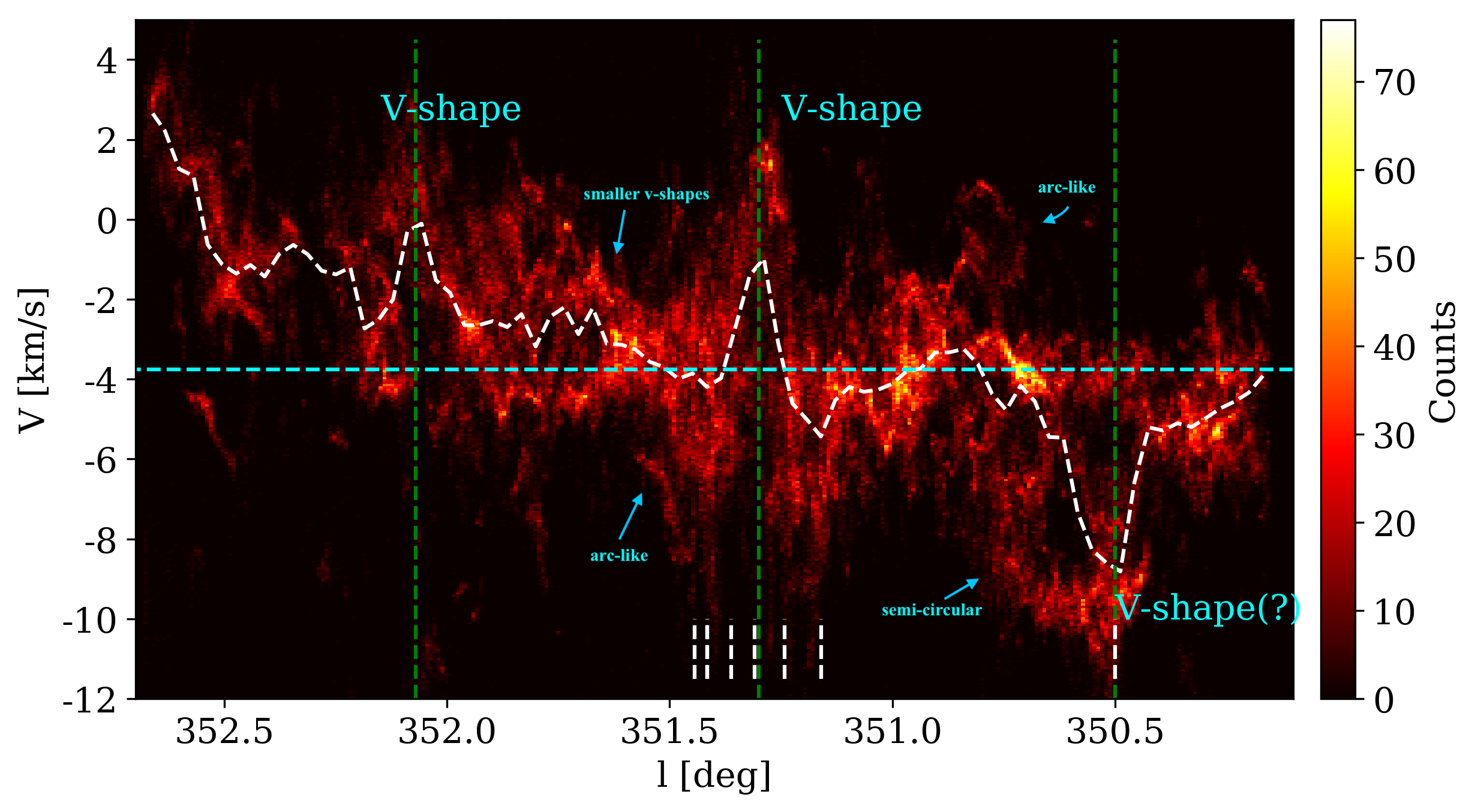}  
    \includegraphics[width=0.7\linewidth]{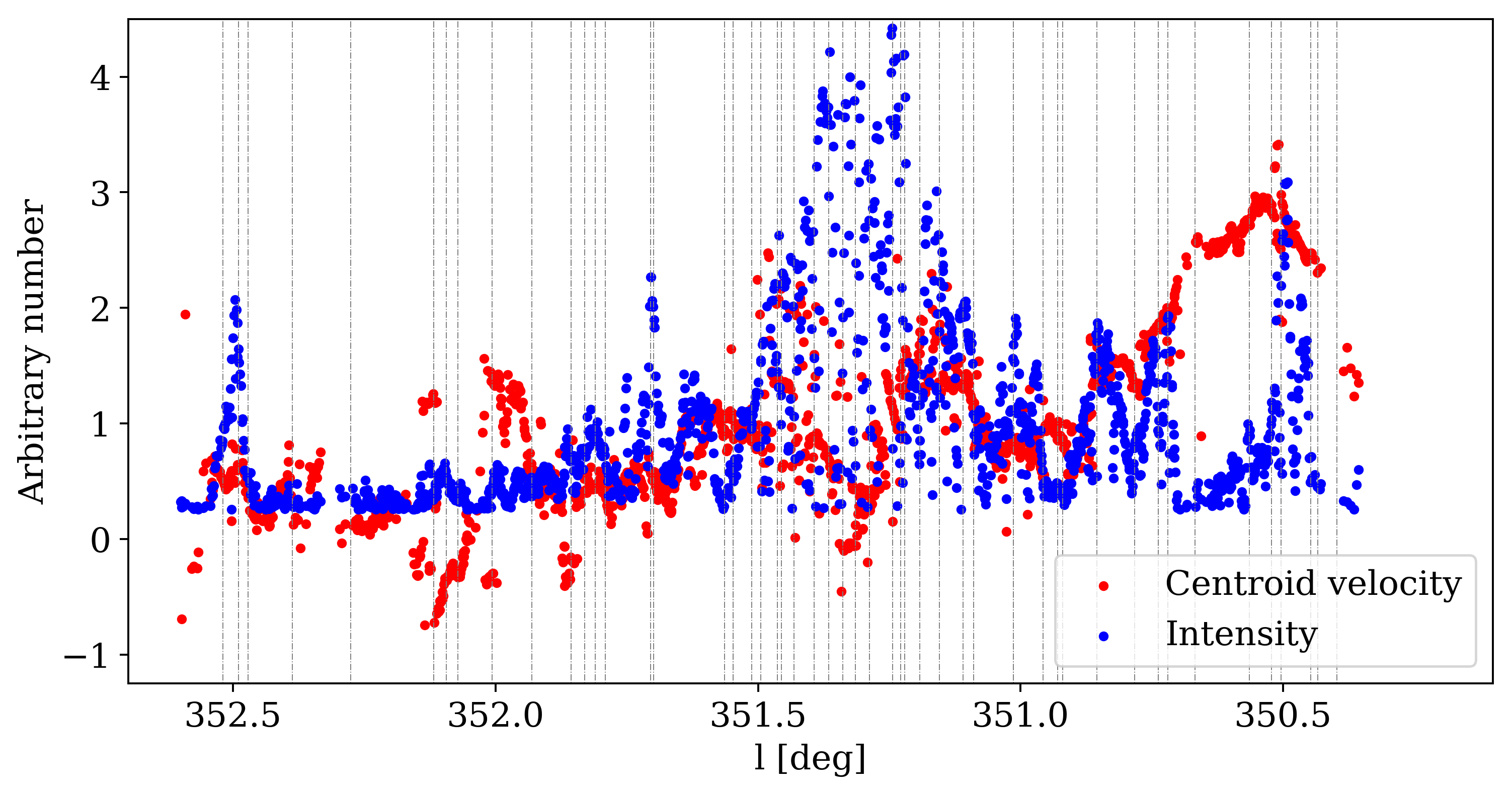}
    \caption{Top : 2D histogram plot of $^{13}$CO velocity components in the full observed region projected along the longitudinal direction from east to west. The bin sizes are 300$\times$300. The color bar shows the counts/density. The dotted white line follows the median velocities at a given longitude. The longitude of the six FIR sources I[N], I, II, III, IV and V, and also the GM24 region is indicated by the vertical dotted whits markers. Also shown are the positions of the V-shaped velocity gradients (labelled in cyan color). Bottom: Velocity and intensity variation along the longitude within the PV-sliced region in Figure \ref{fig:agal+pvmap}. The values in the y-axis are arbitrary numbers normalized by the mean values. 
    The vertical black dotted lines indicate the longitudinal positions of the ATLASGAL clumps in the same region. }
    \label{fig:gaussveldispersion}
\end{figure*}

\subsection{Gas velocity structure}\label{sec:velocitystructure}
\subsubsection{Position-Velocity map}
The middle and lower panels of figure \ref{fig:agal+pvmap} presents the Position-Velocity (PV) maps constructed for the emission within stripe defined by the three red lines in the intensity map (top panel). 

The longitudinal extent of the PV maps is $\sim$2.32 degrees. We find  a $\sim$80 pc long filament with a well connected and coherent velocity structure that has an average velocity of $\sim-$4.0 km s$^{-1}$. 

Since the $^{13}$CO line is mostly optically thin, it is tracing the higher column density  gas regions along the NGC 6334 extended filament. High-velocity emission is not observed in $^{13}$CO. In the $^{12}$CO lines, since it is optically thicker, more diffuse extended emission features at intermediate to high velocities are visible that are not detected in $^{13}$CO. Most of the high-velocity emission is observed mainly from the FIR source along the central filament and GM-24 (indicated by the dashed black markers in Figure \ref{fig:agal+pvmap}). This high-velocity emission is most likely tracing outflows since it shows a clear correspondence with (high-mass) star-forming FIR clumps/cores embedded in the central ridge.

Gas emission at velocities of $\sim$ [$-$25, $-$15] km s$^{-1}$ is observed at  offsets from 0.3 to 1.3 deg. This emission is diffuse and exhibits no clear peaks in the PV map. The origin of this component is at present unknown. In lower excitation lines of CO, this velocity component is seen connected with the main filament (CO $J = 2\rightarrow1$, $1\rightarrow0$ in \citealt{FukuiY2018CC}, and $J = 2\rightarrow1$ in \citealt{Arzoumanian2022}). 

Figure \ref{fig:pvmaps-center} presents PV maps toward two selected slices averaged along the perpendicular direction to the central filament. The slices L1 and L2 indicate same regions as MFS-cold and MFS-warm presented in \citealt{Arzoumanian2022} (see Fig. 12 in the paper). We observe V-shaped (or inverted) velocity structure along the L1 and L2 slices in higher J transitions of $^{13}$CO. In addition, we also present PV-maps (Fig. \ref{fig:pvmaps-center}) toward six FIR sources in the central NGC 6334 ridge. Toward all the sources we observe similar V-shapes in the PV maps along the y-axis. While in FIR source I and II, V-shape appears in both velocity direction, FIR sources III and IV exhibit inverted V-shapes.  \citealt{Arzoumanian2022} interpreted these V-shaped emission features in PV maps as matter flowing within a sheet-like structure compressed by a propagating shock front. We further investigate the velocity structure around H II regions in Sec. \ref{lvbvplots} whether shock compression due to H II regions show similar velocity features in the molecular shells/rings around them.

\subsubsection{Intensity weighted velocity map}\label{sec:mom1}

Figure \ref{fig:mom1} presents the moment 1 map of the $^{13}$CO emission. We used a 7$\sigma$ cutoff in integrated intensity to make the map. The velocity of the gas in the NGC 6334 extended filament shows primarily two components, one at $-$3.9 km s$^{-1}$ and another component at $-$9.2 km s$^{-1}$. Figure \ref{fig:mom1_hist} clearly illustrates double peaked distribution at these velocities. The $-$3.9 km s$^{-1}$ velocity component reveals the coherent velocity structure of the extended filament while the $-$9.2 km s$^{-1}$ velocity component is mainly tracing gas around the GM-24 region. However, it also appears to be tracing the `bridge' structures. We see these features extending south from the FIR source I[N] and I and north-south from the central filament (from FIR sources IV and V, see Fig. \ref{fig:mommaps}). Widespread morphological existence of these two velocity components indicates that the molecular gas in NGC~6334 has at least two origins. 

We also observe signs of velocity gradients in different regions. In the central filament, two `bridge' features are at bluer velocities compared to the trunk of the central ridge. In the central ridge itself, we observe west to east velocity gradient with velocities becoming redder toward the east. Regions EF1, EF2 as well as the SWF filament also show signs of velocity gradients along the filament. A velocity gradient toward the central FIR sources III from both east and west of the central filamentary ridge is seen, which has also been reported by \cite{Zernickel2013}  HCO$^{+}$ $J = 3\rightarrow2$ line data and interpreted as a sign of global collapse.

\subsubsection{Multi Gaussian fitting to the $^{13}$CO line cube}
Given the complex velocity structure of the region, we also  fit the $^{13}$CO line cube using the \texttt{Gausspyplus} package (\citealt{Gausspyplus2019}). \texttt{Gausspyplus} is an automated fitting routine developed and tested to fit complex spectral profiles with a high degree of accuracy that was used in the analysis of the Galactic Ring Survey (GRS)(\citealt{Jackson2006GRS}) (see \citealt{Gausspyplus2019} for more detail). In this routine the spatial coherence is taken into account while fitting multiple Gaussian components. We applied a signal-to-noise ratio (S/N) of 3.0 to constrain the minimum value of $^{13}$CO peak intensity. Figure \ref{fig:gausscomp} presents the number of Gaussian components fitted to the $^{13}$CO spectra within the NGC 6334 extended region. We find that to fit  profiles of the brighter CO emission lines arising from the denser regions along the filaments require multiple Gaussian components to fit the line profiles. 

Figure \ref{fig:gaussveldispersion} (top panel) presents the resulting projected line of sight velocity components along the filament going from east to west in longitudinal direction (from now on called $lv$-plots). The velocities within the PV-sliced region shown in Figure \ref{fig:agal+pvmap} is presented in the bottom panel. The median velocities for each longitude bin are also shown in dashed white contours. In general, the projected velocities decrease from the east of the mapped region to the western direction. Gas velocities in the east of the extended NGC 6334 region are redder (0 to 4 km s$^{-1}$) while in the western region (near GM24) gas at bluer velocities ($-$3 to $-$12 km s$^{-1}$) are found. The velocity gradient over the full filament length ($\sim$80 pc) is much smaller than 1 km/s/pc, which also illustrates the velocity coherence of the filament. 

\begin{figure}[htbp!]
    \centering
    \includegraphics[width=0.85\linewidth]{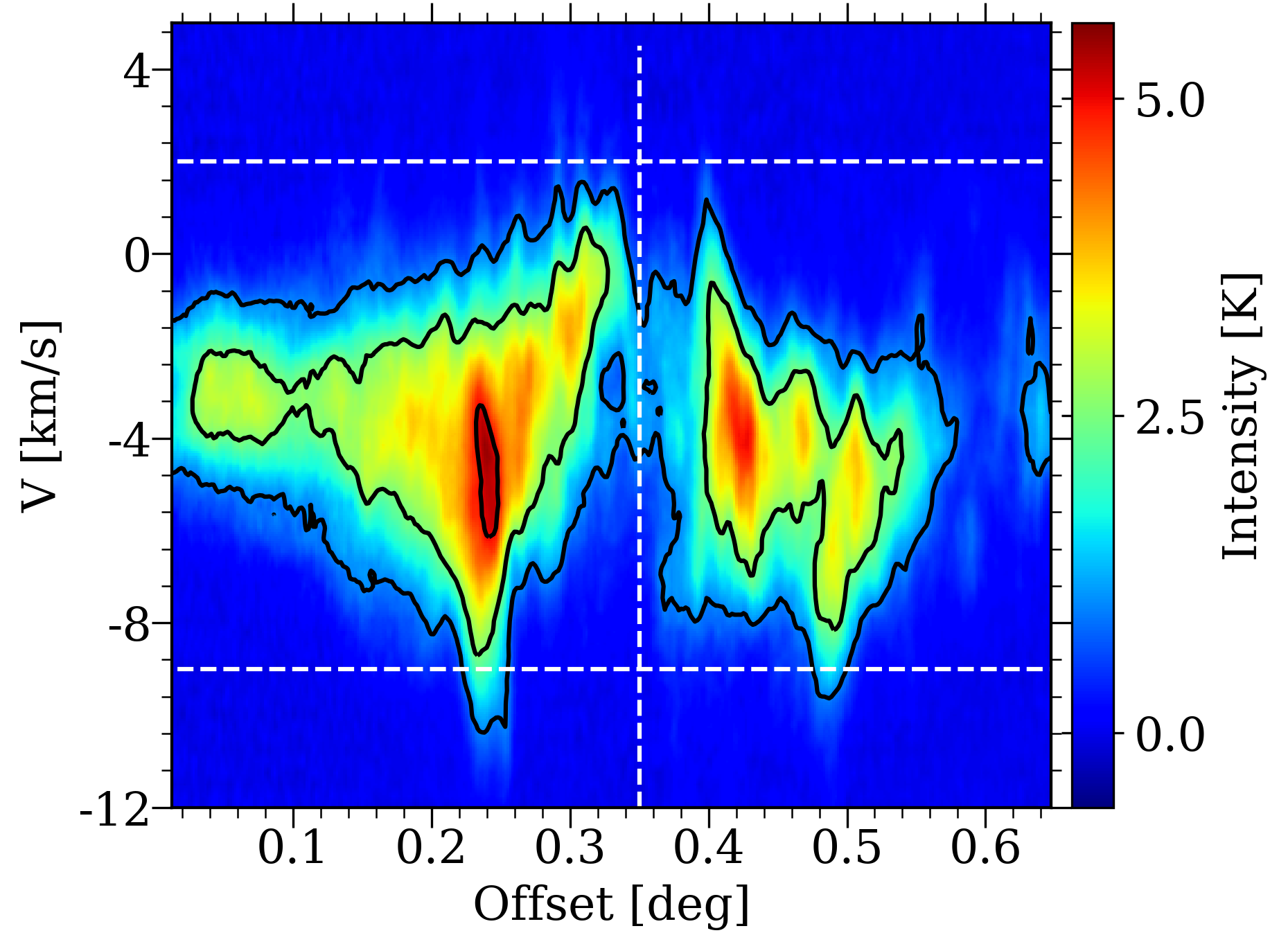}\\
    \includegraphics[width=0.85\linewidth]{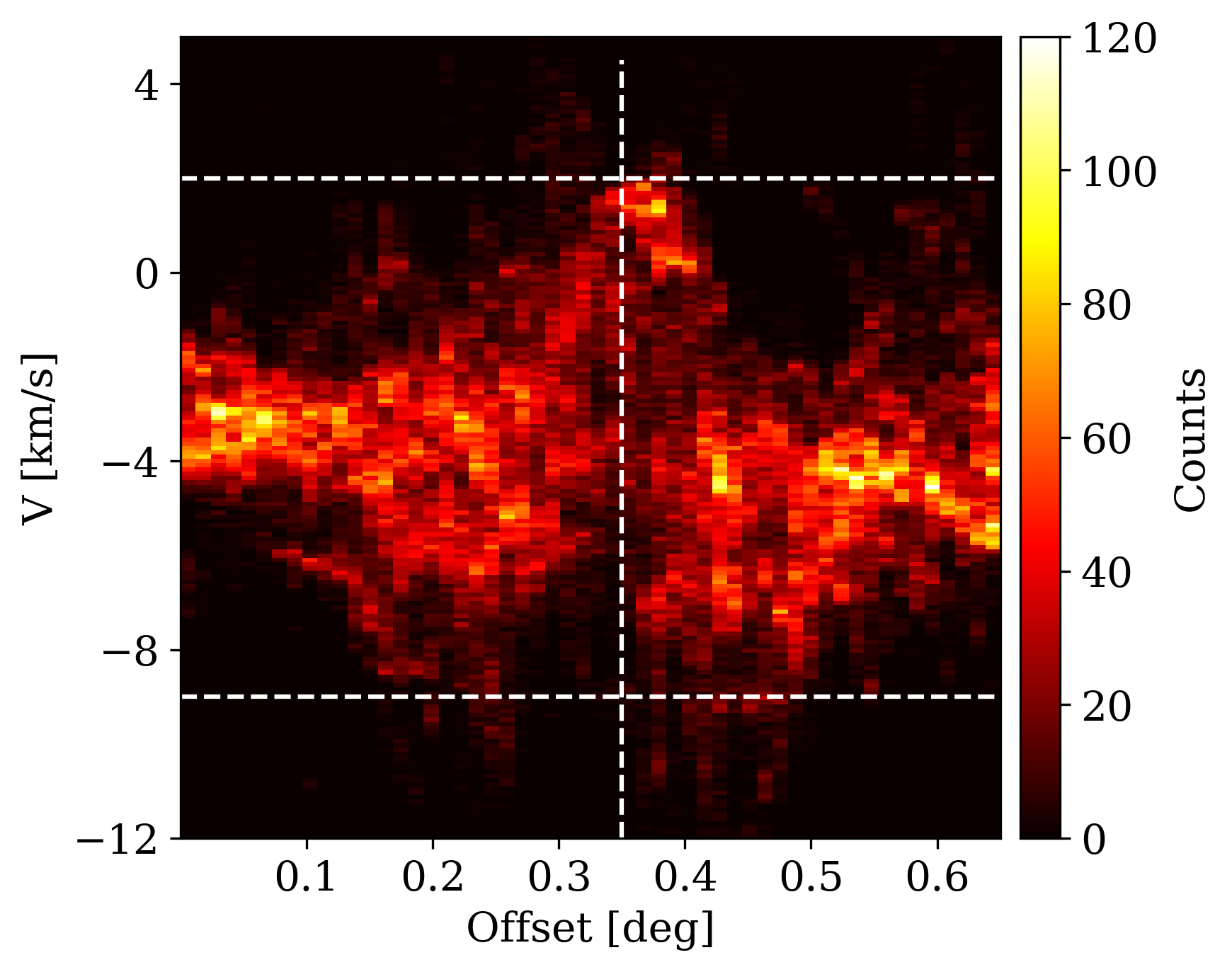}
    \caption{PV-map (top panel) and $lv$-map (bottom panel) toward the NGC 6334 central region (path along $l = 351.0^\circ$ to 351.65$^\circ$ at $b = 0.67^\circ$). The PV-map uses intensity weighting for estimating the velocities while the $lv$-map is made using Gaussian velocity components of $^{13}$CO spectra.}
    \label{fig:pv_lv_center}
\end{figure}

Various features can be identified in the $lv$-plots, for example, smaller V-shapes, arch-like and semi-arch like shapes and semi-circular shapes highlighting the complex gas velocity structure in the region. Many of these features are possibly related to the feedback from the large number of H II regions located in the region. Indeed, complex velocity features around the H II regions are seen in $lv$ and $bv$ plots which will be discussed more in Section \ref{lvbvplots}. We emphasize here that the amount of details of the velocity information one can see in the $lv$-plot, in which velocities are obtained from multi Gaussian fit to the spectra, is superior to the commonly used pv-diagram (see Figure  \ref{fig:pv_lv_center}). 

The intensity and velocity along the dense ridge of the NGC~6334 extended filament show fluctuations or so called `wiggles' at smaller spatial scales and also some broader V-shaped features (see Fig. \ref{fig:gaussveldispersion}, bottom panel). Similar oscillatory velocity fluctuations are also observed toward other Galactic filaments, for example, in $^{13}$CO(1$\rightarrow$0) observations toward the California nebula molecular filament (\citealt{Guo2021}).

Three broad V-shapes (or inverted V-shapes) in the $lv$-plot are indicated by the vertical dotted green lines in the figure at longitudes 352.1, 351.3 and 350.5 degrees. The base length of these V-shapes are broader than 3 pc. The V-shape around $l=352.1^\circ$ is located in the eastern filament (EF1). Two others correspond to the central filament (toward FIR source III) and GM-24 region, respectively. In the GM-24 region, a closer look to the map shows that the velocity structure is rather complex and V-shape is only appearing in median velocity contour. Toward eastern filament (EF1) and central filament, we also observe similar V-shape in intensity variation but phase shifted with respect to the velocity structure. Typically the V-shapes in position velocity diagrams indicate gas compression (due to collision or due to H I/H II bubbles) (e.g., \citealt{Inoue-Fukui2013, Arzoumanian2022}) or a global infall/collapse if observed with phase shifted intensity and velocity gradients (e.g., \citealt{Hacar2011,Zhou2023_1}). 
We will discuss these features in Section \ref{sec:globalcollapse} in conjunction with the observed CO emission morphologies in the channel maps.

\subsection{Channel maps}\label{sec:channelmap}
\begin{figure*}[htbp!]
    \centering
    \includegraphics[width=0.99\linewidth]{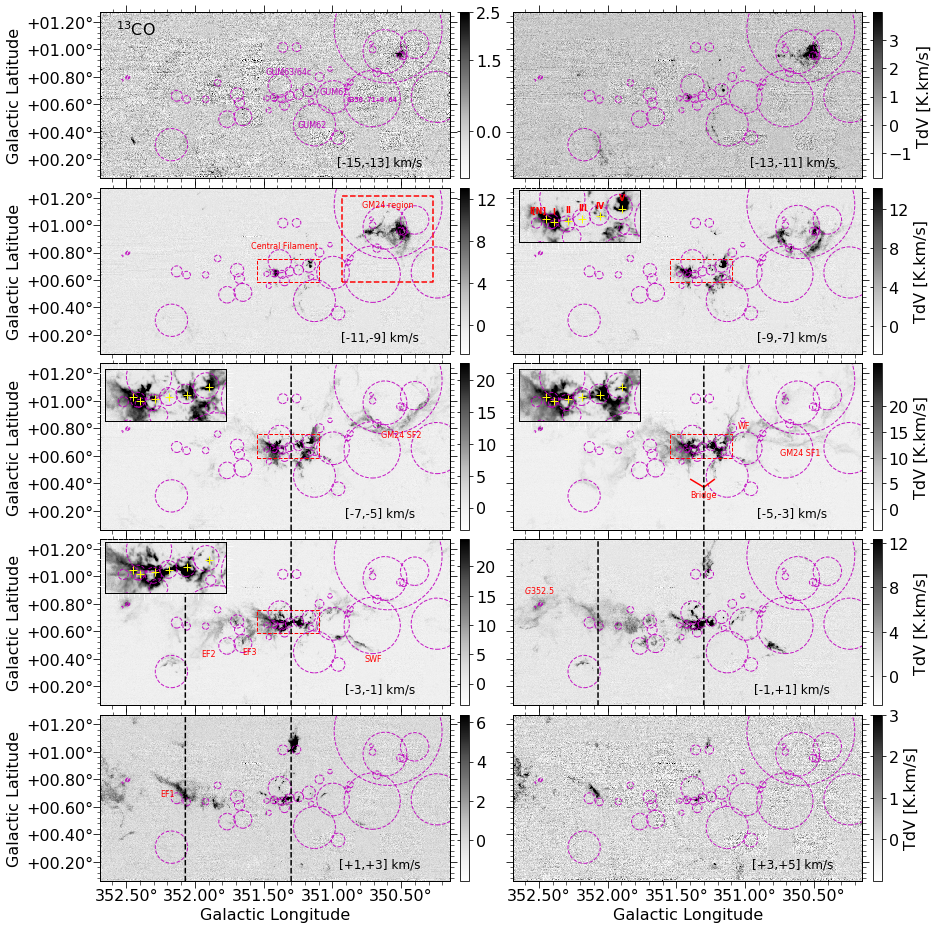}
    \caption{Channel maps of $^{13}$CO emission in the velocity range $-$15 to +5 km s$^{-1}$ with steps of 2 km s$^{-1}$. Magenta dashed circles indicate H II bubbles from \cite{Anderson2014}. Vertical black lines indicate positions of the broad V-shapes observed in $lv$-plot in Figure \ref{fig:gaussveldispersion}. Labels indicating various emission features in the maps are same as in Fig. \ref{fig:mommaps}. Some optical H II regions (same as in Fig. \ref{fig:3color}) discussed in the text are also labelled in the top left channel map for a reference. For a clearer view of the channel maps, zoomed maps toward the central filament, GM-24 and G352 are presented in Fig. \ref{fig:centeronly} to \ref{fig:13CO_g352}.}
    \label{fig:chanelmap13CO}
\end{figure*}

\begin{figure*}[htbp!]
    \centering
    \includegraphics[width=0.99\linewidth]{{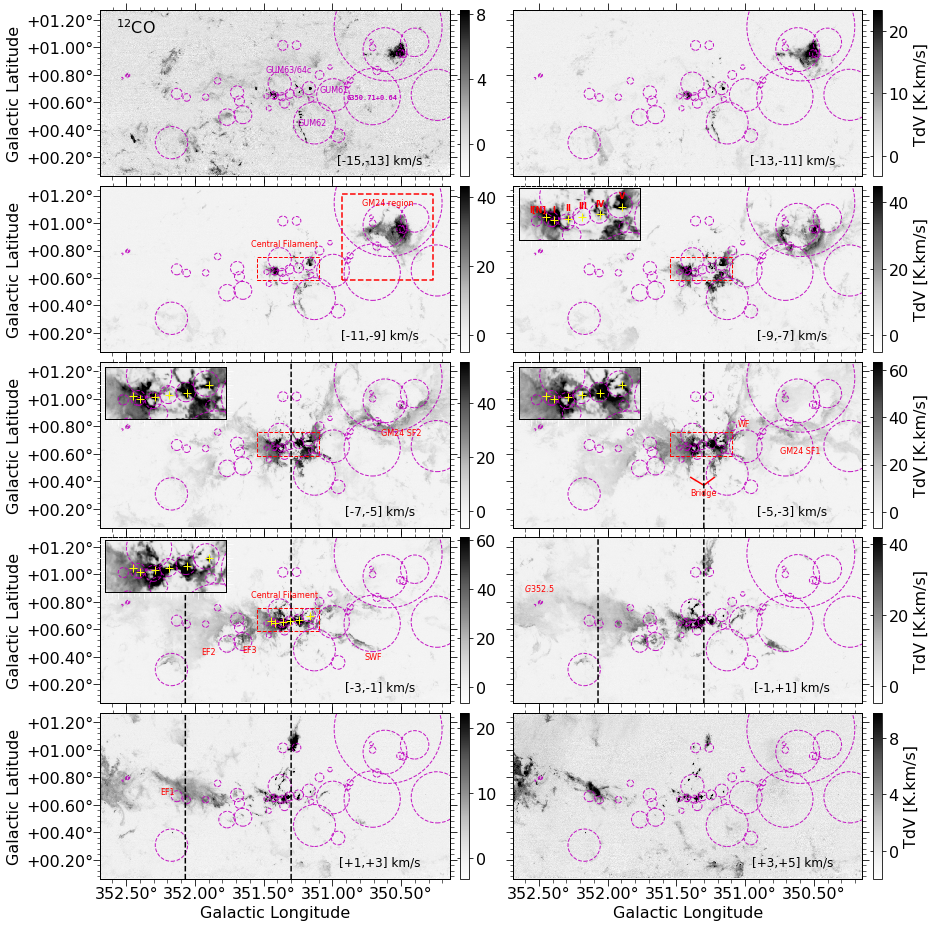}}
    \caption{Same as in Fig. \ref{fig:chanelmap13CO} for $^{12}$CO emission.}
    \label{fig:chanelmap12CO}
\end{figure*}

Figure \ref{fig:chanelmap13CO} and \ref{fig:chanelmap12CO} present channel maps of $^{13}$CO and $^{12}$CO in the velocity range $-$15 to +5 km s$^{-1}$, respectively, with steps of 2 km s$^{-1}$. The dotted circles in magenta 
indicate the H II regions and corresponding bubbles from \cite{Anderson2014}. In addition, we have drawn three vertical dotted lines in black at longitudes 352.1$^\circ$, 351.3$^\circ$ and 350.5$^\circ$ in different sub-plots. These are the positions of V-shapes features observed in the longitude-velocity plot of Figure \ref{fig:gaussveldispersion}. Text labels for various emission features (same as in Fig. \ref{fig:mommaps}) are also indicated in the channel maps. Also presented in Figures \ref{fig:centeronly} to \ref{fig:13CO_g352} are the zoomed velocity channel maps toward the central filament, GM-24 region and G352 region.

The channel maps allow a more detailed, velocity resolved view of the gas  emission morphology. For example, at velocities $-$10 to $-$5 km s$^{-1}$ in both CO lines, the Eastern (near I(N) and I) and Western (near IV and V) parts of the NGC~6334 central filament show bright emission with no connecting ridge structure indicating that from $-$10 to $-$5 km s$^{-1}$, the two sides are disconnected. Note that these velocities are blue-shifted  with respect to the average filament velocity of $-$3.9 km s$^{-1}$. From $-$5, $+$3 km s$^{-1}$, we start seeing emission features connecting the two sides of the central filament. Only at velocities between $-$3 and $+$1 km s$^{-1}$, the connecting ridge is bright in both CO lines. One notable feature is that at these velocities and beyond on the redder side of the emission, the bright emission feature from FIR source V located in the Western part of the central filament is no longer visible. 

Two parallel emission features, that we named the `bridges' in the moment0 map in Figure \ref{fig:mom1}, extend south from the central ridge at velocities $-$15 to $-$3  km s$^{-1}$ in $^{13}$CO. Both features are clearly visible and appear more extended in $^{12}$CO. 
Toward these bridge features, we observe CO emission in both the $-$3.9 and $-$9.2 km s$^{-1}$ velocity components (see Figure \ref{fig:mom1_hist}). The spatial coexistence of the two gas components hints at a possible collision or a merger of the clouds in the central filament. The Eastern bridge extends south from FIR-sources I[N] and I, and is more apparent in velocities from $-$9 to $-$3 km s$^{-1}$. The western bridge extends south from FIR sources IV and V and is also visible at redder velocities up to $-$1 km s$^{-1}$ in $^{12}$CO. The latter emission feature also extends northward from source V up to latitude of $\sim$1 deg toward the North-West. 

The GM24 region presents a clear example of bubbles within bubbles (see Fig. \ref{fig:3color}). In the channel maps this region is mostly bright in the $-$13 to $-$7 km s$^{-1}$ velocity range. The connecting filaments (WF, SWF, SF1,SF2) located to the west of the central filament however are visible at velocities $-$7 to $-$1 km s$^{-1}$.

Regions located east of the central filament (EF1, EF2, EF3 and G352.5) are seen at velocities larger than $-5$ km s$^{-1}$. The  $^{13}$CO  emission at this side of the filament is noticeably weaker. A bright elongated filamentary feature at around $l$ = 352.0 deg, $b$ =0.7 deg is observed both in $^{13}$CO and $^{12}$CO at the velocities from $-$1 to +5 km s$^{-1}$. Emission at these velocities are also bright further east close to the NGC~6357 region at $l$ = 352.5 deg and b = 0.8 deg. Finally to the South of the central filament ($<$ 0.25 deg), there is little or no emission in velocities in between $-$15 and +5 km s$^{-1}$.

\section{Analysis}\label{sec:analysis}

\subsection{Impact of H II regions in shaping the molecular gas structure}\label{sec:bubbleimpact}
To study the impact of H II regions on the surrounding molecular gas and its velocity structure, we first visually inspect the gas emission morphology toward the H II regions located in the extended NGC 6334 region (see Fig. \ref{fig:3color}). For  most of the H II regions a molecular emission counterpart is detected. Average $^{13}$CO velocities of the molecular gas toward the H II regions are presented in Table \ref{tab1}. The velocities are obtained using aperture extraction from the Gaussian fit velocities of $^{13}$CO using radii of the H II regions. We observe a variety of emission morphologies around the H II regions such as bubbles exhibiting ring-like, arc-like shapes and/or central holes in the channel maps. Figure \ref{fig:chan_map_hii0} and \ref{fig:chan_map_hii1} show maps of $^{13}$CO emission towards H II regions.

In the NGC 6334 central filament, eight H II regions are located along the ridge, most of which are associated with FIR sources (see Fig. \ref{fig:chanelmap13CO} and \ref{fig:chanelmap12CO}). Channel maps only towards the central region are also presented in Figure \ref{fig:centeronly}. H II region G351.348+0.593 is found in the South of the main ridge and G351.383+0.737 (GUM63/64) North to the ridge above sources I[N] and I. The central part of the ridge is disconnected up to bluer velocities of $-$5 km s$^{-1}$ and exhibits connected filament only at redder velocities of $-$5 km s$^{-1}$ (see Figure \ref{fig:chanelmap13CO}). Note that our identification of a broad V-shape in the $lv$-plot is associated with this region (see Figure \ref{fig:pv_lv_center}). We also find pillar-like structures above FIR source II (associated with GUM63/64C) and toward IV, observable both in $^{13}$CO and $^{12}$CO (see Figure \ref{fig:centeronly}). Such pillars are identified in many PDR and H II regions, and thought to originate from the expansion of H II regions into a turbulent, non homogeneous medium (e.g., \citealt{Tremblin2013}).

To the west of the central filament, two optical H II regions, G350.995+0.654 (GUM61) and G351.130+0.449 (GUM62) are located south of the FIR sources IV and V. These regions are indicated and labelled in some panels in Fig. \ref{fig:chanelmap13CO}. In both CO channel maps in Fig. \ref{fig:chanelmap13CO} and \ref{fig:chanelmap12CO}, we observe gas emission mostly at their edges but no emission is seen in the center indicating that these H II bubbles have already cleared out the gas around them and are at evolved stages. At velocities $-$1 to +3 km s$^{-1}$, the emission at the southern edge of H II region G350.710+00.641 is  bright in both $^{13}$CO and $^{12}$CO. In the velocity integrated intensity (mom0) maps (Fig. \ref{fig:mommaps}), this emission clearly appears filamentary, extending from source V to the South-West as a structure we named South-West Filament (SWF). These observations illustrate that H II shells could indeed play an important role in forming and shaping the morphology of filamentary cloud structures.

At the intersection of the H II regions G350.710+0.641 and G350.995+0.654 (GUM61), two smaller H II regions (G350.871+0.763 and G350.889+0.728) are seen. This is the region in which GM-24 southern filament 1 (GM24 SF1) departs westward from the West filament (WF). The filamentary emissions are clearly seen at velocities $-$5 to $-$1 km s$^{-1}$ in the channel maps (Fig. \ref{fig:chanelmap13CO} and \ref{fig:chanelmap12CO}). 

The GM24 region is a clear case of bubbles within bubbles toward which we observe evidence of bubbles/shells interacting with gas around them and shaping the emission structure in the region. One example of such an interaction is seen where we observe two arc-like gas layers, one facing South and the other to the North at velocities $-$13 to $-$11 km s$^{-1}$ surrounding the central bright emission spot in GUM1-24 (Figure \ref{fig:chanelmap13CO} and \ref{fig:chanelmap12CO}, see zoomed maps of this region in Fig. \ref{fig:13CO_gm124region}). These arch-like features correspond well with the shells of the H II regions G350.710+00.641 and G350.675+00.832 in the South and G350.401+01.037 in the North. At velocities between $-$9 and $-$7 km s$^{-1}$, we observe gas emission in a shell structure centered slightly east to GUM1-24. This shell structure corresponds to the H II region G351.130+0.449, located east to the compact radio source G350.50+0.95 (\citealt{Russeil2016}). In addition, we also find this shell-like gas emission structure confined within the radius drawn for H II bubble G350.594+1.149. Two HII regions G350.710+0.641 and G350.240+0.654 also appear to act on the the gas from the South in GM-24 region. 

We see spiral/arch-like filamentary gas emission at velocities $-$7 to $-$3 km s$^{-1}$, that is associated with the GM24-SF1 and GM24-SF2 filaments (see also zoomed maps in Fig. \ref{fig:13CO_gm124region}), connecting to the Western Filament (WF) and extending from south of the GM-24. At velocities $-$5 to $-$3 km s$^{-1}$, these filaments are more clearly visible in the channel maps (Figure \ref{fig:chanelmap13CO} and \ref{fig:chanelmap12CO}), more so in $^{12}$CO than in $^{13}$CO emission. In fact, these emission features seem to intersect at the far west at ($l$, $b$) = (350.4 deg, 0.8 deg). We note here that these velocity ranges are slightly bluer but similar to the velocity of the main filamentary structure in the NGC 6334 region. Based on the observed gas distribution of these filaments in the channel maps we suggest that the gas structure is shaped by the group H II regions (G350.594+01.149, G350.617+00.984, G350.710+00.641, G350.995+00.654) (see Fig. \ref{fig:chanelmap13CO}, \ref{fig:chanelmap12CO} and zoomed maps in Fig. \ref{fig:13CO_gm124region}). 

East of the central filament, at around $l = 351.5$ deg and at a similar latitude of the central filament, we find multiple H II regions. In particular four H II regions (G351.651+0.510, G351.676+0.610, G351.693+0.671,G351.766+0.492) appear to be connected to the gas at velocities $-$9 to +1 km s$^{-1}$ (Figure \ref{fig:chanelmap13CO} and \ref{fig:chanelmap12CO}). The emission of $^{13}$CO at this side of the filament is noticeably weaker. A bright elongated filamentary feature at around $l = 352.0$ deg, b =0.7 deg is observed both in $^{13}$CO and $^{12}$CO at velocities of $-$1 to +5 km s$^{-1}$. This is the same region (EF1) toward which we also observe the inverted V-shape in the $lv$-plot (see Figure \ref{fig:gaussveldispersion}), the peak of which is at a longitude of $\sim$ 352.1 deg. Zoomed channel maps toward this region are presented in Figure \ref{fig:13CO_g352}.

\begin{figure}[htp!]
    \centering
    \includegraphics[width=0.75\linewidth]{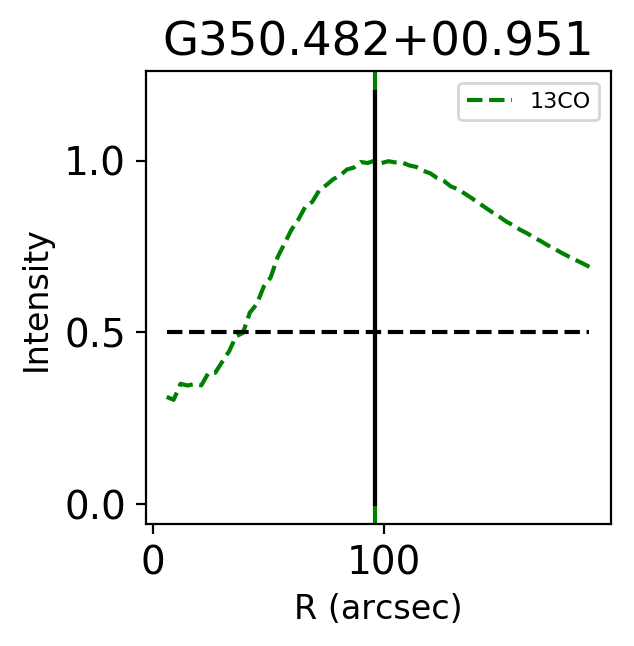}
    \caption{Azimuthally averaged $^{13}$CO radial intensity profile for the H II region G350.482+0.951 for the velocity range [$-$15, +5] km s$^{-1}$. The intensities are normalized by the peak value. Vertical black line indicate the radius of the H II region from \cite{Anderson2014}. }
    \label{fig:radialprofile}
\end{figure}

\subsubsection{Radial profile and contrast parameter}
To perform an unbiased search for shell/ring-like molecular structures around the H II regions, we plotted the azimuthally averaged radial $^{13}$CO intensity profile of the H II regions (using a velocity range of $-$15 to +5 km s$^{-1}$). For this we adopt the position and radius of the H II regions from \cite{Anderson2014} (cols. 3, 4 and 5 in Table \ref{tab1}) and plot the $^{13}$CO intensity profile up to twice these radii. The positions and radii were obtained by 
encircling the WISE mid infrared emission of the H II regions (see \citealt{Anderson2014} for more details). In Figure \ref{fig:radialprofile}, we present, as an example, the radial intensity profile toward H II region G350.482+0.951 associated with GUM1-24. This region exhibits a clear signature of a shell/ring like structure. In Figures \ref{fig:radialprofiles1} and \ref{fig:radialprofiles2}, we present profiles of the $^{13}$CO emission toward all the H II regions located in NGC 6334 extended region.  We individually inspect the emission morphology toward the regions (Fig. \ref{fig:chan_map_hii0} and \ref{fig:chan_map_hii1}) to interpret the radial profiles. We find two general categories of the intensity profiles plotted for full velocity range of $-$15 to +5 km s$^{-1}$.

The first category includes H II regions (22 of 42) that exhibit little emission or a flat emission profile toward their central regions and an increasing intensity profile outwards with respect to the H II radii. Some of these sources clearly exhibit a bumpy emission feature at corresponding H II radii (e.g., G350.482+0.951). In Figures \ref{fig:radialprofiles1} and \ref{fig:radialprofiles11}, radial profiles of these H II regions are presented. The second category of the H II regions (15 of 42) exhibits centrally peaked emission within the adopted H II radii and a decreasing intensity profile outwards. A few H II regions (5 of 42) do not fall into either of these categories and exhibit flat profiles up to twice the adopted radii and only exhibit diffuse $^{13}$CO emission. Intensity profiles of these H II regions are presented in Fig. \ref{fig:radialprofiles2}. 

H II regions with little or flat profile emission toward the center and an increasing intensity profile outwards suggest that they are likely at later evolutionary stages since they appear to have cleared out the molecular gas from the center. More than half of the H II regions (14/22) with an increasing intensity profile exhibit a bumpy feature or an intensity maximum at/near their radii (Fig. \ref{fig:radialprofiles1} and \ref{fig:radialprofiles11}), indicating a molecular shell/ring structure.

On the other hand, centrally peaked $^{13}$CO emission toward H II regions likely reflects an early evolutionary stages. However, it is also possible that these sources have line of sight contamination that caused  their profiles appear centrally peaked. To investigate this we explore the  $^{13}$CO  line profiles for emission averaged over different velocity ranges within $-$15 and +5 km s$^{-1}$ with a step of 5 km s$^{-1}$. Among the 20 H II regions that show either centrally peaked and decreasing or flat profiles for emission averaged over the full $-$15 to +5 km s$^{-1}$ range, we find that the profiles of 14 of them either increase outward  or show a bumpy feature when plotted for different velocity ranges. These profiles are also shown in Figure \ref{fig:radialprofiles2}. In total, we find 36 of 42 H II regions (86\%) that show the signature of molecular gas having been cleared from their center and that exhibit an increasing intensity profile outward or a shell/ring-like molecular structure around them. 

These characteristics of molecular line emission features are noted in col. 8 of Table \ref{tab1}. Centrally peaked and decreasing intensity profiles are annotated by `CP', increasing profiles are annotated by the letter `I', shell/ring-like structure are annotated by the letter `S' and sources with flat profiles are annotated by an `F'. The four 5 km s$^{-1}$ wide velocity ranges within $-$15 to +5 km s$^{-1}$ for which these structures are found are also noted, with V1 to V4.

 We also the estimate sizes of the $^{13}$CO emission associated with the  H II regions from the $^{13}$CO line profiles. For H II regions with clearly decreasing or increasing radial profiles we obtain the sizes at which the intensity is 50\% of the peak emission (indicated by the vertical green lines in Figures \ref{fig:radialprofiles1} to \ref{fig:radialprofiles2}). For H II regions exhibiting a shell/ring like morphology, we adopt the radii where these bumpy or shell-like features are observed. In column 6 of Table \ref{tab1}, the sizes estimated from molecular line emission associated with the H II regions are presented. In general there is a tentative agreement between the sizes that we estimated from $^{13}$CO emission profile and the sizes reported by \cite{Anderson2014}. The mean and standard deviation of the difference in radius are $\sim$15\% and $\sim$40\% with respect to the radius from \cite{Anderson2014}. The  intensity profiles of H II regions with shell/ring like structure begins to increase at a certain inner radius, which varies from source to source (see Figure \ref{fig:radialprofiles1}). We find that the inner radii can have values as small as $\sim$40\% of the shell radii. 

\begin{figure}[htbp!]
    \centering
    \includegraphics[width=0.75\linewidth]{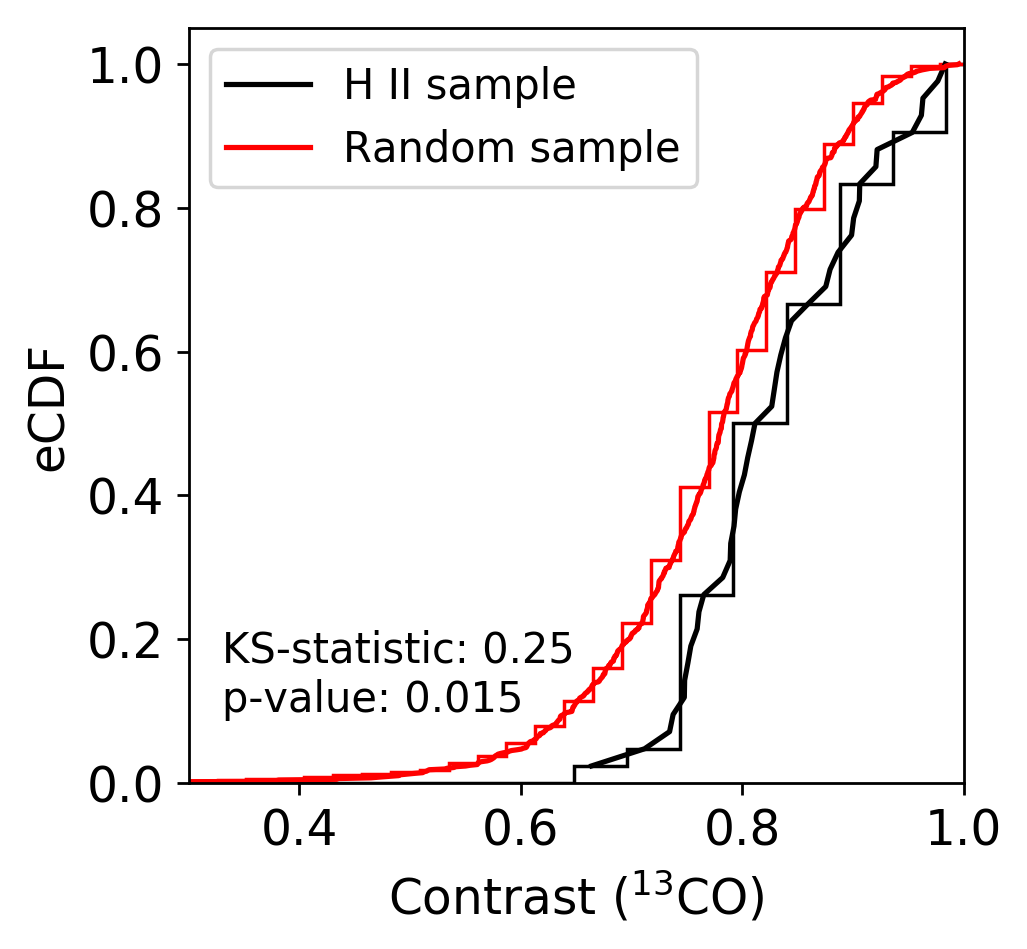}
    \caption{Empirical cumulative distribution function (eCDF) of the contrast values measured for the H II regions (black) and 1000 randomly created rings (red) in the mapped region. The two sample \texttt{ks-test} results are also annotated.}
    \label{fig:ecdf_contrast}
\end{figure}

We employ a contrast measurement method that quantifies the enhancement of the molecular line intensities at the H II region/bubble radii. To do so, we define a contrast parameter \texttt{C} as;
\begin{equation*}
    C = \frac{W_{R_2} - W_{R_1}}{W_{R_2}},
\end{equation*}
where, W$_{R_1}$ and W$_{R_2}$ give sum of the integrated molecular line intensities within the radii of sizes R$_1$ and R$_2$, respectively. R$_1$ represents the size of the inner ring and R$_2$ represents the size of the outer ring. The significance of the parameter for a homogeneous medium is straightforward and proportional to the ratio of the area of the ring to the area of the outer radius considered. 

\begin{table*}[htbp!]
\centering
\footnotesize
\caption{Molecular line emission characteristics of WISE H II regions in NGC 6334.} \label{tab1} \smallskip
\begin{tabular}{lccccccccccl}\hline
WISE H II source		& Type &    l & b & R & R($^{13}$CO)  & V($^{13}$CO) & $^{\dagger}$Emission  &   $C$   & $^{\dagger\dagger}$$C_{max}$&  $^{\ddagger}$N$_{clump}$ &  \\ 
  		                &      &  ($^{\circ}$) & ($^{\circ}$) & ($^{''}$) & ($^{''}$) & km s$^{-1}$ &  profile  &      &    &  &  \\\hline \hline
G350.240+00.654	&	C	&	350.240	&	0.655	&	670	&	498	&	-3.8	&	I,V2	&	---	&	0.78	&	1	\\
G350.401+01.037	&	G	&	350.401	&	1.037	&	372	&	546	&	-8.3	&	I,S	&	0.95	&	0.96	&	5	\\
G350.482+00.951	&	K	&	350.482	&	0.952	&	96	&	96	&	-9.6	&	I,S	&	0.81	&	0.88	&	1	\\
G350.505+00.956	&	K	&	350.505	&	0.957	&	96	&	84	&	-10.0	&	I,V3	&	---	&	--	&	1	\\
G350.594+01.149	&	G	&	350.594	&	1.149	&	1413	&	1314	&	-6.6	&	 I,S	&	---	&	0.84	&	26	\\
G350.617+00.984	&	G	&	350.617	&	0.984	&	574	&	474	&	-8.1	&	I,S	&	0.76	&	0.80	&	14	\\
G350.675+00.832	&	G	&	350.675	&	0.832	&	108	&	156	&	-6.8	&	 I,V3	&	---	&	0.79	&	1	\\
G350.707+00.999	&	C	&	350.707	&	0.999	&	86	&	84	&	-5.0	&	 I,S	&	0.76	&	0.89	&	1	\\
G350.710+00.641	&	G	&	350.710	&	0.642	&	737	&	330	&	-4.2	&	 I,S	&	0.85	&	0.98	&	21	\\
G350.716+01.044	&	Q	&	350.716	&	1.045	&	50	&		&	-4.1	&	 F	&	0.77	&	0.79	&		\\
G350.871+00.763	&	Q	&	350.871	&	0.763	&	86	&	36	&	-1.9	&	 I	&	0.82	&	0.83	&		\\
G350.889+00.728	&	Q	&	350.889	&	0.729	&	76	&	96	&	-2.0	&	 I,V2	&	---	&	0.83	&	1	\\
G350.958+00.358	&	K	&	350.958	&	0.359	&	179	&	135	&		&	 I,S	&	---	&	0.92	&		\\
G350.995+00.654	&	K	&	350.995	&	0.654	&	428	&	330	&	-3.3	&	 I	&	0.90	&	0.92	&	22	\\
G351.017+00.856	&	Q	&	351.017	&	0.857	&	59	&		&		&	 I	&	0.80	&	0.80	&		\\
G351.094+00.800	&	Q	&	351.094	&	0.800	&	116	&		&	-4.5	&	 I,V2	&	---	&	0.79	&		\\
G351.130+00.449	&	K	&	351.130	&	0.450	&	549	&	534	&	-5.1	&	 I,S	&	0.95	&	0.96	&	8	\\
G351.153+00.623	&	Q	&	351.153	&	0.623	&	68	&		&	-4.9	&	 I	&	0.80	&	0.83	&	1	\\
G351.170+00.704	&	K	&	351.170	&	0.704	&	171	&	300	&	-5.1	&	 I,V3	&	---	&	--	&	3	\\
G351.246+00.673	&	K	&	351.246	&	0.673	&	131	&	108	&	-3.3	&	 CP	&	---	&	---	&	4	\\
G351.261+01.016	&	K	&	351.261	&	1.016	&	114	&	111	&	1.4	&	 I,S	&	0.81	&	0.81	&	1	\\
G351.311+00.663	&	K	&	351.311	&	0.663	&	119	&	162	&	-2.0	&	 I,V2	&	---	&	---	&	3	\\
G351.348+00.593	&	G	&	351.348	&	0.593	&	125	&	54	&	-2.5	&	 I,S	&	0.80	&	0.91	&	3	\\
G351.360+01.015	&	K	&	351.360	&	1.015	&	129	&	66	&	1.3	&	 I,V4	&	---	&	---	&		\\
G351.367+00.640	&	K	&	351.367	&	0.641	&	107	&	132	&	-2.9	&	 I,S	&	0.83	&	0.95	&	4	\\
G351.383+00.737	&	K	&	351.383	&	0.737	&	303	&	222	&	-3.6	&	 I,S	&	0.81	&	0.98	&	14	\\
G351.420+00.637	&	Q	&	351.420	&	0.638	&	52	&		&	-5.6	&	 F	&	---	&	---	&	3	\\
G351.424+00.650	&	G	&	351.424	&	0.651	&	99	&		&	-5.0	&	 CP	&	---	&	---	&	4	\\
G351.462+00.556	&	G	&	351.462	&	0.557	&	68	&	42	&	-3.6	&	 I	&	0.84	&	0.88	&		\\
G351.479+00.643	&	Q	&	351.479	&	0.644	&	68	&		&	-4.2	&	 F	&	---	&	0.76	&	2	\\
G351.651+00.510	&	K	&	351.651	&	0.510	&	234	&	96	&	-1.7	&	 I,V2	&	---	&	0.80	&		\\
G351.676+00.610	&	G	&	351.676	&	0.611	&	130	&	162	&	-2.4	&	 I	&	0.89	&	0.90	&		\\
G351.693+00.671	&	K	&	351.693	&	0.672	&	175	&	180	&	-2.1	&	 I,V2	&	---	&	0.79	&		\\
G351.766+00.492	&	G	&	351.766	&	0.493	&	217	&	318	&	-1.1	&	 CP	&	---	&	0.91	&		\\
G351.835+00.756	&	Q	&	351.835	&	0.757	&	87	&		&	-0.2	&	 I	&	0.76	&	0.84	&		\\
G351.922+00.638	&	Q	&	351.922	&	0.638	&	86	&	69	&	-0.3	&	 I,S	&	---	&	0.86	&	1	\\
G352.060+00.636	&	Q	&	352.060	&	0.636	&	99	&		&	1.0	&	 F	&	0.76	&	0.76	&		\\
G352.132+00.663	&	C	&	352.132	&	0.663	&	143	&	72	&	0.8	&	 I	&	0.80	&	0.81	&		\\
G352.171+00.307	&	C	&	352.171	&	0.307	&	423	&	378	&	-1.9	&	 I,S	&	---	&	0.90	&	2	\\
G352.486+00.799	&	C	&	352.486	&	0.799	&	43	&		&	-2.3	&	 I,V4	&	---	&	0.83	&		\\
G352.493+00.793	&	C	&	352.493	&	0.794	&	40	&	42	&	-2.4	&	 CP	&	---	&	---	&		\\
G352.529+00.775	&	Q	&	352.529	&	0.775	&	22	&		&	-1.7	&	 I,V4	&	0.77	&	0.77	&		\\

\hline
\end{tabular}
\tablefoot{Cols. 1 to 5 (H II region name, type, longitude, latitude and radius, respectively) are taken from \citealt{Anderson2014}. The sizes of the molecular shells/rings (R($^{13}$CO)), molecular line velocities (V($^{13}$CO)) and contrast parameter values ($C$) are determined from this work. $^{\dagger}$ $^{13}$CO radial profiles. CP: centrally peaked, I: increasing profile, S: shell-like feature, F: flat profile. V1: [$-$15,$-$10] km s$^{-1}$, V2: [$-$10,$-$5] km s$^{-1}$, V3: [$-$5,0] km s$^{-1}$, V4: [0,+5] km s$^{-1}$. $^{\ddagger\ddagger}$ Maximum value of the contrast in one of the velocity ranges. $^{\ddagger}$ Number of ATLASGAL clumps (from \citealt{Urquhart2018yCat}) located at H II radii range 0.6R-1.2R.}
\end{table*}

Column 9 of Table \ref{tab1} presents the $^{13}$CO contrast parameter values for H II regions computed from the velocity range $-$15 to +5 km s$^{-1}$ for radius ranges of 0.6 to 1.2 times the  radius. For H II regions with shell-like intensity profiles, we find that the inner radii at which the intensity starts increasing is up to 40\% of the shell radii. In addition, since H II regions/bubbles are known to be eccentric (\citealt{Churchwell2007}), these choices of radii between 0.6 R to 1.2 R to measure the contrast parameter incorporate the eccentric nature of the bubbles with eccentricity values of 0.86 to 1. Contrast parameter values larger than the expected value of 0.75 for the radii considered here are presented in col. 9 of Table \ref{tab1}. At these velocity ranges and radii, 21 H II regions show higher contrast parameter than 0.75. 
However, at segmented velocity ranges from $-$15 to +5 km s$^{-1}$ with a step of 5 km s$^{-1}$, we already found that the majority of the H II regions   
show centrally clear or flat profiles that either increase outward or show a bumpy feature at or near the corresponding radii. Therefore, we measured the contrast parameter for velocity ranges from $-$15 to +5 km s$^{-1}$ with a step of 5 km s$^{-1}$. The maximum value of the contrast parameter ($C_{max}$) in these velocity ranges are given in Col. 10 of Table \ref{tab1}. In total, using the contrast measurement method we identify 34 of 42 H II regions (81\%) that exhibit intensity enhancement of the molecular emission at the H II radii.

We performed a two sample \texttt{ks-test} between contrast parameters ($C_{max}$) measured for the H II regions and contrast parameters measured toward a randomly created 1000 shells/rings in the mapped region. Sizes of these randomly created rings range from 20" to 700", similar to the sizes of the H II regions. The \texttt{ks-test statistics} (0.25) and \texttt{p-value} (0.015) suggest that the sample of H II regions is indeed distinct from that representing the randomly created ring structures. Figure \ref{fig:ecdf_contrast} presents the empirical cumulative distribution function (eCDF), which clearly shows that H II regions have a higher probability of getting higher contrast values. Even though most H II regions in our sample show a contrast parameter above the expected value, it is wise to explore the significance of that parameter. To do so we exploited the contrast values derived for the randomly sampled shells/rings and fit a Gaussian distribution with mean and standard deviation of 0.78 and 0.09, respectively (Fig. \ref{random_contrast}). Then we define a contrast threshold above 1$\sigma$ from the mean, which is 0.87. 
14 H II regions show contrast values above 1$\sigma$ threshold in at least one velocity ranges ($-$15 to +5 with a step of 5 km s$^{-1}$) for a shell/ring radii of 0.6 to 1.2 times the H II radius. We repeated the analysis considering smaller shell/ring sizes (0.6R--1.0R and 0.8R--1.2R). In total, we find 22 H II regions (52\%) show a contrast parameter above 1$\sigma$ threshold in at least one radii and velocity ranges.

Caveats of using contrast method are that for the H II regions that are too young to have created a shell/ring like structure and are still embedded in their dense natal gas, or exhibit a  complex morphology, e.g., shaped by a champagne flow, this method may not provide the meaningful results. In addition, both the radial profile and the contrast method applied in this study cannot distinguish the true morphological features if a shell has a full-ring, half-ring or has a clumpy edge. To study the morphology in detail, 2D or 3D emission maps (channel maps) have to be examined. For a larger shell/ring sizes, radii at which contrast is being measured should be carefully selected. Despite the caveats, the contrast method in conjunction with radial profiles can be a useful tool to study H II bubbles/regions, in particular to search for shell-like structures, using large scale Galactic molecular line and continuum surveys. 

\begin{figure}[htbp!]
    \centering
    \includegraphics[width=0.85\linewidth]{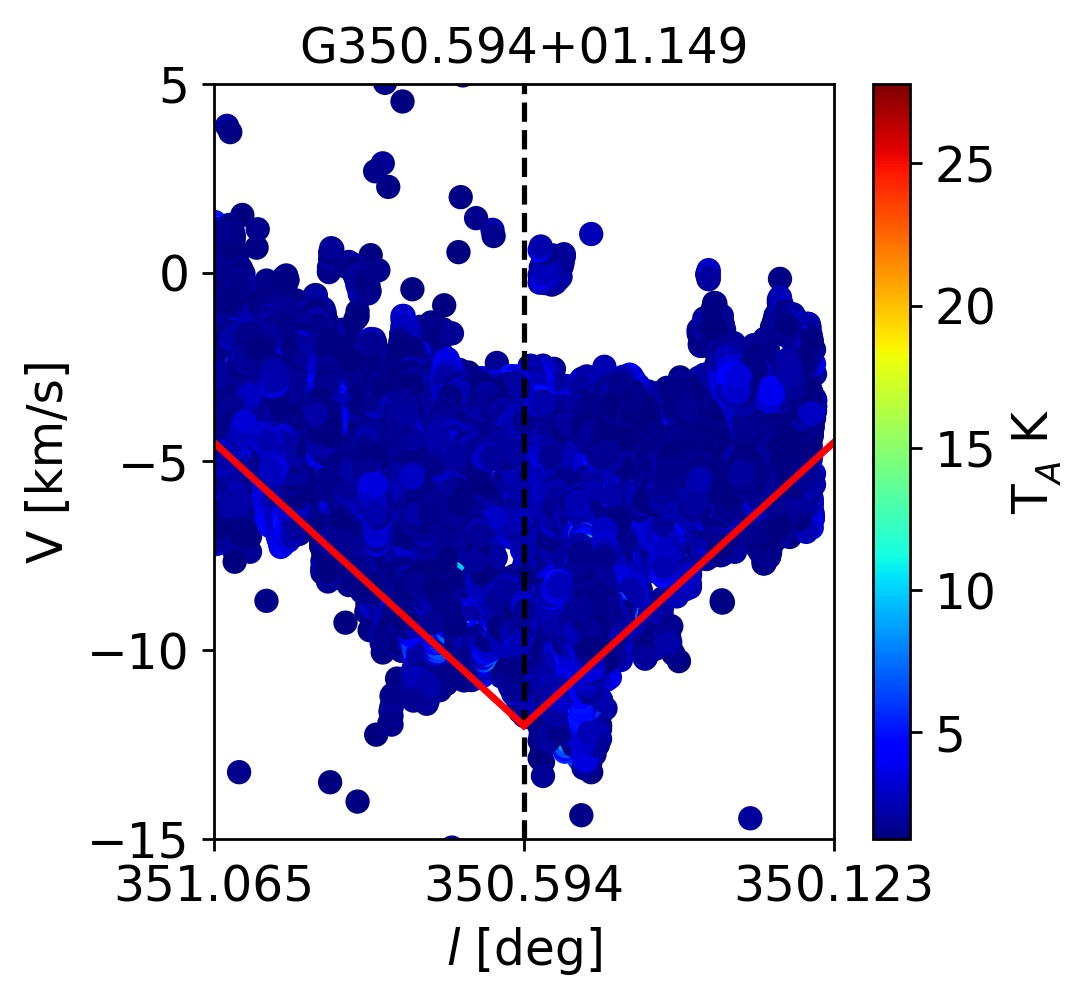}
    \caption{$lv$ plot toward H II region G350.594+1.149. Observed V-shaped emission structure is indicated by the red lines. }
    \label{fig:lvplot_eg}
\end{figure}

\subsubsection{Longitude-velocity ($lv$) and latitude-velocity ($bv$) plots toward H II regions}\label{lvbvplots}
In figures \ref{fig:lvplots_hiisources0} to \ref{fig:lvplots_hiisources9}, we present longitude-velocity ($lv$) and latitude-velocity ($bv$) plots for the H II sources. For the plots, we used the Gauss fitted $^{13}$CO velocities instead of the commonly used intensity averaged pv-plot. We again used the radii range of $\pm$1.2R in both longitudinal and latitudinal direction, where $R$ is the H II region radius. To investigate if we observe higher intensities around the H II regions, we made the plots with intensity color wedges. Toward some H II regions (12/42) we find the velocity structure to have to follow a (partial) elliptical shape either in $lv$ or $bv$ plots. For those, we also present approximate expansion velocities inferred from visually fitted ellipses in Table \ref{tab:vexp}. We also observe broader V-shaped velocity feature, with open cavities or with filled emission. An example of the $lv$ plot toward H II region G350.594+1.149 is shown in figure \ref{fig:lvplot_eg} in which the observed V-shaped emission structure is also indicated by the red lines. Toward 15 H II regions (in $lv$ or $bv$ plots), we observed such features (Fig. \ref{fig:lvplots_hiisources0} to \ref{fig:lvplots_hiisources9}). In particular these features were observed toward the sources located in and around the filaments (see \ref{lvplots_added}). We interpret these velocity structures to represent shock compressed gas layers driven by H II regions in the filament or located nearby. To further investigate the features around these H II regions, we made $pv$ maps along different slices (L3 to L12) shown in \ref{lvplots_added}. These $pv$ maps are constructed along perpendicular direction of the filament axis or the clumpy regions located at the edges of the H II regions. In all slices (except for L7 and L12) we observe V-shaped emission velocity feature. While in slices L7 and L12, we only observe a velocity gradient. Overall the $lv$ and $bv$ plots highlight complex velocity structure around H II regions and in particular their role in shaping the gas velocity structure around them. These results further illustrate multiple sites of gas compression due to H II regions in the extended NGC 6334 region. 

\begin{figure}[htbp!]
    \centering
    \includegraphics[width=0.75\linewidth]{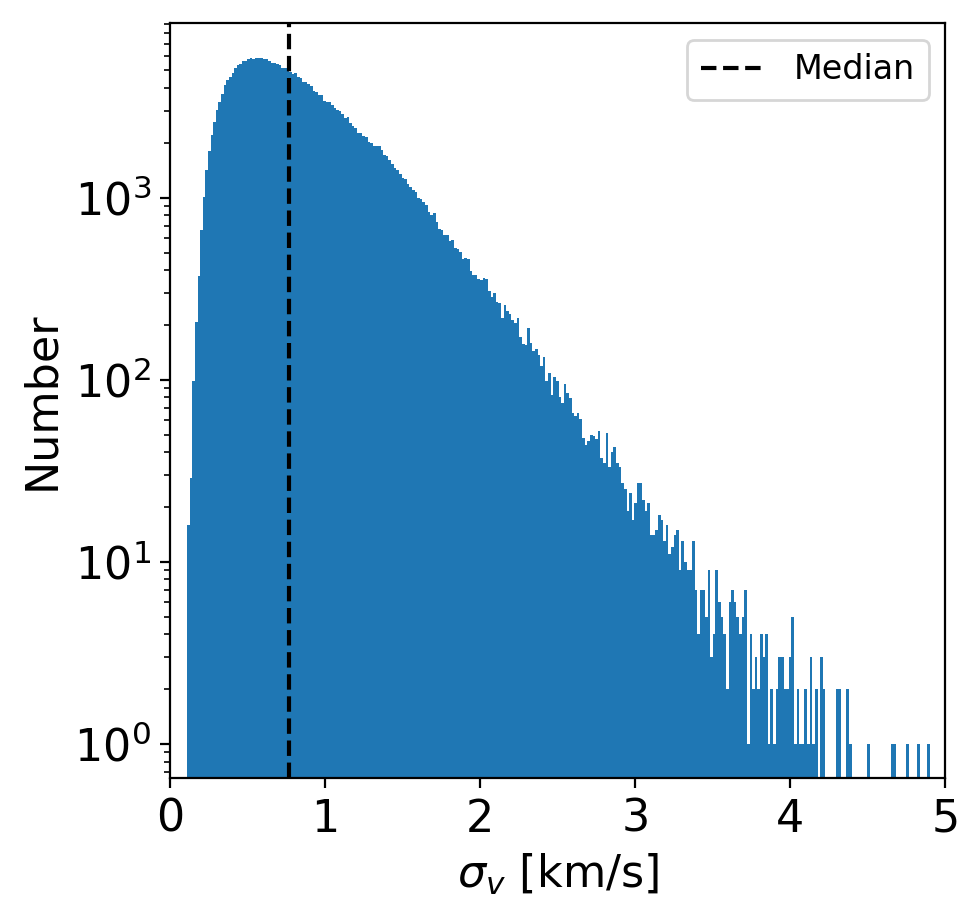}
    \caption{Histogram of velocity dispersion ($\sigma_{v}$)  values determined from the Gaussian fit to the $^{13}$CO line profiles.}
    \label{fig:histveldispersion}
\end{figure}

\begin{figure*}[htbp!]
    \centering
    \includegraphics[width=0.8\linewidth]{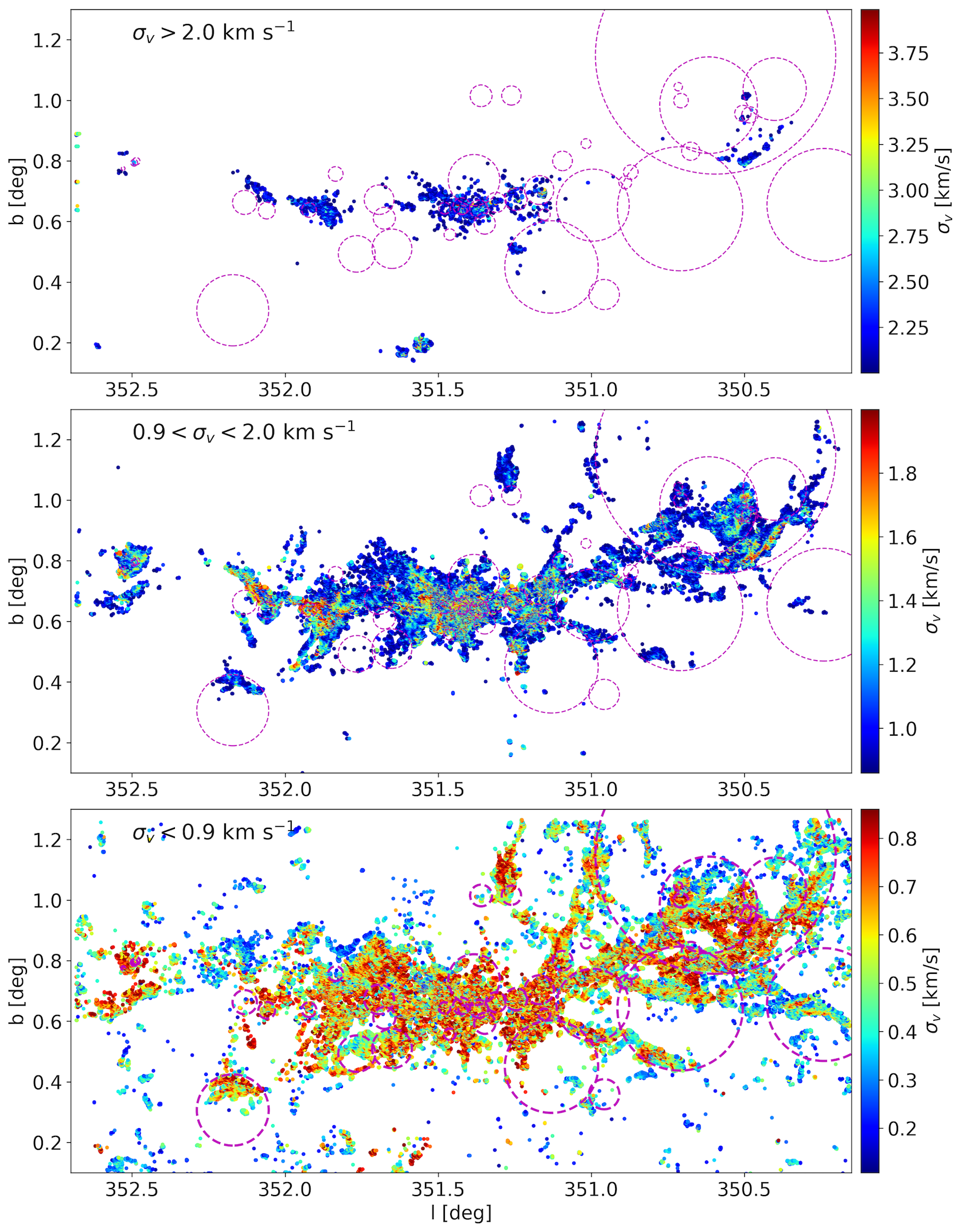}
    \caption{Velocity dispersion measured from $^{13}$CO Gaussian fit. Dispersion velocity ranges are presented on top left corner of each plot and also shown are the H II sources from the catalog of \citep{Anderson2014}.}
    \label{fig:velocitydispersion}
\end{figure*}

\subsection{Velocity dispersion in NGC 6334 extended region}
Figure \ref{fig:histveldispersion} presents a histogram of the velocity dispersion ($\sigma_{v}$) determined from the Gaussian fit to the $^{13}$CO line profiles toward the mapped region. Mean and median values of the distribution are 0.9 and 0.8 km s$^{-1}$, respectively. We then made maps for different velocity dispersion ranges $\sigma_{v}\ge2$, $2.0>\sigma_{v}\ge$0.9 and $0.9\le~\sigma_{v}$ km s$^{-1}$ (top to bottom panels in Figure \ref{fig:velocitydispersion}). The gas with velocity dispersions lower that 0.9 km s$^{-1}$ (mean value), represents the extended regions. At an increased dispersion level from 0.9 to 2 km s$^{-1}$, the emission mostly corresponds to the dense ridges along the filament while three regions (EF1, central ridge and GM24-SF2) exhibit velocity dispersions higher than 2 km s$^{-1}$. Except toward a few H II regions, we do not observe significantly higher velocity dispersions associated with the H II shells. This could mean that  H II regions play a minor role in injecting turbulence to the large scale gas distribution. In the NGC 6334 extended region, we found that the velocity dispersion is well correlated with the dense gas structure along the filament. This indicates that the origin of the velocity dispersion is likely related to formation 
of the filaments themselves. The large values of dispersion toward the NGC 6334 central ridge and filament EF1, $\sigma_{v}\ge2$, may reflect  global collapse of gas onto the filaments. Both of these regions have a broad V-shape velocity structure in the $lv$-plot which will be discussed in Section \ref{sec:globalcollapse}.

\section{Discussion}\label{sec:disc}
\subsection{Velocity coherence and fluctuations}\label{velfluctuations}
The NGC 6334 filamentary structure was studied in lower excitation lines of CO by \cite{Arzoumanian2022}. Based on their position-velocity map, these authors reported a $\sim$50 pc sized velocity coherent structure between  $l = 350.4^{\circ}$ and 352.6$^{\circ}$). In this study, with the higher resolution data and improved method for creating position velocity map by following the dense gas ridge along the filament, we investigated the velocity structure in the NGC 6334 extended region ($l = 350.15^{\circ}$ to 352.65$^{\circ}$) (see Section \ref{sec:velocitystructure}). We found that the extent of the filament in which we observe the velocity coherent structure is $\sim$ 80 pc in length and thus longer than previously thought. The velocity gradient along the entire filament ($\sim$80 pc) is much smaller than 1 km/s/pc which quantitatively illustrates its velocity coherency. Such coherency in large scale structures is also found in the simulations of galactic filaments (e.g., \citealt{DaurteDobbs2016, Smith2020}). In a recent review of the filamentary ISM, \cite{Hacar2023} suggest that velocity coherency could simply be a necessary feature of survival of the large filaments since larger gradients would otherwise lead to their rapid destruction. 

We also observed smaller scale velocity and intensity fluctuations (or so called `wiggles') along the filament (see Section \ref{sec:velocitystructure} and Figure \ref{fig:gaussveldispersion}). These fluctuations or oscillatory features in the velocity centroid and intensity (with phase shift) are thought to trace core-forming flows or local density enhancement in the filament (e.g., \citealt{Hacar2011, Guo2021, Henshaw2020}).   

\subsection{Multiple gas compression, global collapse and infall}\label{sec:globalcollapse}
We studied both localized and the large scale velocity structure of the NGC 6334 filament using position velocity diagrams (see Sect. \ref{sec:velocitystructure}). In addition to the commonly used position-velocity plots, which are intensity weighted, we used longitude/latitude velocity ($lv$ or $bv$) plots using the Gauss fitted velocities from $^{13}$CO line profiles. We investigated the velocity structure toward the FIR sources in the central filament and also toward the H II sources located in the extended region. Toward all the FIR sources we observe V-shaped (inverted V-shape) velocity structures latitude-velocity plots (see Fig. \ref{fig:pvmaps-center}). Toward the H II regions the $lv$ and $bv$ plots exhibited a variety of complex velocity features (see Fig. \ref{fig:lvplots_hiisources0} to \ref{fig:lvplots_hiisources9}). A more common feature is again the V-shape (inverted V-shape) velocity structure particularly toward sources that are located in or adjacent to the filament. The V-shape velocity features are thought to trace gas compression due to propagating shock fronts or colliding flows (for example, \citealt{Inoue-Fukui2013, FukuiY2018CC,Arzoumanian2022}). Formation of molecular clouds after multiple compressions in interacting shells or bubbles has been proposed in theoretical studies (e.g., \citealt{Inutsuka2015}). Our observations of multiple gas compression features observed toward H II shells/radii, as well as toward FIR sources in the central ridge support this scenario.

On a larger scale, we have identified broader V-shape velocity structures toward the NGC 6334 central filament, the GM-24 region and EF 1 (the G352 region) in the median velocity contours projected along longitude direction (see Figure \ref{fig:gaussveldispersion} in Section \ref{sec:velocitystructure}). We discuss these three regions individually here.

\begin{figure*}[htbp!]
    \centering
    \includegraphics[width=0.9\linewidth]{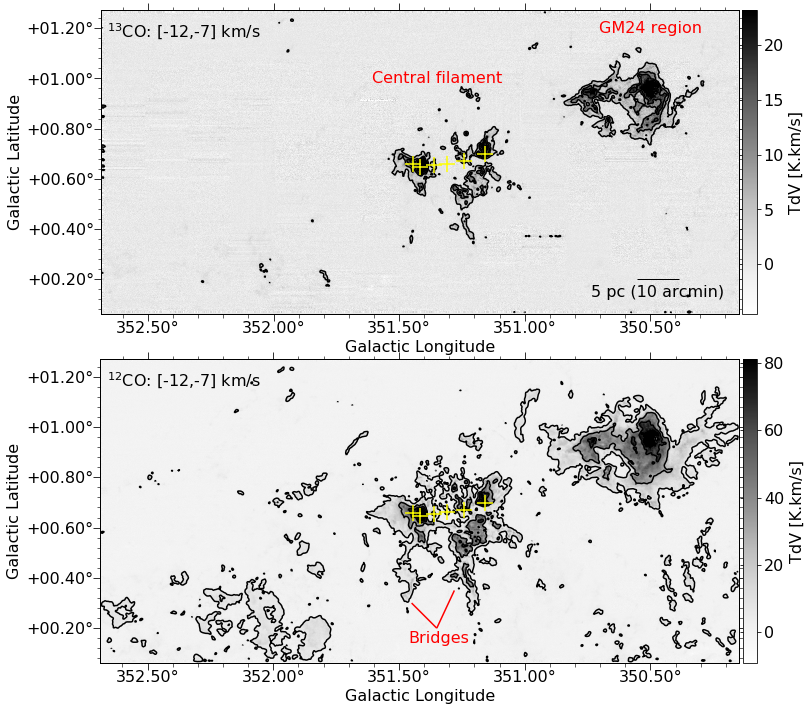}\\
    \caption{Maps of $^{13}$CO (top) and $^{12}$CO (bottom) integrated over the velocity range from $-$12 to $-$7 km s$^{-1}$. Labels and markers are the same as in Fig. \ref{fig:mommaps}}
    \label{fig:bridges}
\end{figure*}

\subsubsection{NGC 6334 central filament:}
The NGC 6334 central filament is approximately 10 pc long and runs almost parallel to the Galactic plane (see Figure \ref{fig:mommaps}). Three distinct properties of the gas velocity structure are observed toward this region. 

First, the east and west side of the central ridge exhibit bright emission at blue-shifted velocities with respect to the average velocity of the NGC 6334 central filament $-$3.9 km s$^{-1}$. This is evident from the channel maps presented in Figure \ref{fig:chanelmap13CO} and \ref{fig:chanelmap12CO}. The case for a global collapse scenario has already been made by \cite{Zernickel2013} for this  filament. Our observations provide further evidence of the globally collapsing gas in the NGC 6334 central ridge. 

Second, we observed a broad inverted V-shape structure in the $lv$-plot (see Figure \ref{fig:gaussveldispersion}). The base length of the V-shape is $\sim$6 pc (0.2 deg at 1.7 kpc) and it is located between FIR sources II and IV. Additionally, we also observe the intensity fluctuation phase shifted with respect to the observed velocity structure (Figure \ref{fig:gaussveldispersion}, bottom panel). Such V-shapes indicate gas compression (due to collision or due to H I/H II bubbles) (e.g., \citealt{Inoue-Fukui2013, Arzoumanian2022}) or a global infall/collapse if observed with phase shifted intensity and velocity gradients (e.g., \citealt{Hacar2011,Zhou2023_1}). The velocity gradient inferred from the arms of the V-shape is 1.3 km s$^{-1}$ pc$^{-1}$. Assuming the observed velocity gradient is due to free fall and using the extension of V-shape (R$\sim$3 pc), we estimate a kinetic mass ($M \approx 2R^{3} \nabla V^{2}/G$; see Eqn. 1 in \citealt{Zhou2023_1}) of the central filament of $\sim2\times$10$^{4}$ M$_\odot$. The mass estimate from the free fall assumption is consistent with the reports of line mass per unit length of 1000 M$_\odot$ pc$^{-1}$ toward the central NGC 6334 filament by \cite{andre2016}.

\begin{figure*}[htbp!]
    \centering
    \includegraphics[width=0.9\linewidth]{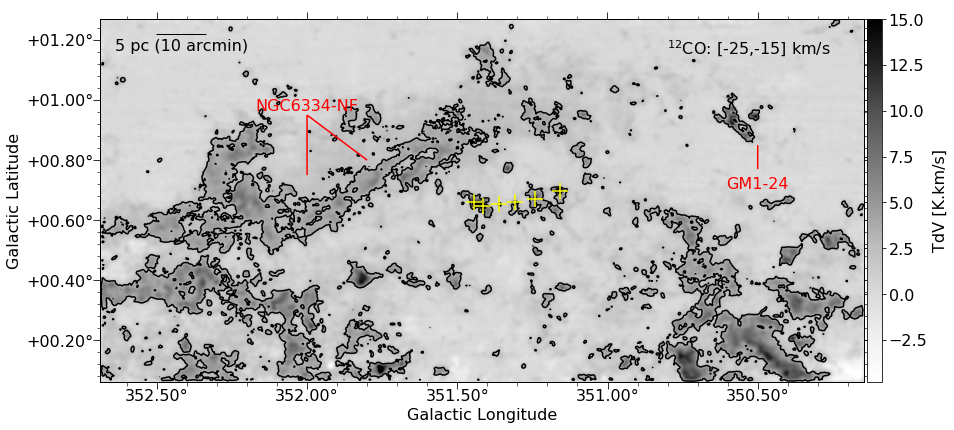}
    \caption{Map of $^{12}$CO integrated emission integrated over the velocity range from $-25$ to $-$15 km s$^{-1}$ the contour line represening the 5$\sigma$ level. The emission north to the NGC 6334 central region is labelled as the northern filament (NF). Yellow plus markers denote the FIR-sources (I[N], I to V from east to west) in the central filament.}
    \label{fig:northernfilament}
\end{figure*}

Third, toward both longitudinal ends of the central filament, we observe two `bridge' structures (see Figure \ref{fig:bridges}). The gas emission in the bridges correspond to the bluer velocity component ($-$9.2 km s$^{-1}$) shown in Figure \ref{fig:mom1_hist}.  From $^{12}$CO and $^{13}$CO emission properties, it is clear that these are prominent dense gas structures, filamentary in nature, and running almost perpendicular to the central ridge of the main filament. The same gas velocity component is also associated with GM-24 region. From the central ridge, the eastern bridge feature in $^{12}$CO (3-2) emission extends approximately 7 pc to the south while the western bridge is about 15 pc long extending both toward south and north passing from FIR source V. The south extension of the west bridge from sources V is $\sim$8 pc. Along the central ridge both the $-$3.9 km s$^{-1}$ and the $-$9.2 km s$^{-1}$ velocity components coexist spatially. This hints mixing of the gas either due to collision or merger of the clouds. 

\cite{FukuiY2018CC} presented a cloud-cloud collision scenario in which the $-$20 km s$^{-1}$ filamentary cloud is in collision with the NGC 6334 filament. In fact, we also observe at velocities [$-$25, $-$15] km s$^{-1}$ a feature in  $^{12}$CO emission,  but not in $^{13}$CO. The emission morphology presented in Figure \ref{fig:northernfilament} at this velocity range runs north-east to the central filament (with reference to the source I[N]). This emission region extends from (351.5 deg,0.8 deg) to (352.2 deg, 0.5 deg). We refer to this as the Northern Filament and have labeled it "NGC6334-NF" in the figure. This emission feature is also detected in CO (2-1) and CO (1-0) and is connected in the position velocity map with the central NGC 6334 filament (see \citealt{FukuiY2018CC,Arzoumanian2022}). The detection of higher excitation $J = 3-2$ lines of CO in this work indeed indicates that it also contains relatively denser gas but likely at quiescent phase since no ATLASGAL clumps are detected toward this region. According to \cite{FukuiY2018CC}, cloud-cloud collision scenario explains the observations of the "bridge" features. In addition they also postulate that the cloud collision is happening at different locations at different time scales and, therefore, giving rise to the different evolutionary phases of the NGC 6334 extended filament from west to east. GM-24 region located in the west is an evolved phase, the central ridge is in an active phase of star formation, and, the filament to the east (G352 region) is in a quiescent phase. We suggest that the spatial and kinematic connection of the NGC6334-NF filament and the `bridge' features with the main gas velocity structure should be considered in explaining star formation in the NGC 6334 filamentary complex.

\subsubsection{GM-24 region:}
A broad inverted V-shape is observed toward the GM-24 region in the median velocity contour in the $lv$-plot (see Fig. \ref{fig:gaussveldispersion}). However, the GM-24 region contains gas emission at both $-$3.9 km s$^{-1}$ and $-$9.2 km s$^{-1}$. Therefore the broad V-shape in median velocity contours shown in Figure \ref{fig:gaussveldispersion} does not exactly illustrate the velocity structure. 
\cite{Fukui2018_GM24} suggested cloud-cloud collision scenario toward this region based on the complimentary distribution of gas emission at two velocities ($-$10 and $-$6 km s$^{-1}$) and the V-shape observed in position velocity map of $^{12}$CO 2-1 at spatial resolution of 90$\arcsec$. Our LAsMA observations with 20$\arcsec$ spatial resolution do not confirm complimentary distribution of the gas. We find that the observed gas emission morphology and velocity structure toward GM-24 region can be explained by the multiple H II bubbles and shell-like structures. In addition, we observe that the larger bubbles located south to the GM-24 region are shaping the filamentary gas emission structure. 

\subsubsection{Eastern Filament (EF1): a hub-filament in formation?}
Filament EF1 is located around $l = 352$ deg and $b = 0.7$ deg (see Figure \ref{fig:mom1} and \ref{fig:13CO_g352}). The filament corresponds to the dark lane observed in the infrared three color map in Figure \ref{fig:3color} caused by  dust absorption. In addition, \cite{Arzoumanian2022} have reported velocity coherent filamentary structure (VCF47) toward this filament. 

In the longitude velocity plots, we observe a broad V-shape toward this filament. These velocity fluctuations are accompanied by the intensity fluctuations with phase shift (Fig. \ref{fig:gaussveldispersion}, bottom panel). The velocity gradient is also visible in the moment 1 map of $^{13}$CO (Fig. \ref{mom1g352}). We suggest that this is consistent with  the global collapse scenario in which  gas is infalling toward the filament EF1. This filament harbours a few ATLASGAL clumps (see Fig. \ref{fig:13CO_g352}). The velocity gradient inferred from the arms of the V-shape is 1.3 km s${-1}$ pc$^{-1}$. Assuming a free fall velocity corresponding to the observed velocity gradient and extension of V-shape (R$\sim$2 pc), we estimate that the kinetic mass ($\approx 2R^{3} \nabla V^{2}/G$) of the filament $\sim$5$\times$10$^{3}$ M$_\odot$. The channel maps in the [+1, +5] km s$^{-1}$ also show multiple gas streams that are more clearly visible in $^{12}$CO (Fig. \ref{fig:13CO_g352}). The global infall and filamentary gas streams in EF1 indicate that this region is a hub-filament system in formation.  

\subsection{Supersonic velocity dispersion}
We observed supersonic dispersion velocities along the dense gas ridge in NGC 6334 filament. At the spatial resolution of our observations (beam size of 20$\arcsec$), the Mach number, corresponding to $\sigma_{v}$ of 0.9 to 3.0 km s$^{-1}$ are 3 to 11, values inferred from sound speed at an  average temperature of 20 K. Supersonic velocity dispersions are commonly observed toward giant filaments (see review by \citealt{Hacar2023}). Theories on the origin of the velocity dispersion in molecular clouds and filaments are under intense discussion both in theoretical and observational studies if it reflects the gravitational or turbulent origin (e.g., \citealt{BallesterosParedes2011, Heitsch2013, Arzoumanian2013, Smith2016,Clarke2017, Mattern2018, Xu2019, Vazquez-semadeni2019,Hacar2023}).

Observationally, it is clear that determinations of velocity dispersionss depend on the spatial resolution. Resolved clouds and filaments seem to exhibit dispersions close to the sonic speed (e.g., \citealt{Hacar2011,Arzoumanian2013}). In NGC 6334 itself, high resolution observations with ALMA have revealed subsonic and transonic velocity dispersions in filaments and cores (NGC 6334S: \citealt{Shanghuo2020}, NGC 6334 I[N] and I: \citealt{Sadaghiani2020}). \cite{Hacar2023} investigated relations between velocity dispersion, filament length and line mass and showed that these scaling relations are followed by various types of filaments, which indicates that at larger scales non-thermal motions govern the gas dynamics. Whether the non-thermal motions originate from a turbulence cascade, core forming flows in the filament, global gravitational collapse or cloud collisions requires further investigation. 

\subsection{Role of H II regions in the star formation processes in NGC 6334 extended region}
Various feedback processes act upon molecular clouds, providing mechanical and radiative energy input. Supernova explosions of dying high-mass stars ($> 8 M_{\odot}$), stellar winds, stellar jets and outflows provide mechanical feedback while ionizing and non-ionizing radiation provide radiative energy inputs (\citealt{Klessen-Glover2016}). Toward the NGC 6334 region, only one supernova remnant (G351.7+0.8) has been detected (\citealt{Green2019}), which is located North-East of the central filament. However, based on optical extinction data, the distance of this supernova remnant was found to be 3.4 kpc (\citealt{wangsu2020}), twice the accepted  distance to NGC 6334 
($\sim$ 1.7 kpc; \citealt{Russeil2012}). A large number of OB stars and the H II regions/bubbles created by them drive the feedback in the NGC 6334 extended region.  

As shown in Figure \ref{fig:3color}, OB stars are clustered in the GM-24 region while they are found in association in the NGC 6334 central region (see \citealt{Russeil2020OBcat}). A total of 42 H II regions/bubbles are found in the extended region that are visually identified in the mid-infrared with varying sizes from 0.2 to 12 pc (\citealt{Anderson2014}). At least eight H II regions, associated with present star formation, are found in and around the NGC 6334 central ridge (see \citealt{persitapia2008}). The GM-24 region has bubbles within bubbles (see Figure  \ref{fig:3color}). Only a few H II regions, mostly of smaller size, are found toward the eastern filaments (near EF1, EF2 and G352.5, see Figure  \ref{fig:3color}, \ref{fig:chanelmap13CO}).  
Of the 42 H II regions from \cite{Anderson2014}, most H II regions (40 of 42) show gas velocities derived from $^{13}$CO that are consistent with the velocity of the bulk gas emission (see Table \ref{tab1}). This confirms that they are indeed part of the NGC 6334 region. From a visual inspection of the channel maps of the CO emission structure around the H II regions  also using quantitative methods, we find that a significant fraction {($>$80\%)} of H II regions have an impact on the surrounding gas. Impact here either means clear shell-like/arc-like structure in emission morphology and/or high-contrast values in the CO intensities at bubble edges. Visually, clear signatures of H II bubbles interacting with the filamentary structure are seen toward GM-24 region in which filaments extending west from the central ridge pass through the edges of the bubbles (see Figure \ref{fig:3color} and \ref{fig:chanelmap13CO}). We suggest that these bubbles, typically larger in size, have induced accumulation of gas in the filament and reached pressure equilibrium conditions with the environment. \cite{Barnes2020} have reported that the maximum size of an H II region is set by pressure equilibrium with the ambient ISM, consistent with our interpretation. Similar to our findings, the  role of H II bubbles in the formation of a molecular filament has also been suggested for the case of RCW 120 (\citealt{zavango2020}). 

Hundred and sixty-seven ATLASGAL clumps are found in the extended NGC 6334 region, most of which are embedded in the filamentary structure. We searched for their location to investigate possible associations with the H II bubbles. A total of 56 (of 167) ATLASGAL clumps are located on the edges of the H II bubbles (0.8-1.2$\times R_{H II}$). We also found that 26 of 42 H II regions have at least one ATLASGAL clump located at the bubble radii. This result hints that H II regions may play an important role in formation of the filaments in which the ATLASGAL clumps are embedded. \citealt{Elmegreen1998} has highlighted the important role of the H II region in the formation of stars at their edges/shells (see also \citealt{Inutsuka2015}). Cases of positive feedback from  H II bubbles are reported towards multiple star forming regions, for example, toward the G305 complex `collect and collapse' model of triggered star formation is reported by \cite{Mazumdar2021b}.

\section{Summary}\label{sec:summary}
We conducted observations of the $^{13}$CO and $^{12}$CO $3\rightarrow2$ molecular lines toward the extended NGC 6334 filament using the LAsMA instrument on the APEX telescope, with  a spectral resolution of 0.25 km s$^{-1}$ and a spatial resolution of $\sim$20 arcsec (0.16 pc at 1.7 kpc). In this paper, we focused on studying the emission morphology and velocity structure of the gas in the filament traced by carbon monoxide. The results are summarized here:
\begin{itemize}
\item The CO traced gas in the NGC 6334 extended region is filamentary, and extends over 80 pc parallel to the Galactic plane. The central NGC~6334 filament exhibits bright CO emission tracing the dense gas reservoir that extends over 10 pc scale. While $^{13}$CO traces denser regions, $^{12}$CO exhibits a more extended emission morphology.
\item We fitted the $^{13}$CO line profiles using an automated Gauss fit algorithm \texttt{Gausspyplus}. Multiple velocity components were required to fit the observed line profiles toward the denser regions indicating the complex gas velocity structure of  the filament. Overall, NGC 6334 exhibits two distinct gas components at velocities $-$3.9 and $-$9.2 km s$^{-1}$. A third component at $\sim$$-$20 km s$^{-1}$ is kinematically connected to the extended region.
\item We observed velocity and intensity fluctuations (so called `wiggles') along the dense ridge of the filament.  We suggest that such fluctuations are likely to be associated with the local density enhancements and gravitational infall onto the filament. 
\item We found that the dense gas along the filament shows velocity dispersions of $2 > \sigma_{v} >0.9$ km s$^{-1}$. Higher velocity dispersions ($> 2$ km s$^{-1}$ were observed toward NGC 6334 central filament and an Eastern Filament (EF1). The velocity dispersion in NGC 6334 filament is supersonic at the spatial resolution of our observations. 
\item We investigated the molecular gas structure around the infrared H II regions identified by \cite{Anderson2014} using azimuthally averaged radial  $^{13}$CO  intensity profiles and measured the line intensity enhancement using the contrast method. Toward most H II regions, we detected molecular line emission and reported the systemic velocities. We found 36 of 42 H II regions to show the signature of molecular gas clearance from the center. These sources exhibit little emission or a flat emission profile toward the central region with an intensity increasing outward or a bumpy feature near or at the H II radii. Using a contrast measurement method, we found intensity enhancements toward 34 of 42 H II regions. In addition, we found that many H II regions (26 of 42) have at least one ATLASGAL clump located at their shell radii.

\item A visually clear signature of H II bubble shells emanating from the filamentary structure is observed in particular toward the GM-24 region, in which filaments extending west from the central ridge are located on the edges of the bubbles. We suggest that these bubbles, typically evolved and larger in size, have assisted the formation of gas filament by accumulating gas at their edges and have reached pressure equilibrium with the environment.

\item We investigated the gas velocity structure around six FIR sources (I[N], I to V) in the central NGC 6334 filament using position velocity diagrams and found evidence for gas compression toward all of them.
In addition, we studied the gas velocity structure around H II regions using $lv$ and $bv$ plots in which velocities were obtained from Gauss fitting of $^{13}$CO line profiles. Toward a third of the H II regions we observed a V-shape emission indicating multiple gas compression in the NGC 6334 extended filament. 

\item In the longitude-velocity ($lv$) plot for the entire mapped region (Figure \ref{fig:gaussveldispersion}), we observed broad V-shaped (or inverted V-shaped) velocity structure toward NGC 6334 central filament and the eastern filament ($l\sim$352.1 deg). 
Toward the NGC 6334 central filament and eastern filament EF1 ($l\sim$352.1 deg), the velocity gradient inferred from the arms of the V-shapes were consistent with a global infall scenario. 
We conclude that east filament EF1 ($l\sim$352.1 deg) is a hub-filament system in formation. 

\item Finally, we studied the kinematic connection of  `bridge' features and the northern filament (NGC 6334-NF) to the main gas component in  NGC 6334. We found NGC 6334-NF contains relatively quiescent gas and does not harbour any star forming clumps traced by submillimeter emission from dust. We suggest that the `bridge' features are possibly linked to cloud-cloud collisions in NGC 6334-NF and the NGC 6334 main filament. 
\end{itemize}

In summary, our observations revealed a complex gas velocity structure in the NGC 6334 filament that extends over $\sim$80 pc. Located in the west, GM24 region exhibits bubbles within bubbles and is at relatively evolved stage of star formation. The NGC 6334 central ridge is undergoing global gas infall and exhibits two gas `bridge' features possibly indicating cloud-cloud collisions in the NGC 6334-NF and the NGC 6334 main gas component. The relatively quiescent eastern filament (EF1 - G352.1) also shows the kinematic signature of global gas infall onto the filament. We detected molecular emission around most infrared H II regions and found that most H II regions have already cleared out the molecular gas from the center and that many have shell/ring like molecular structure around them. We also observed multiple gas compression signatures around the H II regions and  highlighted their important role in shaping the gas emission and velocity structure in the NGC 6334 extended region and in the overall evolution of this star forming complex.

\section*{Data availability}
The supplementary materials (figures and tables of the Appendix section) are available online at: \url{https://doi.org/10.5281/zenodo.13642114}. 

\begin{acknowledgements}
      We thank anonymous referee for the constructive feedback that helped to improve the manuscript. This publication is based on data acquired with the Atacama Pathfinder Experiment (APEX) under program ID M-0107.F-9518A-2021. APEX has been a collaboration between the Max-Planck-Institut f\"{u}r Radioastronomie, the European Southern Observatory, and the Onsala Space Observatory. This work was supported by the Collaborative Research Council (CRC) 956, sub-project A6, and CRC 1601, sub projected B1, funded by the Deutsche Forschungsge-meinschaft (DFG). G.G acknowledges support by the ANID BASAL project FB210003. 
\end{acknowledgements}

\bibliographystyle{aa}
\bibliography{biblio}

\begin{thebibliography}{53}
\expandafter\ifx\csname natexlab\endcsname\relax\def\natexlab#1{#1}\fi

\bibitem[{{Anderson} {et~al.}(2014){Anderson}, {Bania}, {Balser}, {Cunningham},
  {Wenger}, {Johnstone}, \& {Armentrout}}]{Anderson2014}
{Anderson}, L.~D., {Bania}, T.~M., {Balser}, D.~S., {et~al.} 2014, \apjs, 212,
  1

\bibitem[{{Andr{\'e}} {et~al.}(2016){Andr{\'e}}, {Rev{\'e}ret}, {K{\"o}nyves},
  {Arzoumanian}, {Tig{\'e}}, {Gallais}, {Roussel}, {Le Pennec}, {Rodriguez},
  {Doumayrou}, {Dubreuil}, {Lortholary}, {Martignac}, {Talvard}, {Delisle},
  {Visticot}, {Dumaye}, {De Breuck}, {Shimajiri}, {Motte}, {Bontemps},
  {Hennemann}, {Zavagno}, {Russeil}, {Schneider}, {Palmeirim}, {Peretto},
  {Hill}, {Minier}, {Roy}, \& {Rygl}}]{andre2016}
{Andr{\'e}}, P., {Rev{\'e}ret}, V., {K{\"o}nyves}, V., {et~al.} 2016, \aap,
  592, A54

\bibitem[{{Arzoumanian} {et~al.}(2013){Arzoumanian}, {Andr{\'e}}, {Peretto}, \&
  {K{\"o}nyves}}]{Arzoumanian2013}
{Arzoumanian}, D., {Andr{\'e}}, P., {Peretto}, N., \& {K{\"o}nyves}, V. 2013,
  \aap, 553, A119

\bibitem[{{Arzoumanian} {et~al.}(2022){Arzoumanian}, {Russeil}, {Zavagno},
  {Chun-Yuan Chen}, {Andr{\'e}}, {Inutsuka}, {Misugi}, {S{\'a}nchez-Monge},
  {Schilke}, {Men'shchikov}, \& {Kohno}}]{Arzoumanian2022}
{Arzoumanian}, D., {Russeil}, D., {Zavagno}, A., {et~al.} 2022, \aap, 660, A56

\bibitem[{{Ballesteros-Paredes} {et~al.}(2011){Ballesteros-Paredes},
  {Hartmann}, {V{\'a}zquez-Semadeni}, {Heitsch}, \&
  {Zamora-Avil{\'e}s}}]{BallesterosParedes2011}
{Ballesteros-Paredes}, J., {Hartmann}, L.~W., {V{\'a}zquez-Semadeni}, E.,
  {Heitsch}, F., \& {Zamora-Avil{\'e}s}, M.~A. 2011, \mnras, 411, 65

\bibitem[{{Barnes} {et~al.}(2020){Barnes}, {Longmore}, {Dale}, {Krumholz},
  {Kruijssen}, \& {Bigiel}}]{Barnes2020}
{Barnes}, A.~T., {Longmore}, S.~N., {Dale}, J.~E., {et~al.} 2020, \mnras, 498,
  4906

\bibitem[{{Churchwell}(2002)}]{Churchwell2002}
{Churchwell}, E. 2002, \araa, 40, 27

\bibitem[{{Churchwell} {et~al.}(2007){Churchwell}, {Watson}, {Povich},
  {Taylor}, {Babler}, {Meade}, {Benjamin}, {Indebetouw}, \&
  {Whitney}}]{Churchwell2007}
{Churchwell}, E., {Watson}, D.~F., {Povich}, M.~S., {et~al.} 2007, \apj, 670,
  428

\bibitem[{{Clarke} {et~al.}(2017){Clarke}, {Whitworth}, {Duarte-Cabral}, \&
  {Hubber}}]{Clarke2017}
{Clarke}, S.~D., {Whitworth}, A.~P., {Duarte-Cabral}, A., \& {Hubber}, D.~A.
  2017, \mnras, 468, 2489

\bibitem[{{Duarte-Cabral} \& {Dobbs}(2016)}]{DaurteDobbs2016}
{Duarte-Cabral}, A. \& {Dobbs}, C.~L. 2016, \mnras, 458, 3667

\bibitem[{{Elmegreen}(1998)}]{Elmegreen1998}
{Elmegreen}, B.~G. 1998, in Astronomical Society of the Pacific Conference
  Series, Vol. 148, Origins, ed. C.~E. {Woodward}, J.~M. {Shull}, \&
  J.~{Thronson}, Harley~A., 150

\bibitem[{{Frerking} {et~al.}(1982){Frerking}, {Langer}, \&
  {Wilson}}]{Frerking1982}
{Frerking}, M.~A., {Langer}, W.~D., \& {Wilson}, R.~W. 1982, \apj, 262, 590

\bibitem[{{Fukui} {et~al.}(2018{\natexlab{a}}){Fukui}, {Kohno}, {Yokoyama},
  {Nishimura}, {Torii}, {Hattori}, {Sano}, {Ohama}, {Yamamoto}, \&
  {Tachihara}}]{Fukui2018_GM24}
{Fukui}, Y., {Kohno}, M., {Yokoyama}, K., {et~al.} 2018{\natexlab{a}}, \pasj,
  70, S44

\bibitem[{{Fukui} {et~al.}(2018{\natexlab{b}}){Fukui}, {Kohno}, {Yokoyama},
  {Torii}, {Hattori}, {Sano}, {Nishimura}, {Ohama}, {Yamamoto}, \&
  {Tachihara}}]{FukuiY2018CC}
{Fukui}, Y., {Kohno}, M., {Yokoyama}, K., {et~al.} 2018{\natexlab{b}}, \pasj,
  70, S41

\bibitem[{{Green}(2019)}]{Green2019}
{Green}, D.~A. 2019, Journal of Astrophysics and Astronomy, 40, 36

\bibitem[{{Guo} {et~al.}(2021){Guo}, {Chen}, {Feng}, {Sun}, {Wang}, {Su},
  {Sun}, {Ao}, {Zhang}, {Zhou}, {Yuan}, \& {Yang}}]{Guo2021}
{Guo}, W., {Chen}, X., {Feng}, J., {et~al.} 2021, \apj, 921, 23

\bibitem[{{Hacar} {et~al.}(2023){Hacar}, {Clark}, {Heitsch}, {Kainulainen},
  {Panopoulou}, {Seifried}, \& {Smith}}]{Hacar2023}
{Hacar}, A., {Clark}, S.~E., {Heitsch}, F., {et~al.} 2023, in Astronomical
  Society of the Pacific Conference Series, Vol. 534, Astronomical Society of
  the Pacific Conference Series, ed. S.~{Inutsuka}, Y.~{Aikawa}, T.~{Muto},
  K.~{Tomida}, \& M.~{Tamura}, 153

\bibitem[{{Hacar} \& {Tafalla}(2011)}]{Hacar2011}
{Hacar}, A. \& {Tafalla}, M. 2011, \aap, 533, A34

\bibitem[{{Heitsch}(2013)}]{Heitsch2013}
{Heitsch}, F. 2013, \apj, 769, 115

\bibitem[{{Henshaw} {et~al.}(2020){Henshaw}, {Kruijssen}, {Longmore}, {Riener},
  {Leroy}, {Rosolowsky}, {Ginsburg}, {Battersby}, {Chevance}, {Meidt},
  {Glover}, {Hughes}, {Kainulainen}, {Klessen}, {Schinnerer}, {Schruba},
  {Beuther}, {Bigiel}, {Blanc}, {Emsellem}, {Henning}, {Herrera}, {Koch},
  {Pety}, {Ragan}, \& {Sun}}]{Henshaw2020}
{Henshaw}, J.~D., {Kruijssen}, J.~M.~D., {Longmore}, S.~N., {et~al.} 2020,
  Nature Astronomy, 4, 1064

\bibitem[{{Inoue} \& {Fukui}(2013)}]{Inoue-Fukui2013}
{Inoue}, T. \& {Fukui}, Y. 2013, \apjl, 774, L31

\bibitem[{{Inutsuka} {et~al.}(2015){Inutsuka}, {Inoue}, {Iwasaki}, \&
  {Hosokawa}}]{Inutsuka2015}
{Inutsuka}, S.-i., {Inoue}, T., {Iwasaki}, K., \& {Hosokawa}, T. 2015, \aap,
  580, A49

\bibitem[{{Jackson} {et~al.}(2006){Jackson}, {Rathborne}, {Shah}, {Simon},
  {Bania}, {Clemens}, {Chambers}, {Johnson}, {Dormody}, {Lavoie}, \&
  {Heyer}}]{Jackson2006GRS}
{Jackson}, J.~M., {Rathborne}, J.~M., {Shah}, R.~Y., {et~al.} 2006, \apjs, 163,
  145

\bibitem[{{Klein} {et~al.}(2012){Klein}, {Hochg{\"u}rtel}, {Kr{\"a}mer},
  {Bell}, {Meyer}, \& {G{\"u}sten}}]{Klein2012}
{Klein}, B., {Hochg{\"u}rtel}, S., {Kr{\"a}mer}, I., {et~al.} 2012, \aap, 542,
  L3

\bibitem[{{Klessen} \& {Glover}(2016)}]{Klessen-Glover2016}
{Klessen}, R.~S. \& {Glover}, S. C.~O. 2016, in Saas-Fee Advanced Course,
  Vol.~43, Saas-Fee Advanced Course, ed. Y.~{Revaz}, P.~{Jablonka},
  R.~{Teyssier}, \& L.~{Mayer}, 85

\bibitem[{{Kraemer} \& {Jackson}(1999)}]{Kraemer+Jackson1999}
{Kraemer}, K.~E. \& {Jackson}, J.~M. 1999, \apjs, 124, 439

\bibitem[{{Li} {et~al.}(2020){Li}, {Zhang}, {Liu}, {Beuther}, {Palau},
  {Girart}, {Smith}, {Hora}, {Lin}, {Qiu}, {Strom}, {Wang}, {Li}, \&
  {Yue}}]{Shanghuo2020}
{Li}, S., {Zhang}, Q., {Liu}, H.~B., {et~al.} 2020, \apj, 896, 110

\bibitem[{{Loughran} {et~al.}(1986){Loughran}, {McBreen}, {Fazio},
  {Rengarajan}, {Maxson}, {Serio}, {Sciortino}, \& {Ray}}]{Loughran1986}
{Loughran}, L., {McBreen}, B., {Fazio}, G.~G., {et~al.} 1986, \apj, 303, 629

\bibitem[{{Mattern} {et~al.}(2018){Mattern}, {Kauffmann}, {Csengeri},
  {Urquhart}, {Leurini}, {Wyrowski}, {Giannetti}, {Barnes}, {Beuther},
  {Bronfman}, {Duarte-Cabral}, {Henning}, {Kainulainen}, {Menten}, {Schisano},
  \& {Schuller}}]{Mattern2018}
{Mattern}, M., {Kauffmann}, J., {Csengeri}, T., {et~al.} 2018, \aap, 619, A166

\bibitem[{{Mazumdar} {et~al.}(2021){Mazumdar}, {Wyrowski}, {Urquhart},
  {Colombo}, {Menten}, {Neupane}, \& {Thompson}}]{Mazumdar2021b}
{Mazumdar}, P., {Wyrowski}, F., {Urquhart}, J.~S., {et~al.} 2021, \aap, 656,
  A101

\bibitem[{{Mu{\~n}oz} {et~al.}(2007){Mu{\~n}oz}, {Mardones}, {Garay},
  {Rebolledo}, {Brooks}, \& {Bontemps}}]{Munoz2007}
{Mu{\~n}oz}, D.~J., {Mardones}, D., {Garay}, G., {et~al.} 2007, \apj, 668, 906

\bibitem[{{Persi} \& {Tapia}(2008)}]{persitapia2008}
{Persi}, P. \& {Tapia}, M. 2008, in Handbook of Star Forming Regions, Volume
  II, ed. B.~{Reipurth}, Vol.~5, 456

\bibitem[{{Riener} {et~al.}(2019){Riener}, {Kainulainen}, {Henshaw}, {Orkisz},
  {Murray}, \& {Beuther}}]{Gausspyplus2019}
{Riener}, M., {Kainulainen}, J., {Henshaw}, J.~D., {et~al.} 2019, \aap, 628,
  A78

\bibitem[{{Russeil} {et~al.}(2017){Russeil}, {Adami}, {Bouret}, {Herv{\'e}},
  {Parker}, {Zavagno}, \& {Motte}}]{Russeil2017}
{Russeil}, D., {Adami}, C., {Bouret}, J.~C., {et~al.} 2017, \aap, 607, A86

\bibitem[{{Russeil} {et~al.}(2013){Russeil}, {Schneider}, {Anderson},
  {Zavagno}, {Molinari}, {Persi}, {Bontemps}, {Motte}, {Ossenkopf},
  {Andr{\'e}}, {Arzoumanian}, {Bernard}, {Deharveng}, {Didelon}, {Di
  Francesco}, {Elia}, {Hennemann}, {Hill}, {K{\"o}nyves}, {Li}, {Martin},
  {Nguyen Luong}, {Peretto}, {Pezzuto}, {Polychroni}, {Roussel}, {Rygl},
  {Spinoglio}, {Testi}, {Tig{\'e}}, {Vavrek}, {Ward-Thompson}, \&
  {White}}]{Russeil2013}
{Russeil}, D., {Schneider}, N., {Anderson}, L.~D., {et~al.} 2013, \aap, 554,
  A42

\bibitem[{{Russeil} {et~al.}(2016){Russeil}, {Tig{\'e}}, {Adami}, {Anderson},
  {Schneider}, {Zavagno}, {Samal}, {Amram}, {Guennou}, {Le Coarer}, {Walsh},
  {Longmore}, \& {Purcell}}]{Russeil2016}
{Russeil}, D., {Tig{\'e}}, J., {Adami}, C., {et~al.} 2016, \aap, 587, A135

\bibitem[{{Russeil} {et~al.}(2012){Russeil}, {Zavagno}, {Adami}, {Anderson},
  {Bontemps}, {Motte}, {Rodon}, {Schneider}, {Ilmane}, \&
  {Murphy}}]{Russeil2012}
{Russeil}, D., {Zavagno}, A., {Adami}, C., {et~al.} 2012, \aap, 538, A142

\bibitem[{{Russeil} {et~al.}(2020){Russeil}, {Zavagno}, {Nguyen}, {Figueira},
  {Adami}, \& {Bouret}}]{Russeil2020OBcat}
{Russeil}, D., {Zavagno}, A., {Nguyen}, A., {et~al.} 2020, \aap, 642, A21

\bibitem[{{Sadaghiani} {et~al.}(2020){Sadaghiani}, {S{\'a}nchez-Monge},
  {Schilke}, {Liu}, {Clarke}, {Zhang}, {Girart}, {Seifried}, {Aghababaei},
  {Li}, {Ju{\'a}rez}, \& {Tang}}]{Sadaghiani2020}
{Sadaghiani}, M., {S{\'a}nchez-Monge}, {\'A}., {Schilke}, P., {et~al.} 2020,
  \aap, 635, A2

\bibitem[{{Smith} {et~al.}(2016){Smith}, {Glover}, {Klessen}, \&
  {Fuller}}]{Smith2016}
{Smith}, R.~J., {Glover}, S. C.~O., {Klessen}, R.~S., \& {Fuller}, G.~A. 2016,
  \mnras, 455, 3640

\bibitem[{{Smith} {et~al.}(2020){Smith}, {Tre{\ss}}, {Sormani}, {Glover},
  {Klessen}, {Clark}, {Izquierdo}, {Duarte-Cabral}, \& {Zucker}}]{Smith2020}
{Smith}, R.~J., {Tre{\ss}}, R.~G., {Sormani}, M.~C., {et~al.} 2020, \mnras,
  492, 1594

\bibitem[{{Tig{\'e}} {et~al.}(2017){Tig{\'e}}, {Motte}, {Russeil}, {Zavagno},
  {Hennemann}, {Schneider}, {Hill}, {Nguyen Luong}, {Di Francesco}, {Bontemps},
  {Louvet}, {Didelon}, {K{\"o}nyves}, {Andr{\'e}}, {Leuleu}, {Bardagi},
  {Anderson}, {Arzoumanian}, {Benedettini}, {Bernard}, {Elia}, {Figueira},
  {Kirk}, {Martin}, {Minier}, {Molinari}, {Nony}, {Persi}, {Pezzuto},
  {Polychroni}, {Rayner}, {Rivera-Ingraham}, {Roussel}, {Rygl}, {Spinoglio}, \&
  {White}}]{Tige2017}
{Tig{\'e}}, J., {Motte}, F., {Russeil}, D., {et~al.} 2017, \aap, 602, A77

\bibitem[{{Tremblin} {et~al.}(2013){Tremblin}, {Minier}, {Schneider}, {Audit},
  {Hill}, {Didelon}, {Peretto}, {Arzoumanian}, {Motte}, {Zavagno}, {Bontemps},
  {Anderson}, {Andr{\'e}}, {Bernard}, {Csengeri}, {Di Francesco}, {Elia},
  {Hennemann}, {K{\"o}nyves}, {Marston}, {Nguyen Luong}, {Rivera-Ingraham},
  {Roussel}, {Sousbie}, {Spinoglio}, {White}, \& {Williams}}]{Tremblin2013}
{Tremblin}, P., {Minier}, V., {Schneider}, N., {et~al.} 2013, \aap, 560, A19

\bibitem[{{Urquhart} {et~al.}(2018{\natexlab{a}}){Urquhart}, {Koenig},
  {Giannetti}, {Leurini}, {Moore}, {Eden}, {Pillai}, {Thompson}, {Braiding},
  {Burton}, {Csengeri}, {Dempsey}, {Figura}, {Froebrich}, {Menten}, {Schuller},
  {Smith}, \& {Wyrowski}}]{Urquhart2018yCat}
{Urquhart}, J.~S., {Koenig}, C., {Giannetti}, A., {et~al.} 2018{\natexlab{a}},
  VizieR Online Data Catalog, J/MNRAS/473/1059

\bibitem[{{Urquhart} {et~al.}(2018{\natexlab{b}}){Urquhart}, {K{\"o}nig},
  {Giannetti}, {Leurini}, {Moore}, {Eden}, {Pillai}, {Thompson}, {Braiding},
  {Burton}, {Csengeri}, {Dempsey}, {Figura}, {Froebrich}, {Menten}, {Schuller},
  {Smith}, \& {Wyrowski}}]{Urquhart2018}
{Urquhart}, J.~S., {K{\"o}nig}, C., {Giannetti}, A., {et~al.}
  2018{\natexlab{b}}, \mnras, 473, 1059

\bibitem[{{van der Tak} {et~al.}(2007){van der Tak}, {Black}, {Sch{\"o}ier},
  {Jansen}, \& {van Dishoeck}}]{RADEX2007}
{van der Tak}, F.~F.~S., {Black}, J.~H., {Sch{\"o}ier}, F.~L., {Jansen}, D.~J.,
  \& {van Dishoeck}, E.~F. 2007, \aap, 468, 627

\bibitem[{{V{\'a}zquez-Semadeni} {et~al.}(2019){V{\'a}zquez-Semadeni}, {Palau},
  {Ballesteros-Paredes}, {G{\'o}mez}, \&
  {Zamora-Avil{\'e}s}}]{Vazquez-semadeni2019}
{V{\'a}zquez-Semadeni}, E., {Palau}, A., {Ballesteros-Paredes}, J.,
  {G{\'o}mez}, G.~C., \& {Zamora-Avil{\'e}s}, M. 2019, \mnras, 490, 3061

\bibitem[{{Wang} {et~al.}(2020){Wang}, {Zhang}, {Jiang}, {Zhao}, {Chen},
  {Chen}, {Gao}, \& {Liu}}]{wangsu2020}
{Wang}, S., {Zhang}, C., {Jiang}, B., {et~al.} 2020, \aap, 639, A72

\bibitem[{{Wilson} \& {Rood}(1994)}]{wilson-rood1994}
{Wilson}, T.~L. \& {Rood}, R. 1994, \araa, 32, 191

\bibitem[{{Xu} {et~al.}(2019){Xu}, {Ji}, \& {Lazarian}}]{Xu2019}
{Xu}, S., {Ji}, S., \& {Lazarian}, A. 2019, \apj, 878, 157

\bibitem[{{Zavagno} {et~al.}(2020){Zavagno}, {Andr{\'e}}, {Schuller},
  {Peretto}, {Shimajiri}, {Arzoumanian}, {Csengeri}, {Figueira}, {Fuller},
  {K{\"o}nyves}, {Men'shchikov}, {Palmeirim}, {Roussel}, {Russeil},
  {Schneider}, \& {Zhang}}]{zavango2020}
{Zavagno}, A., {Andr{\'e}}, P., {Schuller}, F., {et~al.} 2020, \aap, 638, A7

\bibitem[{{Zernickel} {et~al.}(2013){Zernickel}, {Schilke}, \&
  {Smith}}]{Zernickel2013}
{Zernickel}, A., {Schilke}, P., \& {Smith}, R.~J. 2013, \aap, 554, L2

\bibitem[{{Zhou} {et~al.}(2023){Zhou}, {Wyrowski}, {Neupane}, {Urquhart},
  {Evans}, {V{\'a}zquez-Semadeni}, {Menten}, {Gong}, \& {Liu}}]{Zhou2023_1}
{Zhou}, J.~W., {Wyrowski}, F., {Neupane}, S., {et~al.} 2023, \aap, 676, A69

\end{thebibliography}

\appendix
\onecolumn

\section{rms noise and Gaussian components: $^{13}$CO (3-2)}

\begin{figure*}[htbp!]
    \centering
    \includegraphics[width=0.95\linewidth]{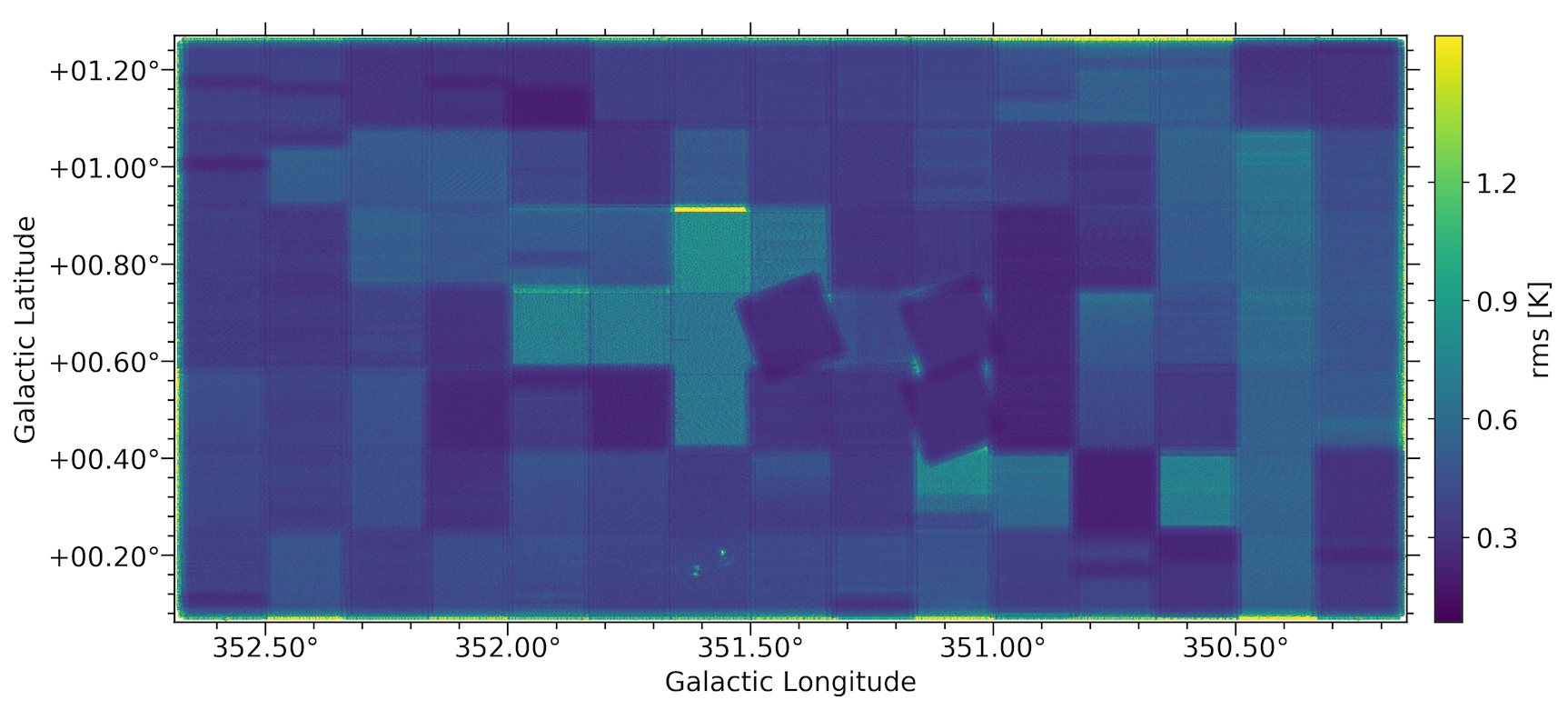}
    \caption{rms noise map of $^{13}$CO toward the NGC 6334 mapped region.}
    \label{fig:rmsnoisemap}
\end{figure*}

\begin{figure*}[htbp!]
    \centering
    \includegraphics[width=0.95\linewidth]{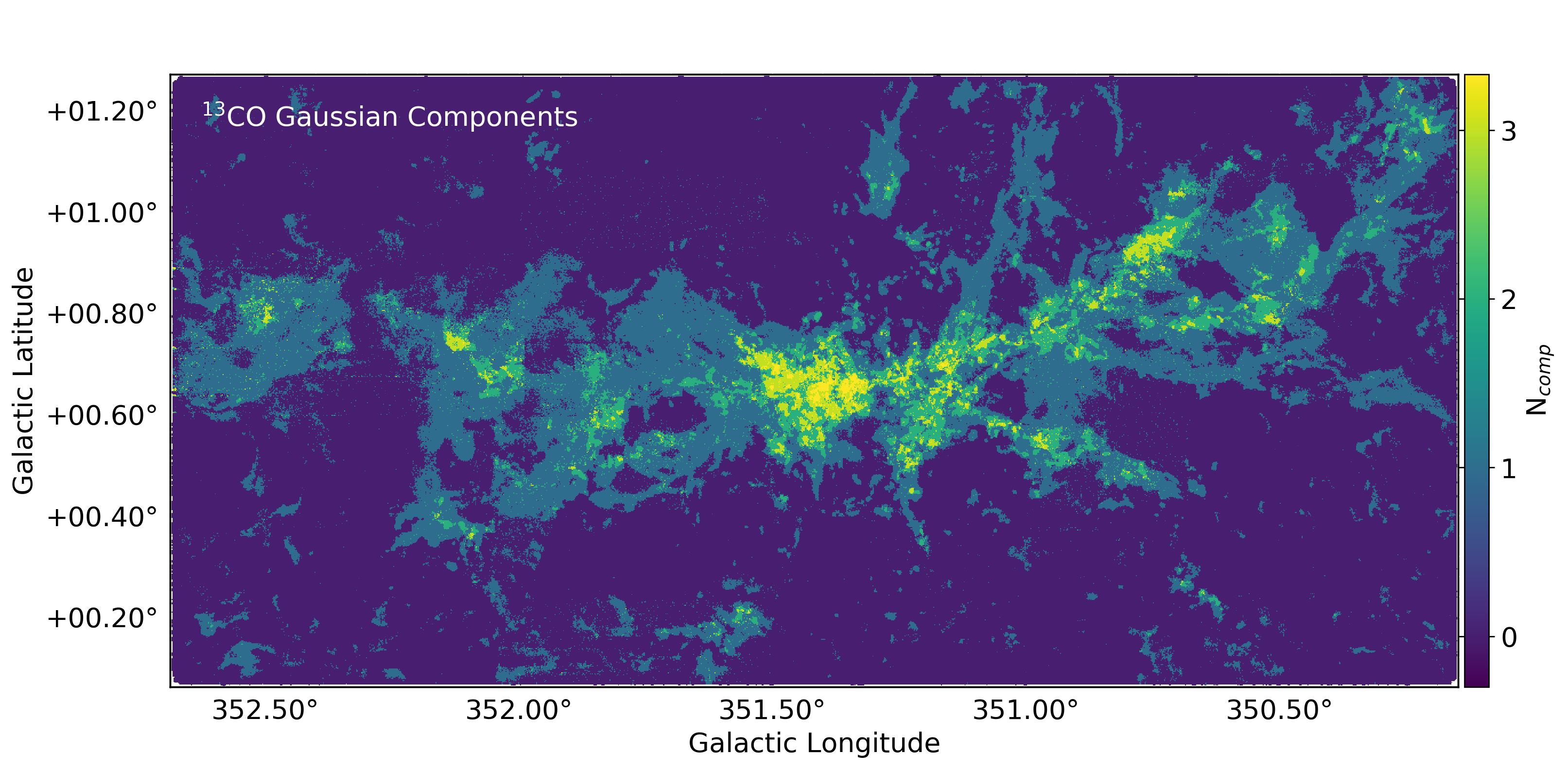}
    \caption{Number of Gauss components per pixel fitted to the $^{13}$CO line profile toward the NGC 6334 region.}
    \label{fig:gausscomp}
\end{figure*}

\clearpage
\section{PV maps toward selected regions and FIR sources in the NGC 6334 central filament}

\begin{figure*}[htbp!]
        \centering
    \includegraphics[width=0.65\linewidth]{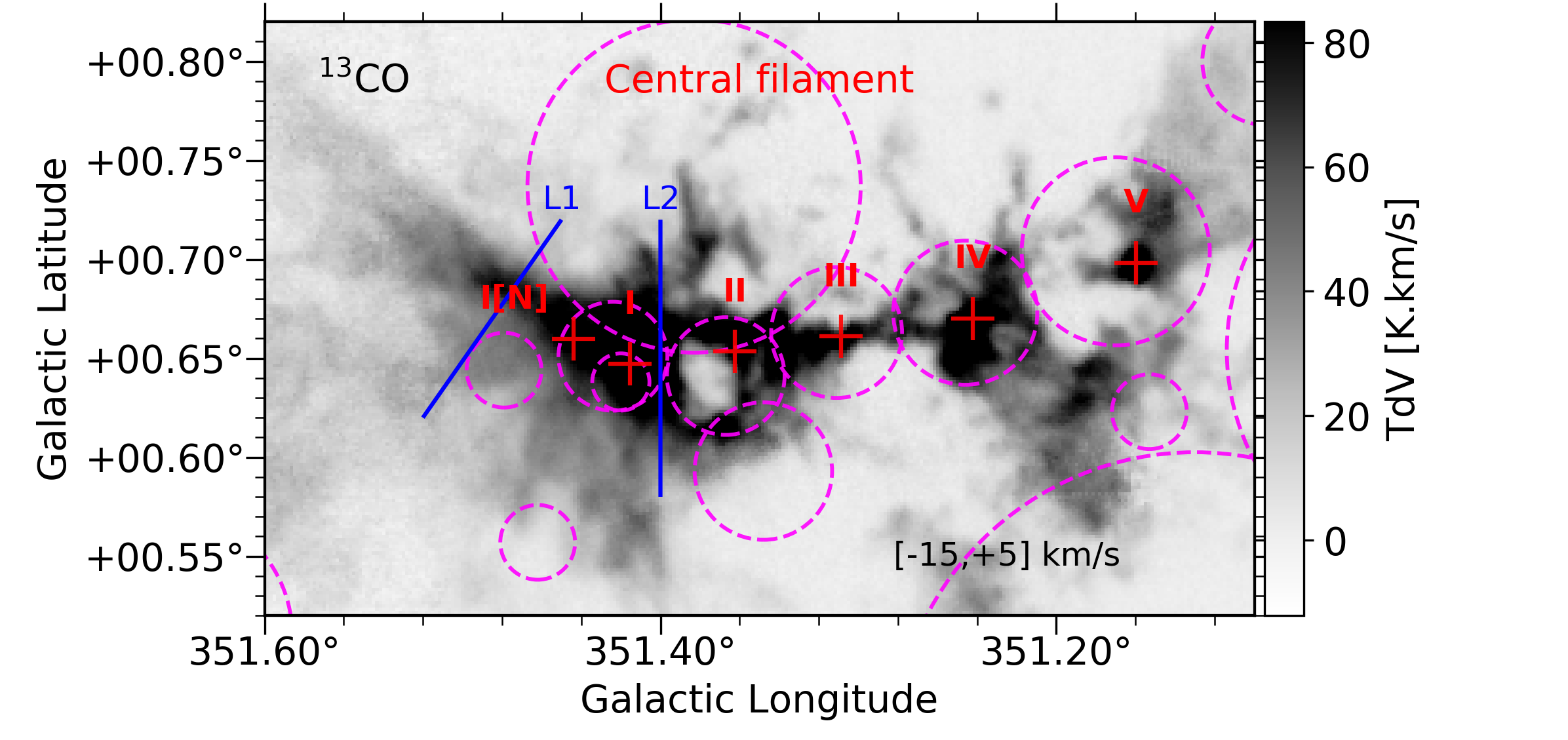}\\
    \includegraphics[width=0.31\linewidth]{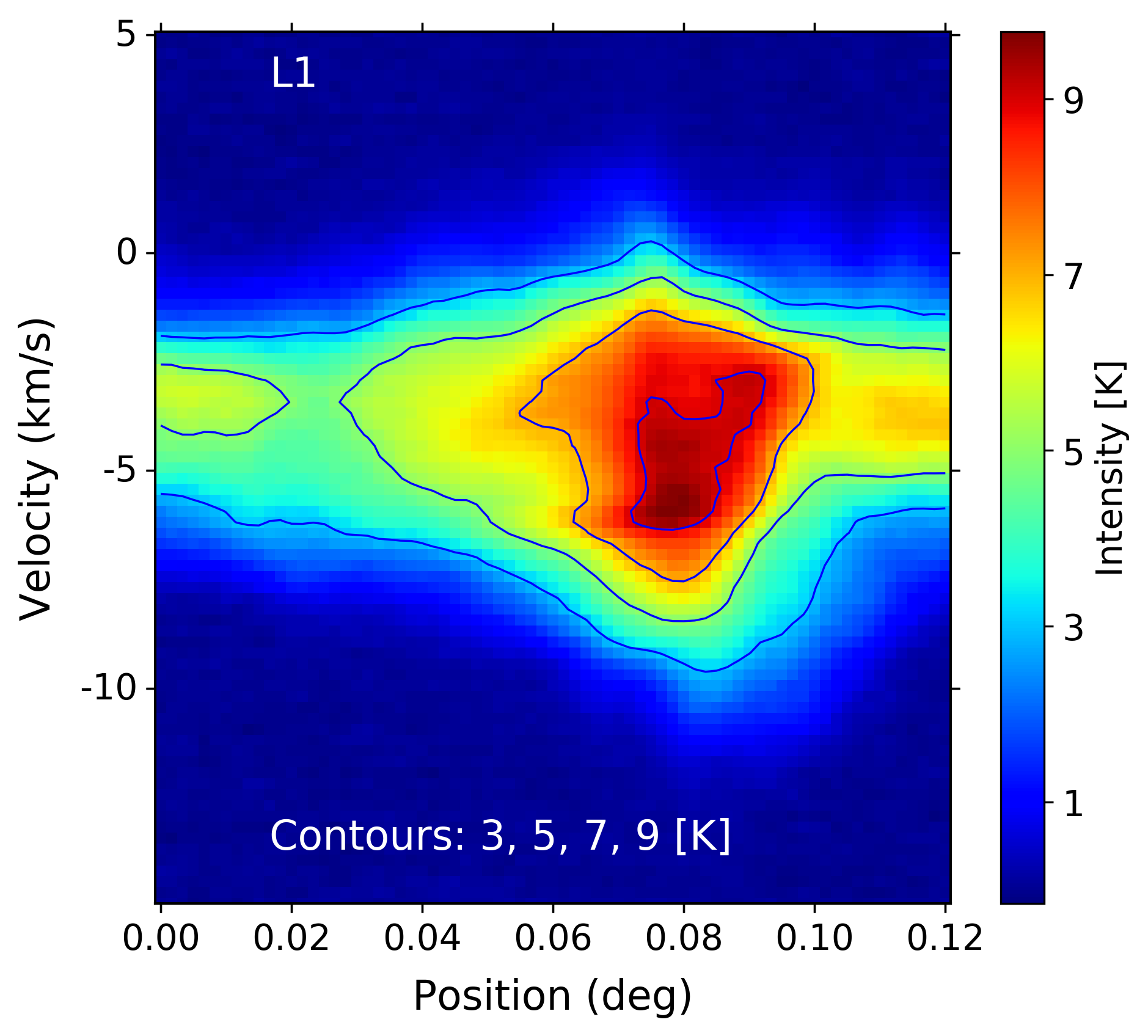}
    \includegraphics[width=0.31\linewidth]{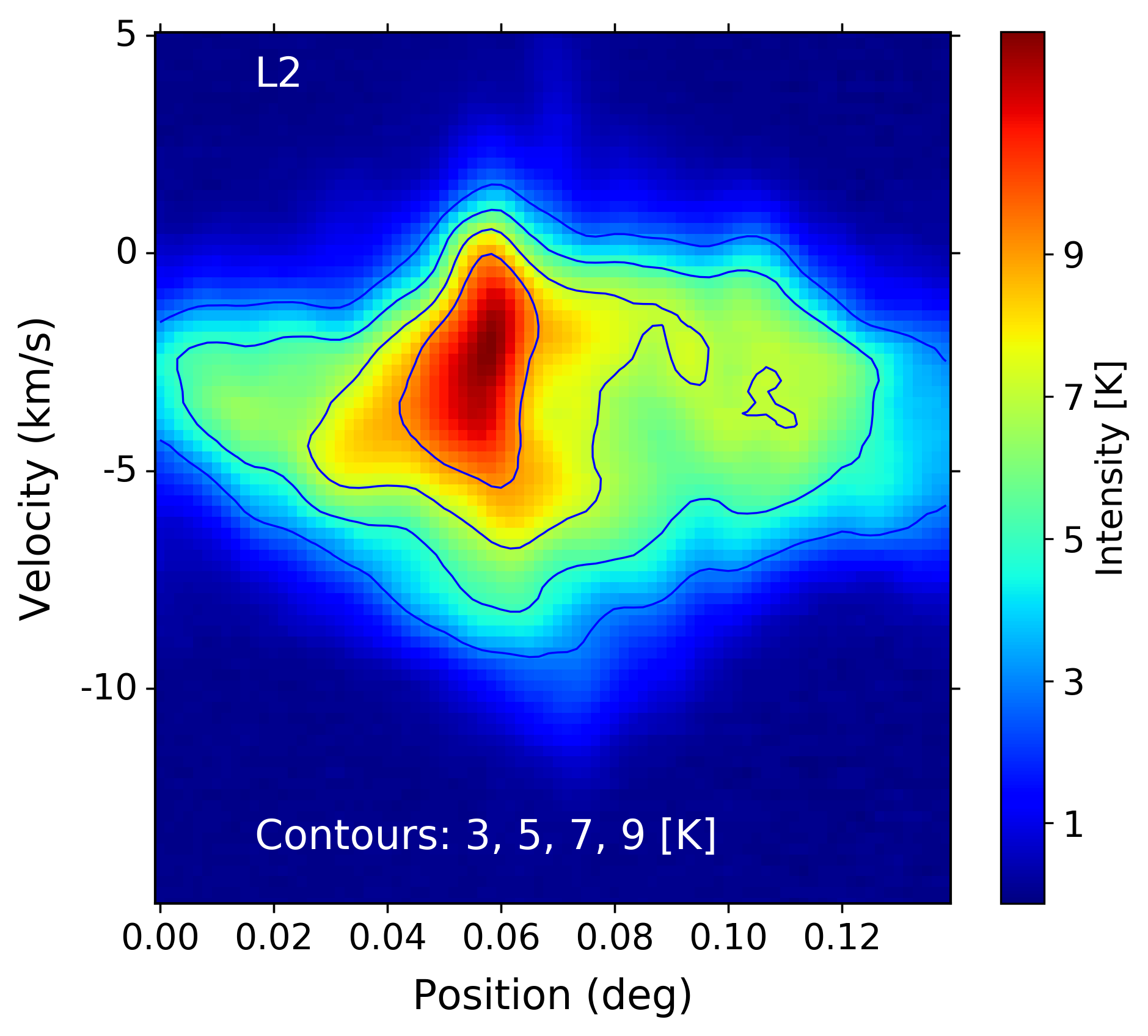}\\
    \includegraphics[width=0.31\linewidth]{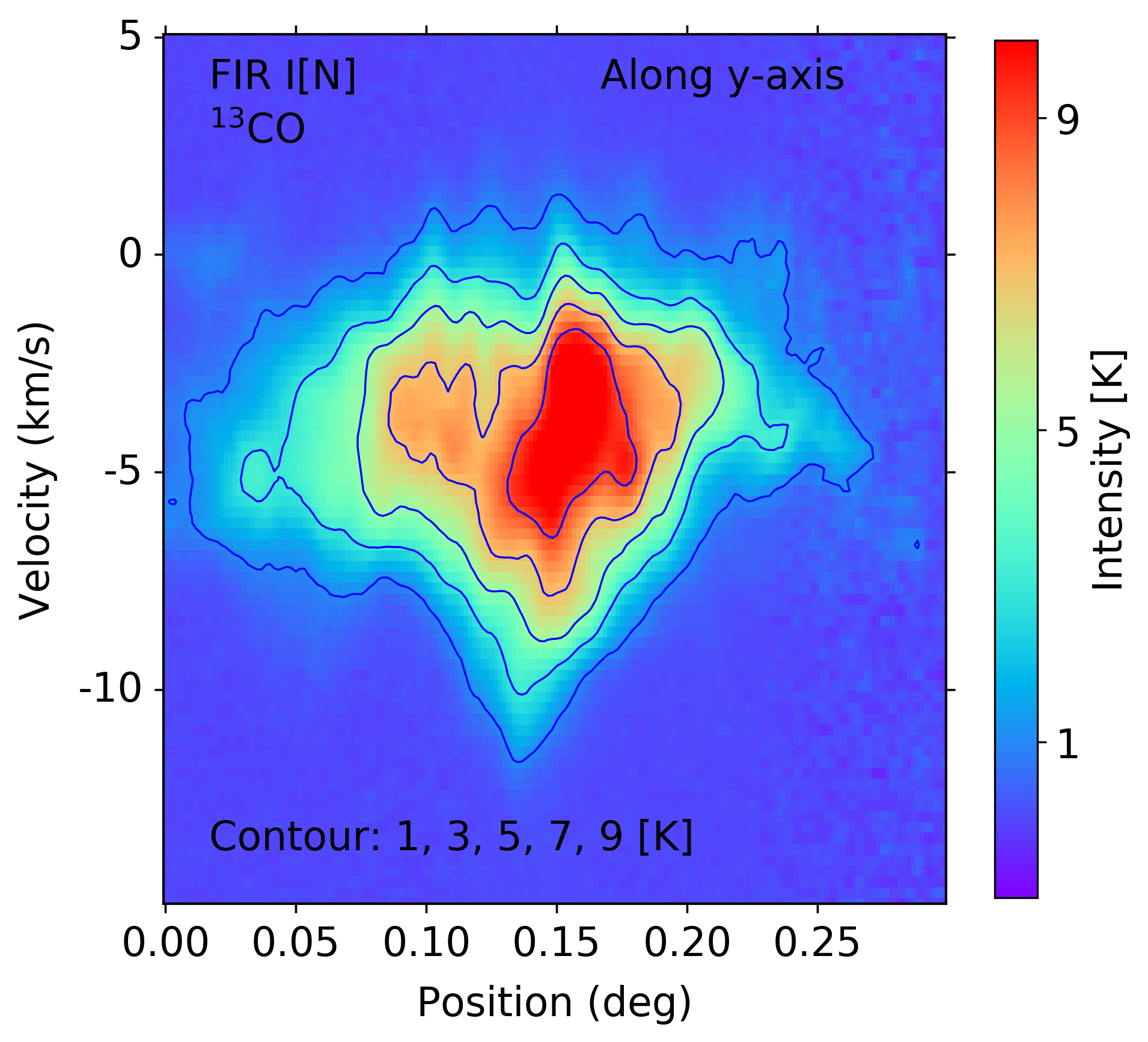}
    \includegraphics[width=0.31\linewidth]{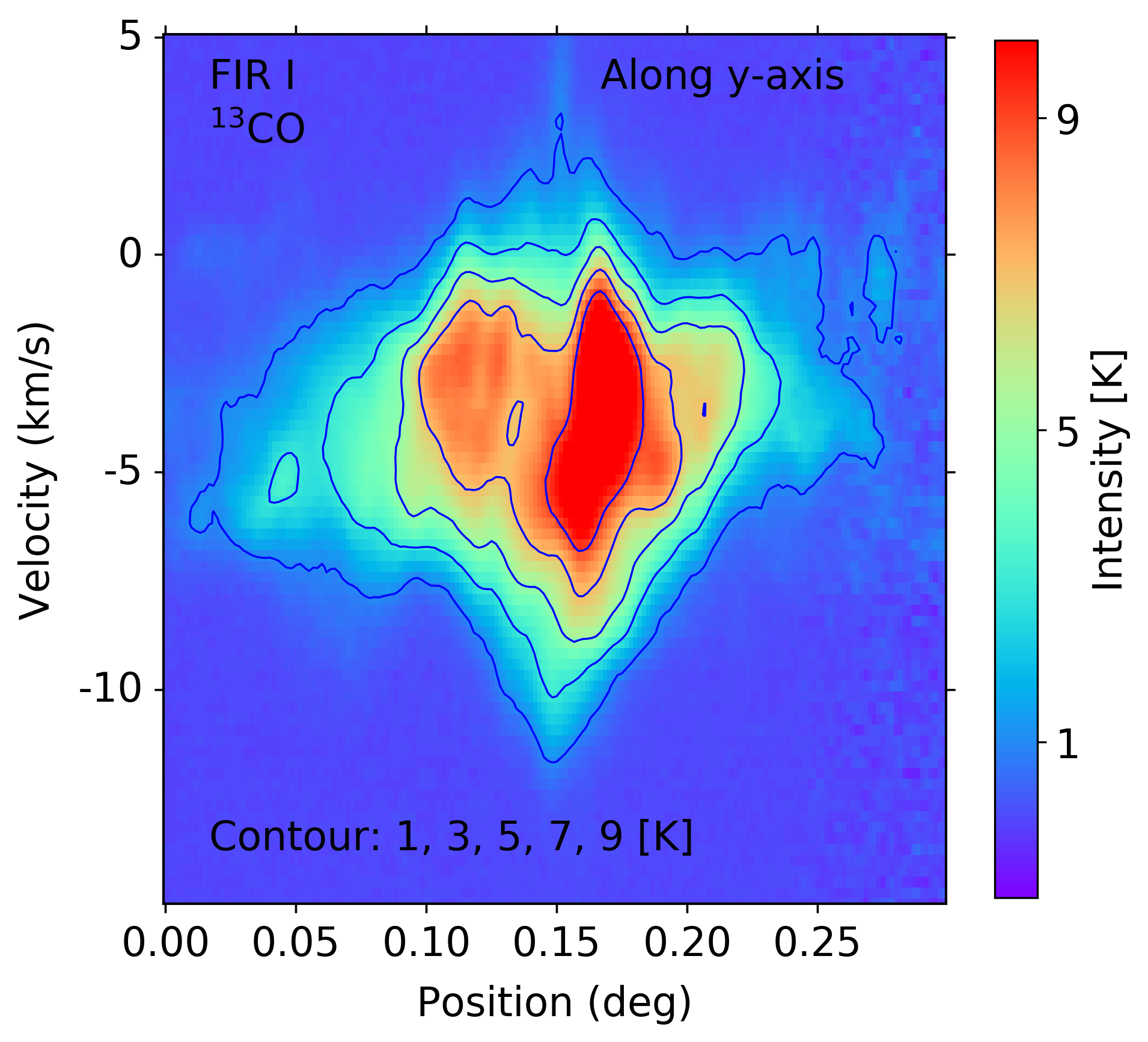}
    \includegraphics[width=0.31\linewidth]{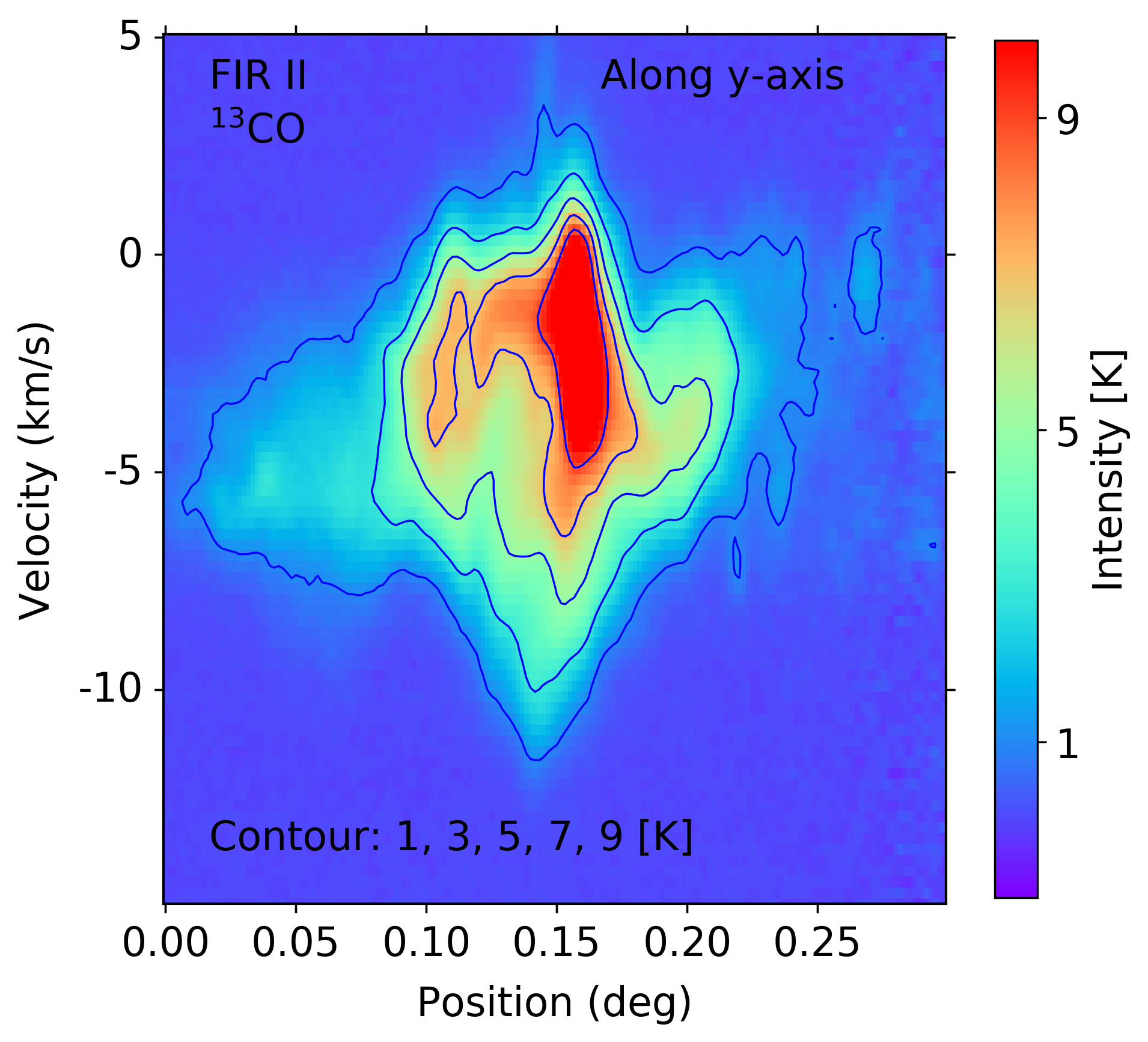}\\
    \includegraphics[width=0.31\linewidth]{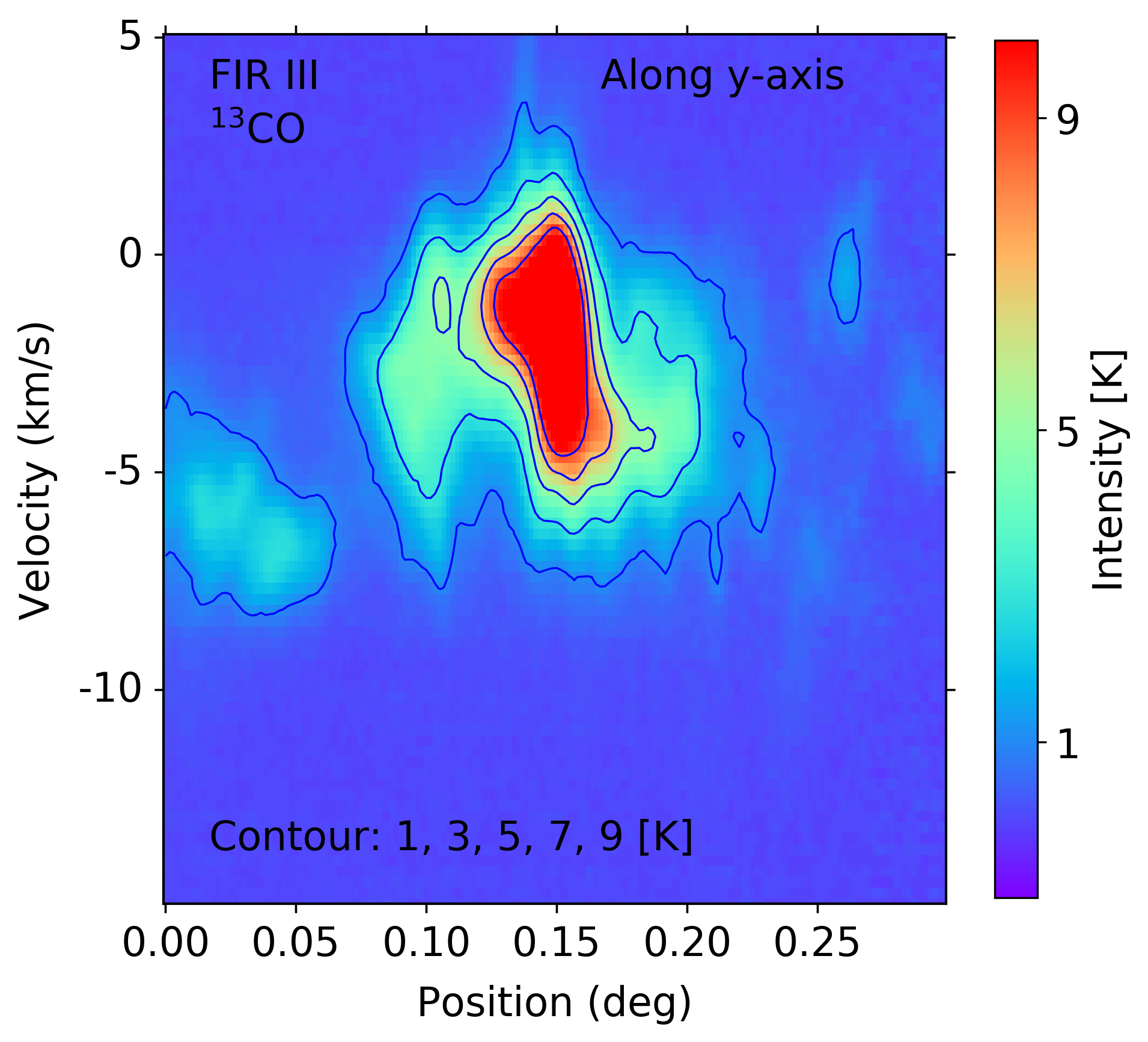}
    \includegraphics[width=0.31\linewidth]{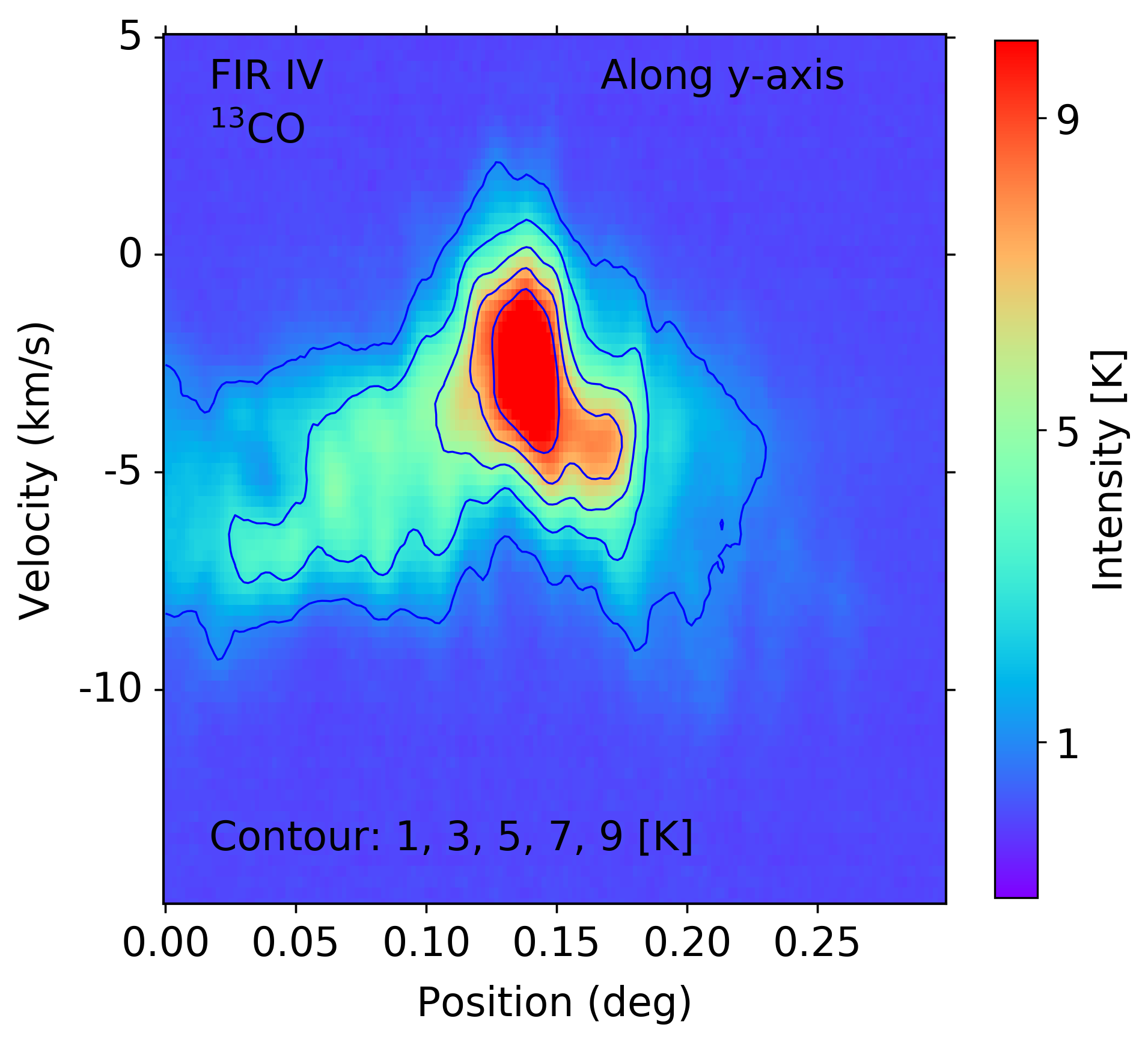}
    \includegraphics[width=0.31\linewidth]{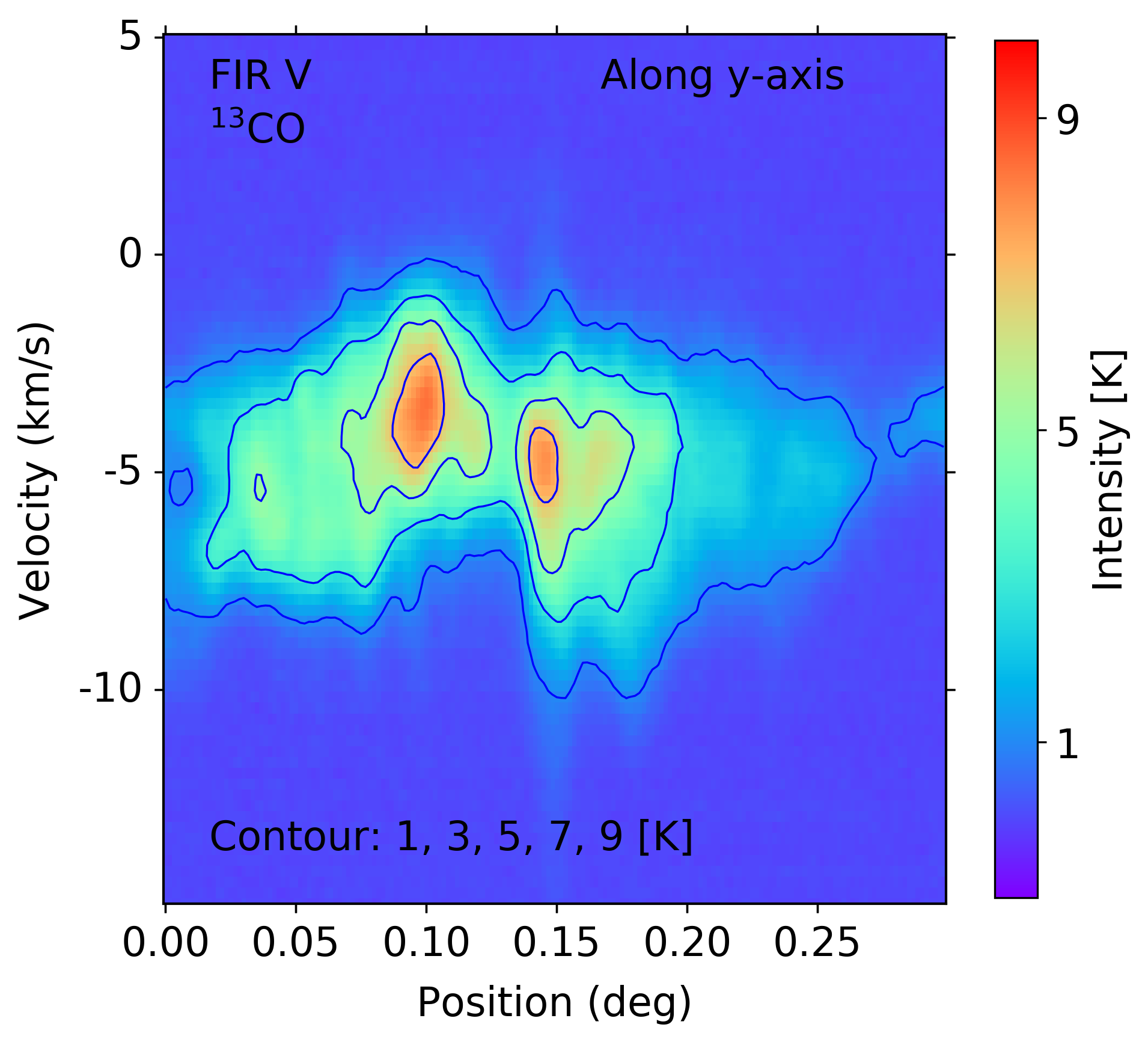}
    \caption{Position velocity maps (bottom panels) toward three selected slices (L1 and L2 indicated in top panel) perpendicular to the central filament and toward FIR-sources (I[N], I to V indicated by red plus markers in top panel) along y-axis (latitude). Magenta circles indicate H II regions from \citealt{Anderson2014}. The slices L1 and L2 show same regions, MFS cold and MFS warm, presented in Figure 12 of \citealt{Arzoumanian2022}.}
    \label{fig:pvmaps-center}

\end{figure*}

\clearpage\clearpage
\section{Channel maps of selected regions}
\begin{figure*}[htbp!]
    \centering
    \includegraphics[width=0.62\linewidth]{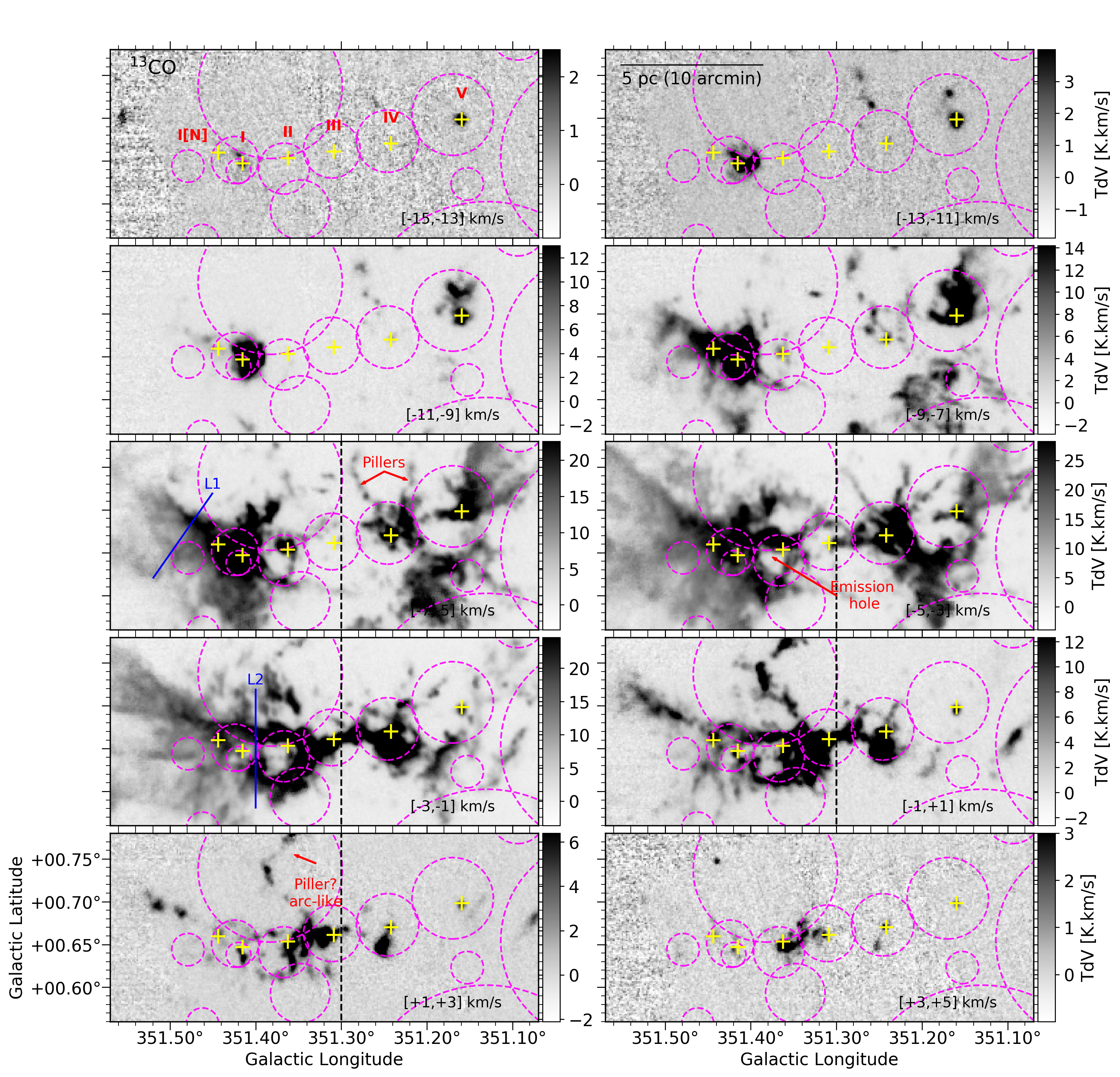}\\
    \includegraphics[width=0.62\linewidth]{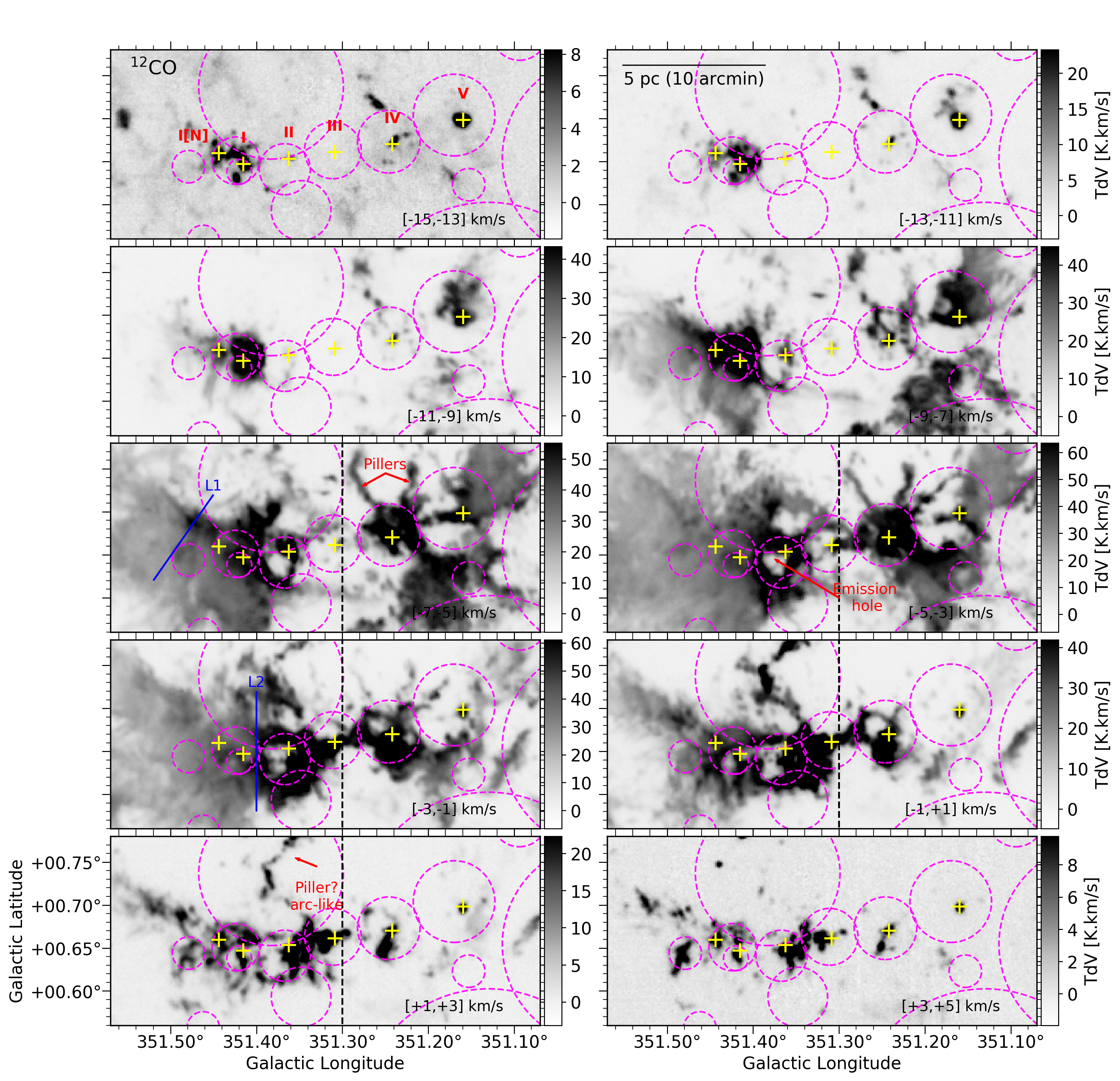}
    \caption{Channel map of $^{13}$CO and $^{12}$CO toward NGC 6334 central filament region. The vertical dotted line indicates the longitude where we observe the V shape in the $lv$ plot (in Fig \ref{fig:lvplot_eg}). Magenta/red circles indicate H II regions from \cite{Anderson2014}. Blue lines indicate along which the pv-maps are shown in Fig. \ref{fig:pvmaps-center}.}
    \label{fig:centeronly}
\end{figure*}

\begin{figure*}[htbp!]
    \centering
    \includegraphics[width=0.497\linewidth]{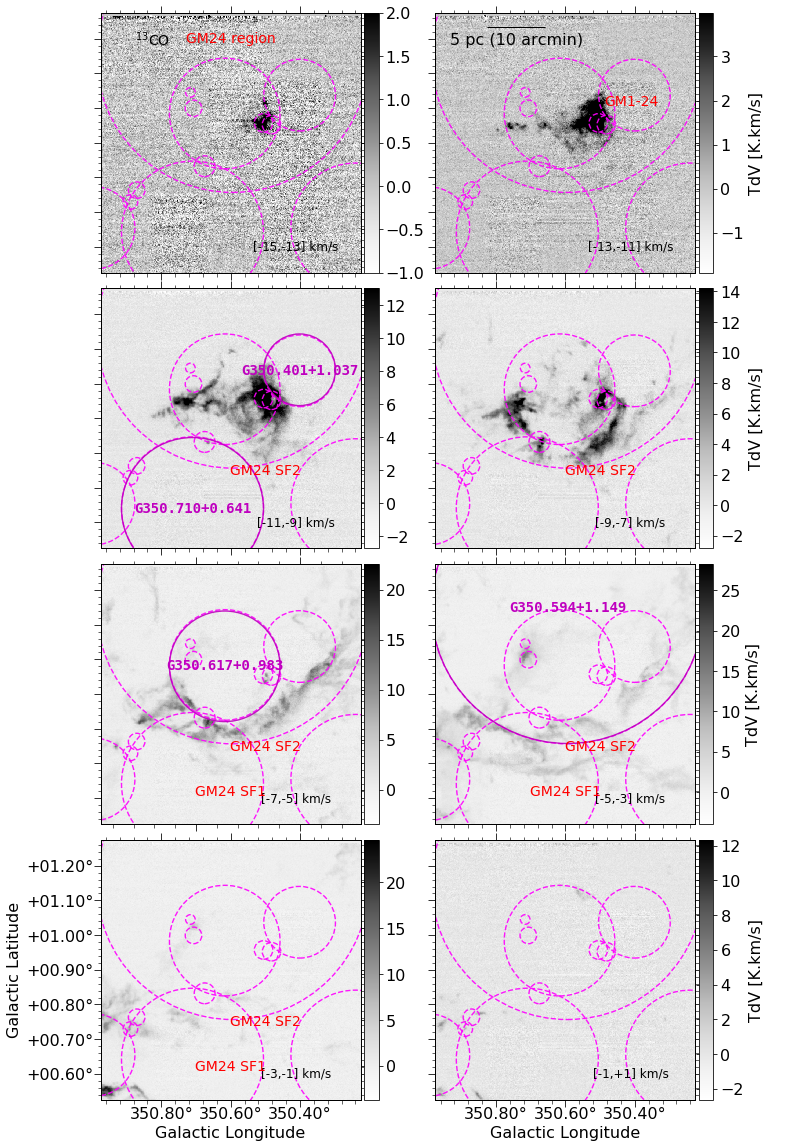}
    \includegraphics[width=0.497\linewidth]{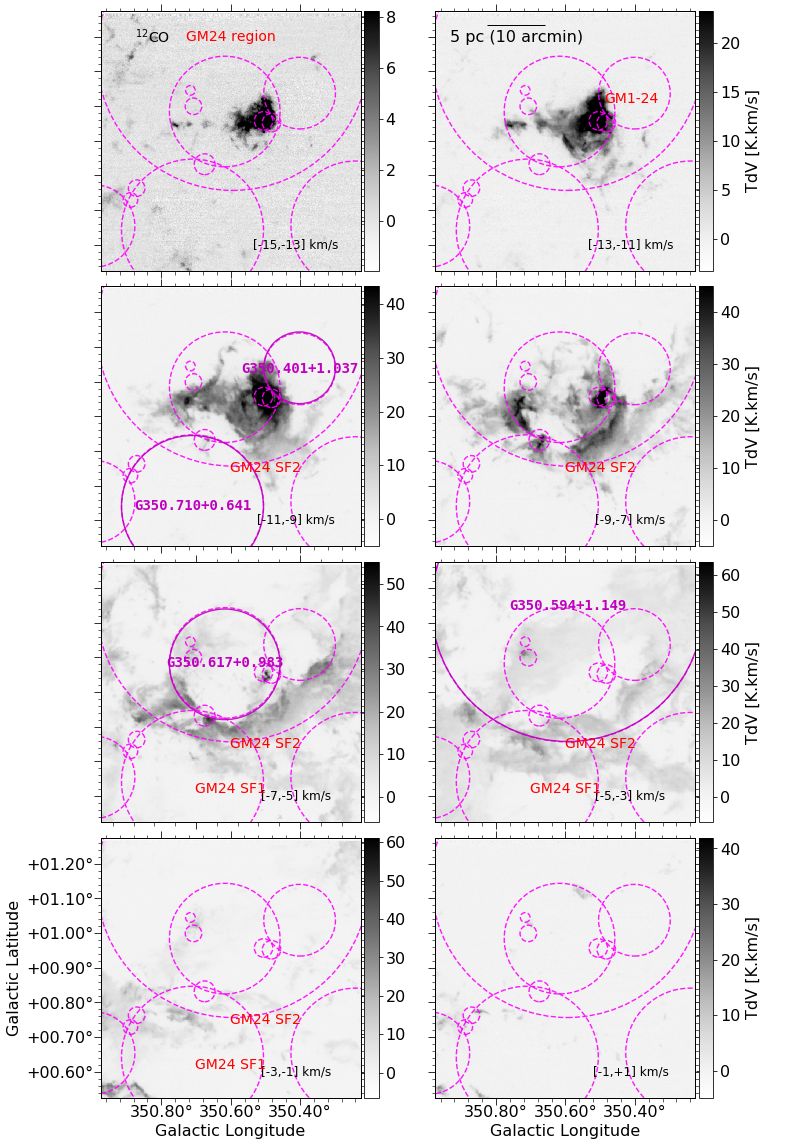}
    \caption{Channel map of $^{13}$CO and $^{12}$CO toward GM-24 region. Magenta/red circles indicate H II regions from \cite{Anderson2014}.}
    \label{fig:13CO_gm124region}
\end{figure*}

\begin{figure*}[htbp!]
    \centering
    \includegraphics[width=0.497\linewidth]{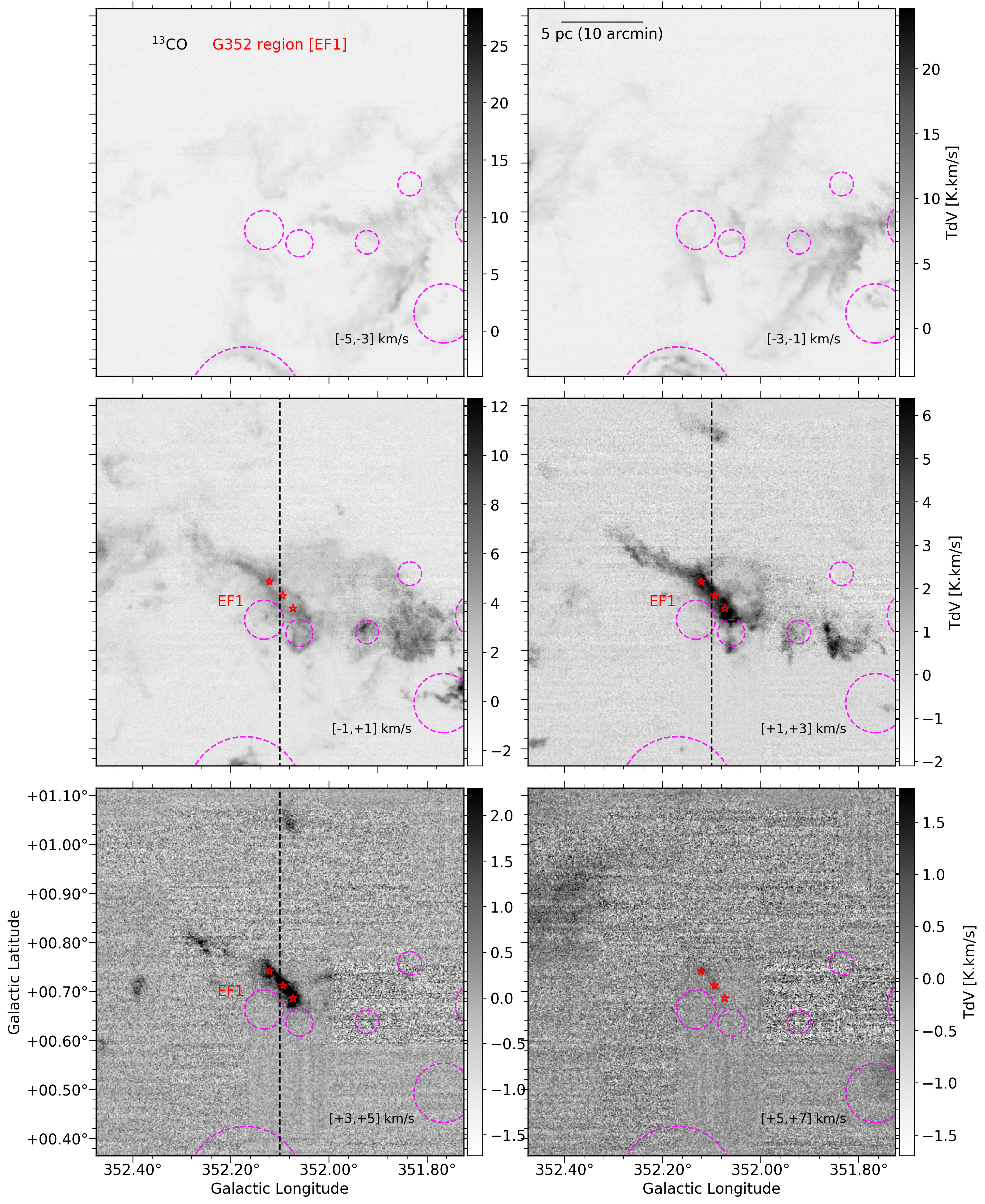}
    \includegraphics[width=0.497\linewidth]{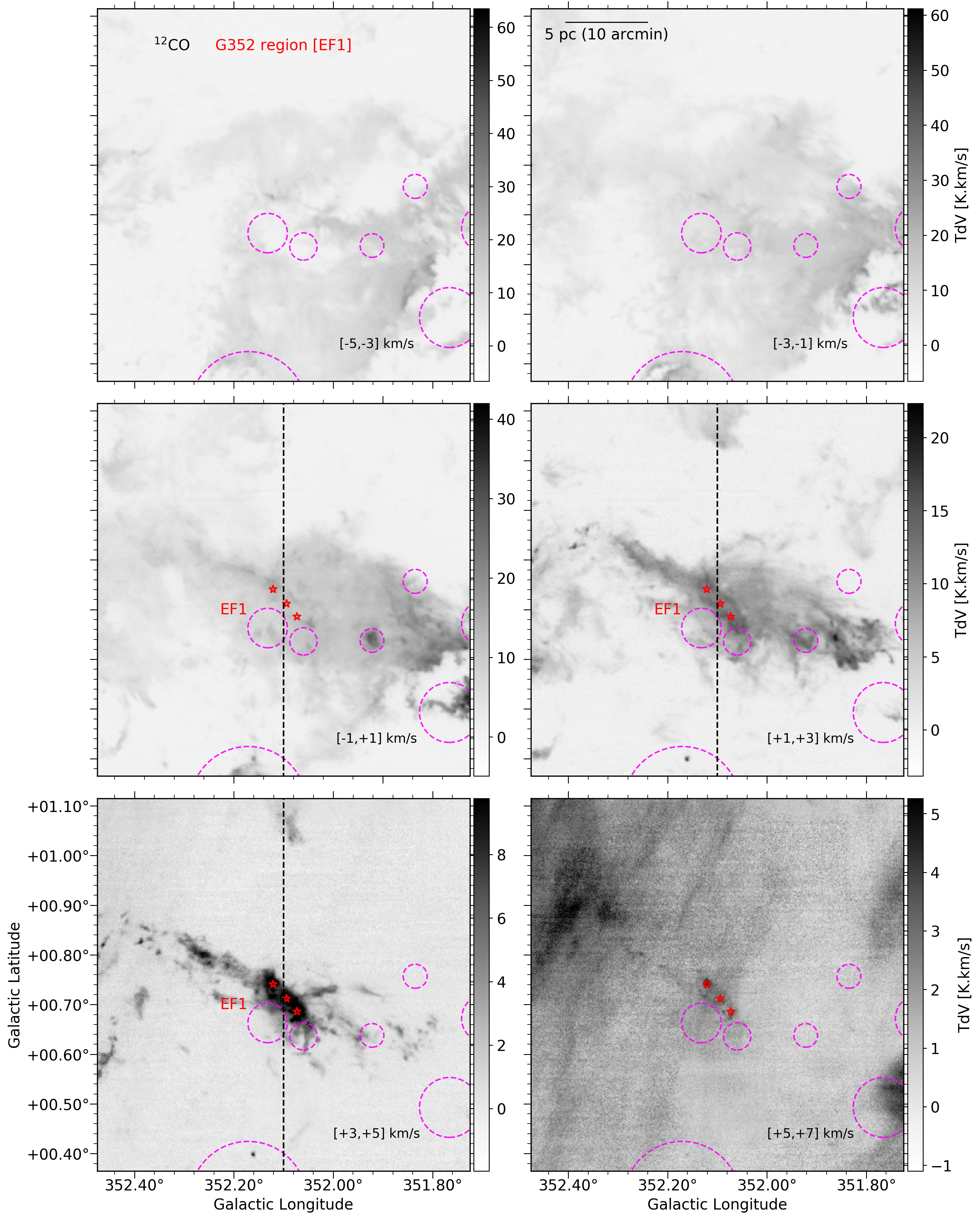}
    \caption{Channel map of $^{13}$CO and $^{12}$CO toward G352.1 region. The vertical dotted line indicates the longitude where we observe the V shape in the $lv$ plot. Magenta circles indicate H II regions from \cite{Anderson2014}. Red star symbols show the location of ATLASGAL clumps found in the filament. }
    \label{fig:13CO_g352}
\end{figure*}

\clearpage

\section{$^{13}$CO emission maps of H II regions}
\begin{figure}[htbp!]
\centering
\includegraphics[width=0.22\linewidth]{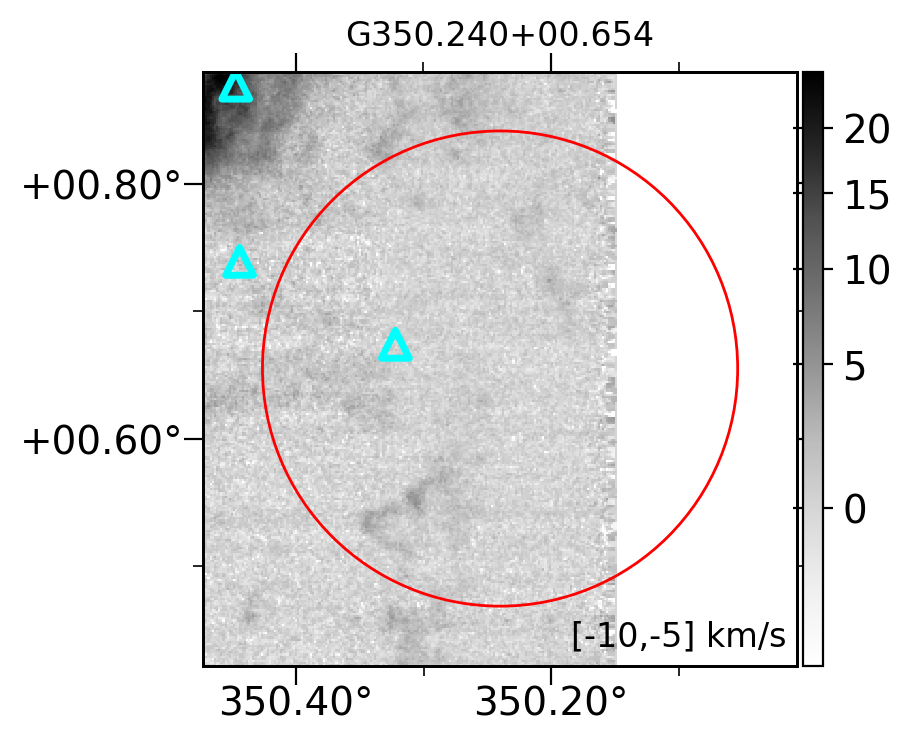}
\includegraphics[width=0.22\linewidth]{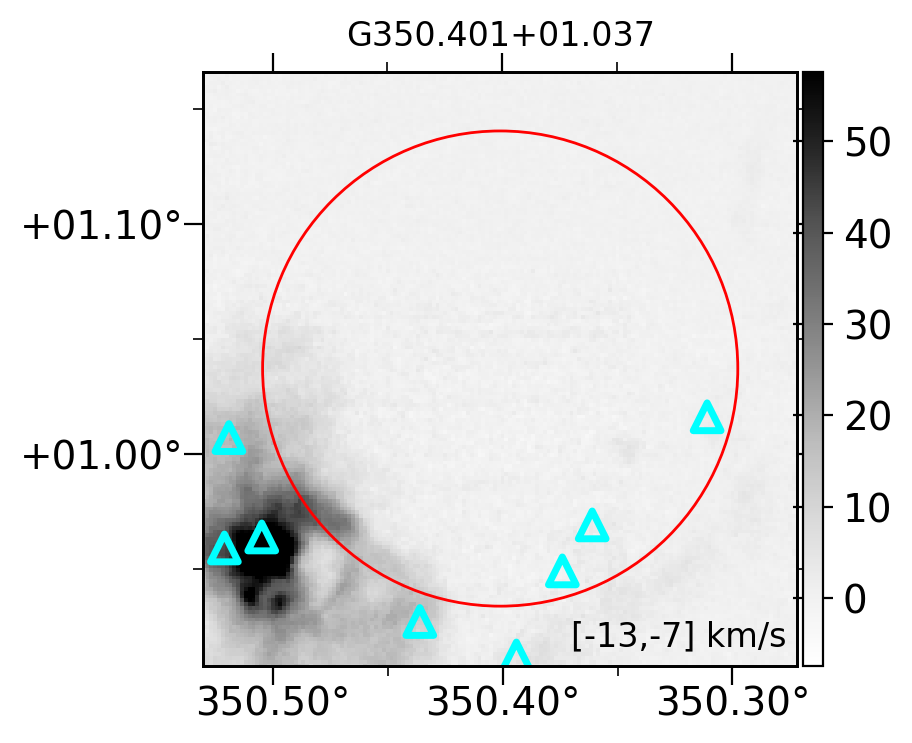}
\includegraphics[width=0.22\linewidth]{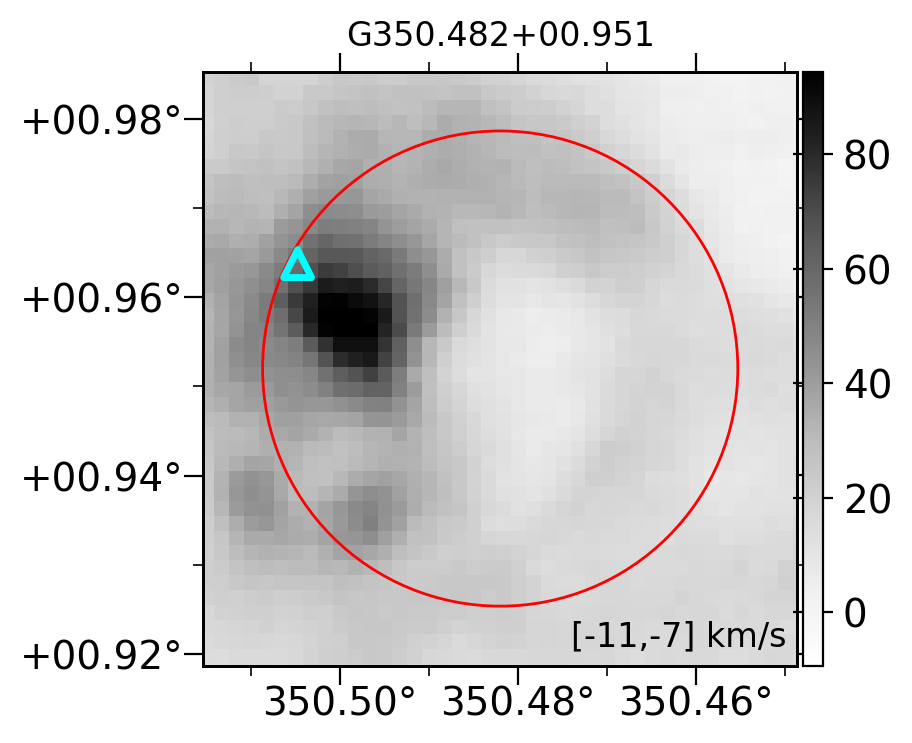}
\includegraphics[width=0.22\linewidth]{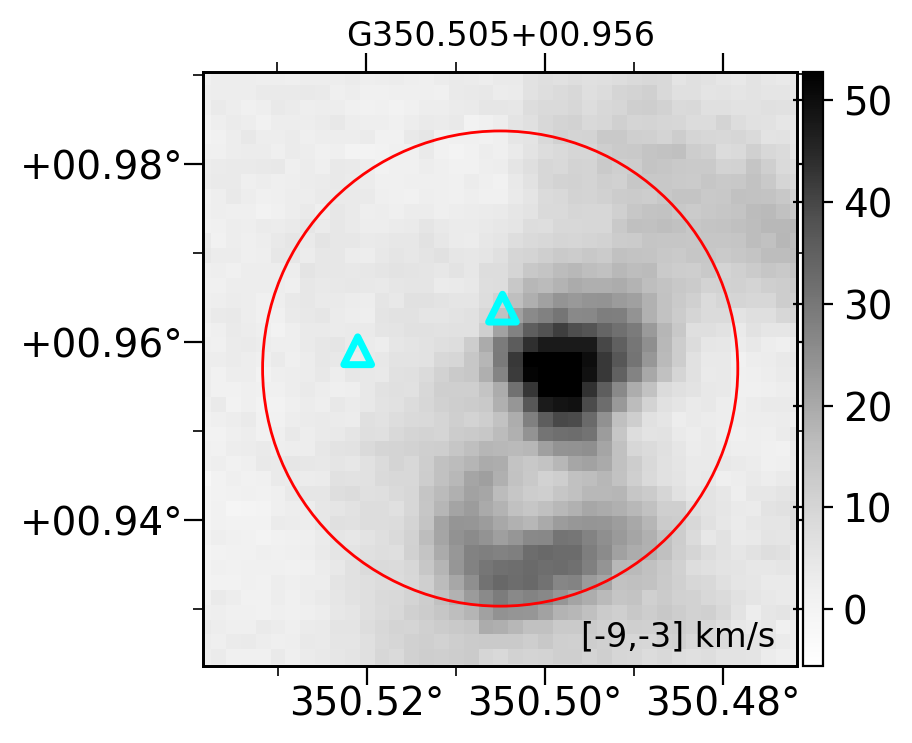}\\
\includegraphics[width=0.22\linewidth]{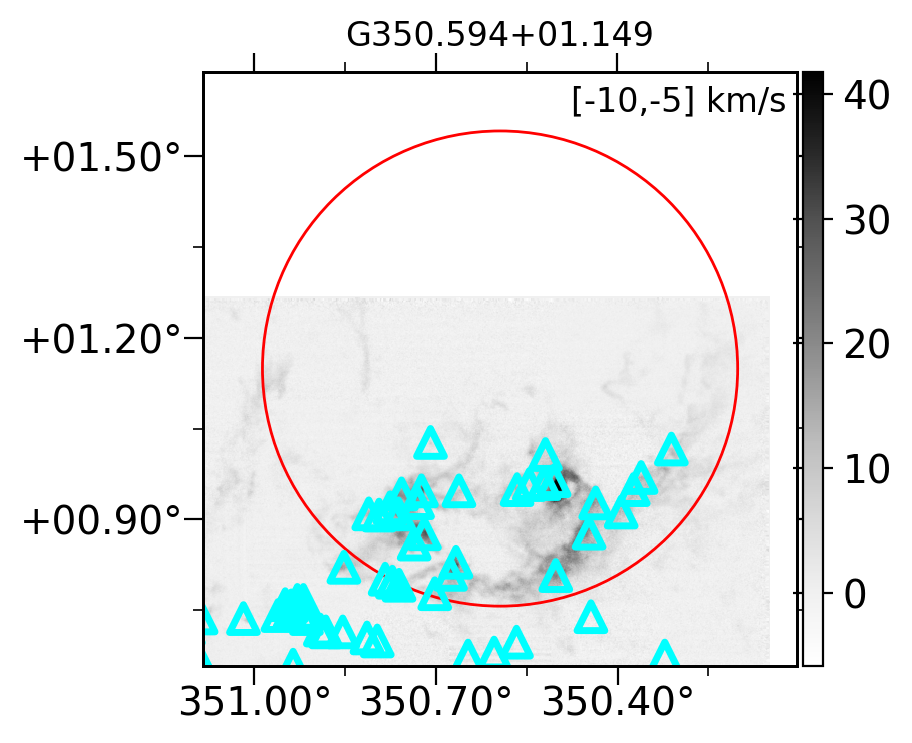}
\includegraphics[width=0.22\linewidth]{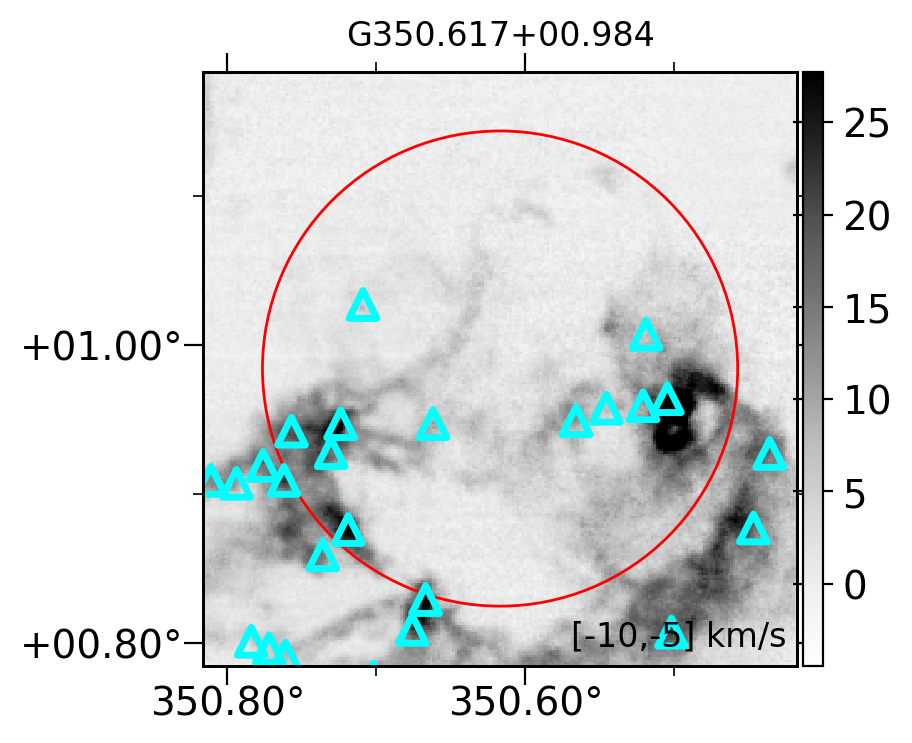}
\includegraphics[width=0.22\linewidth]{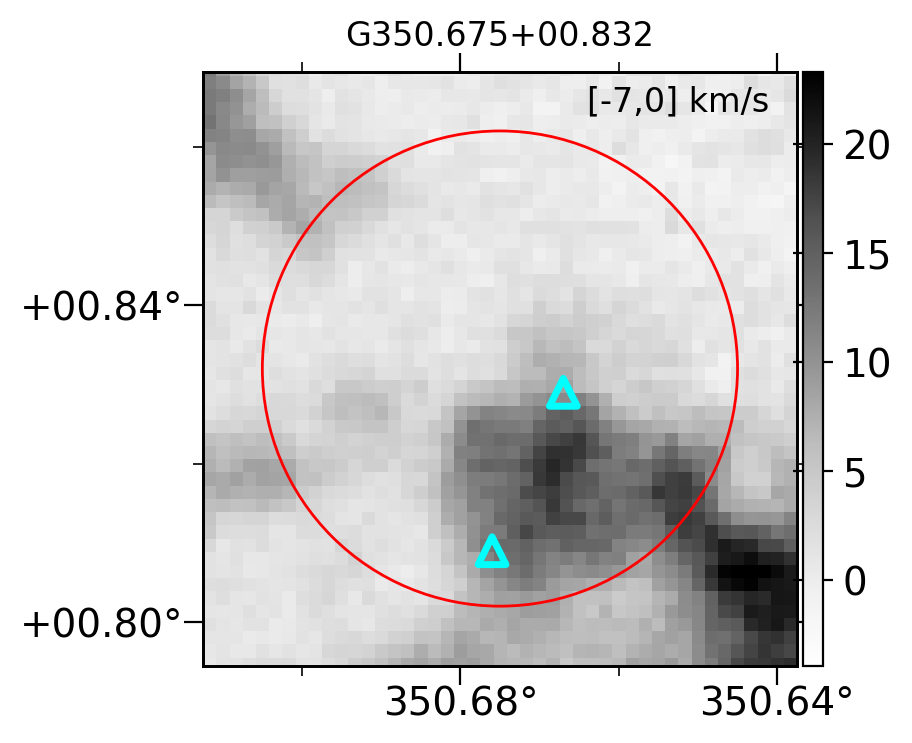}
\includegraphics[width=0.22\linewidth]{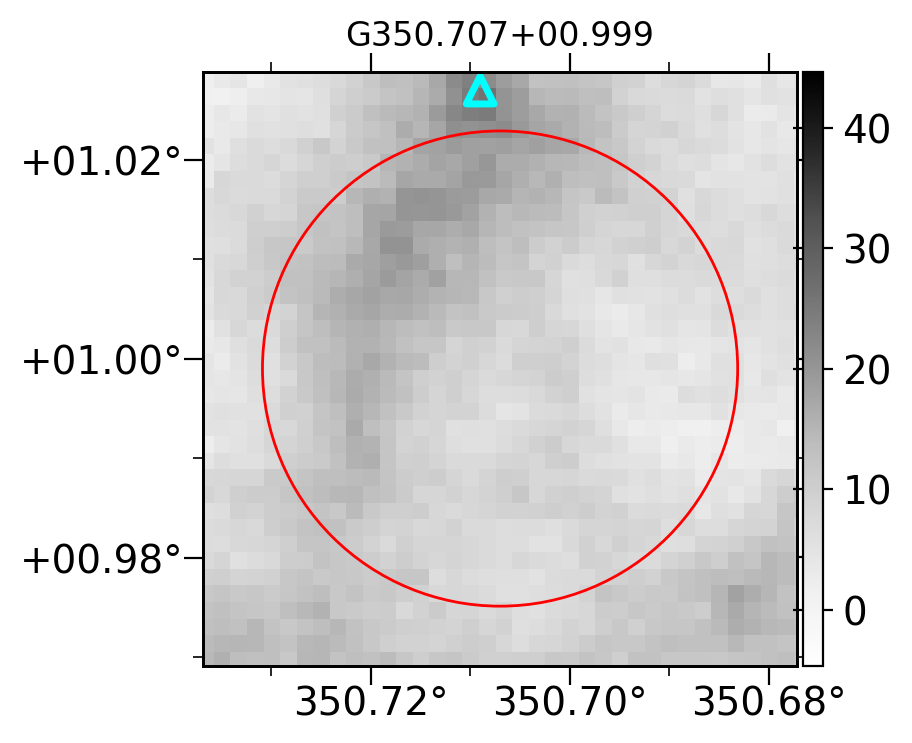}\\
\includegraphics[width=0.22\linewidth]{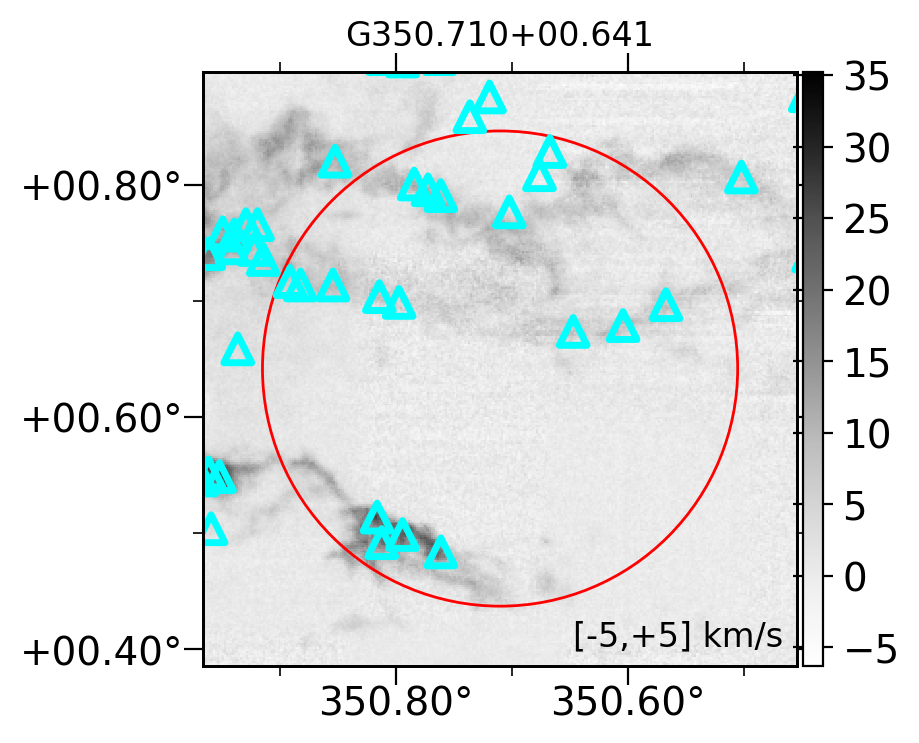}
\includegraphics[width=0.22\linewidth]{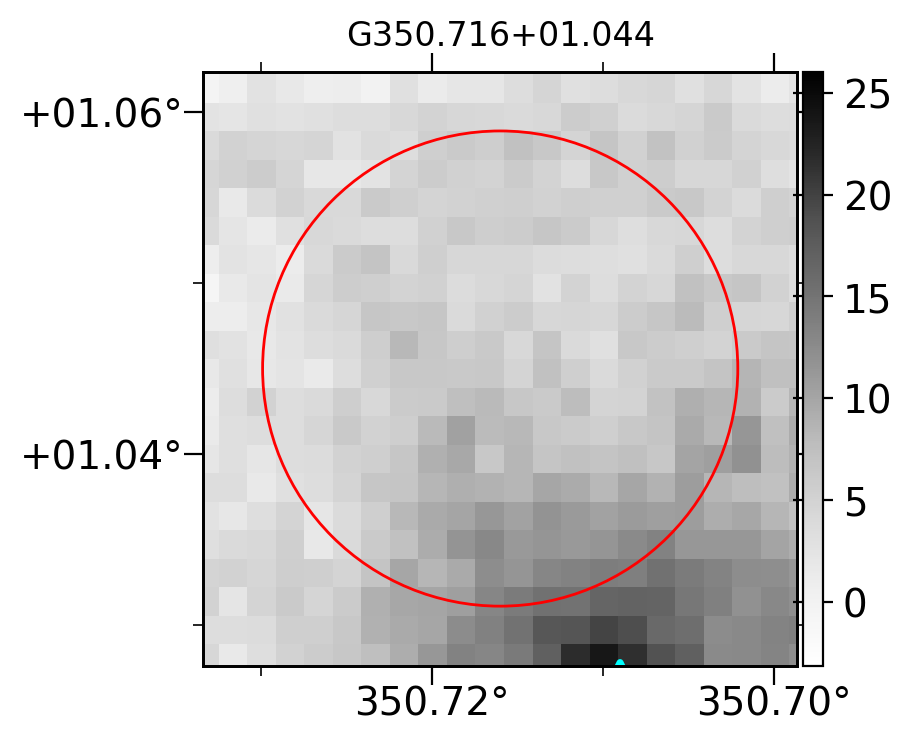}
\includegraphics[width=0.22\linewidth]{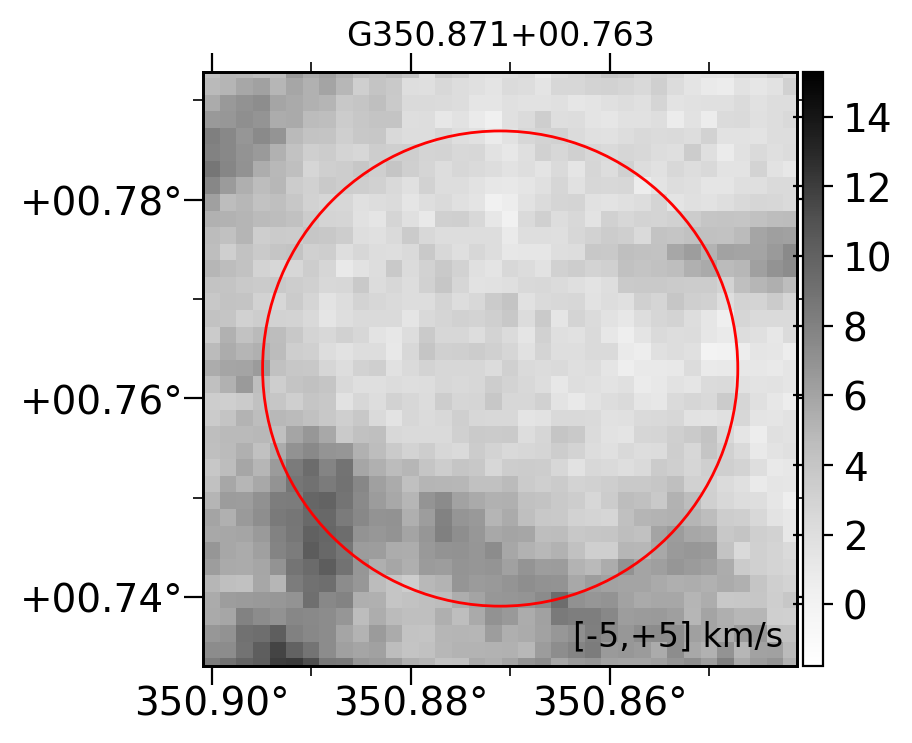}
\includegraphics[width=0.22\linewidth]{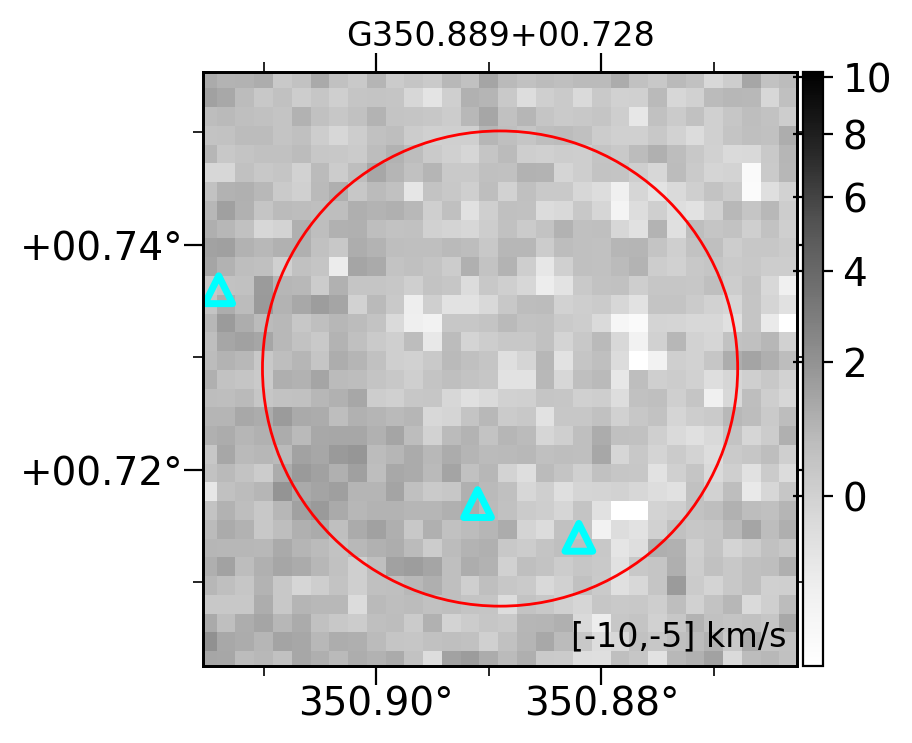}\\
\includegraphics[width=0.22\linewidth]{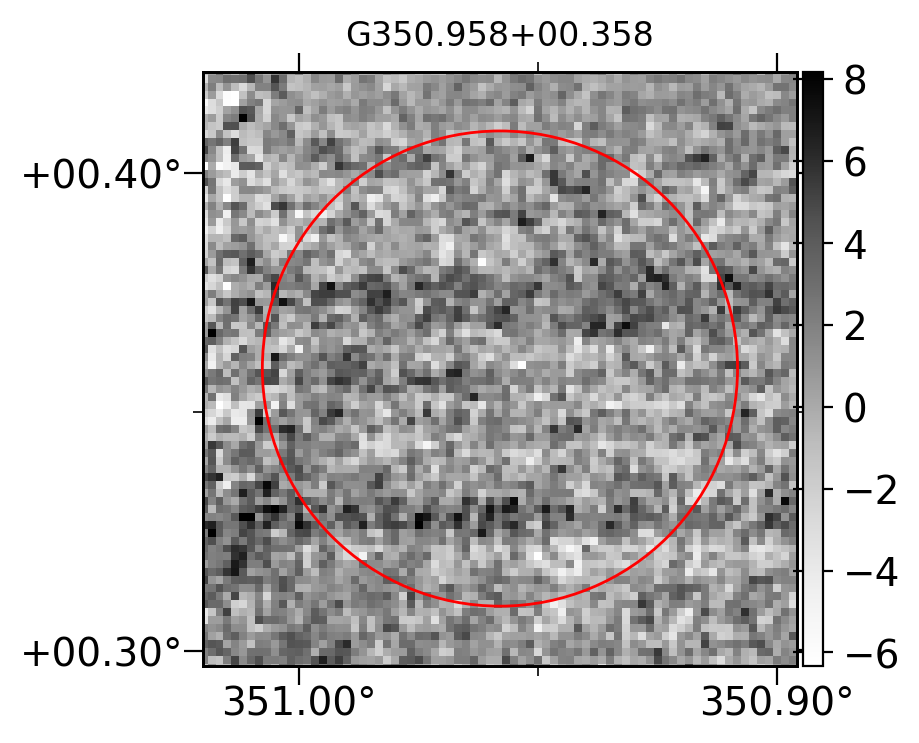}
\includegraphics[width=0.22\linewidth]{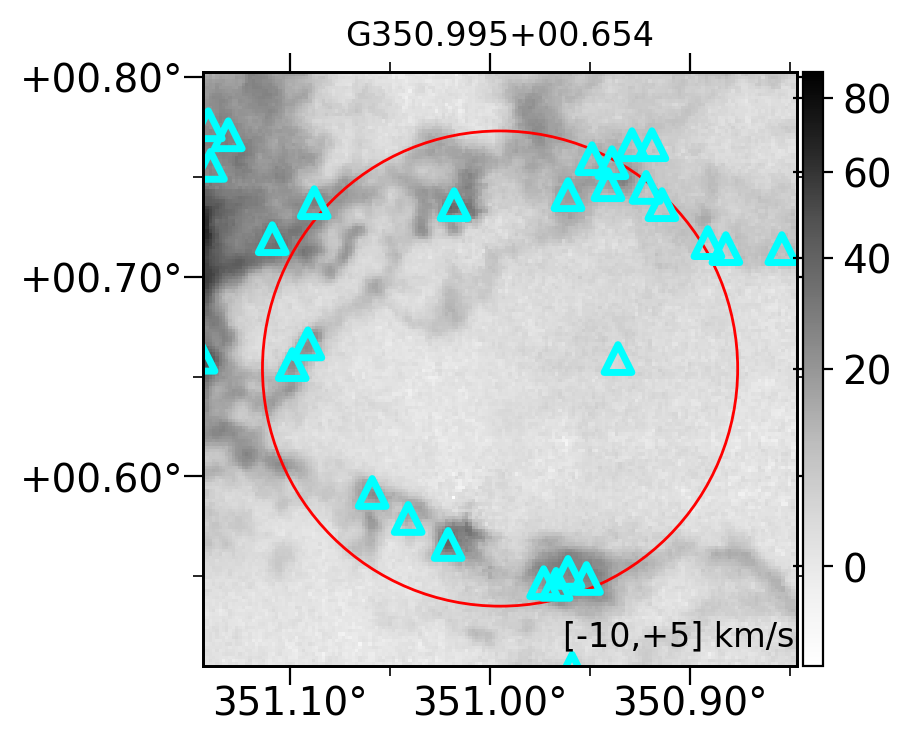}
\includegraphics[width=0.22\linewidth]{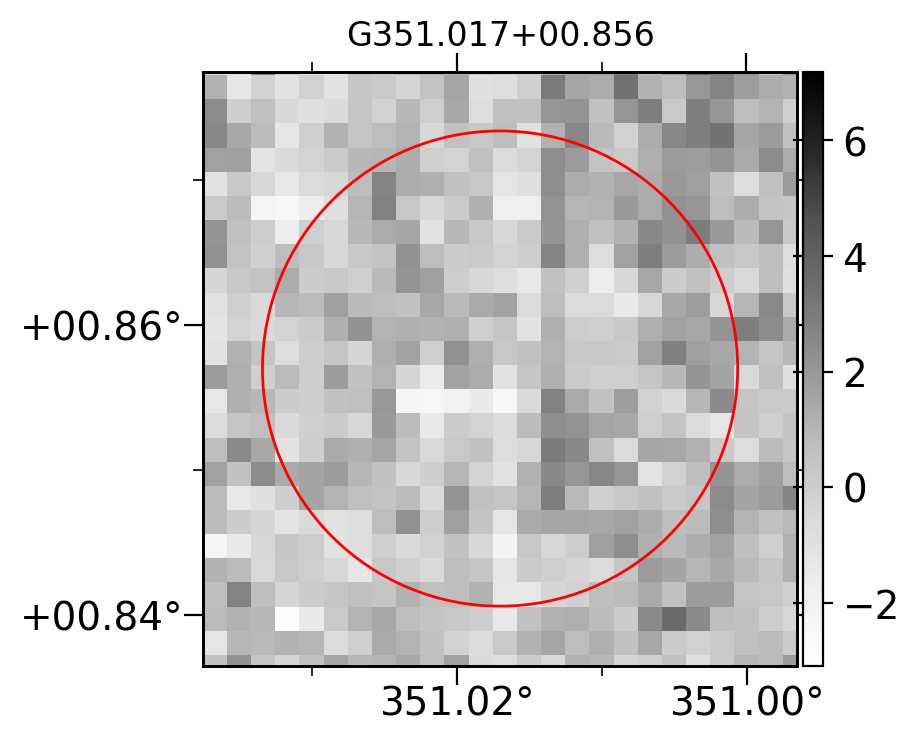}
\includegraphics[width=0.22\linewidth]{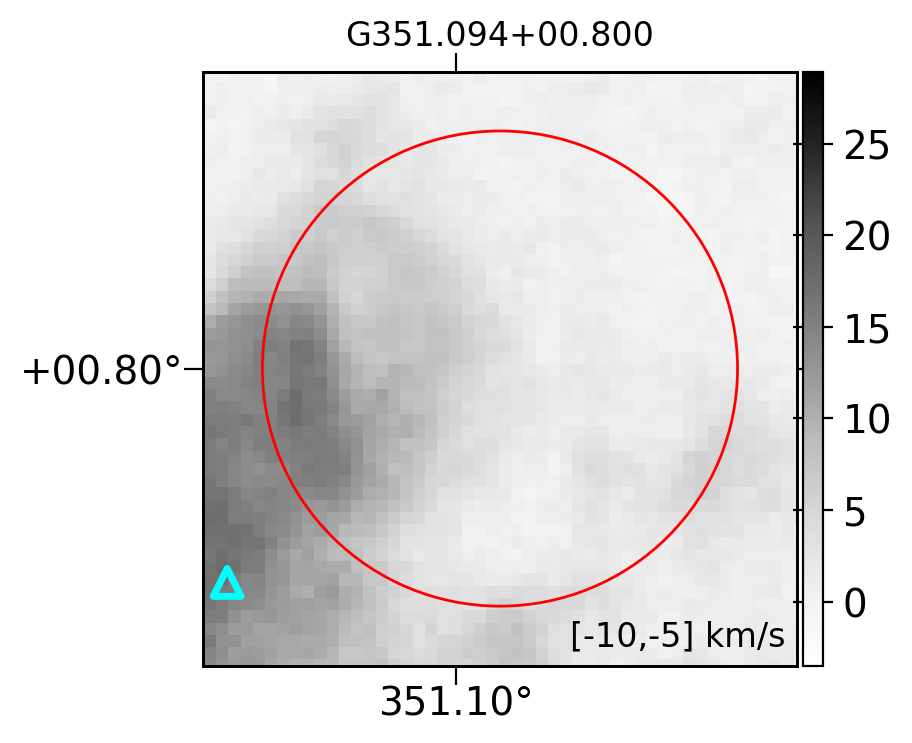}\\
\includegraphics[width=0.22\linewidth]{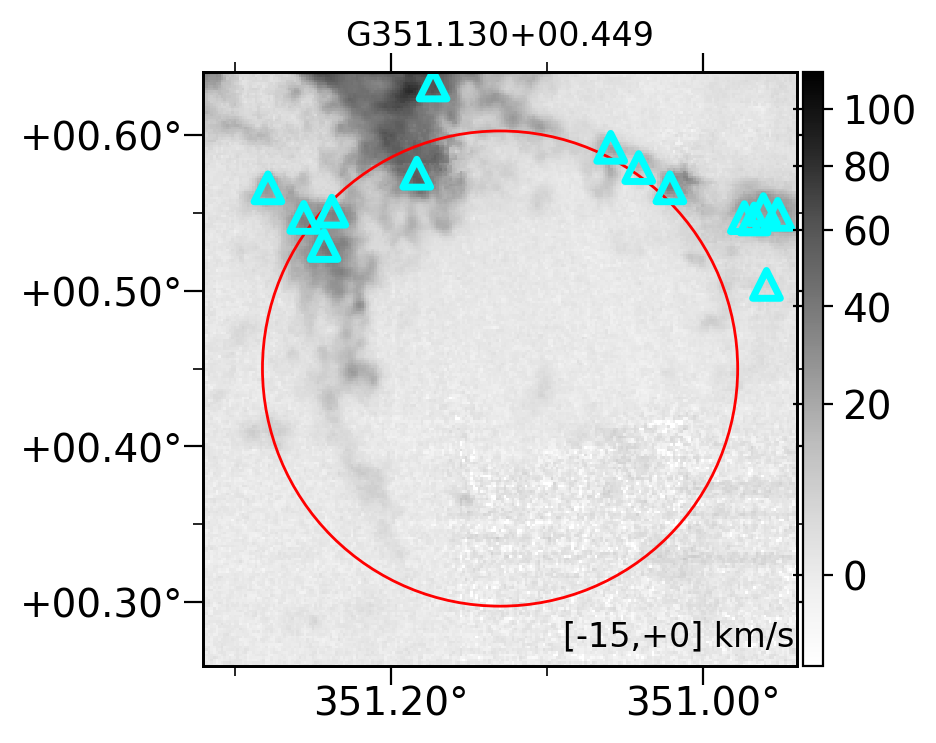}
\includegraphics[width=0.22\linewidth]{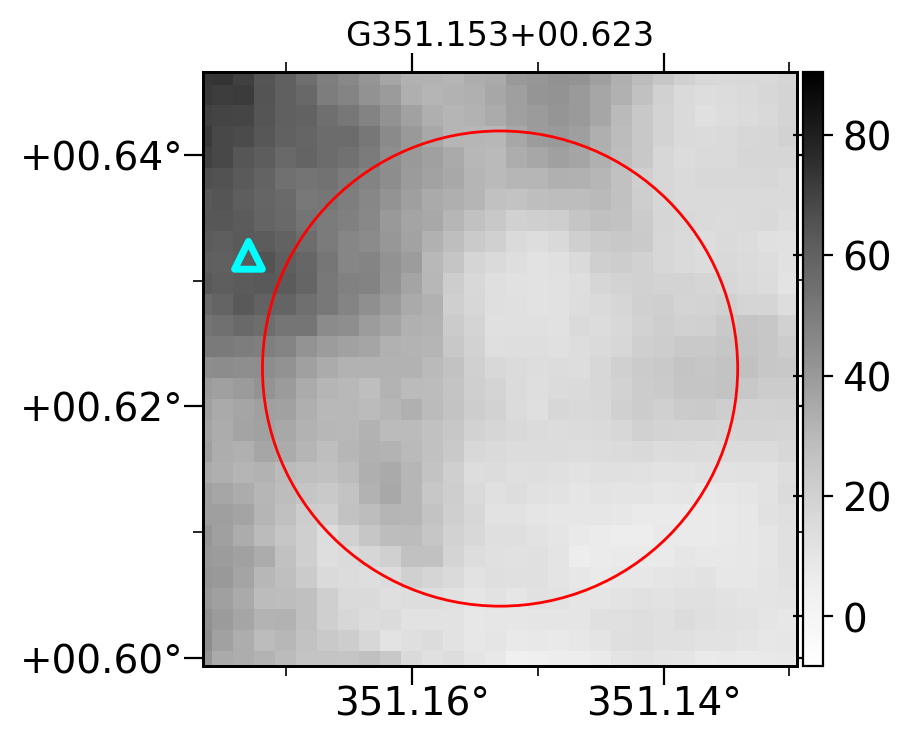}
\includegraphics[width=0.22\linewidth]{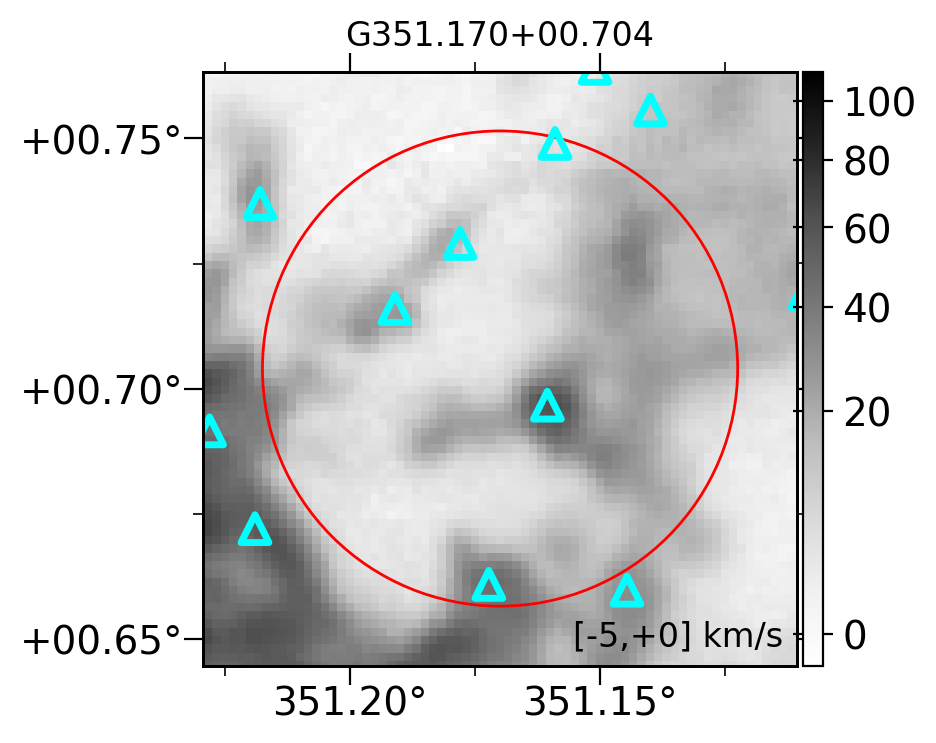}
\includegraphics[width=0.22\linewidth]{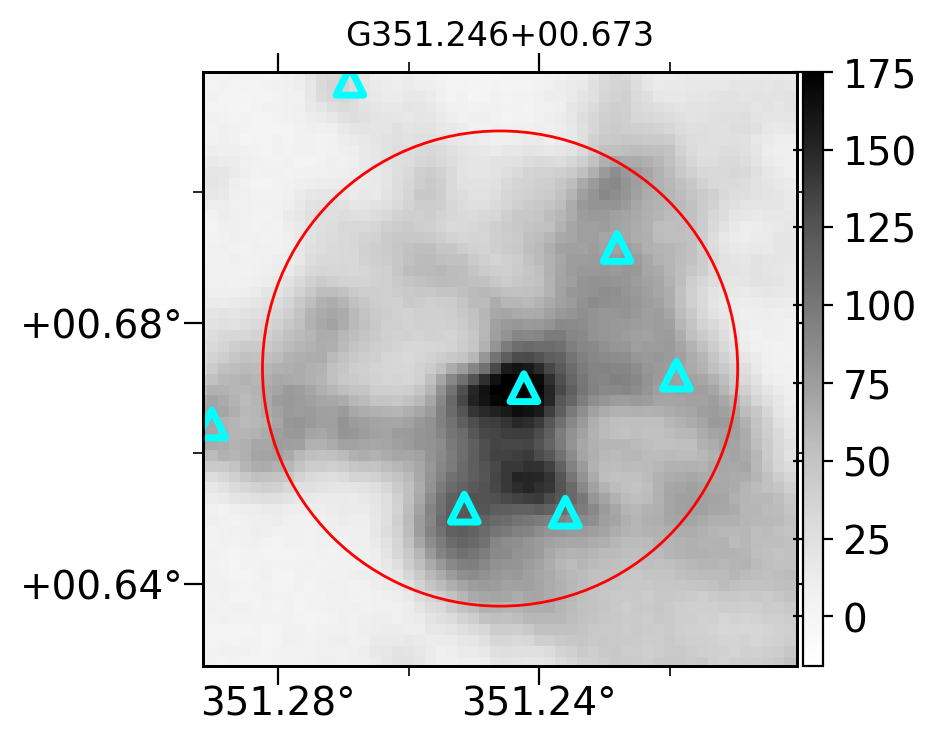}\\
\includegraphics[width=0.22\linewidth]{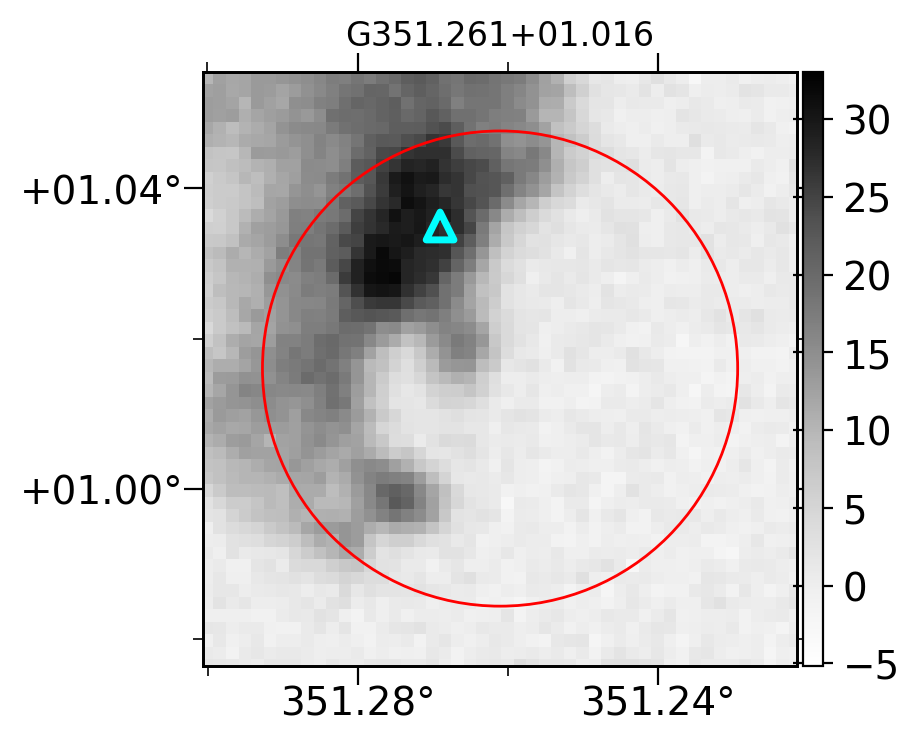}
\includegraphics[width=0.22\linewidth]{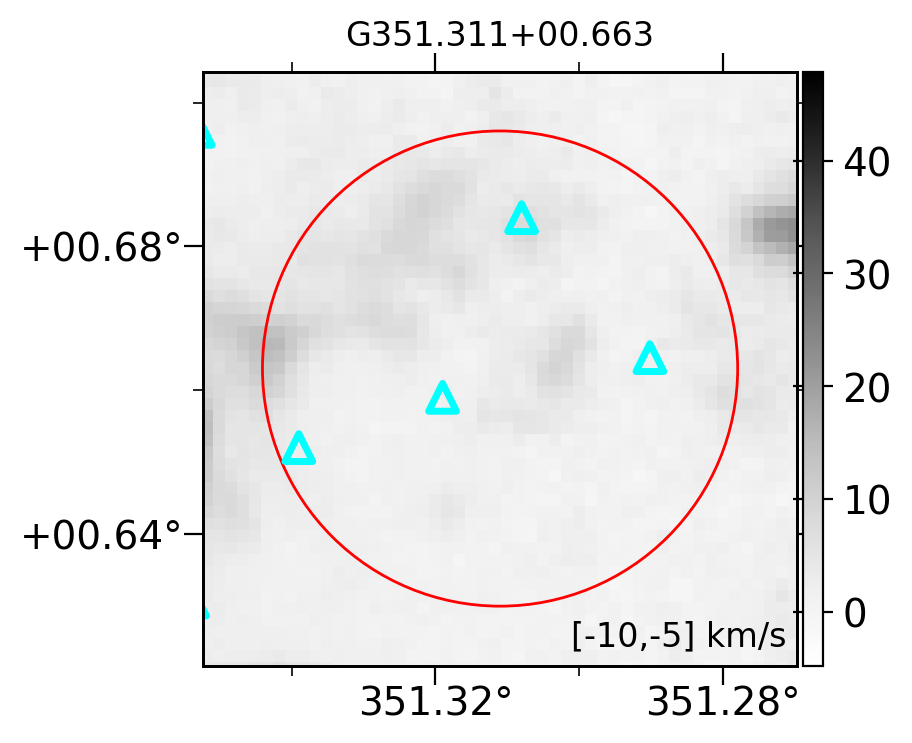}
\includegraphics[width=0.22\linewidth]{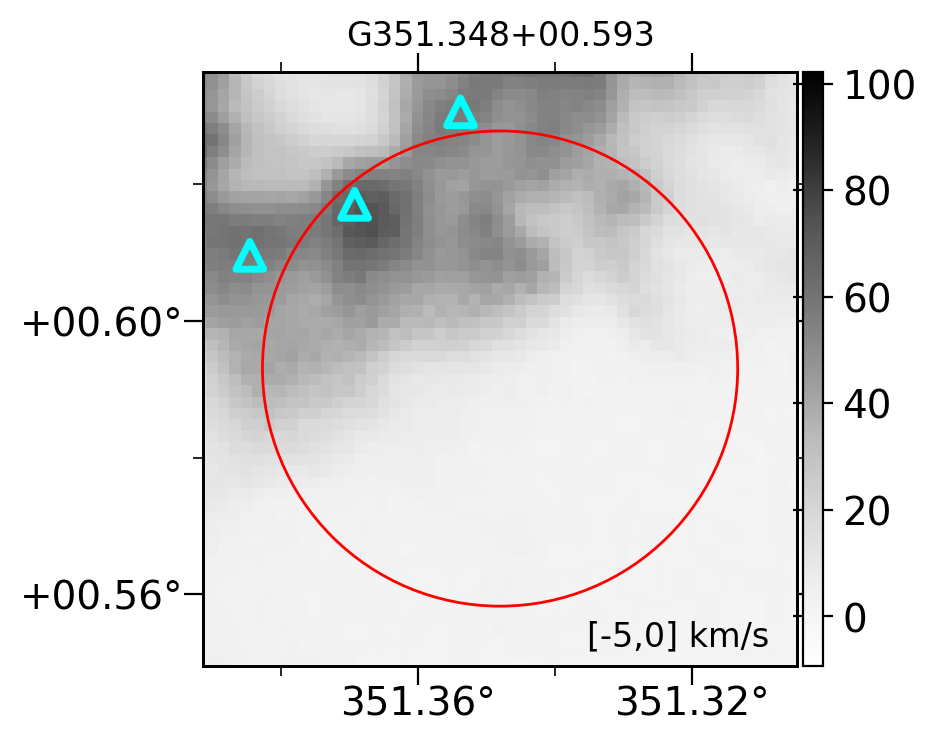}
\includegraphics[width=0.22\linewidth]{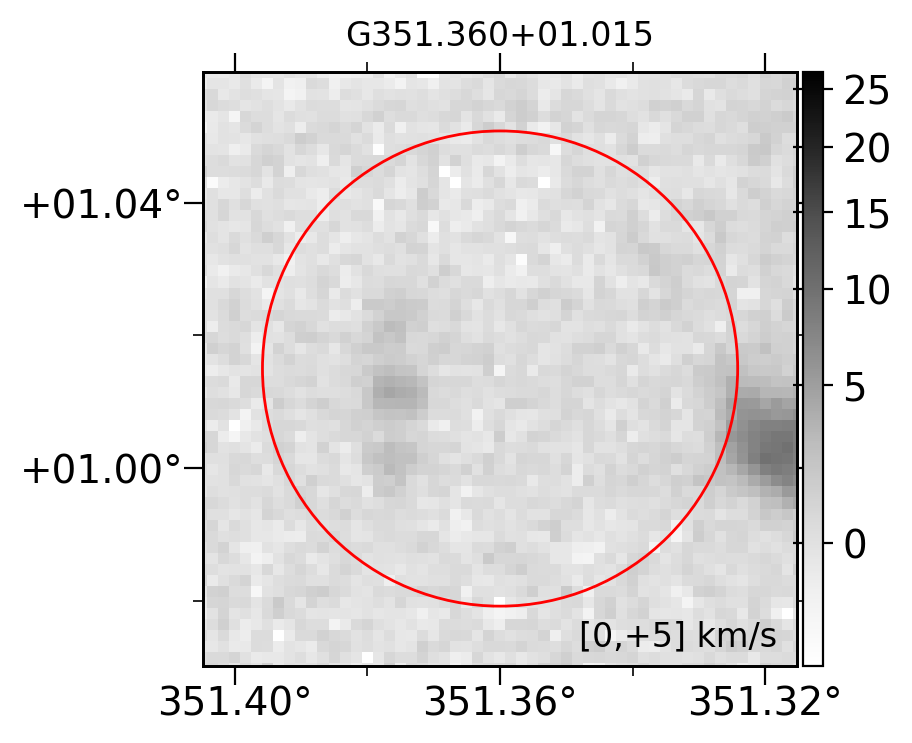}\\
    \caption{$^{13}$CO emission maps of H II regions integrated over $-$15 to +5 km s$^{-1}$, otherwise indicated inside each maps. Color wedges indicate TdV in K km s$^{-1}$. Cyan triangles indicate positions of the ATLASGAL clumps.}
    \label{fig:chan_map_hii0}
\end{figure}

\begin{figure}[htbp!]
\centering
\includegraphics[width=0.22\linewidth]{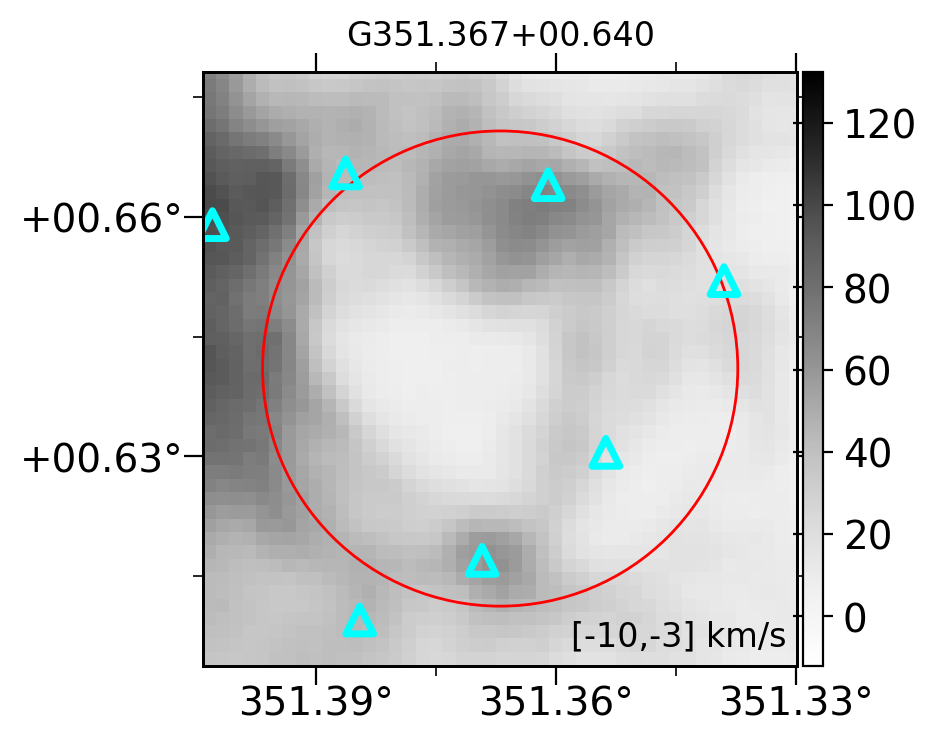}
\includegraphics[width=0.22\linewidth]{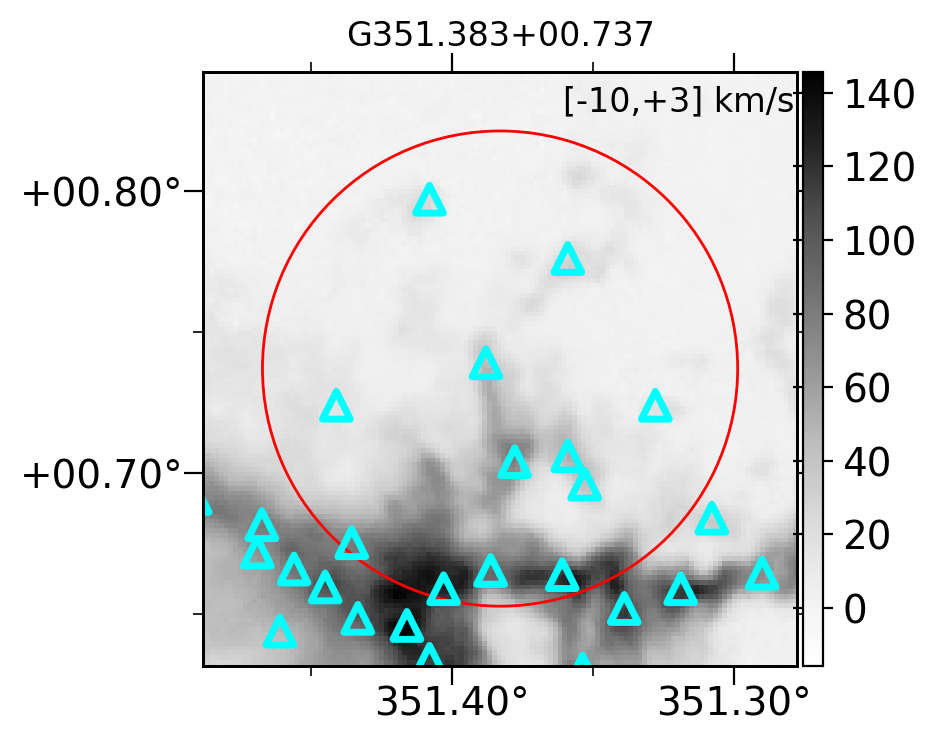}
\includegraphics[width=0.22\linewidth]{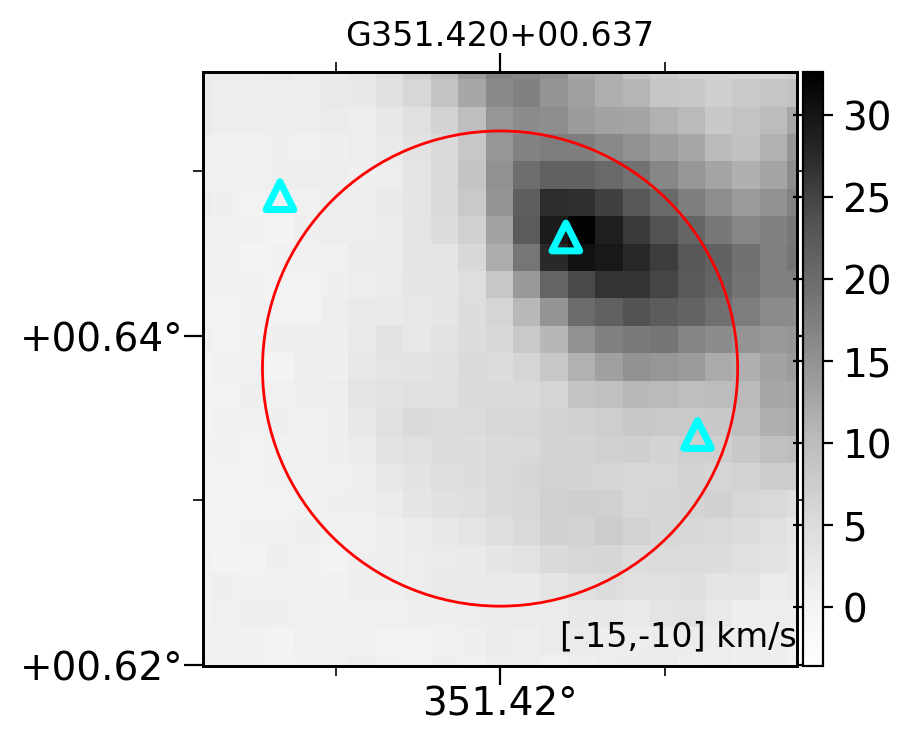}
\includegraphics[width=0.22\linewidth]{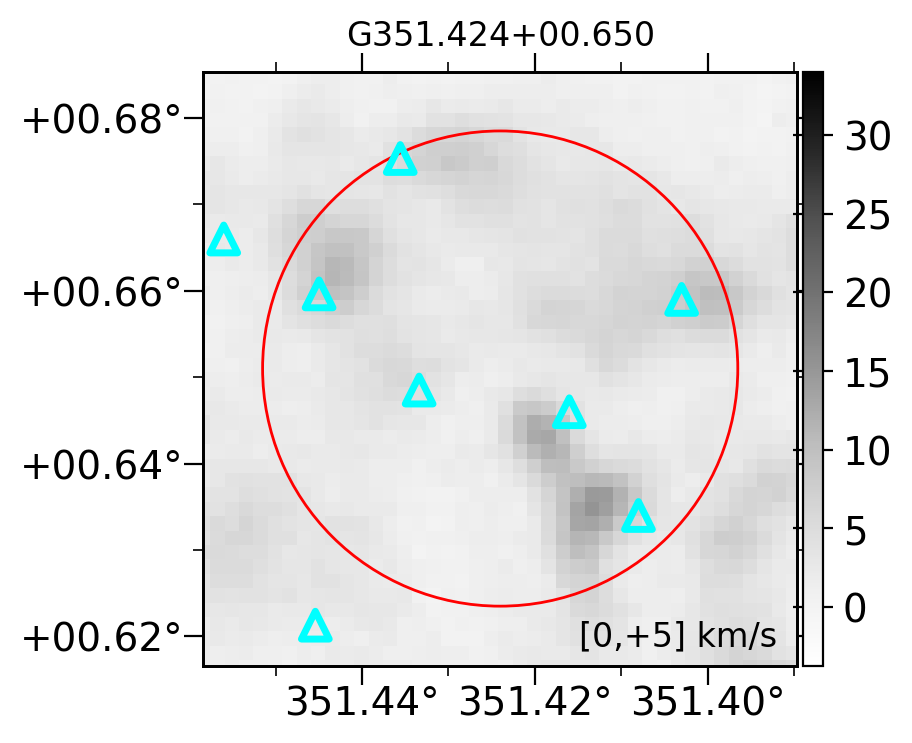}\\
\includegraphics[width=0.22\linewidth]{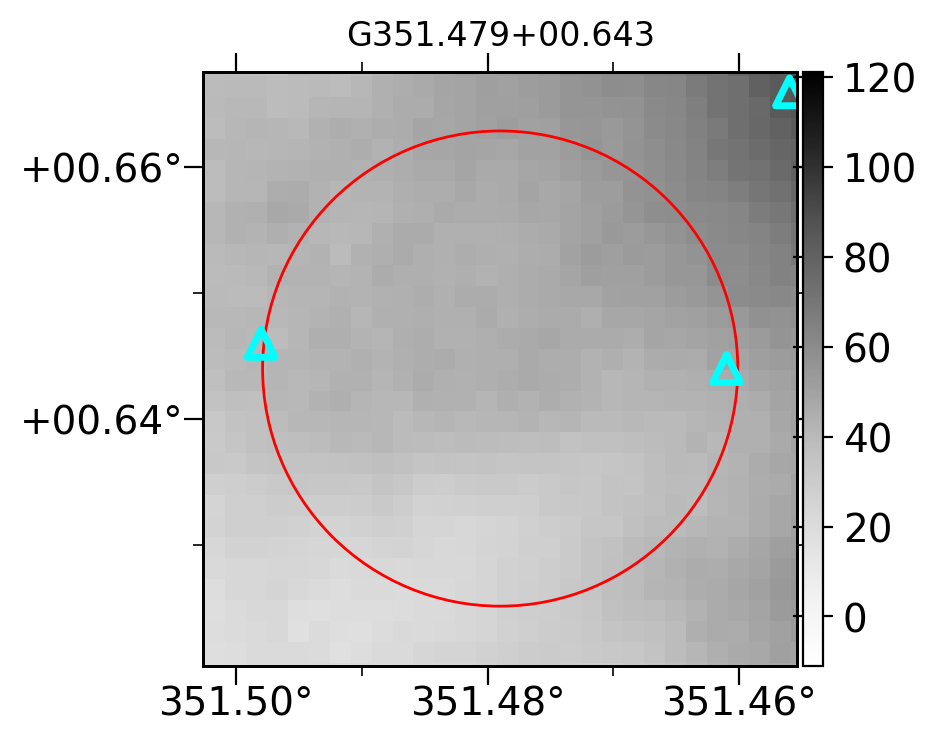}
\includegraphics[width=0.22\linewidth]{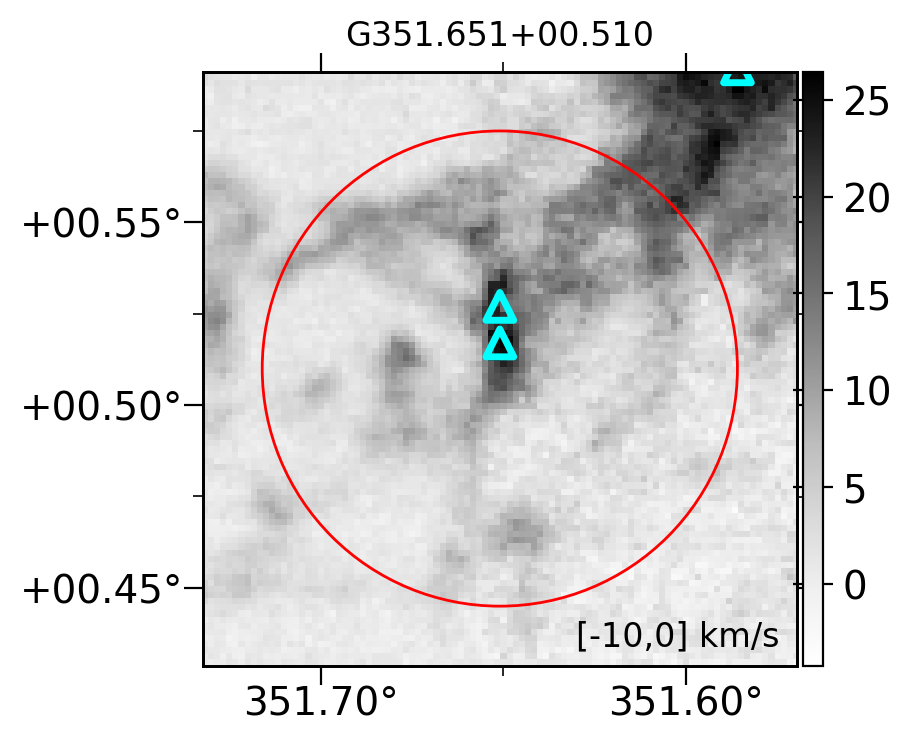}
\includegraphics[width=0.22\linewidth]{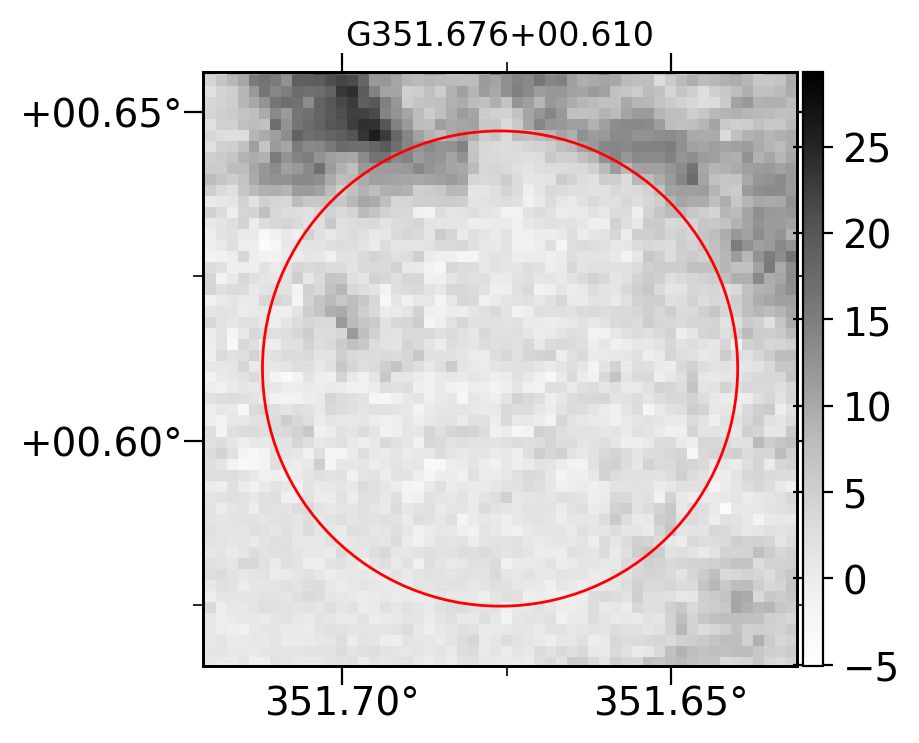}
\includegraphics[width=0.22\linewidth]{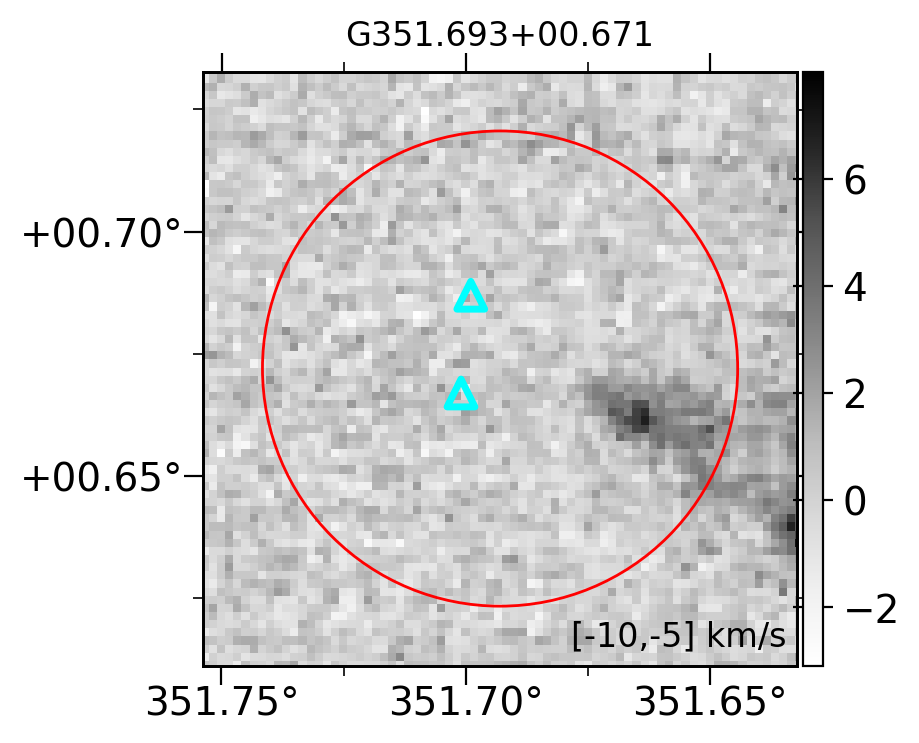}\\
\includegraphics[width=0.22\linewidth]{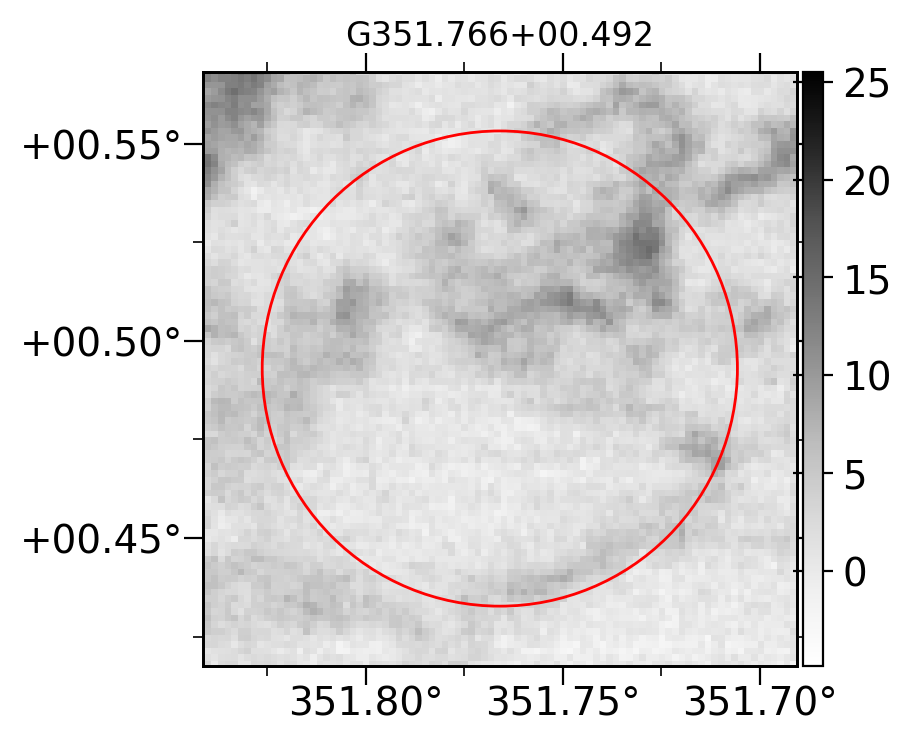}
\includegraphics[width=0.22\linewidth]{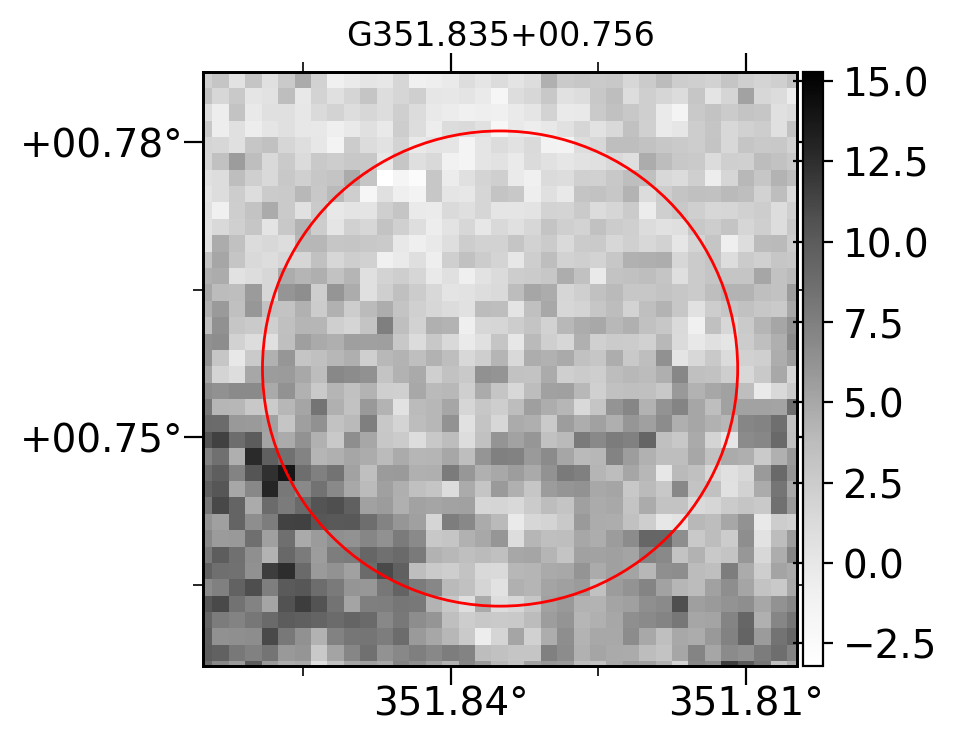}
\includegraphics[width=0.22\linewidth]{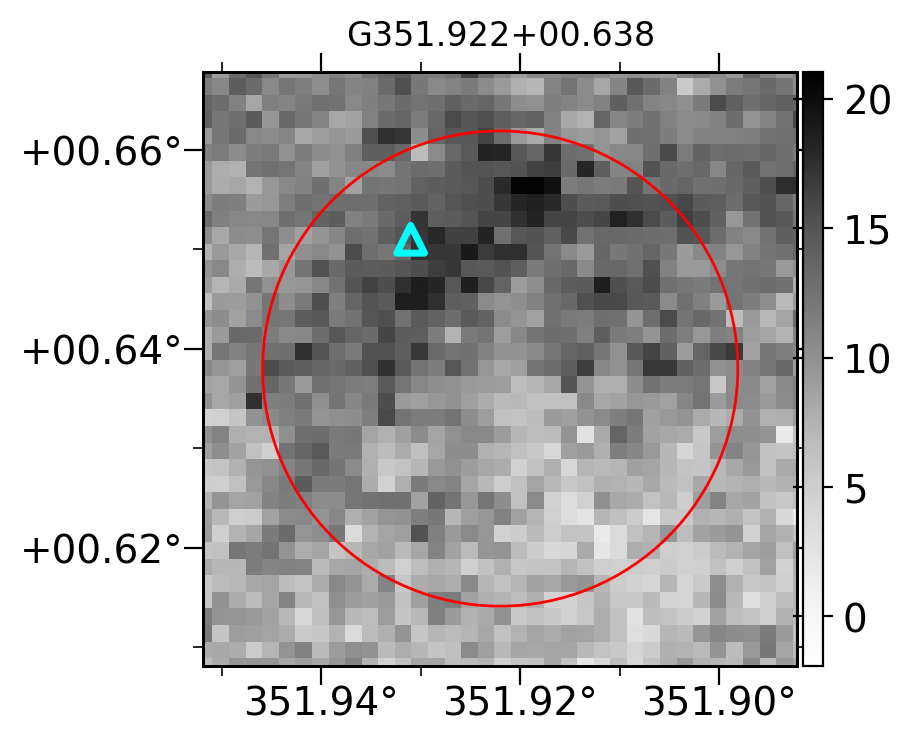}
\includegraphics[width=0.22\linewidth]{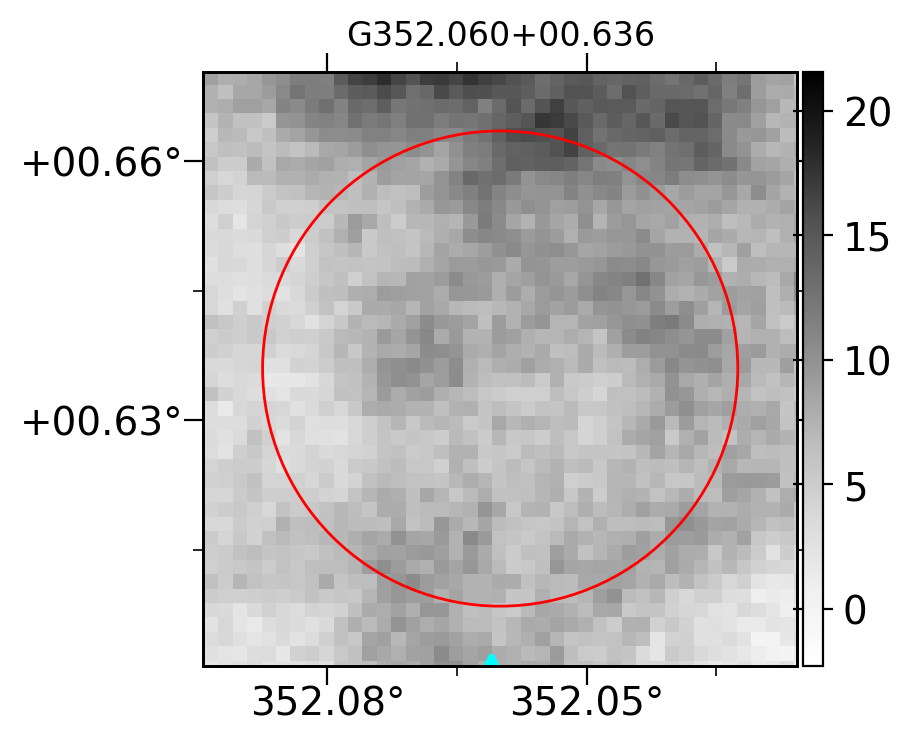}\\
\includegraphics[width=0.22\linewidth]{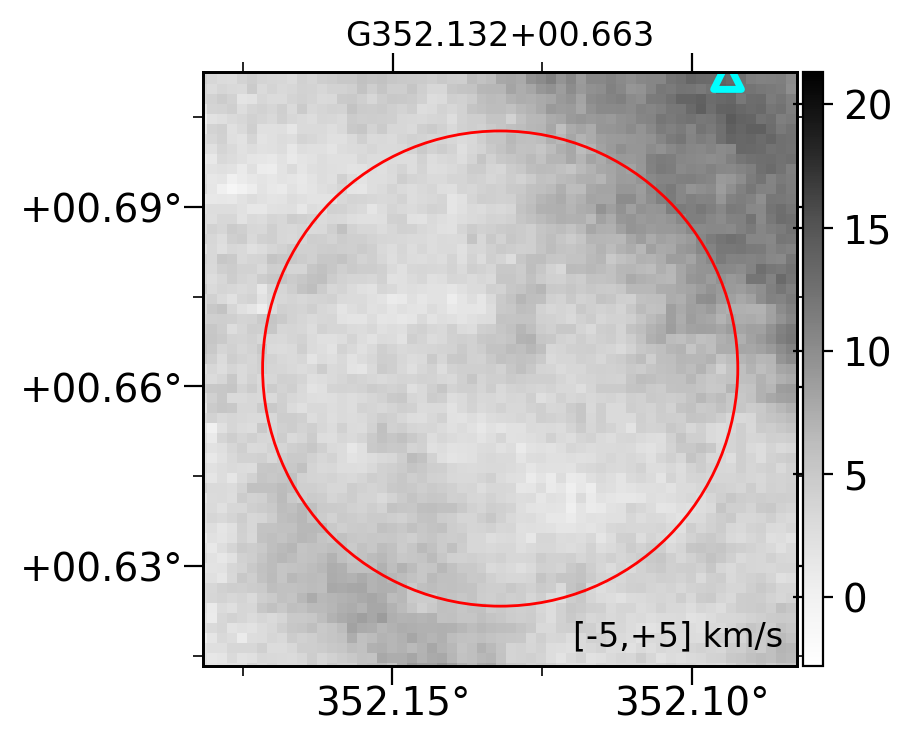}
\includegraphics[width=0.22\linewidth]{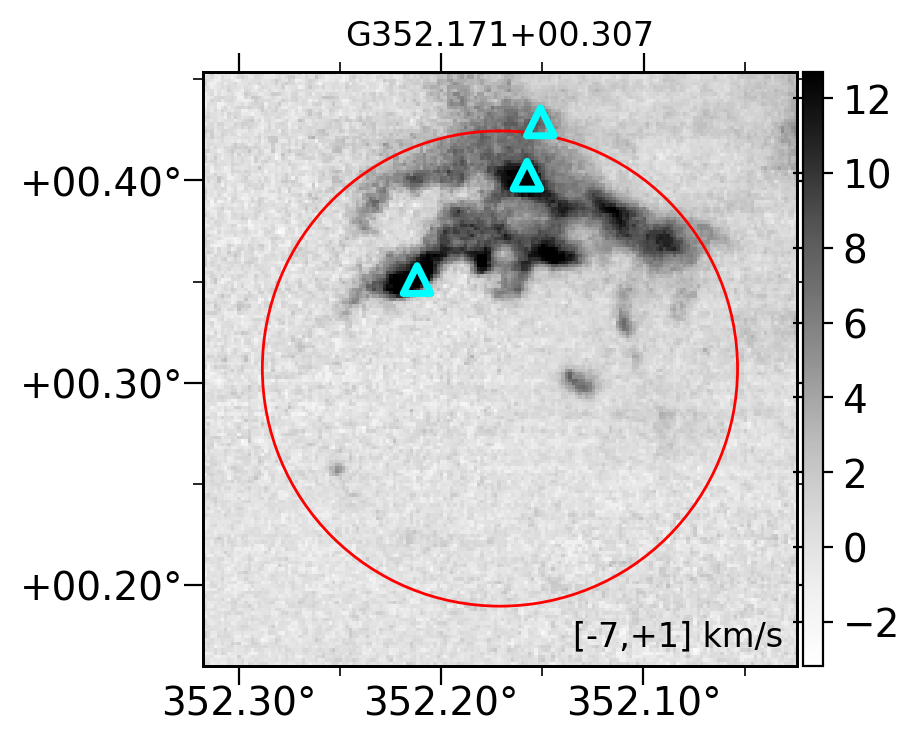}
\includegraphics[width=0.22\linewidth]{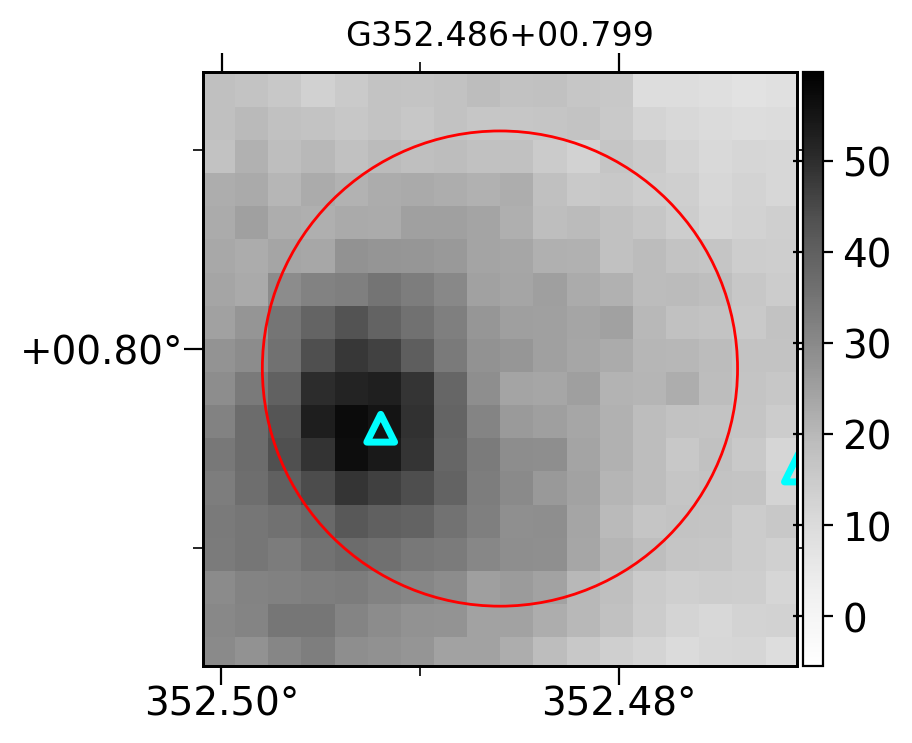}
\includegraphics[width=0.22\linewidth]{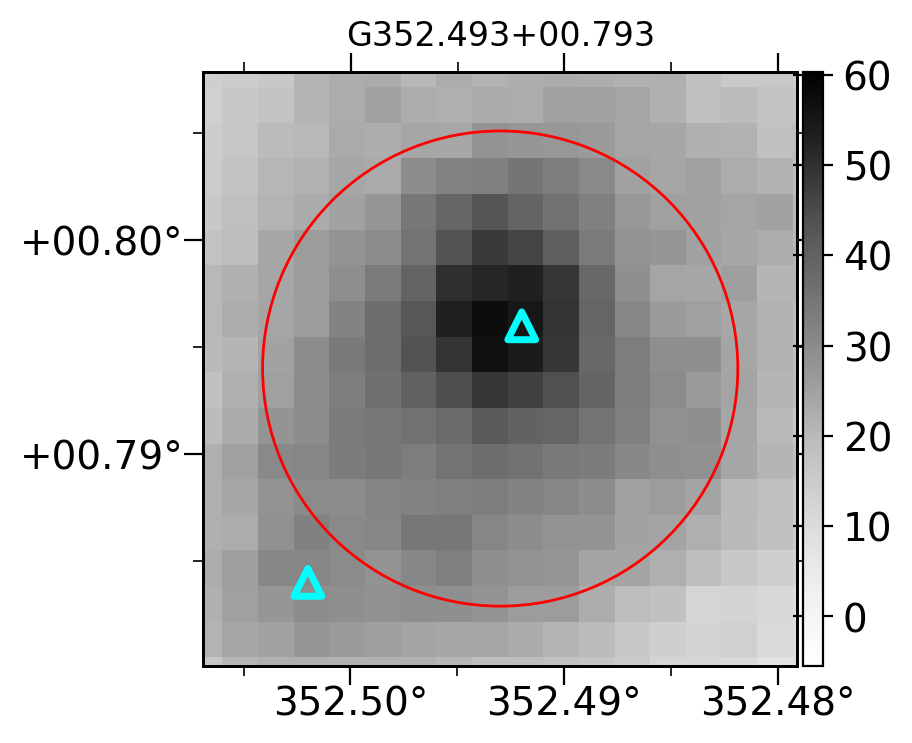}\\
\includegraphics[width=0.22\linewidth]{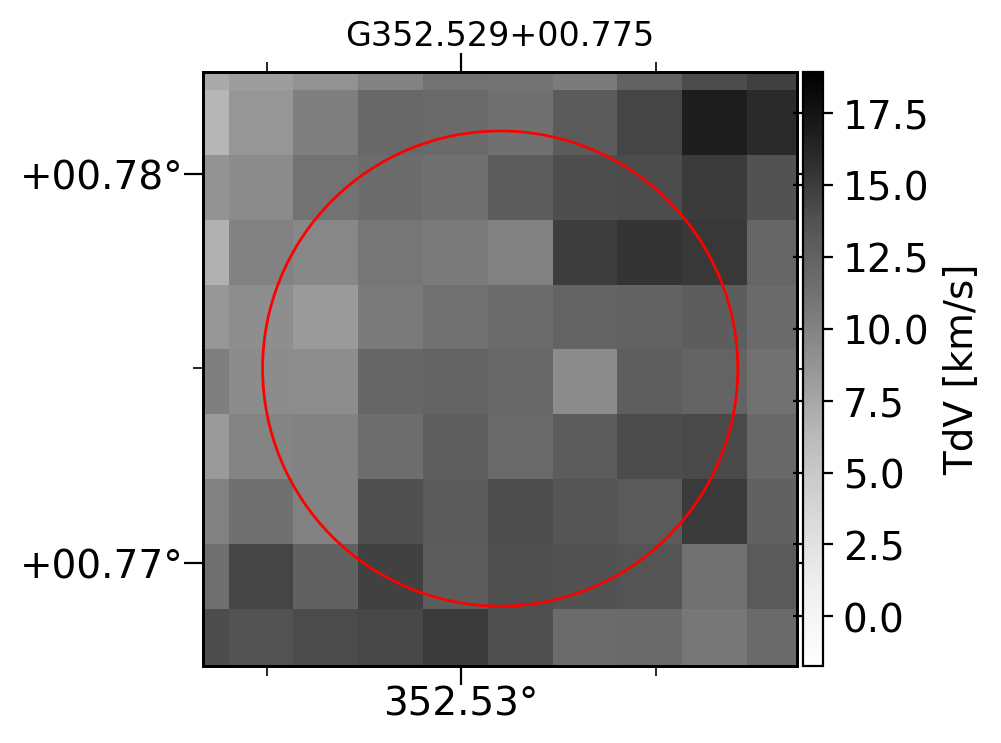}
    \caption{Same as Figure \ref{fig:chan_map_hii0}}
    \label{fig:chan_map_hii1}
\end{figure}

\clearpage
\section{Radial profiles and contrast parameter}
\begin{figure}[htbp!]
\centering
\includegraphics[width=0.22\linewidth]{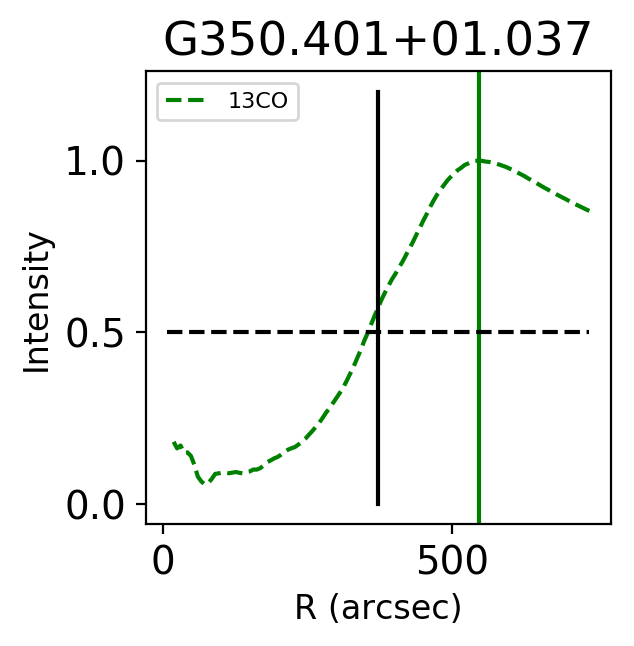}
\includegraphics[width=0.22\linewidth]{rprofiles/2_rprofile_13coalone_v3.png}
\includegraphics[width=0.22\linewidth]{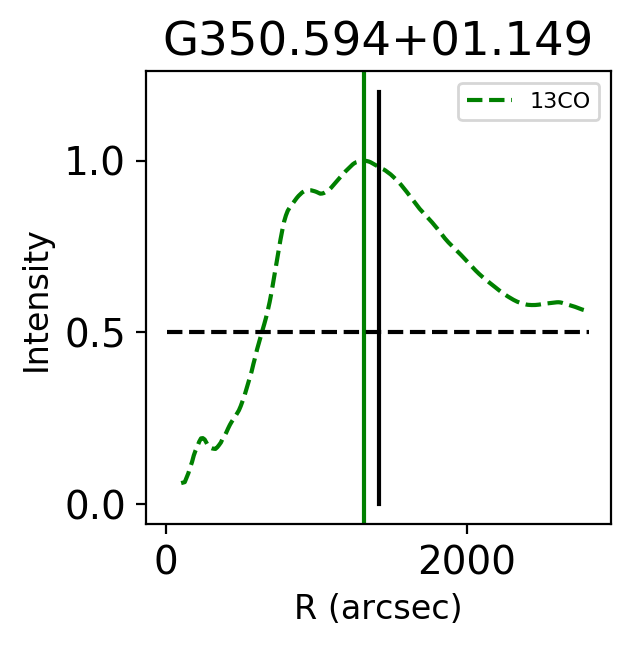}
\includegraphics[width=0.22\linewidth]{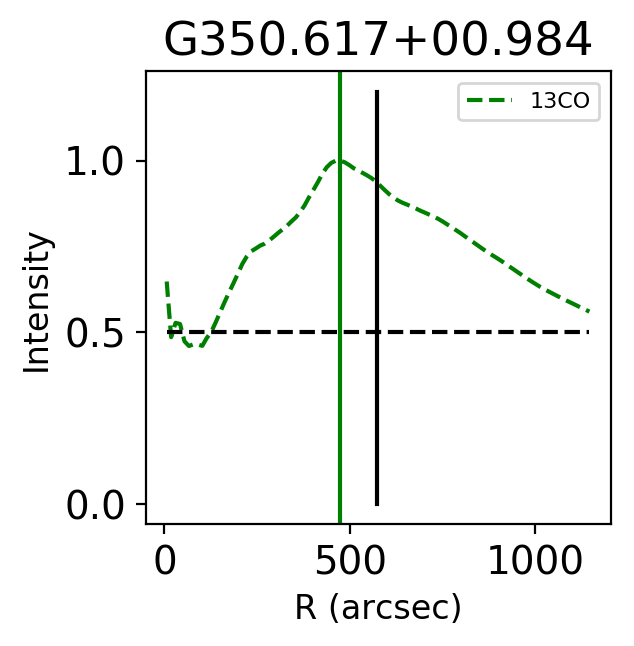}\\
\includegraphics[width=0.22\linewidth]{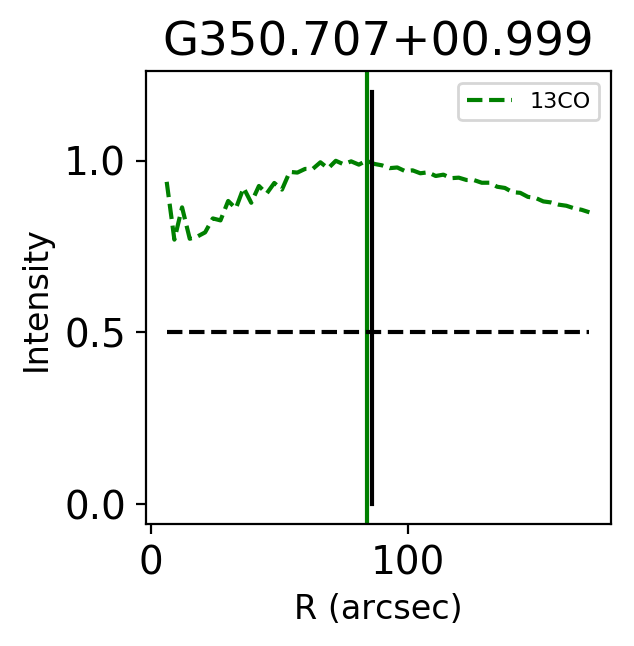}
\includegraphics[width=0.22\linewidth]{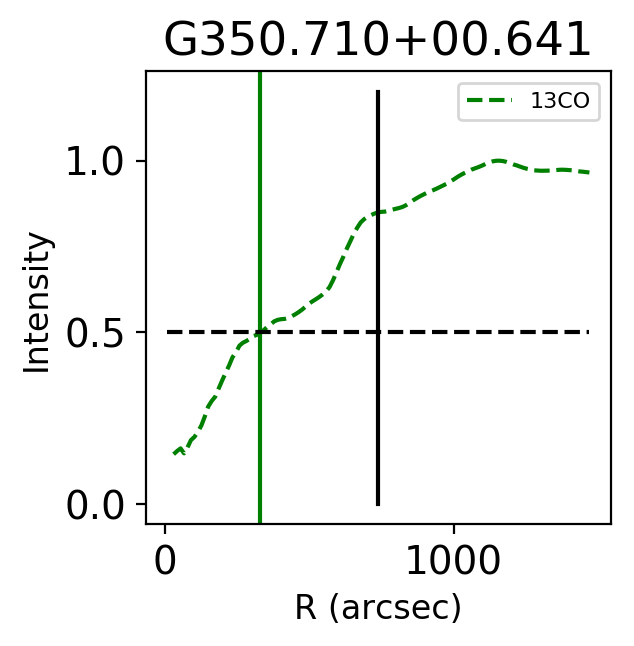}
\includegraphics[width=0.22\linewidth]{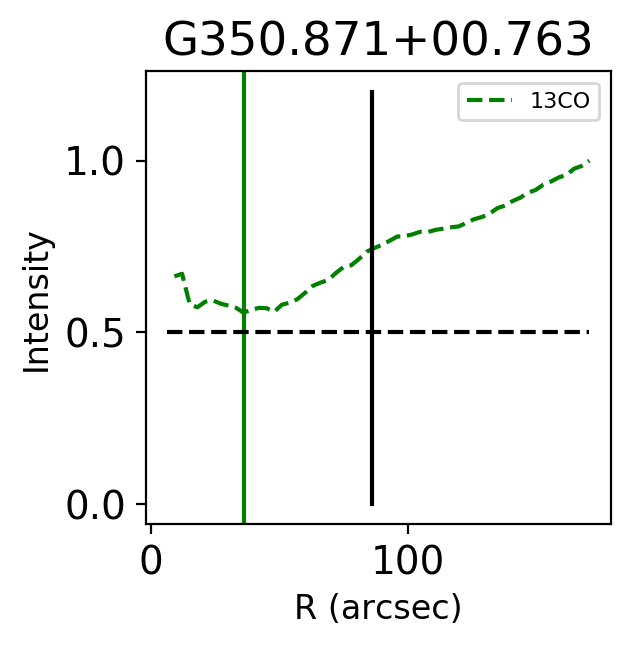}
\includegraphics[width=0.22\linewidth]{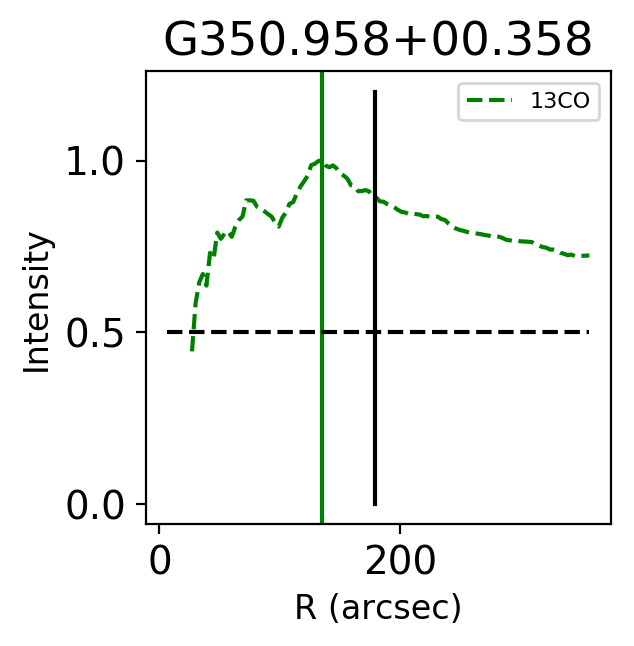}\\
\includegraphics[width=0.22\linewidth]{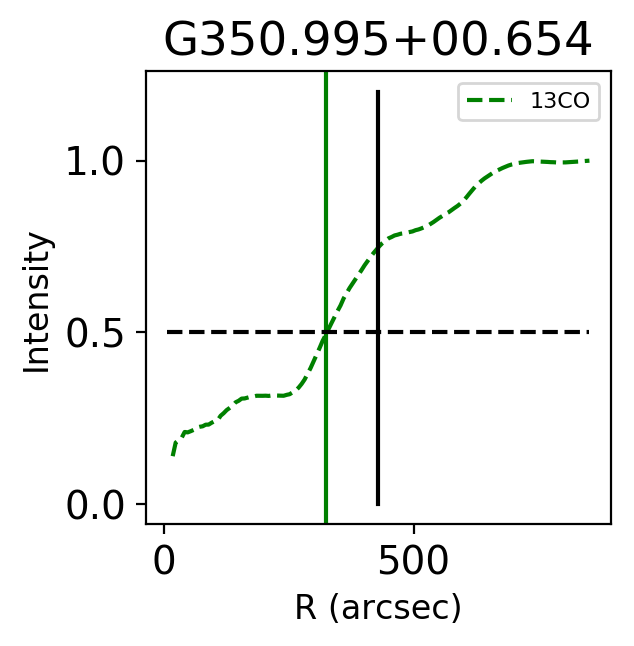}
\includegraphics[width=0.22\linewidth]{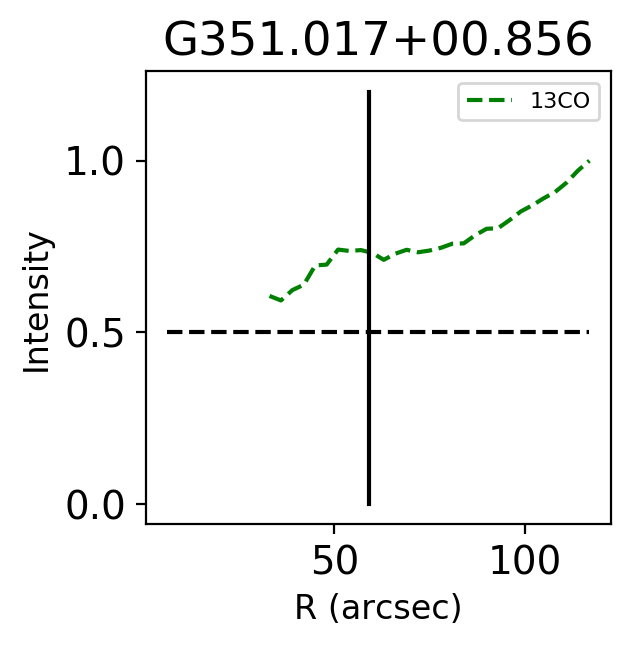}
\includegraphics[width=0.22\linewidth]{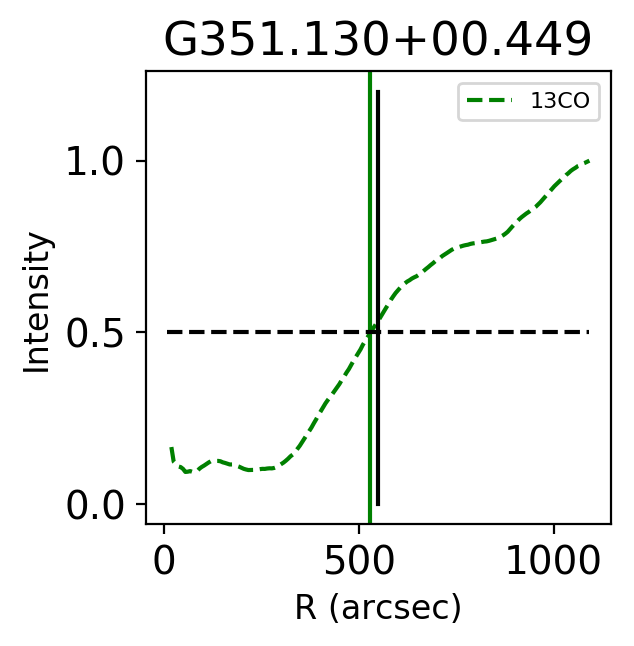}
\includegraphics[width=0.22\linewidth]{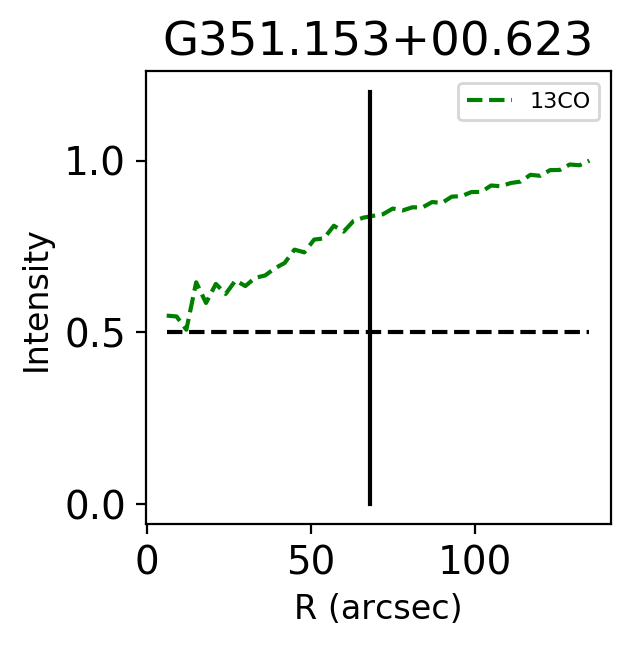}
\includegraphics[width=0.22\linewidth]{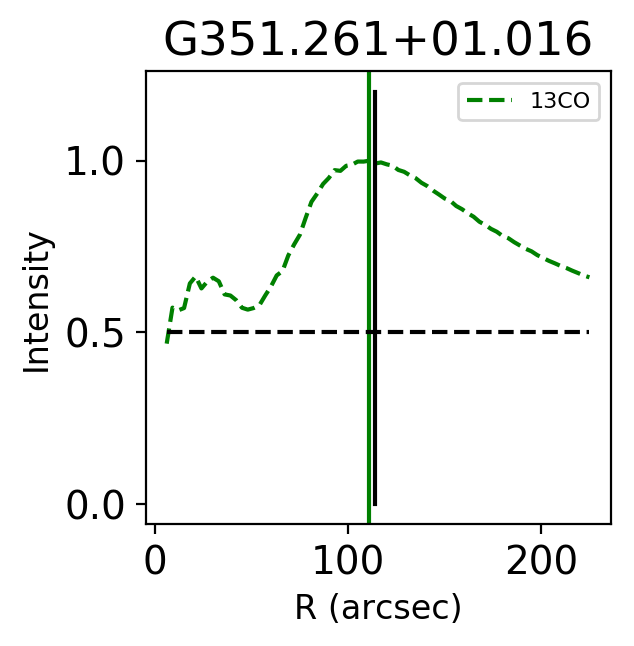}
\includegraphics[width=0.22\linewidth]{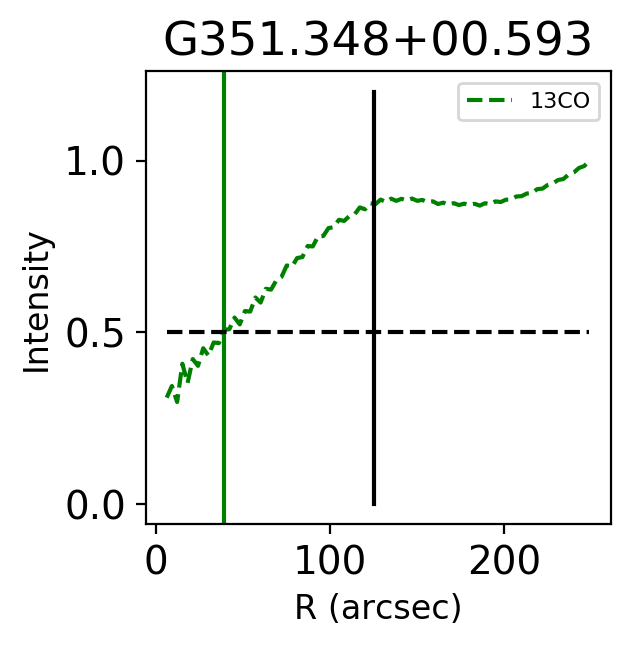}
\includegraphics[width=0.22\linewidth]{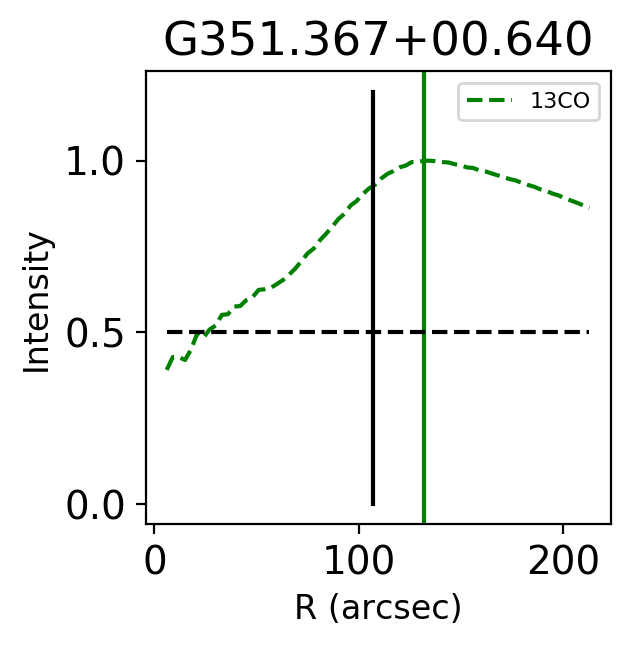}
\includegraphics[width=0.22\linewidth]{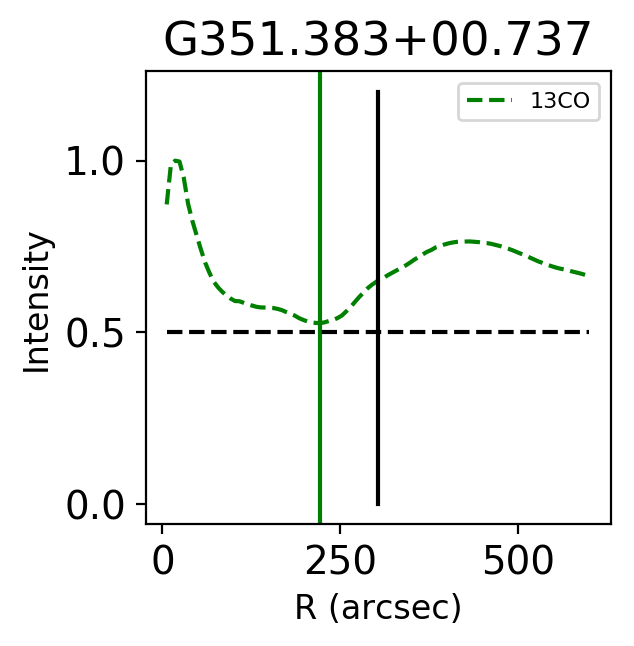}
\includegraphics[width=0.22\linewidth]{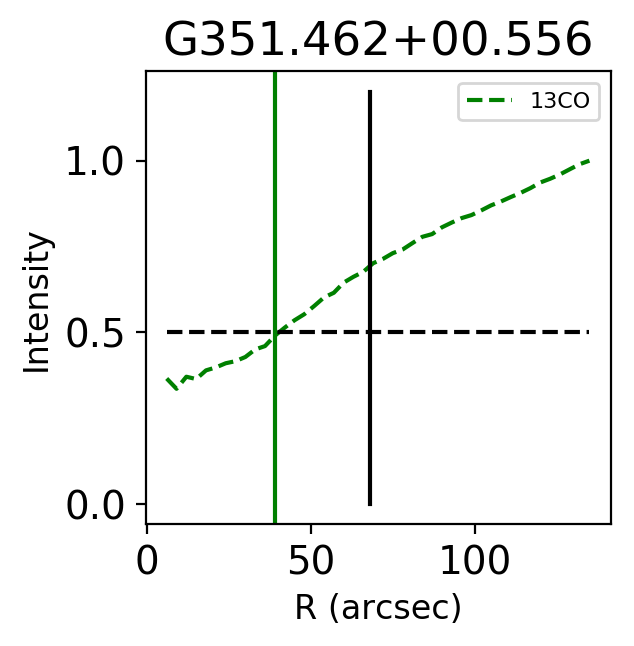}
\includegraphics[width=0.22\linewidth]{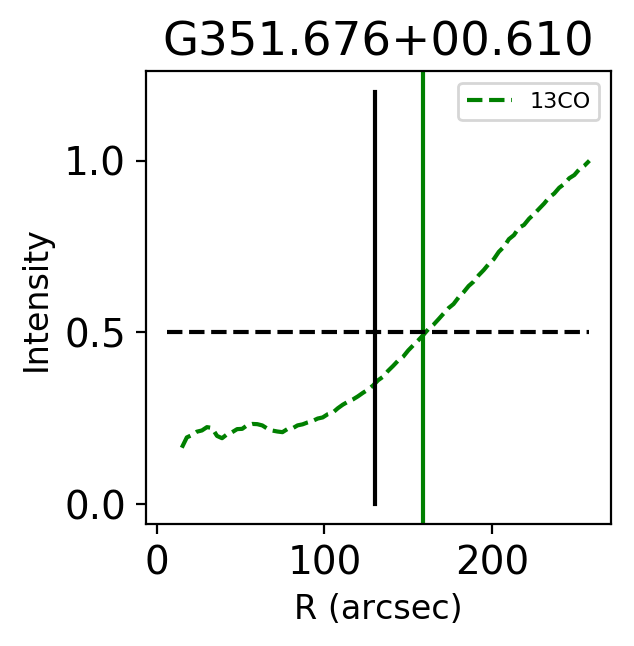}
\includegraphics[width=0.22\linewidth]{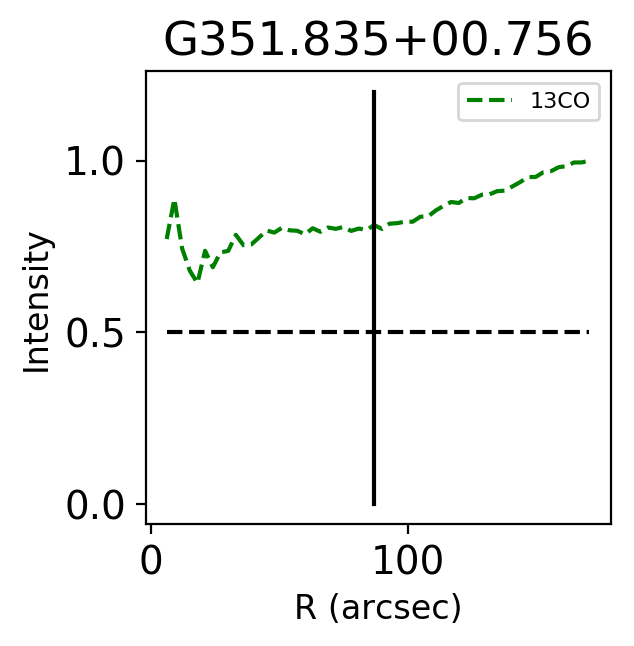}
\includegraphics[width=0.22\linewidth]{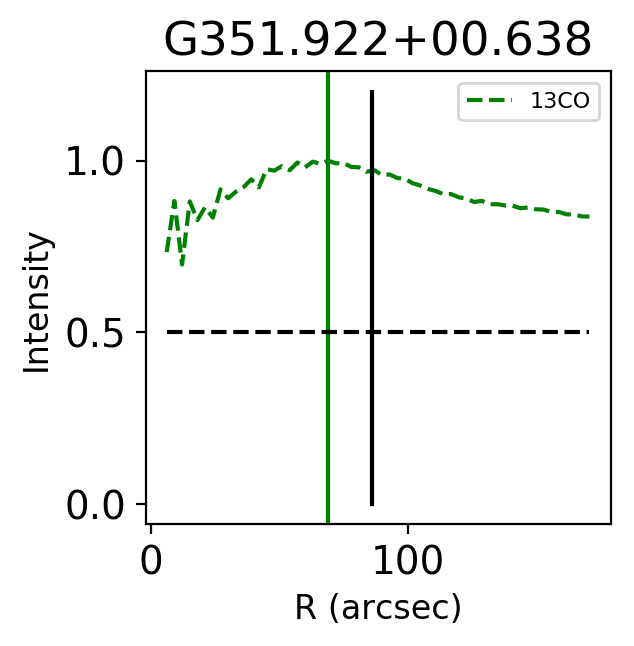}
    \caption{Azimuthally averaged $^{13}$CO emission profiles (shown in green color) toward H II regions exhibiting increasing radial intensity profiles, with or without the shell/ring like features, computed from velocity range $-$15 to +5 km s$^{-1}$. Vertical black line indicate the radii from \citealt{Anderson2014} and vertical green line indicate sizes estimated from $^{13}$CO emission profiles. Horizontal black line indicates 50\% value of the intensity maxima.}
    \label{fig:radialprofiles1}
\end{figure}

\begin{figure}[htbp!]
\centering
\includegraphics[width=0.2\linewidth]{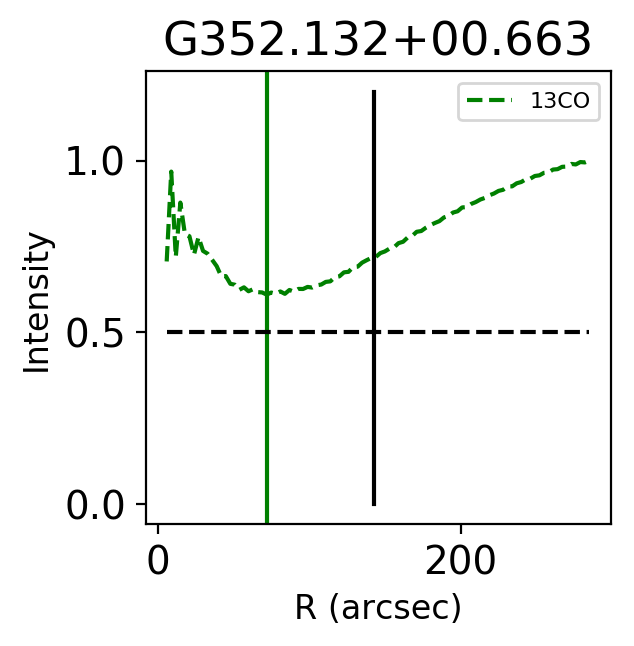}
\includegraphics[width=0.2\linewidth]{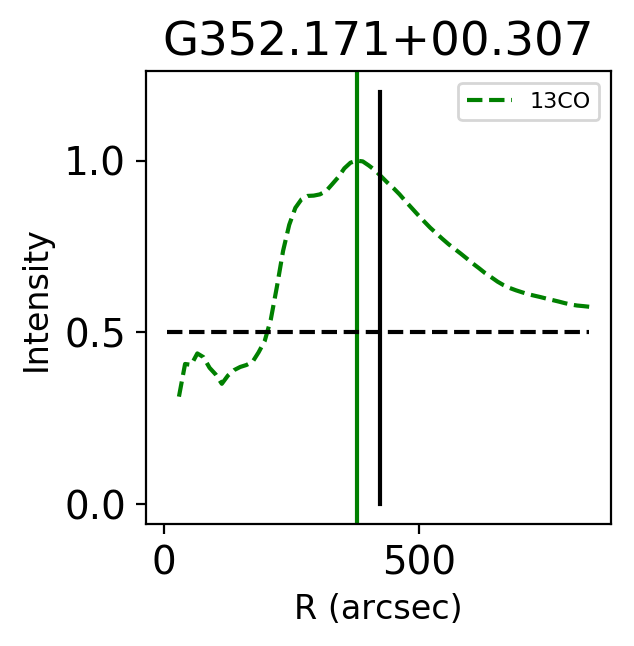}
    \caption{Same as Fig. \ref{fig:radialprofiles1}. \label{fig:radialprofiles11}}
\end{figure}

\begin{figure}[htbp!]
\centering
\includegraphics[width=0.22\linewidth]{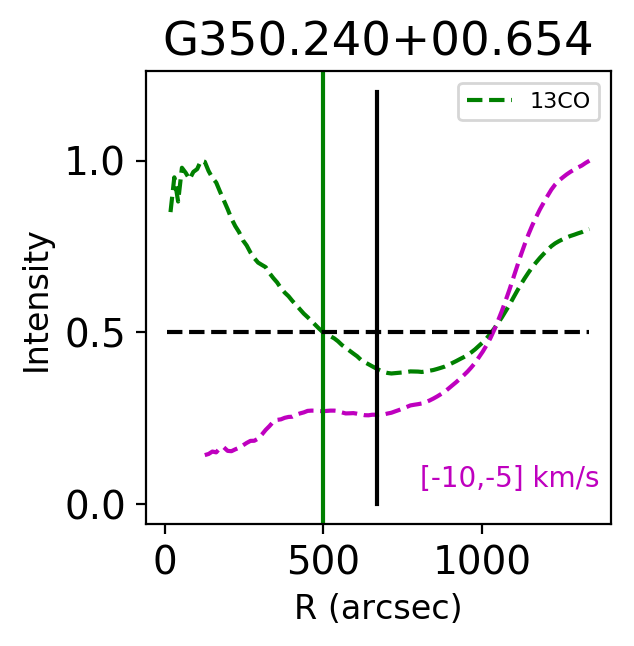}
\includegraphics[width=0.22\linewidth]{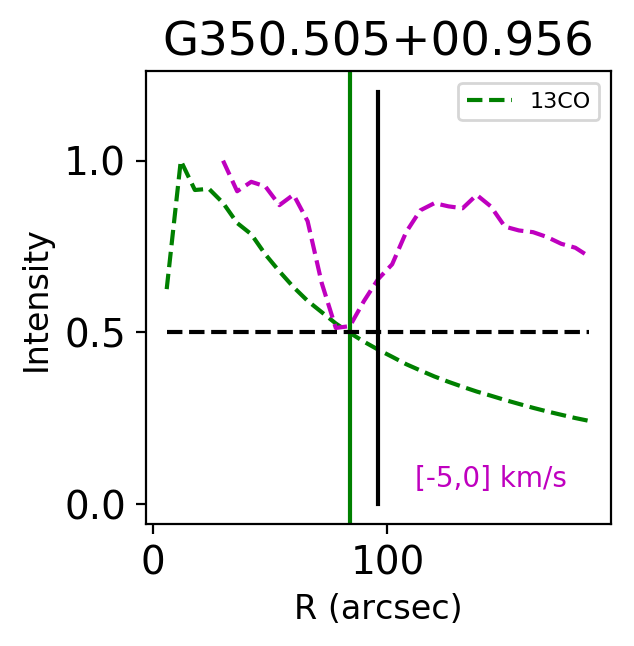}
\includegraphics[width=0.22\linewidth]{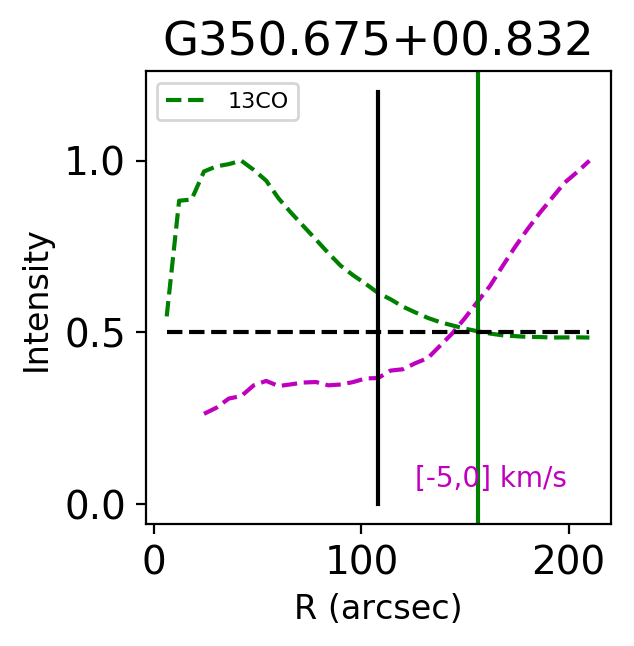}
\includegraphics[width=0.22\linewidth]{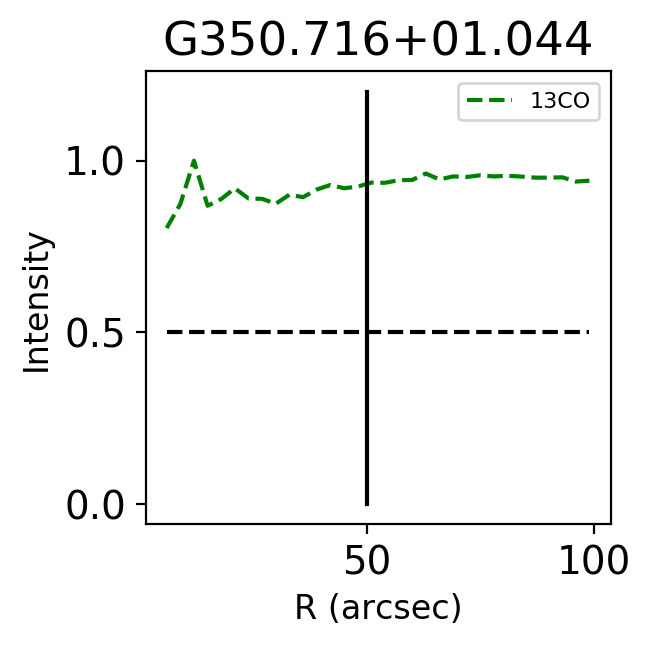}
\includegraphics[width=0.22\linewidth]{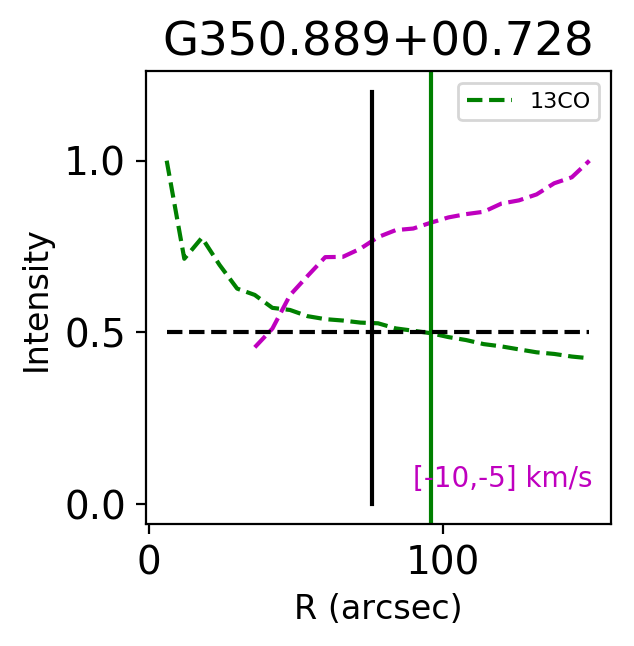}
\includegraphics[width=0.22\linewidth]{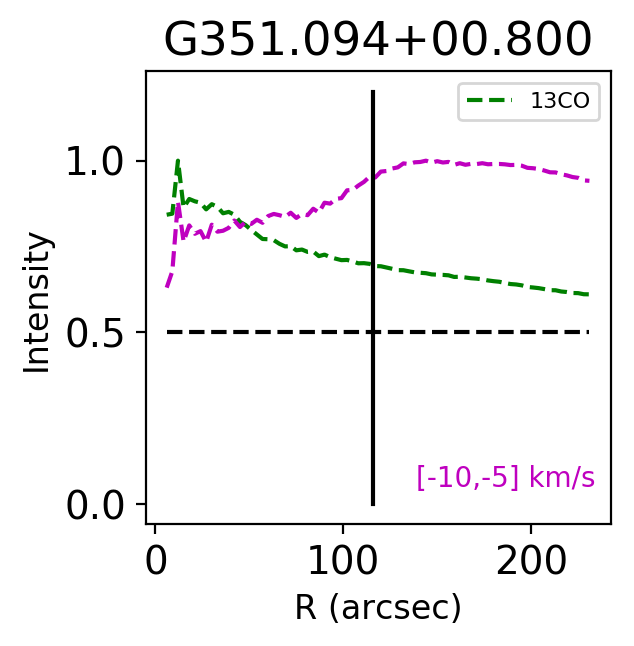}
\includegraphics[width=0.22\linewidth]{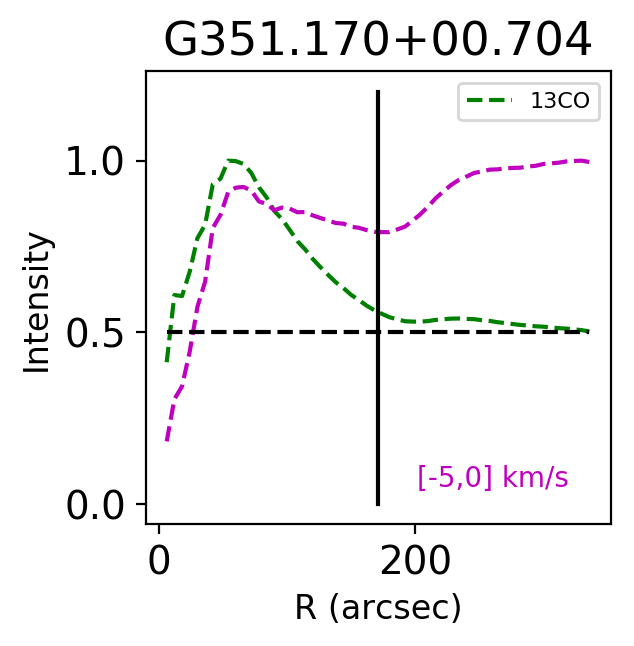}
\includegraphics[width=0.22\linewidth]{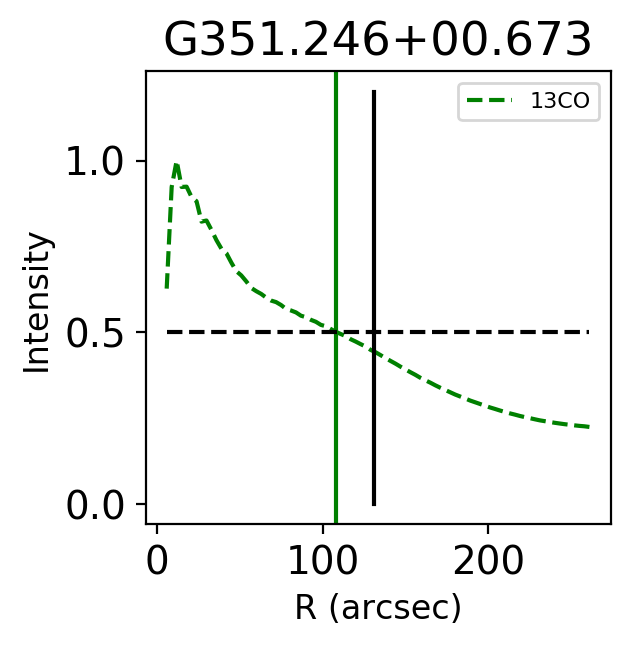}
\includegraphics[width=0.22\linewidth]{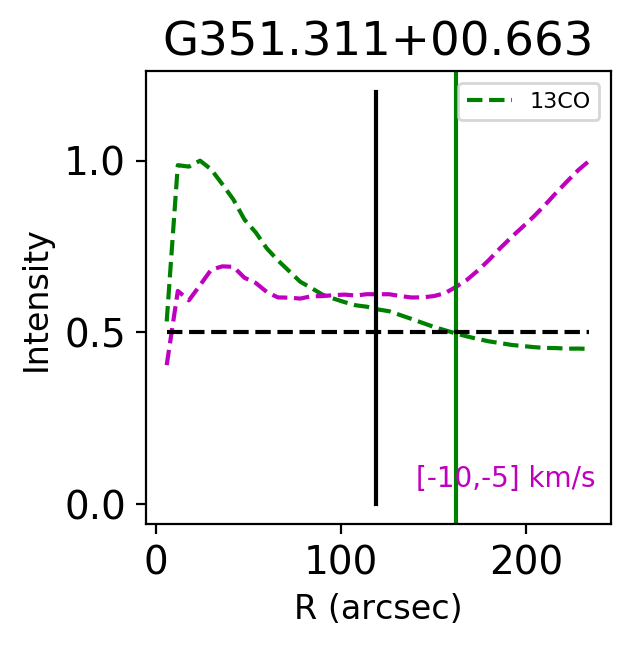}
\includegraphics[width=0.22\linewidth]{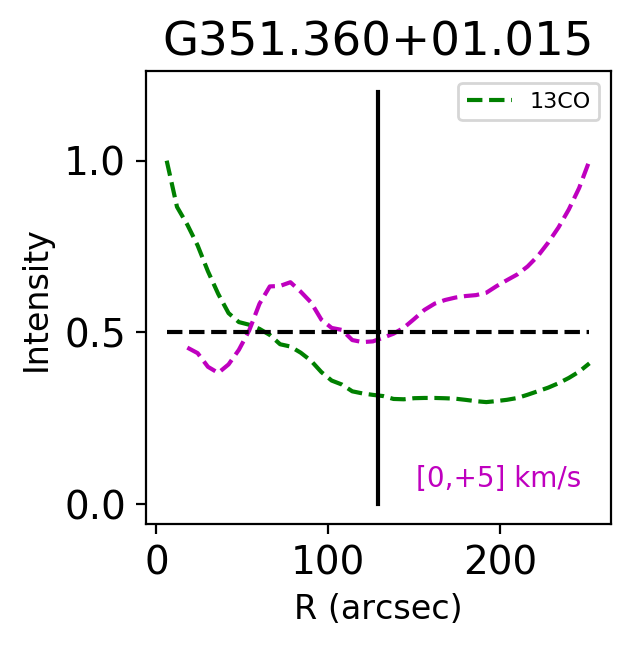}
\includegraphics[width=0.22\linewidth]{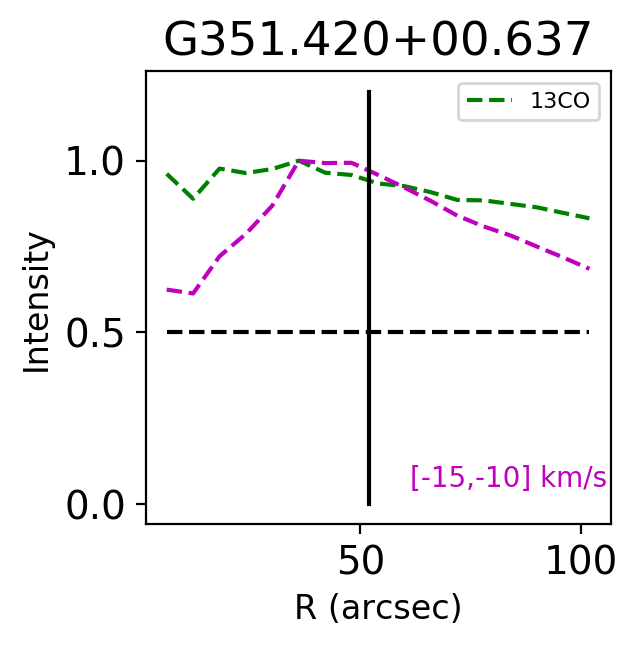}
\includegraphics[width=0.22\linewidth]{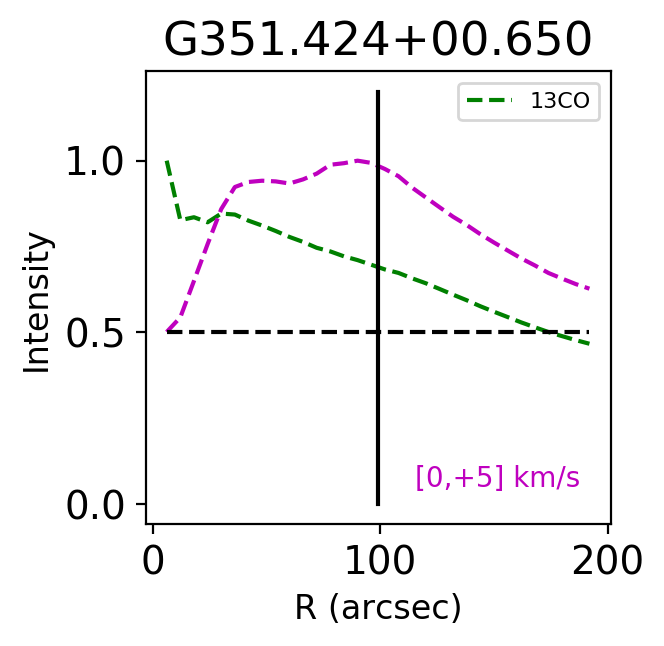}
\includegraphics[width=0.22\linewidth]{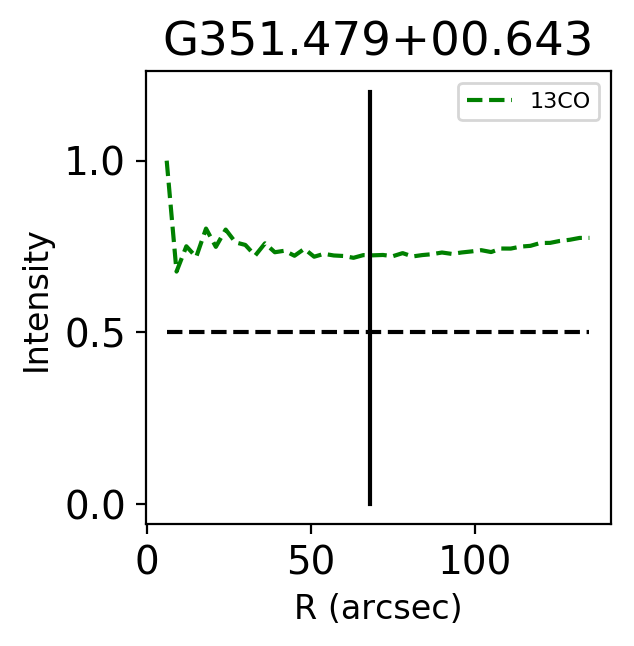}
\includegraphics[width=0.22\linewidth]{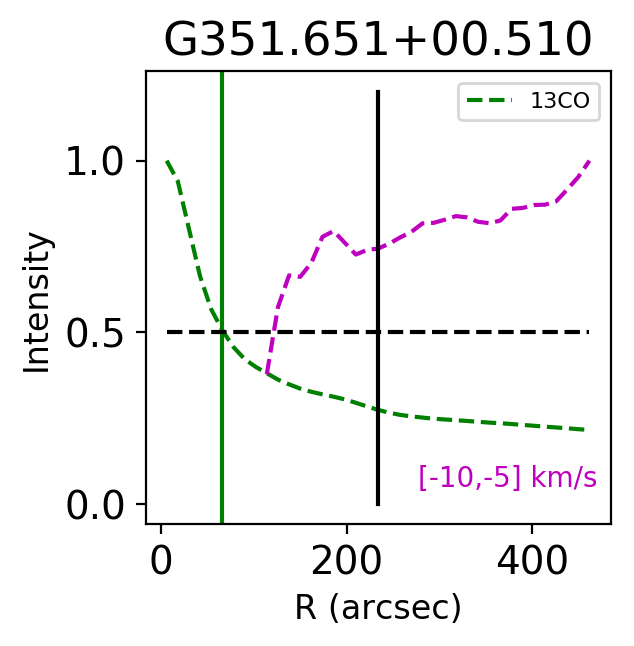}
\includegraphics[width=0.22\linewidth]{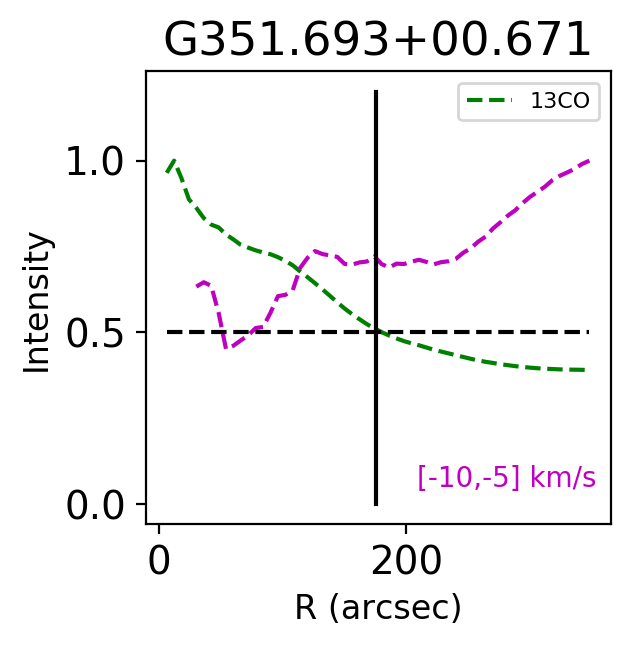}
\includegraphics[width=0.22\linewidth]{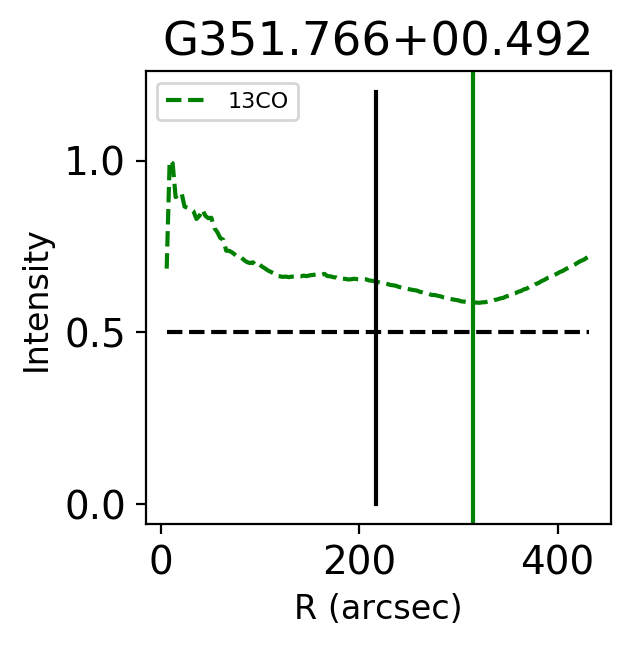}
\includegraphics[width=0.22\linewidth]{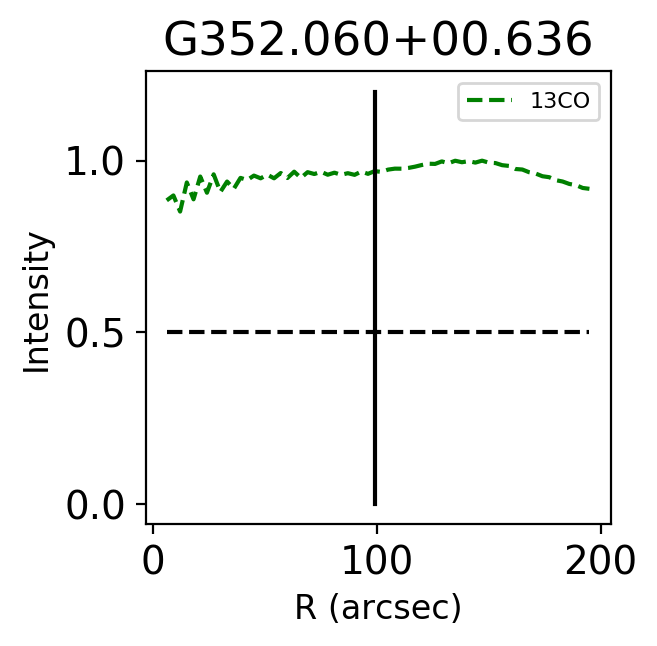}
\includegraphics[width=0.22\linewidth]{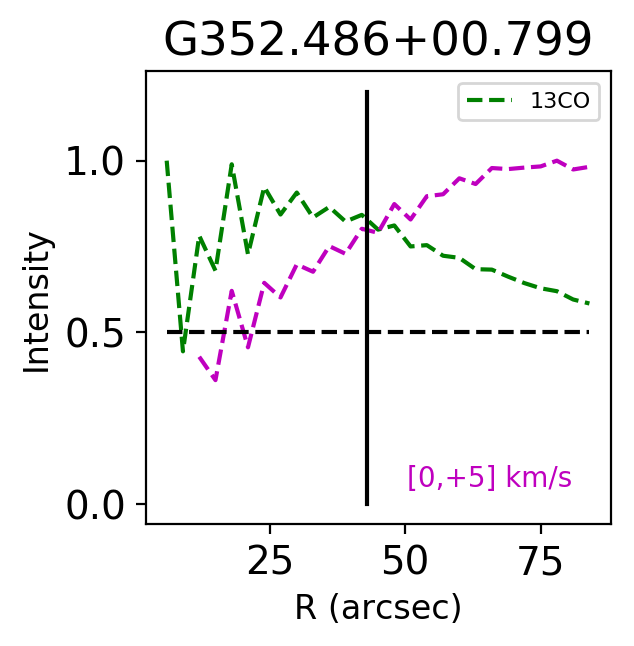}
\includegraphics[width=0.22\linewidth]{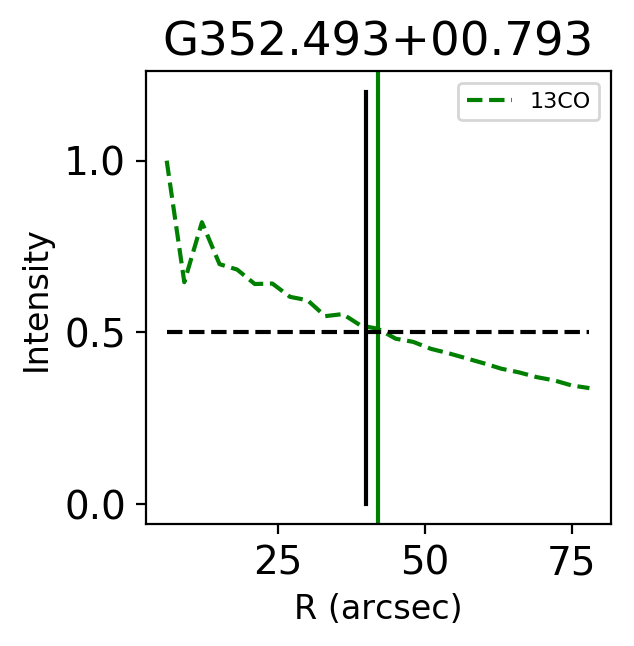}
\includegraphics[width=0.22\linewidth]{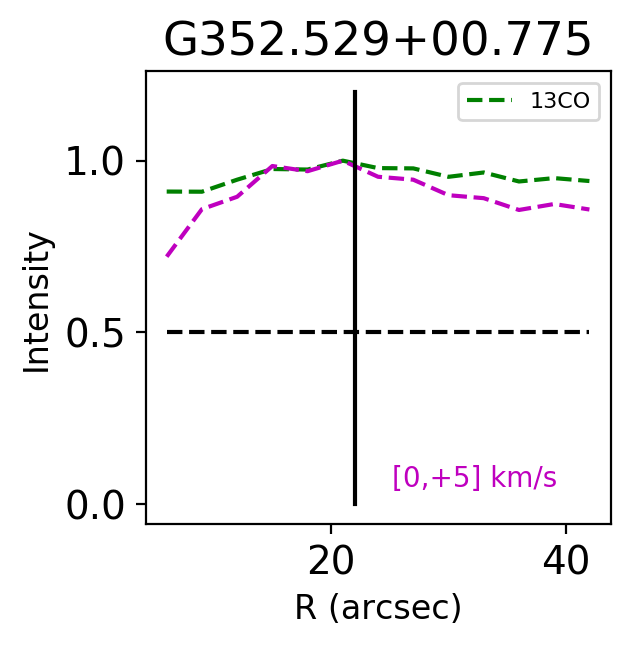}
    \caption{Same as Fig. \ref{fig:radialprofiles1} for H II regions exhibiting centrally peaked emission with decreasing intensity profiles or flat radial profiles. In addition, magenta radial line profiles are shown for particular velocity ranges labelled at the bottom left corner of each plots. }
    \label{fig:radialprofiles2}
\end{figure}

\clearpage
\begin{figure}[htbp!]
    \centering
    \includegraphics[width=0.45\linewidth]{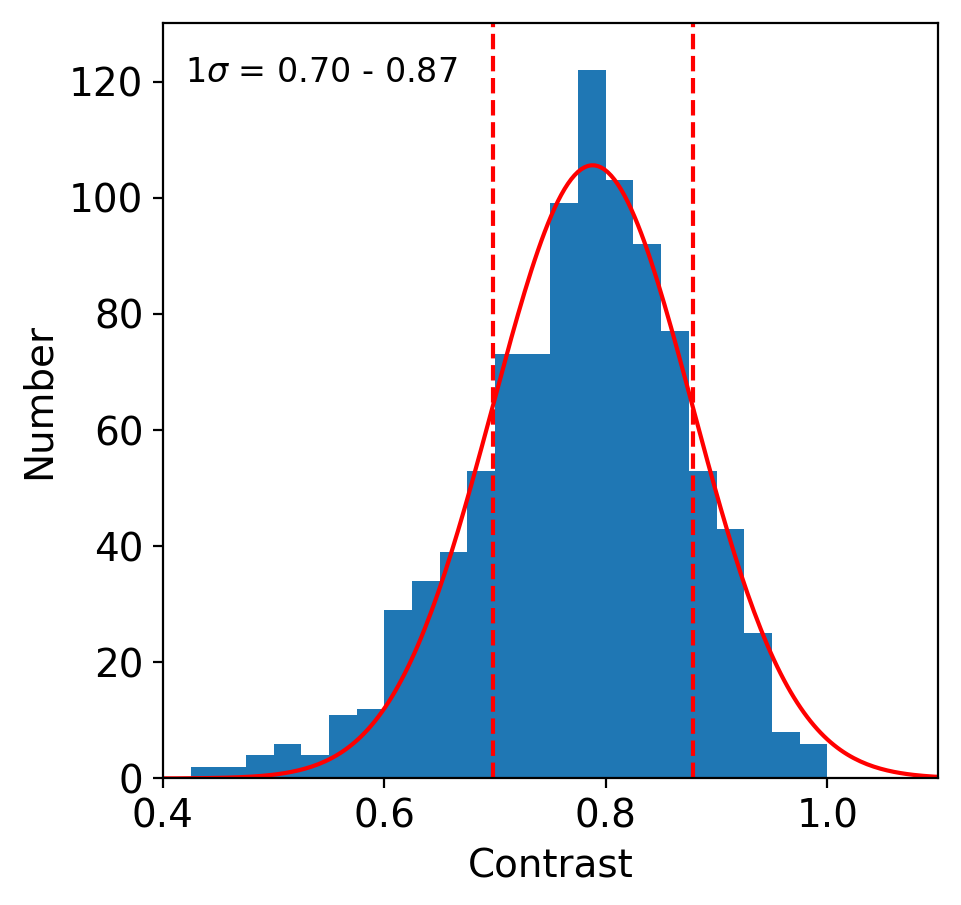}
    \caption{Contrast values of 1000 randomly sampled shells/rings obtained from $^{13}$CO moment map for velocity ranges $-$15 to +5 km s$^{-1}$ and radii 0.6R to 1.2R, where R is radius of randomly selected region. The radius of randomly created rings range from 20" to 700", similar to the sizes of H II regions found in NGC 6334 extended region. 1$\sigma$ range of contrast values are indicated by the dotted red lines.}
    \label{random_contrast}
\end{figure}

\clearpage
\section{Longitude-velocity ($lv$) and latitude-velocity ($bv$) plots}
\begin{figure}[htbp!]
    \centering
    \includegraphics[width=0.45\linewidth]{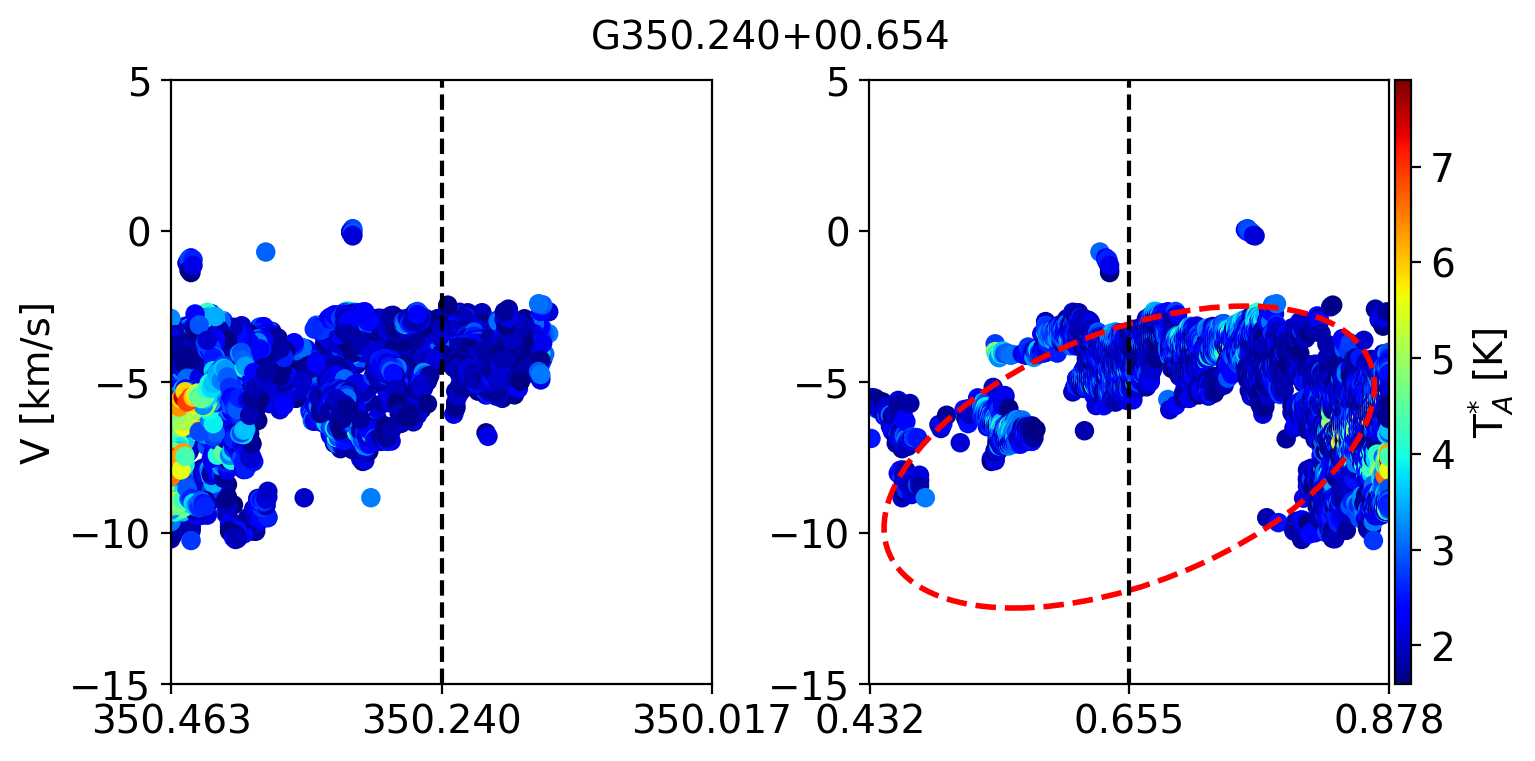}
    \includegraphics[width=0.45\linewidth]{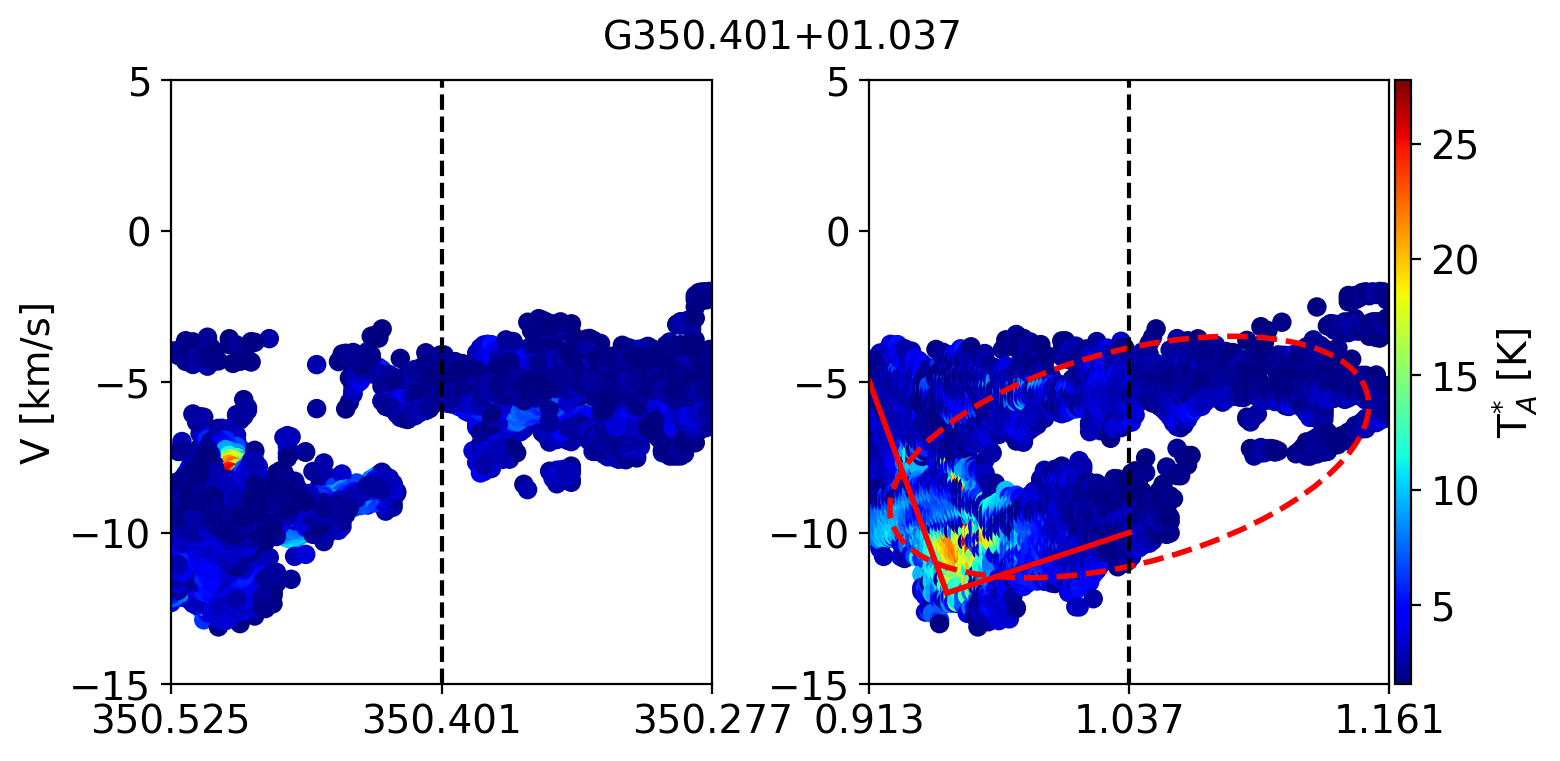}
    \includegraphics[width=0.45\linewidth]{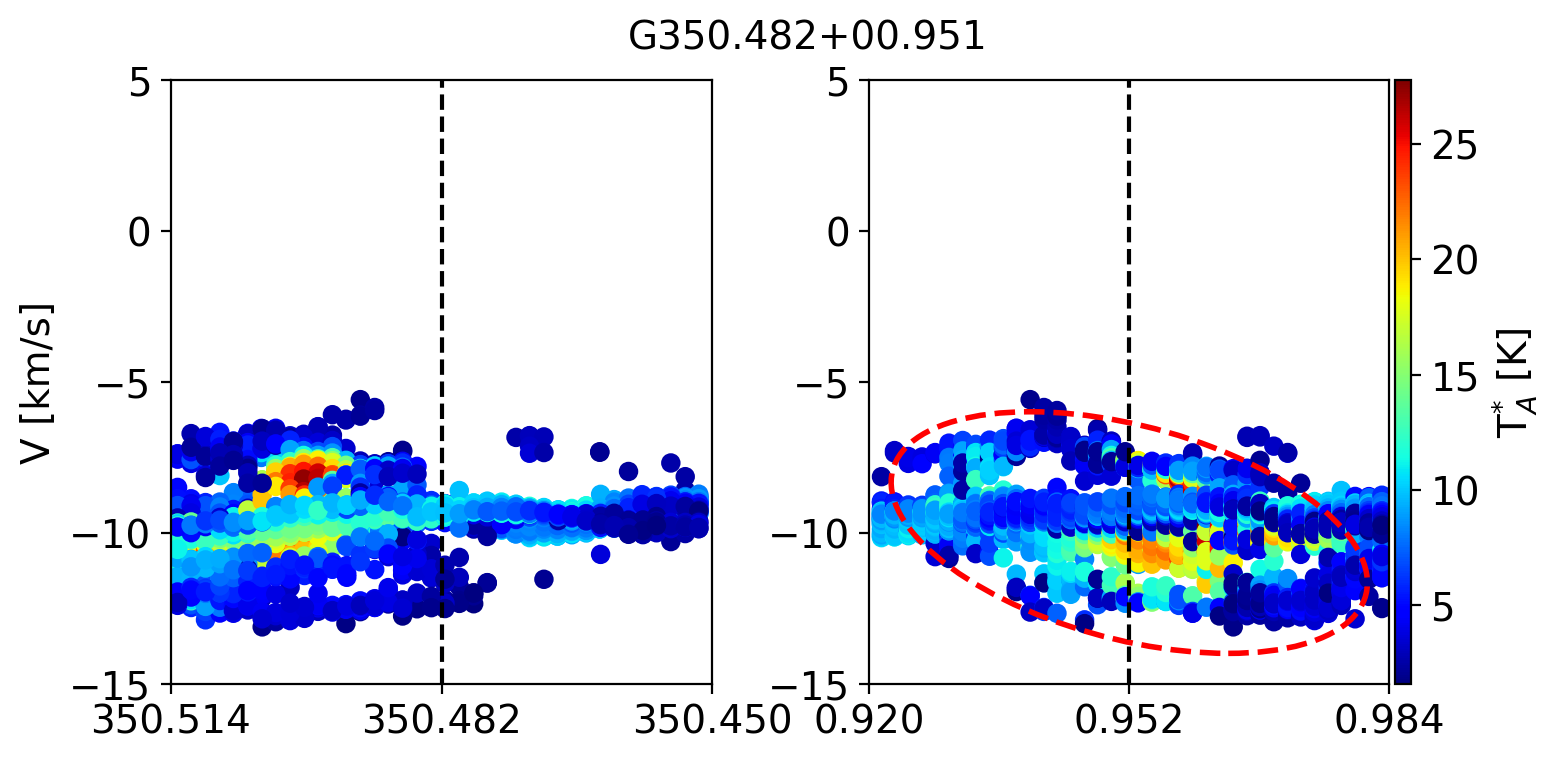}
    \includegraphics[width=0.45\linewidth]{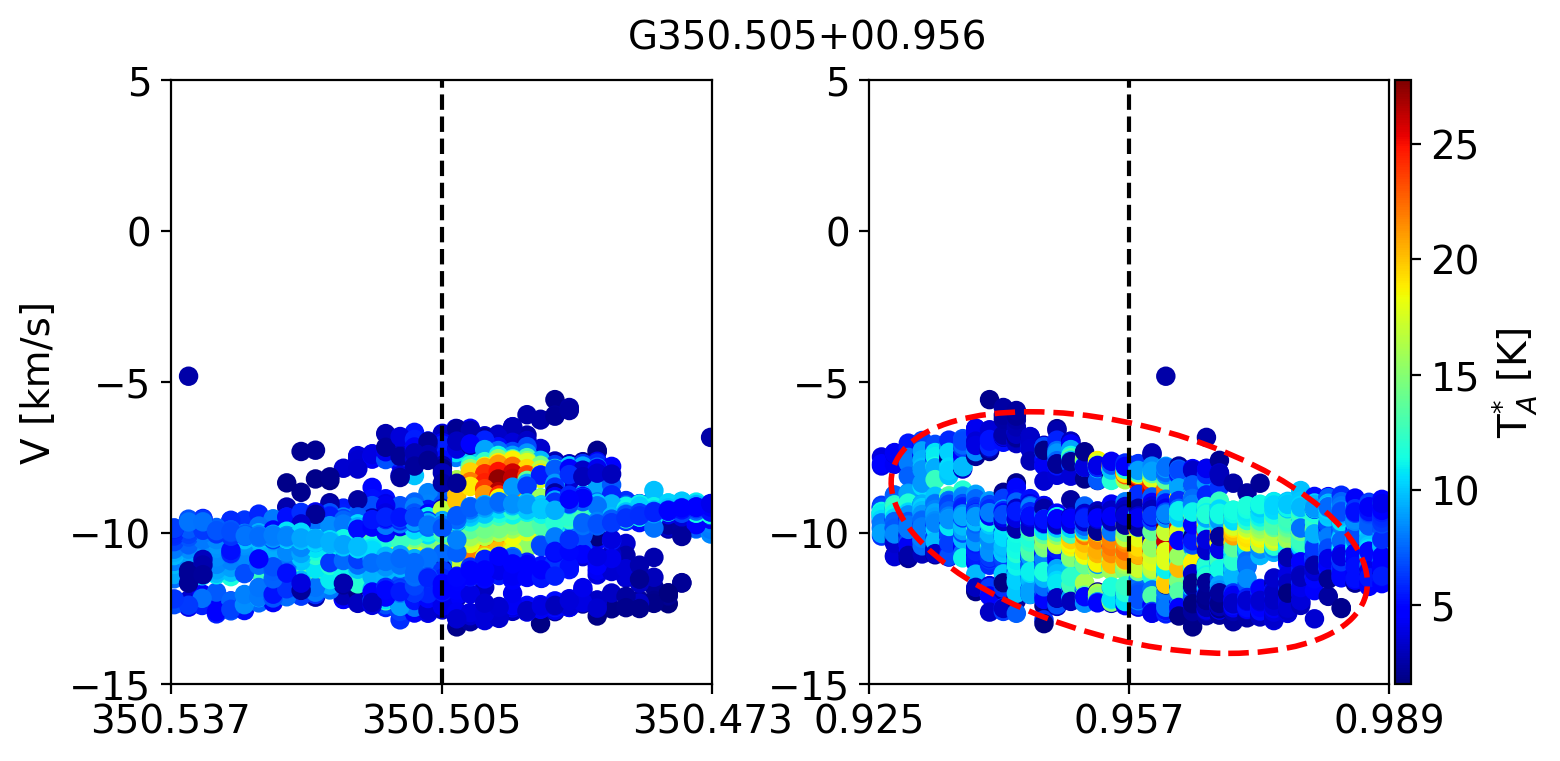}
    \includegraphics[width=0.45\linewidth]{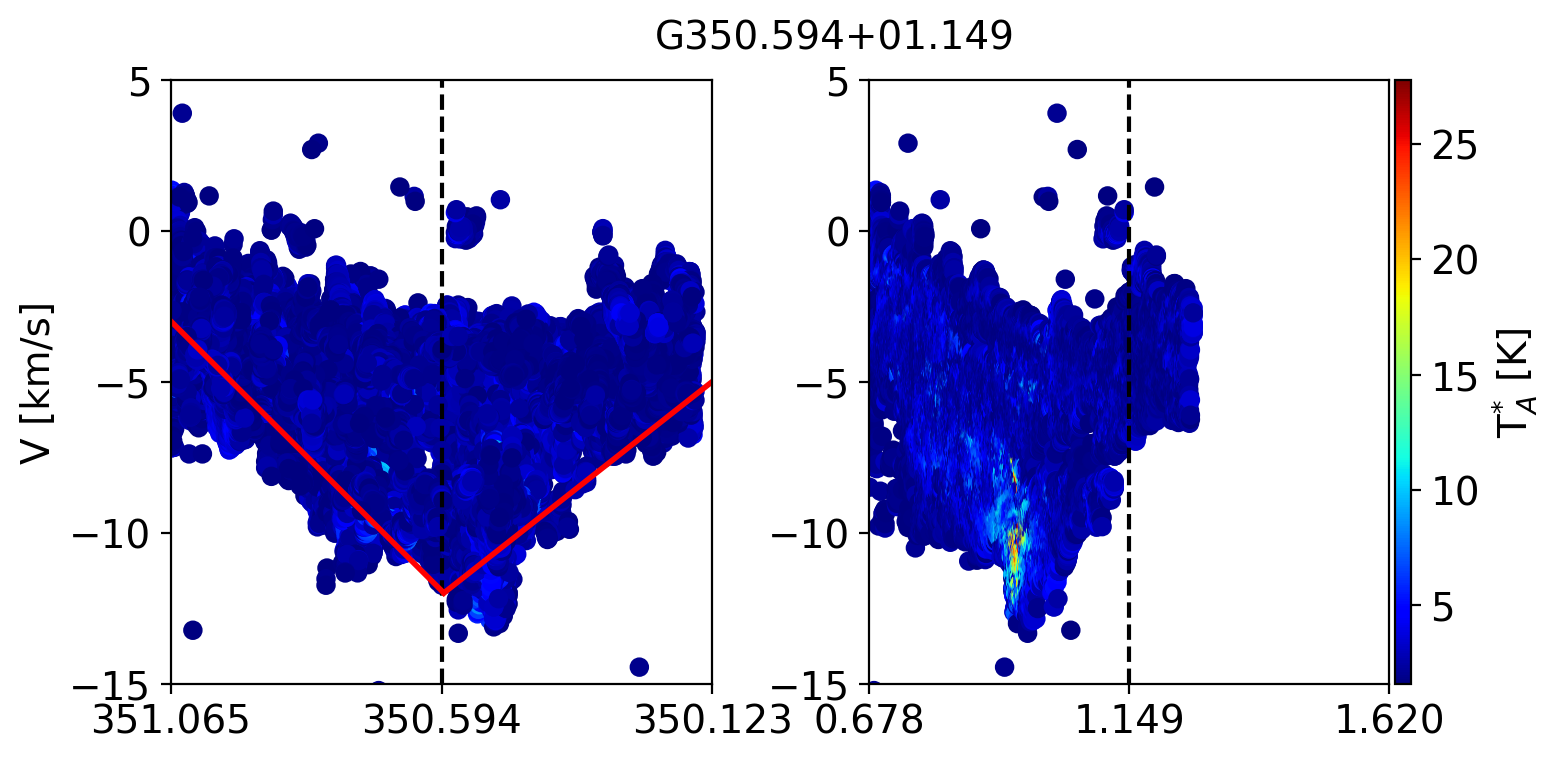}
    \includegraphics[width=0.45\linewidth]{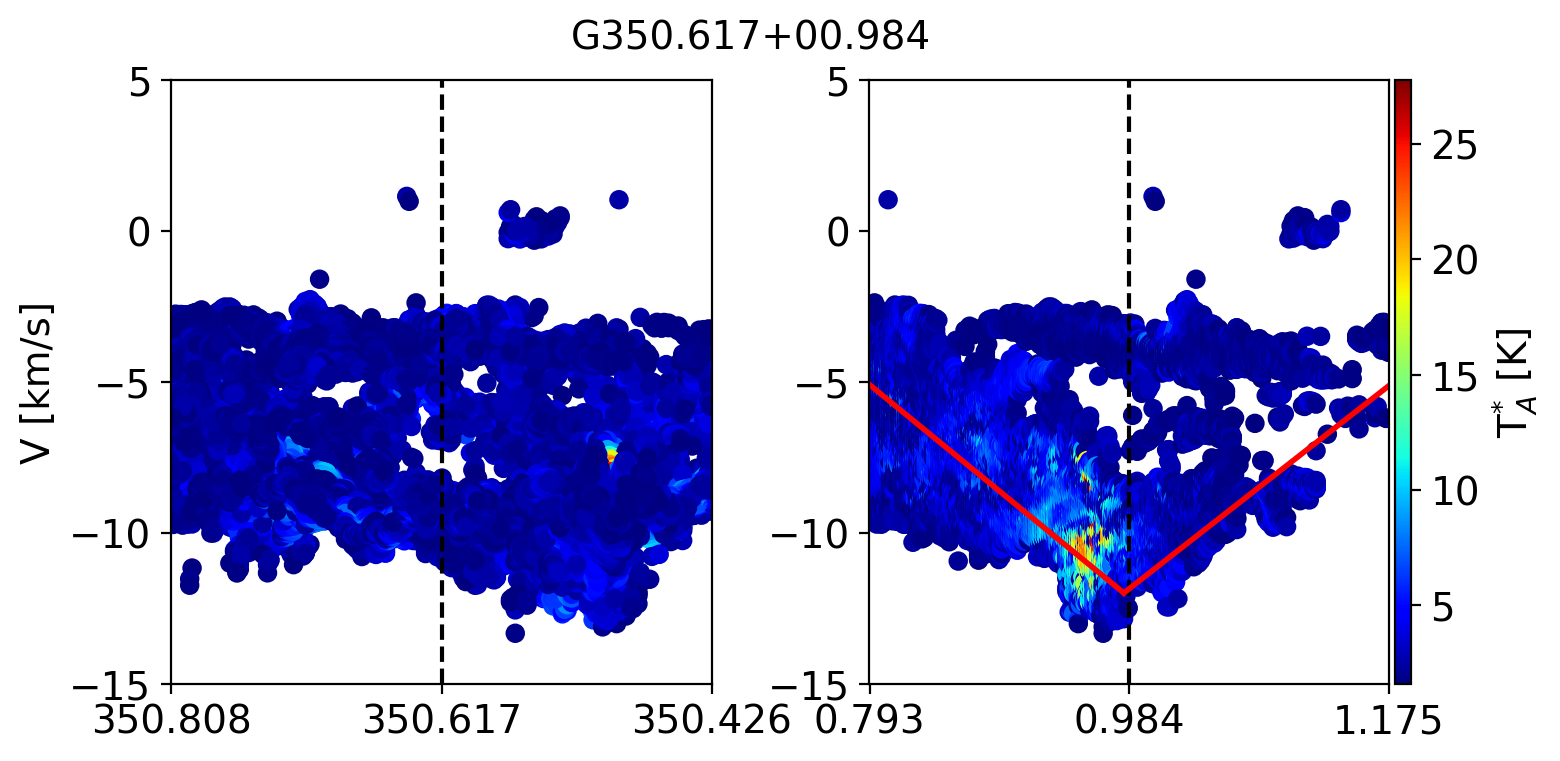}
    \includegraphics[width=0.45\linewidth]{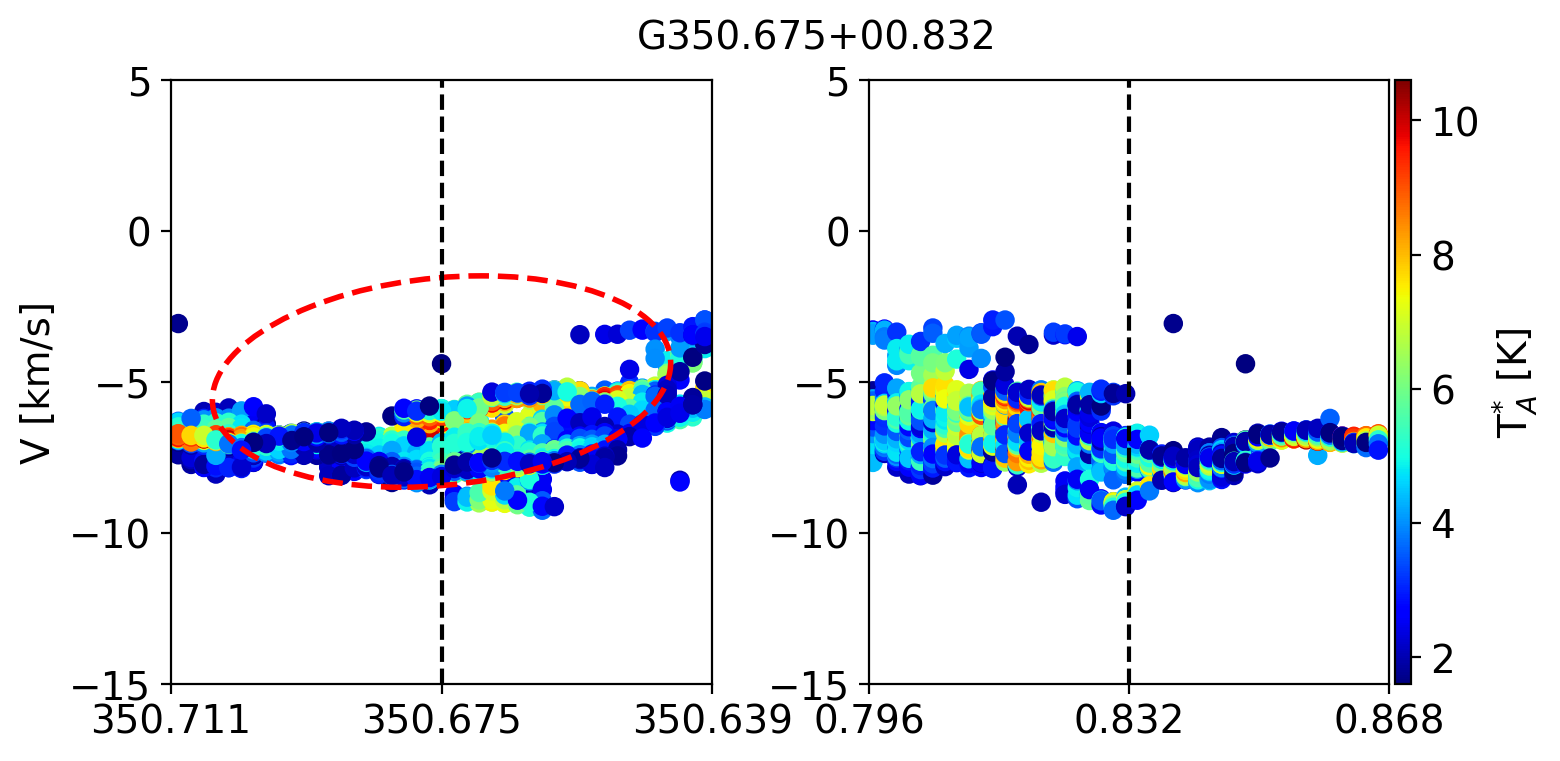}
    \includegraphics[width=0.45\linewidth]{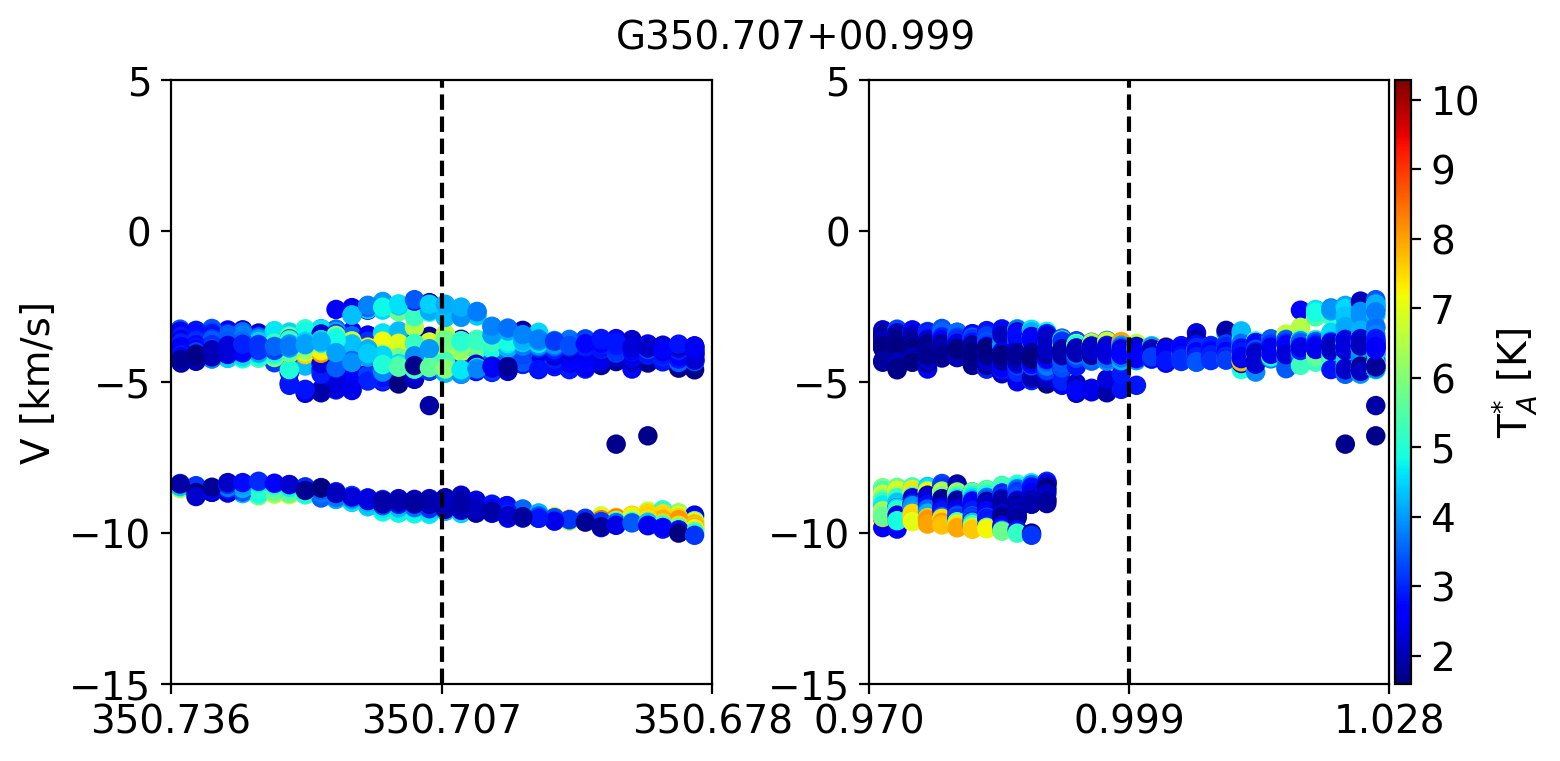}
    \includegraphics[width=0.45\linewidth]{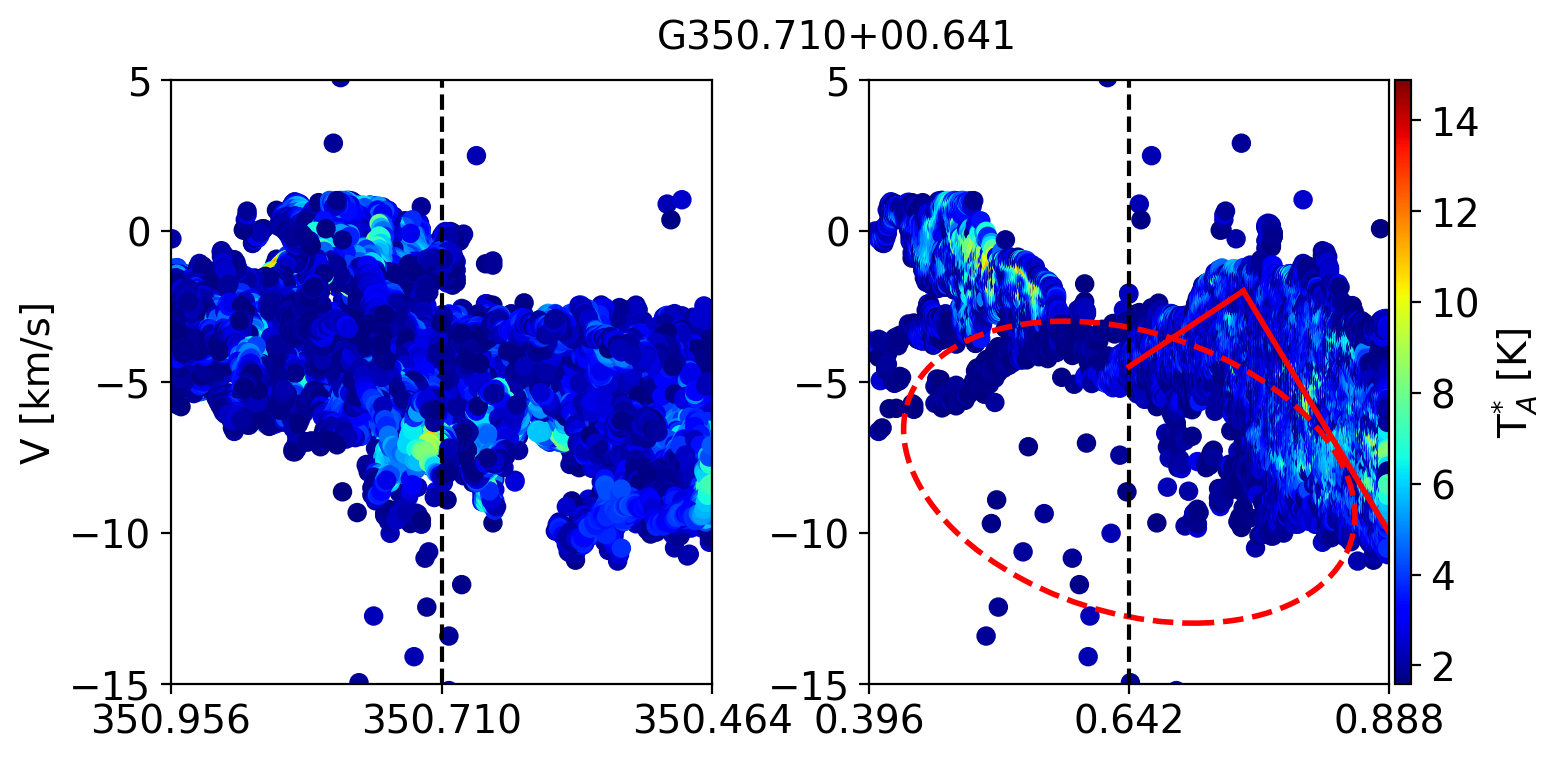}
    \includegraphics[width=0.45\linewidth]{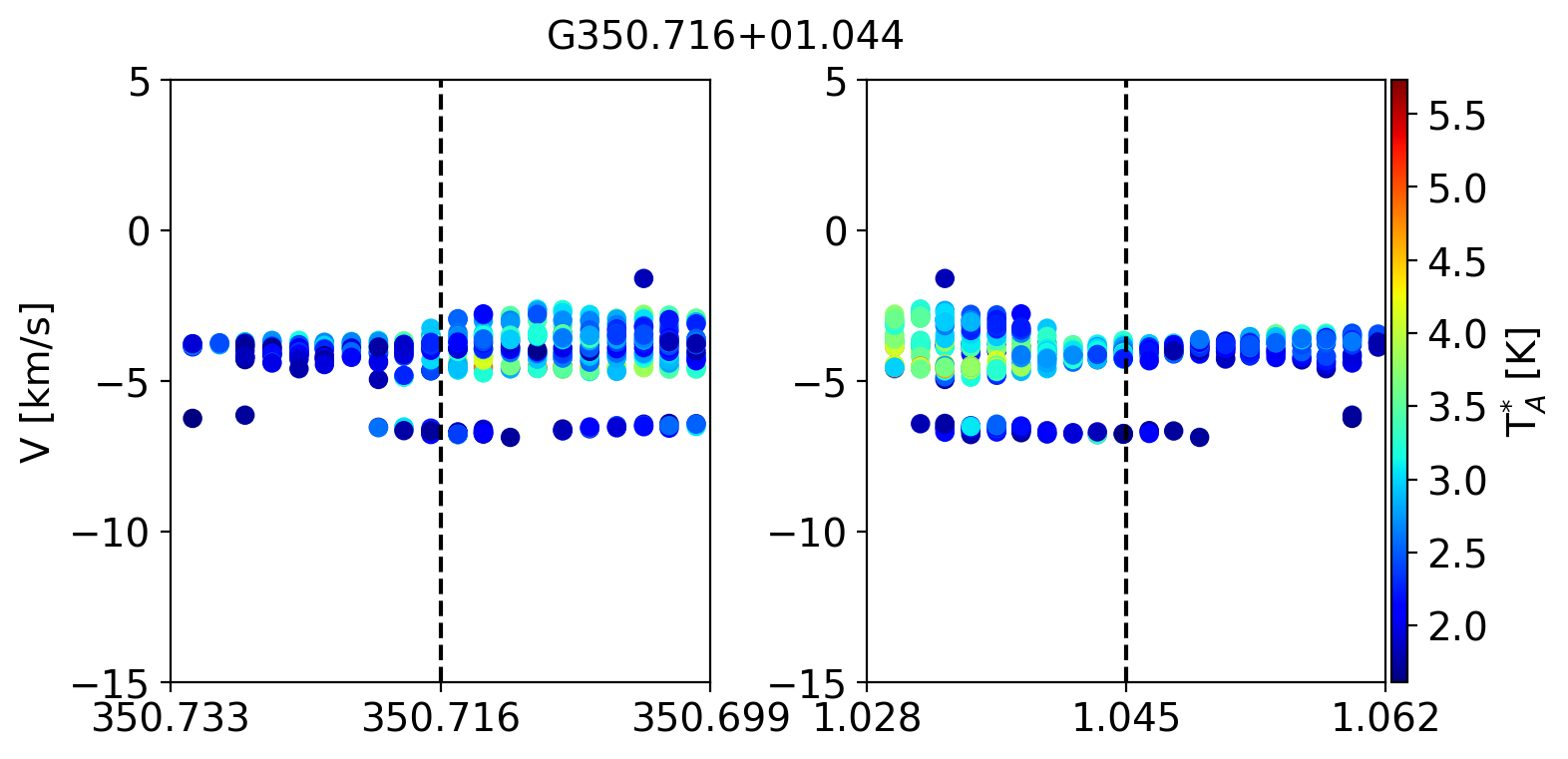}
    \caption{Longitude-velocity ($lv$) and latitude-velocity ($bv$) plots toward the H II sources. Color bars shows Gaussian peak intensity (T$^{*}_{A}$ [K]). Ellipse in red color (if present) indicate visual fit to the velocity structure to infer expansion velocities. Red lines are used to indicate tentative V-shapes in the velocity structure. Name of the H II regions are given on top of each pair of maps.}
    \label{fig:lvplots_hiisources0}
\end{figure}

\begin{figure}[htbp!]
    \centering
    \includegraphics[width=0.45\linewidth]{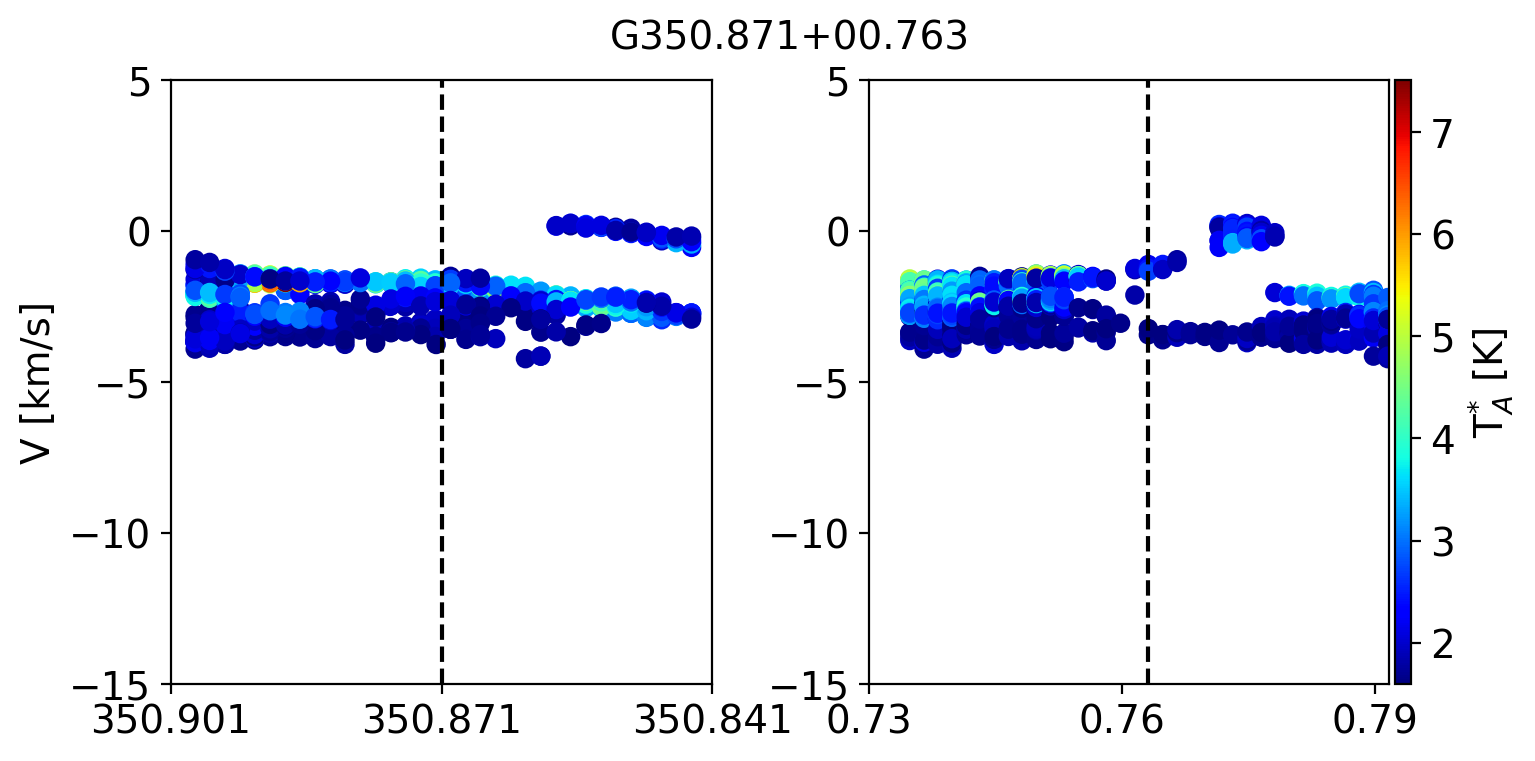}
    \includegraphics[width=0.45\linewidth]{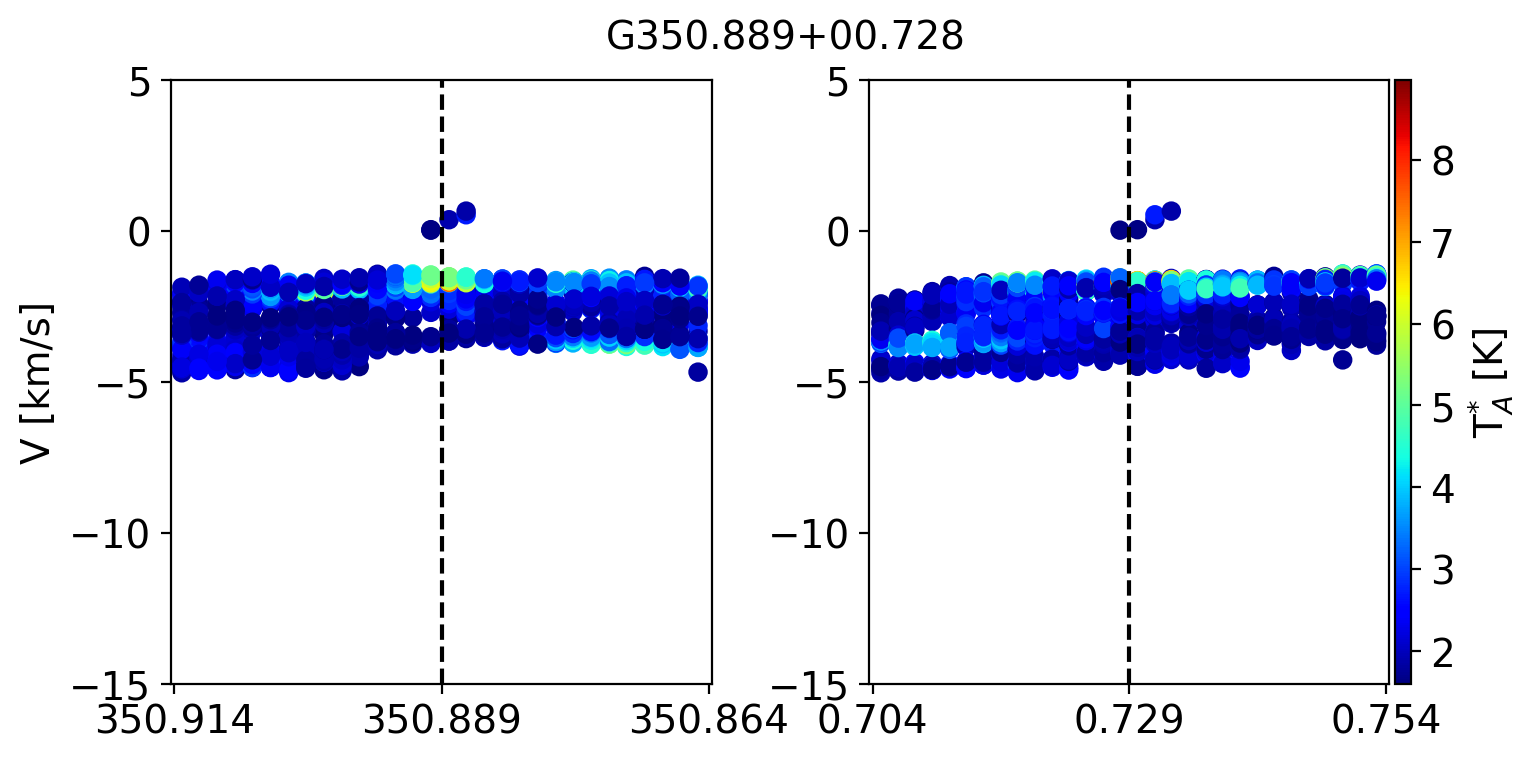}
    \includegraphics[width=0.45\linewidth]{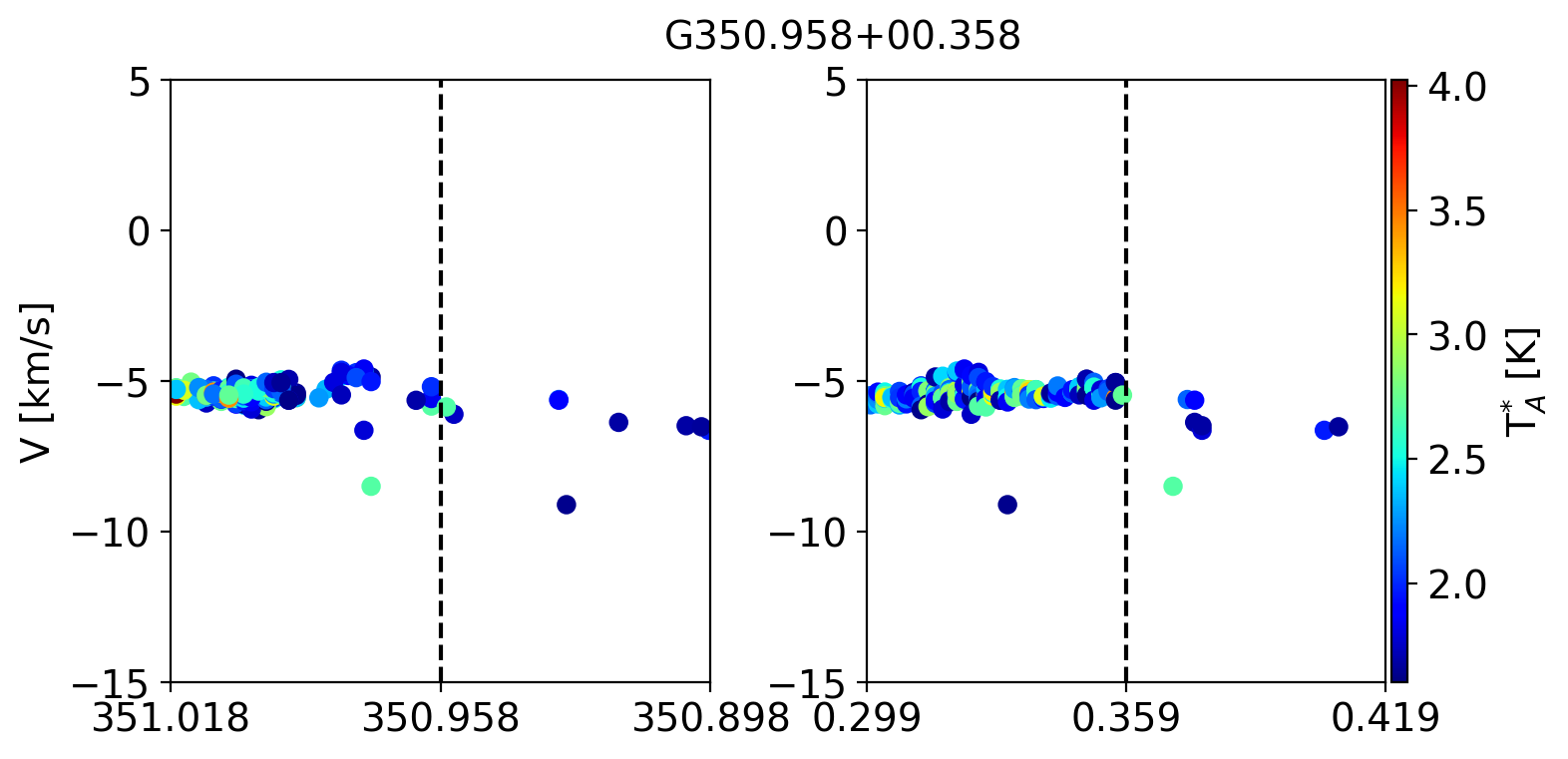}
    \includegraphics[width=0.45\linewidth]{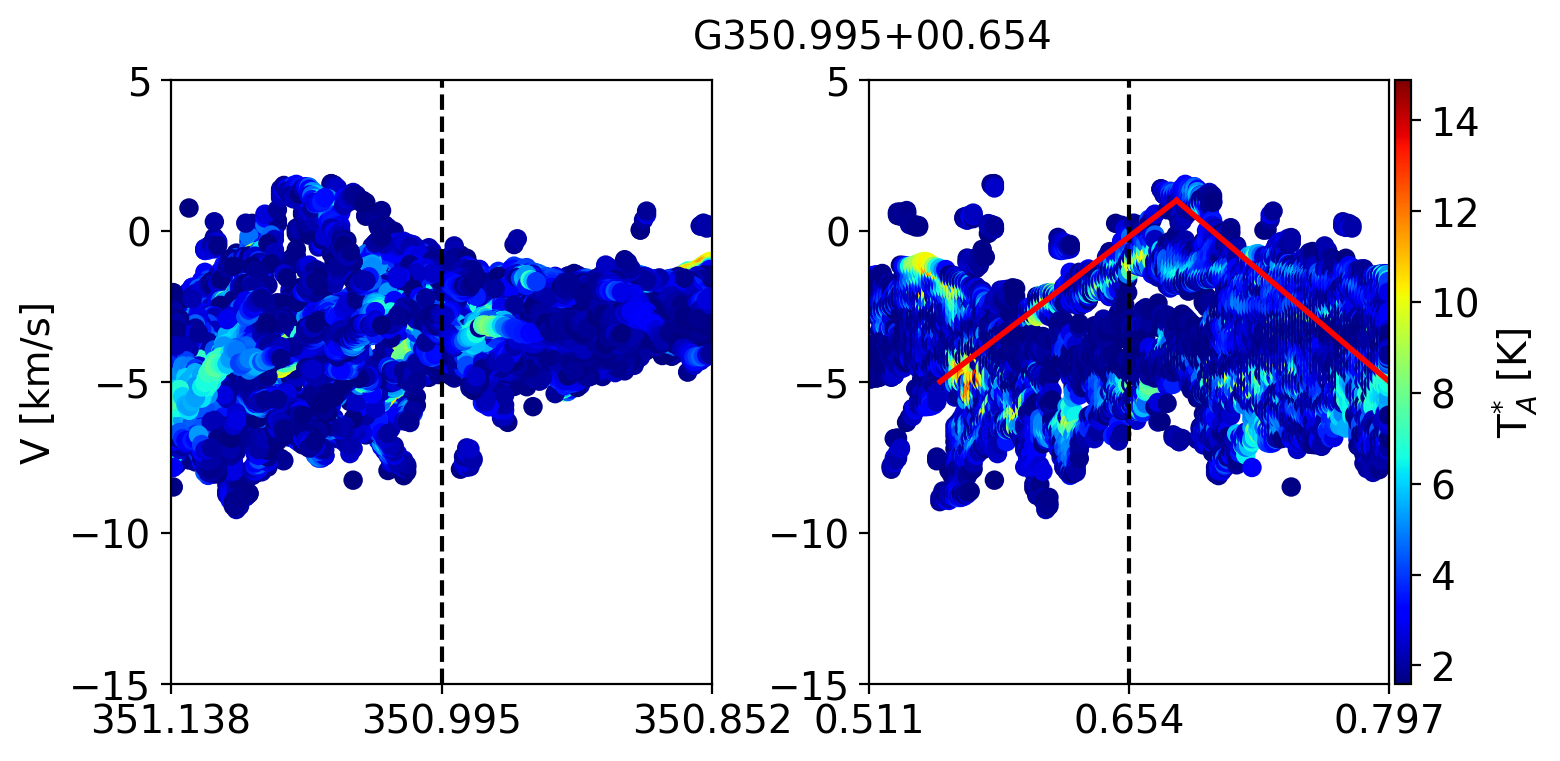}
    \includegraphics[width=0.45\linewidth]{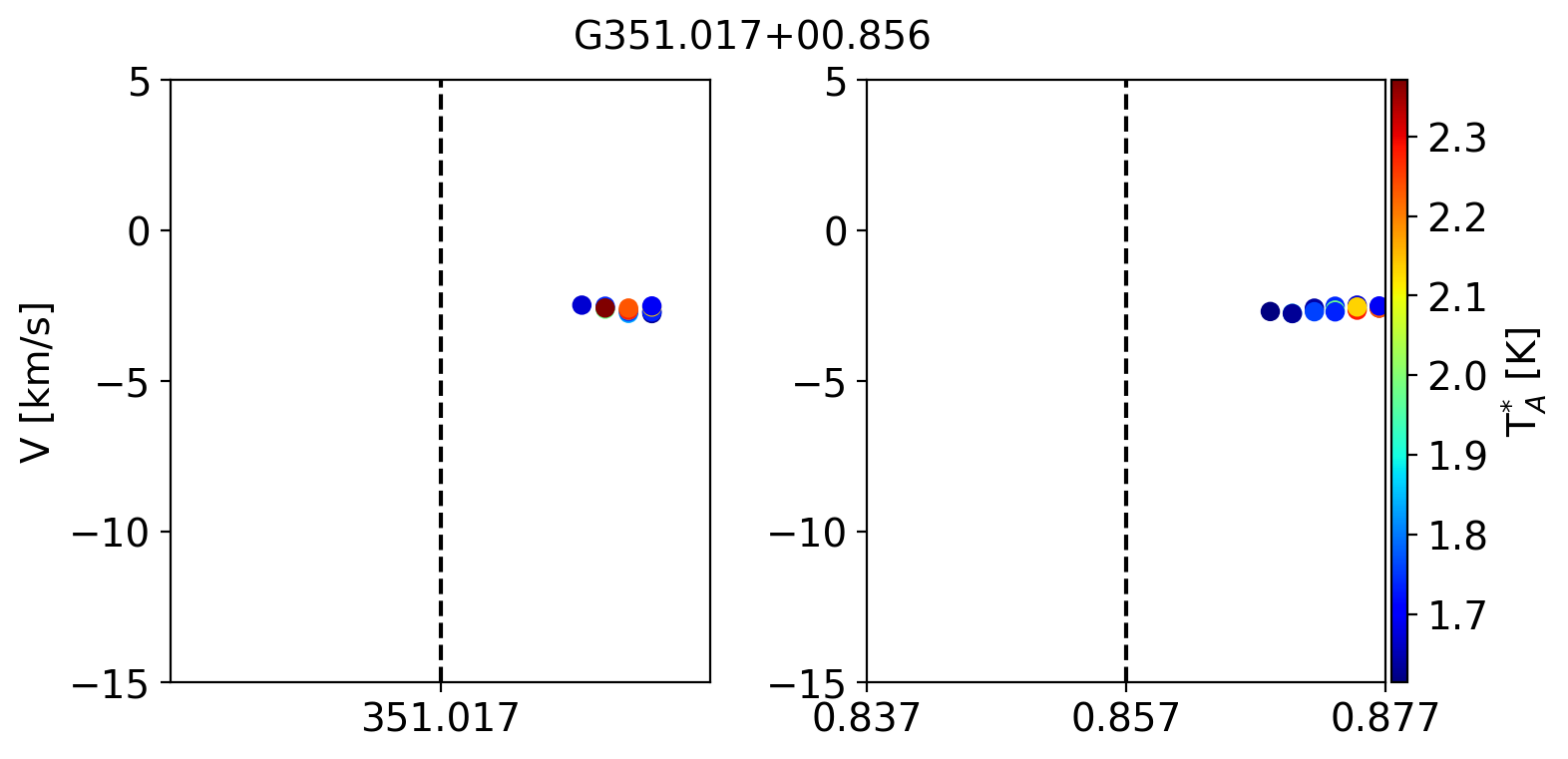}
    \includegraphics[width=0.45\linewidth]{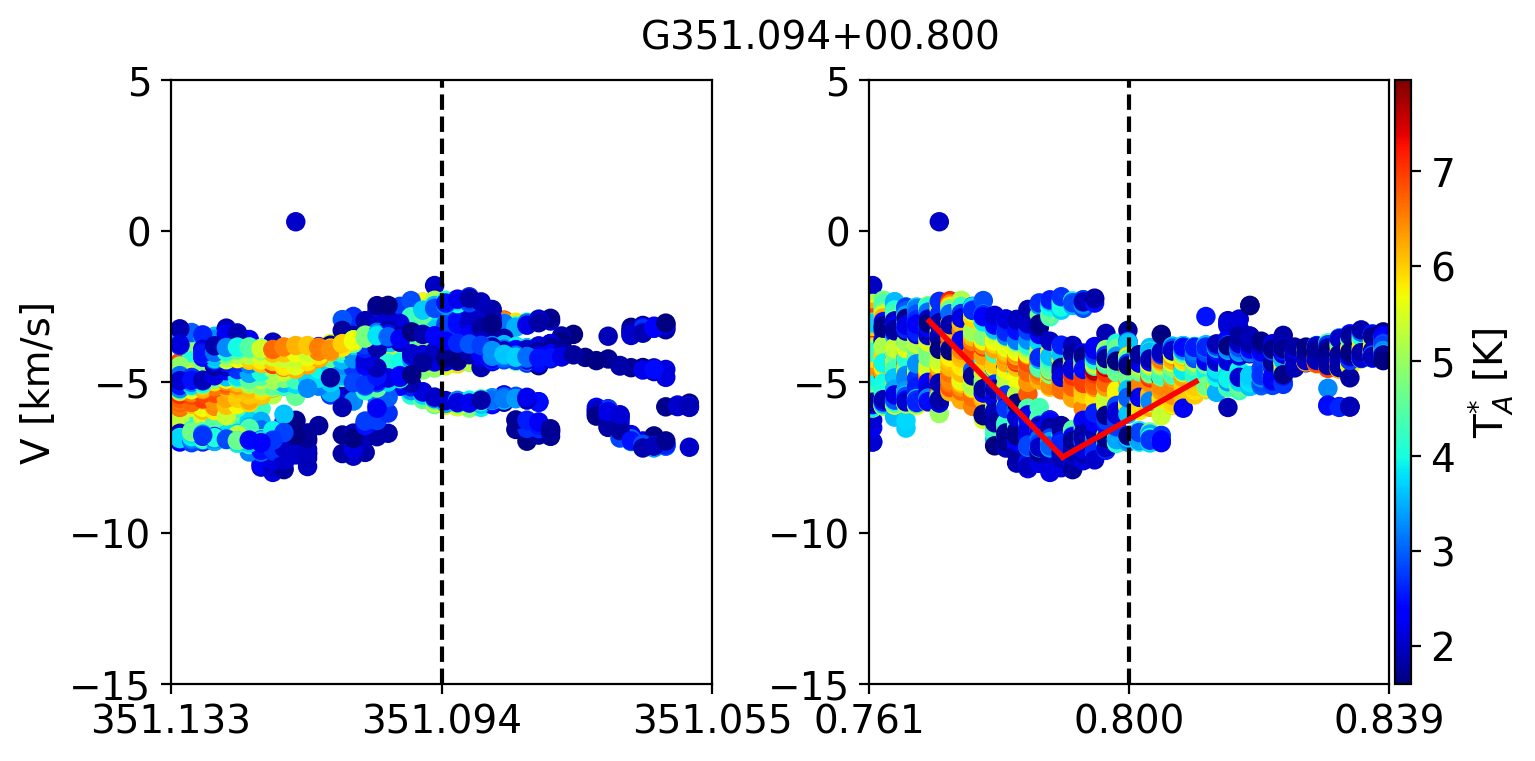}
    \includegraphics[width=0.45\linewidth]{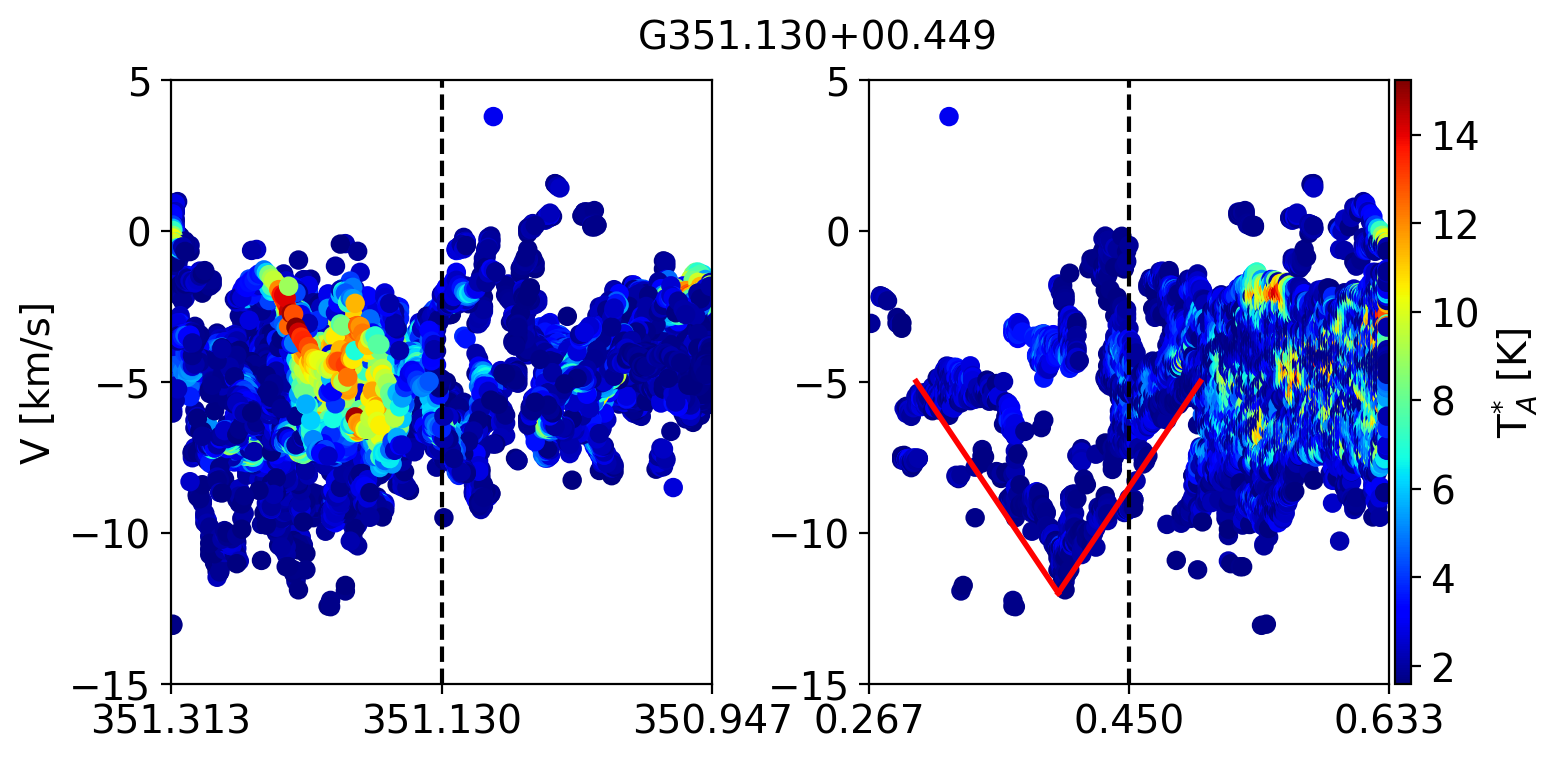}
    \includegraphics[width=0.45\linewidth]{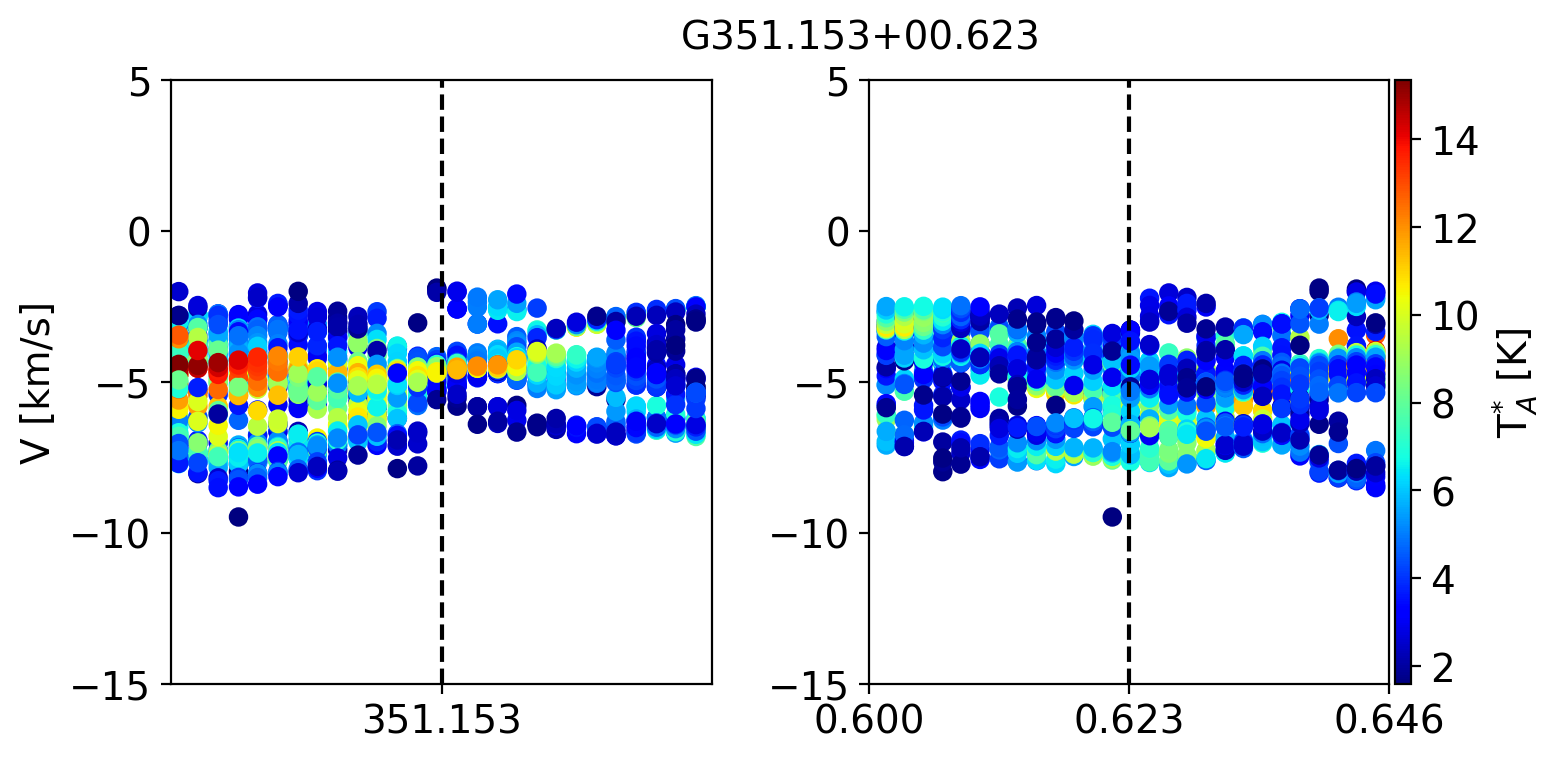}
    \includegraphics[width=0.45\linewidth]{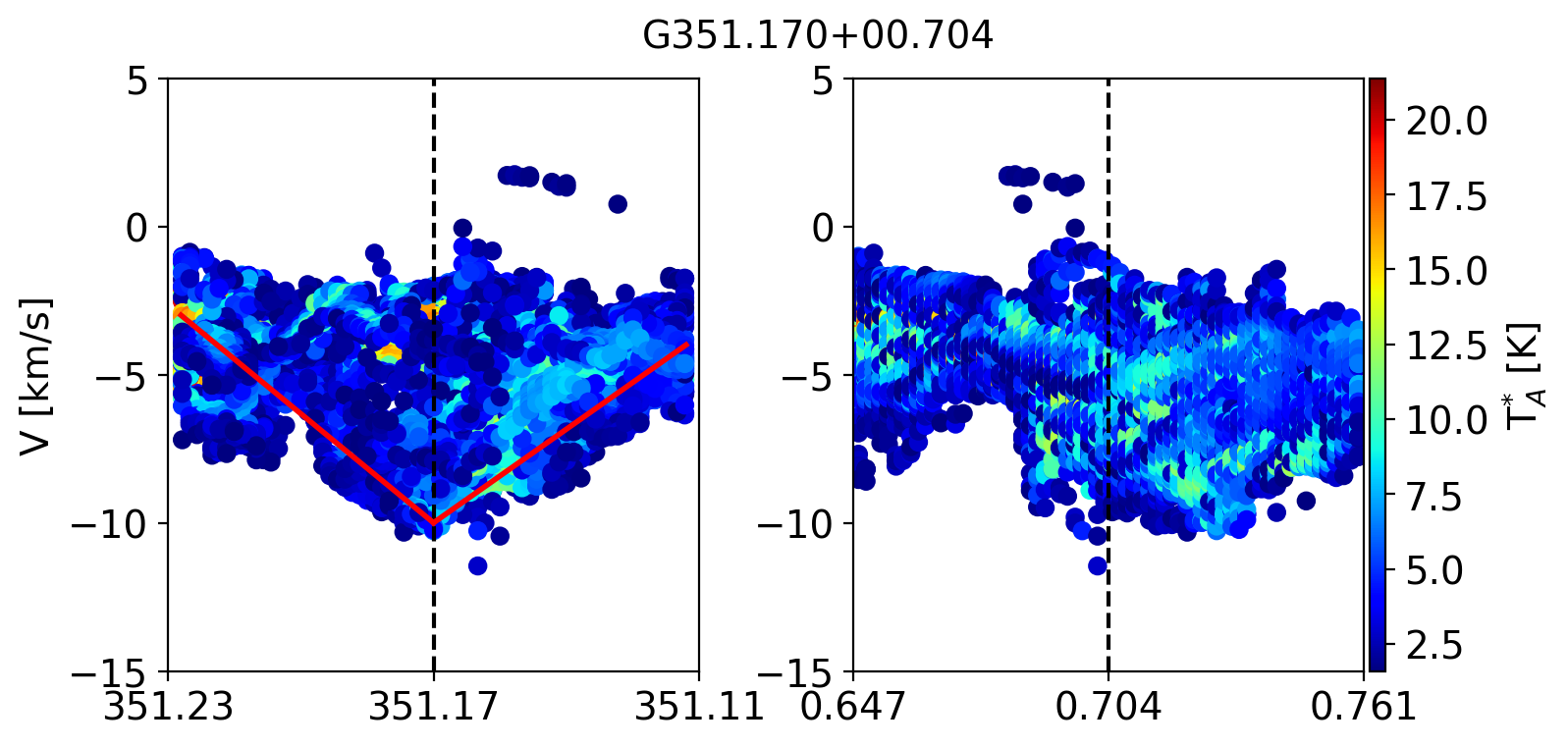}
    \includegraphics[width=0.45\linewidth]{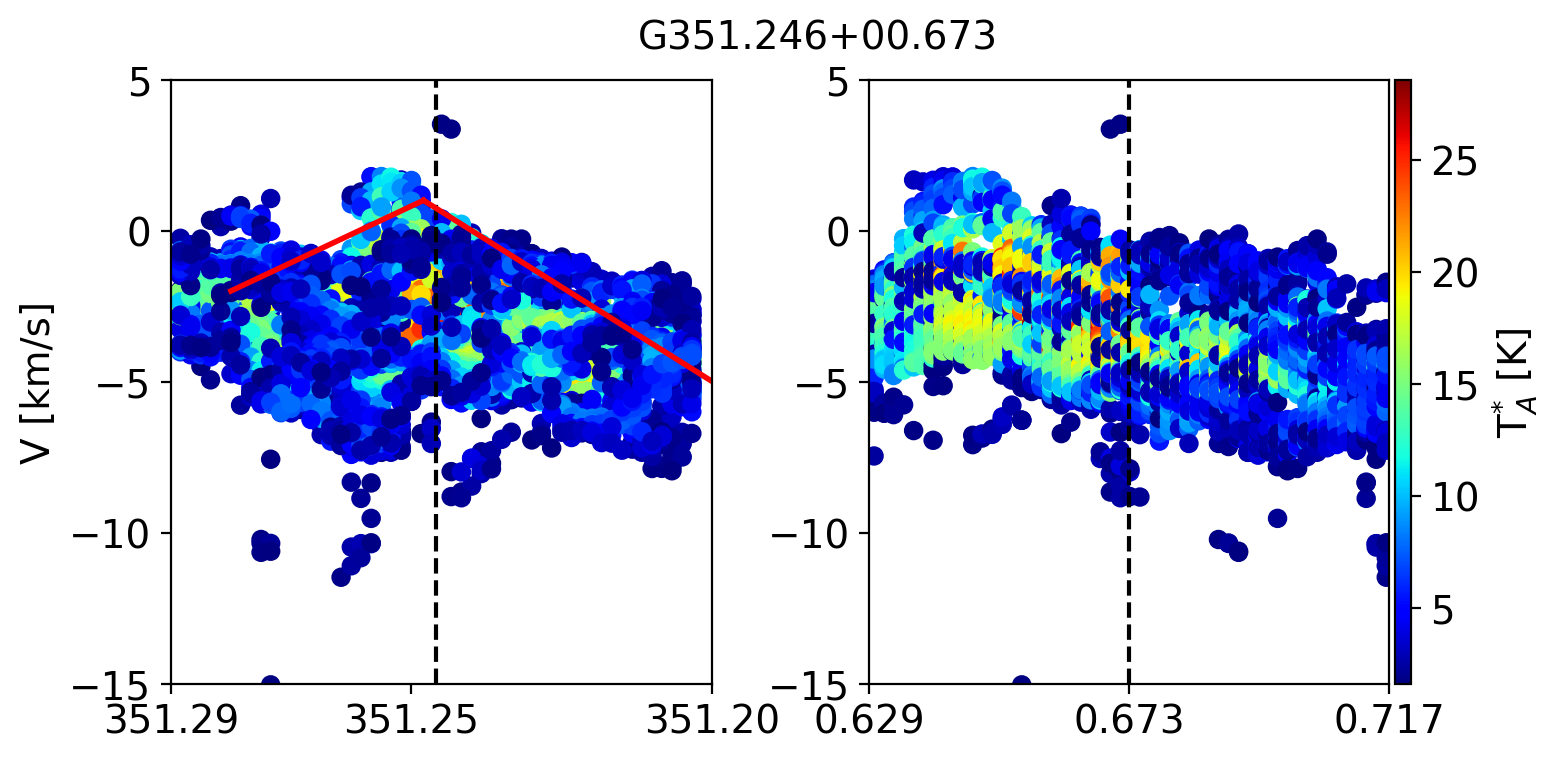}
    \caption{Same as Figure \ref{fig:lvplots_hiisources0}}
    \label{fig:lvplots_hiisources4}
\end{figure}

\begin{figure}[htbp!]
    \centering
    \includegraphics[width=0.45\linewidth]{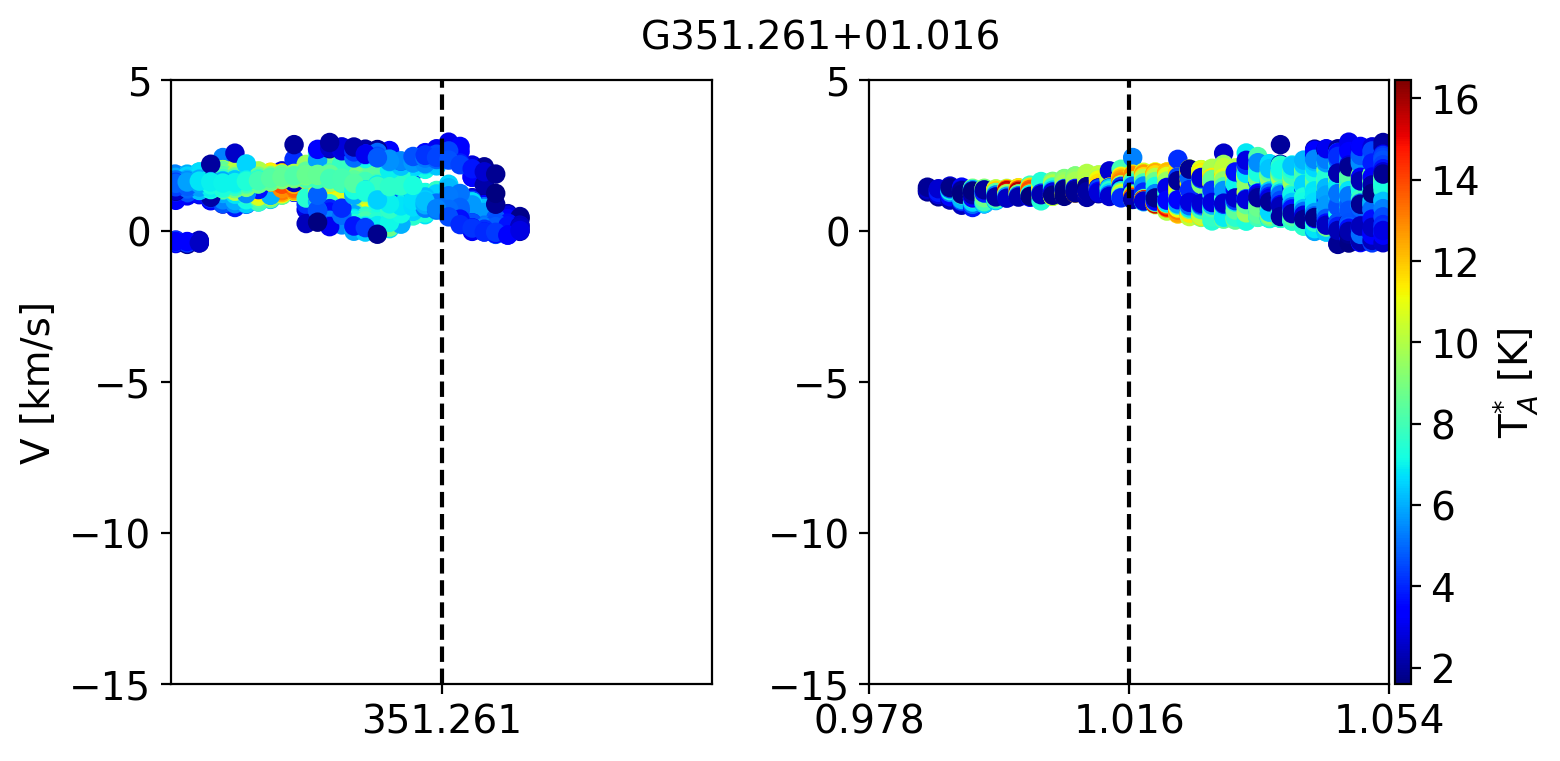}
    \includegraphics[width=0.45\linewidth]{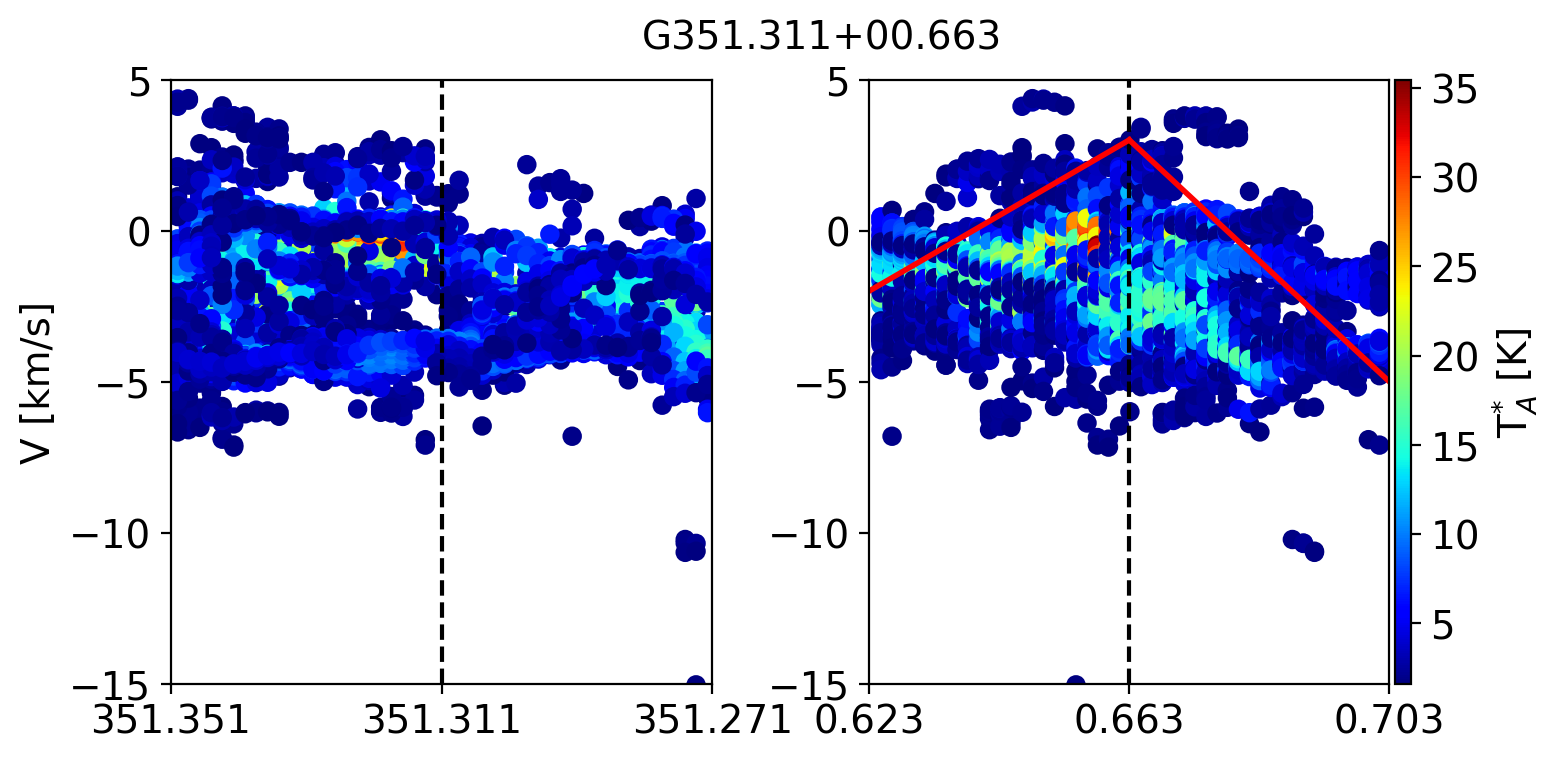}
    \includegraphics[width=0.45\linewidth]{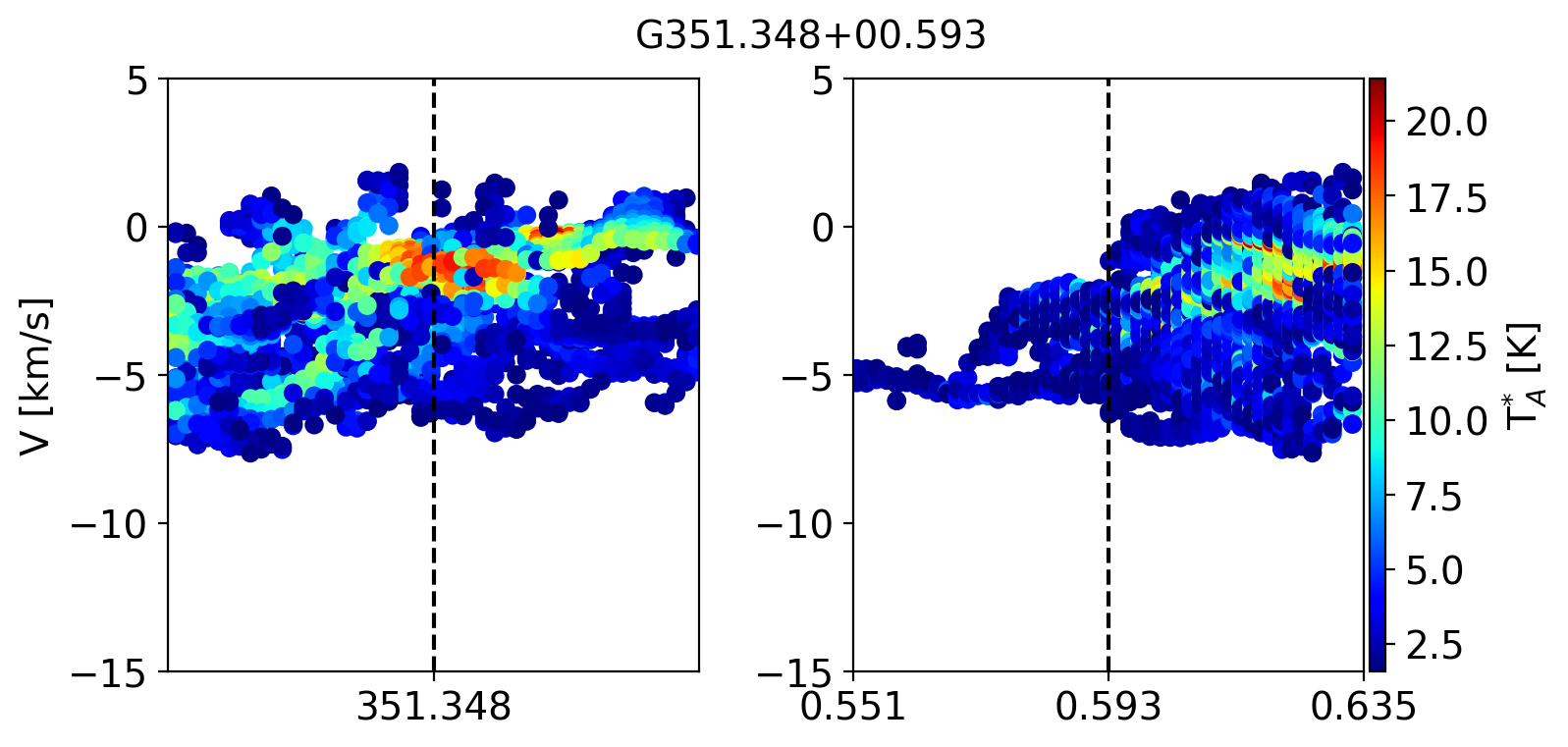}
    \includegraphics[width=0.45\linewidth]{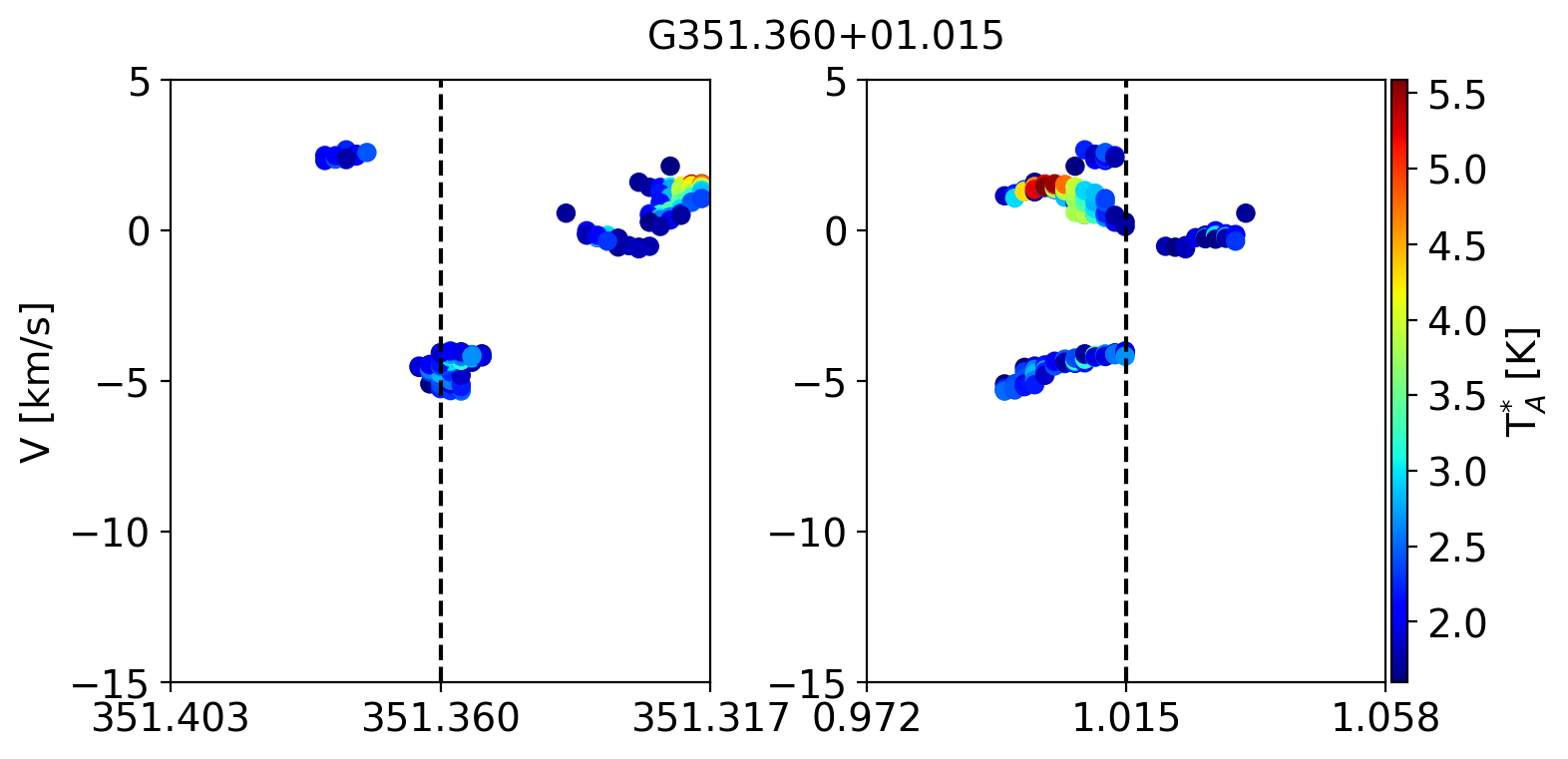}
    \includegraphics[width=0.45\linewidth]{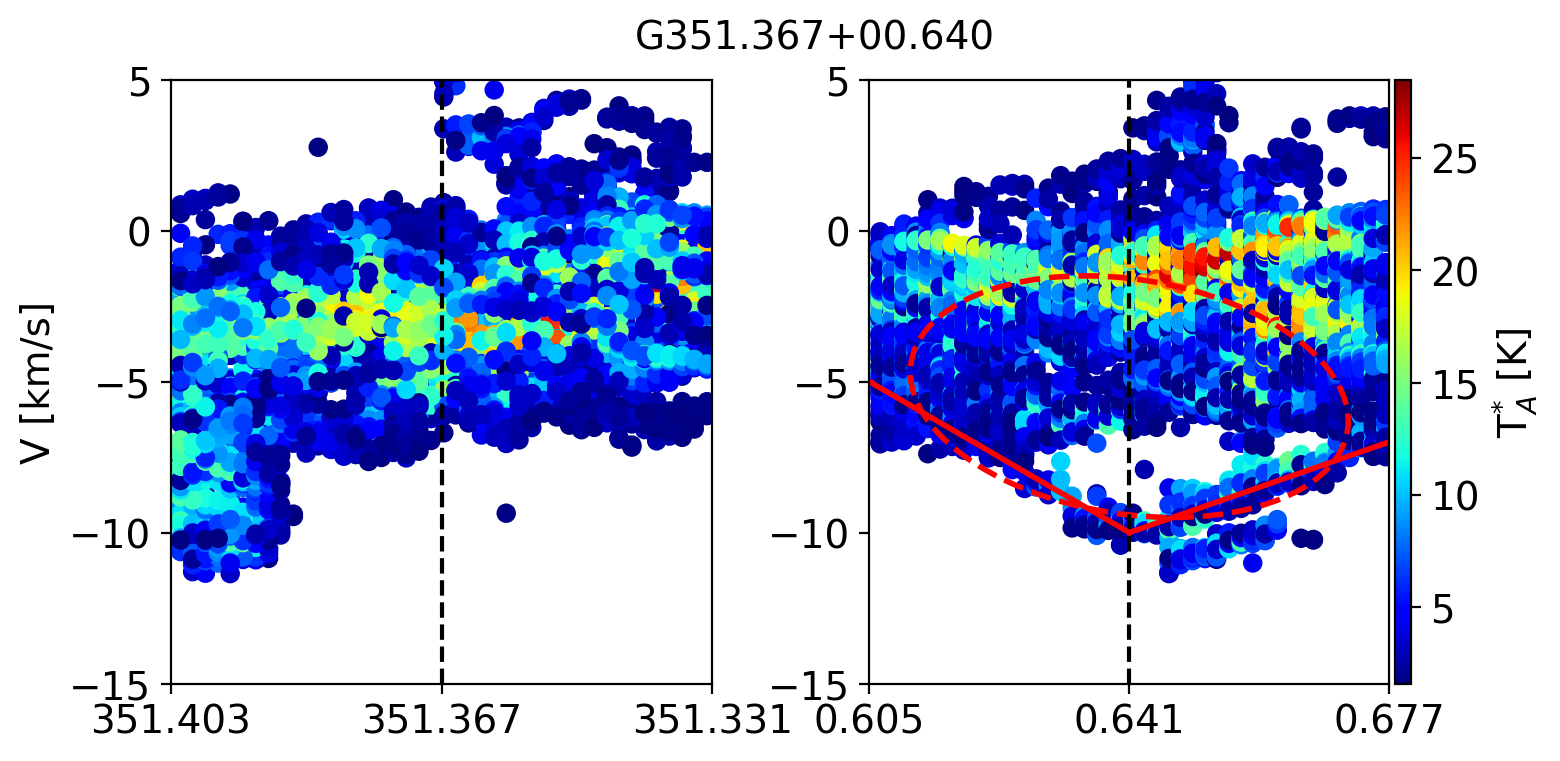}
    \includegraphics[width=0.45\linewidth]{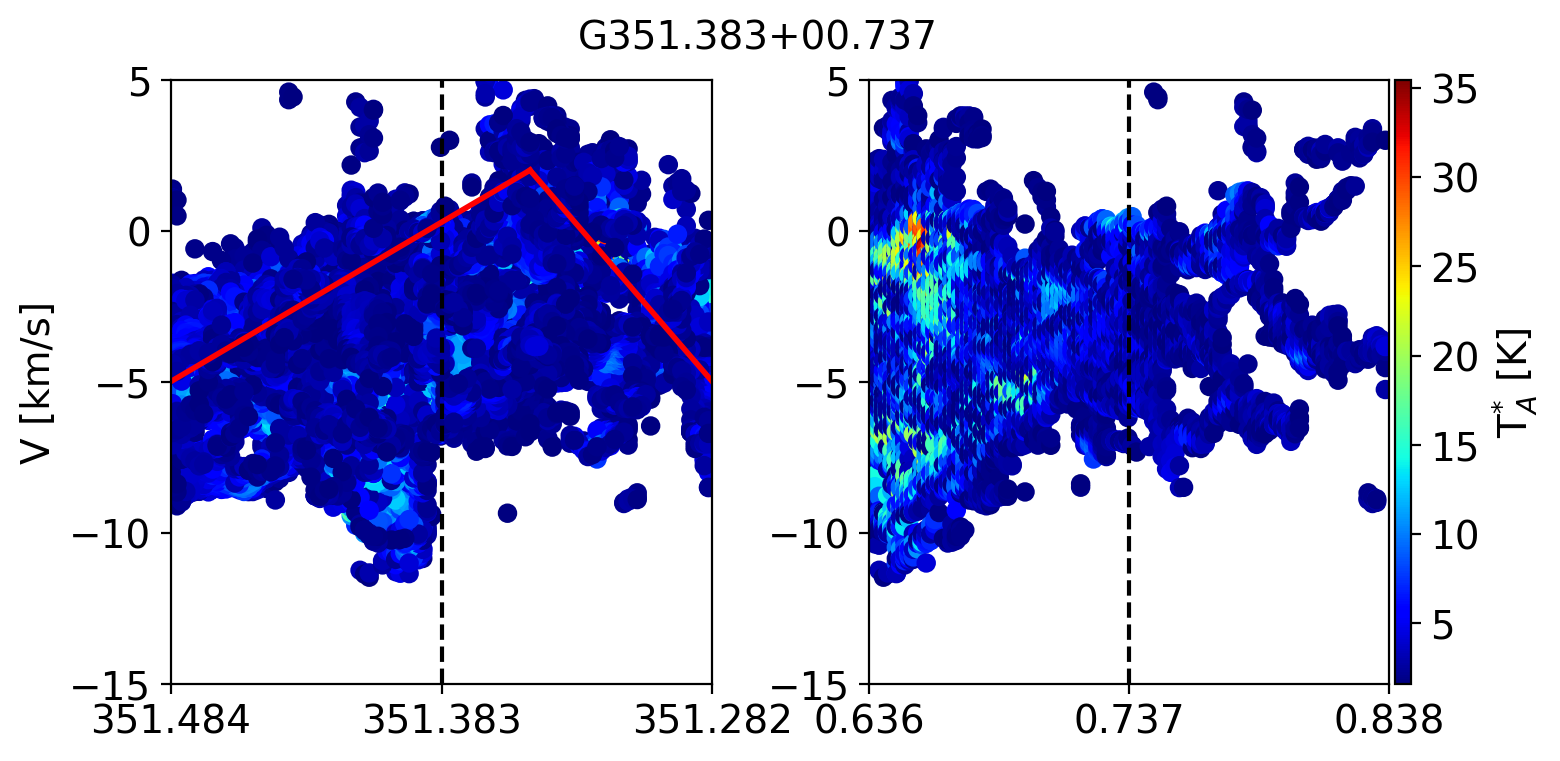}
    \includegraphics[width=0.45\linewidth]{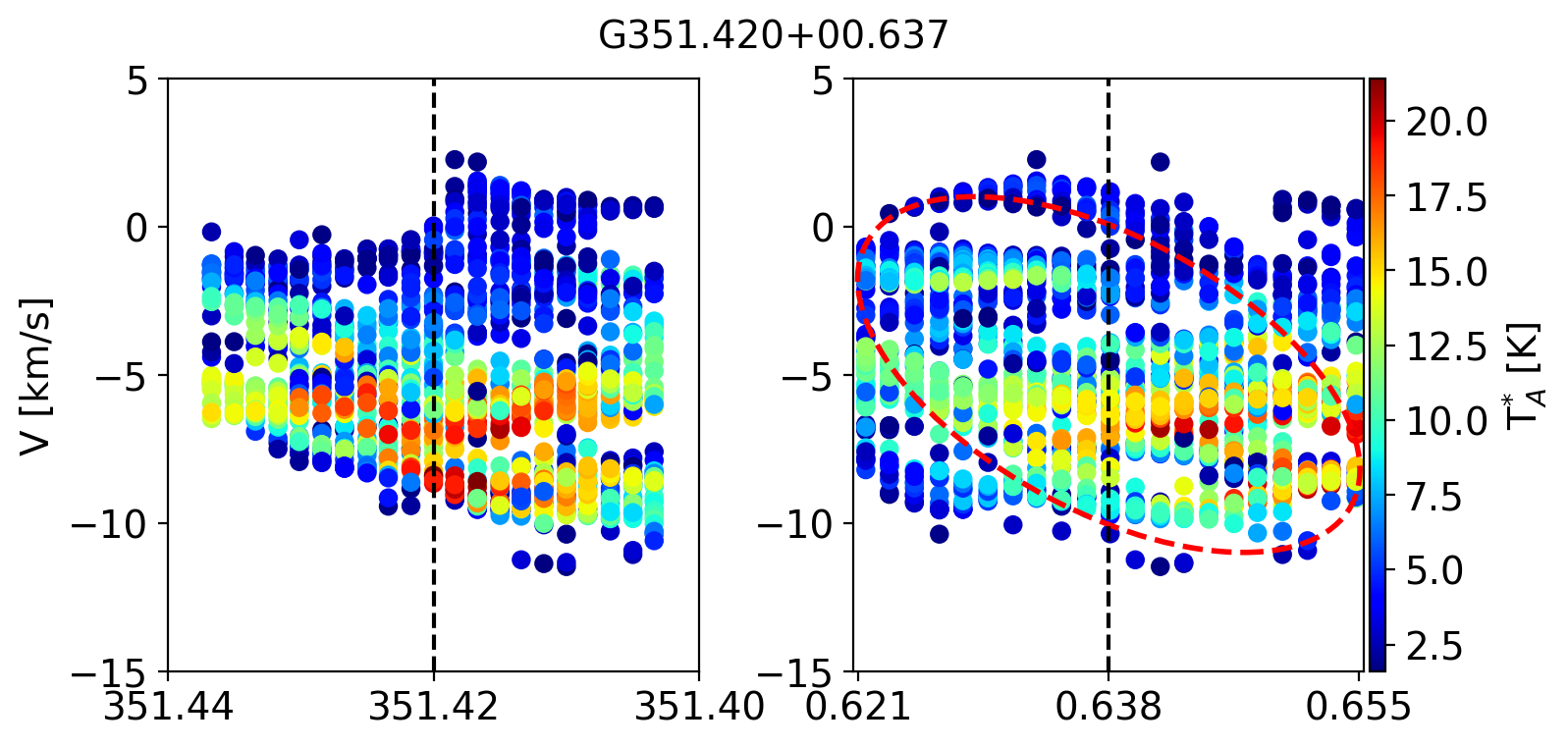}
    \includegraphics[width=0.45\linewidth]{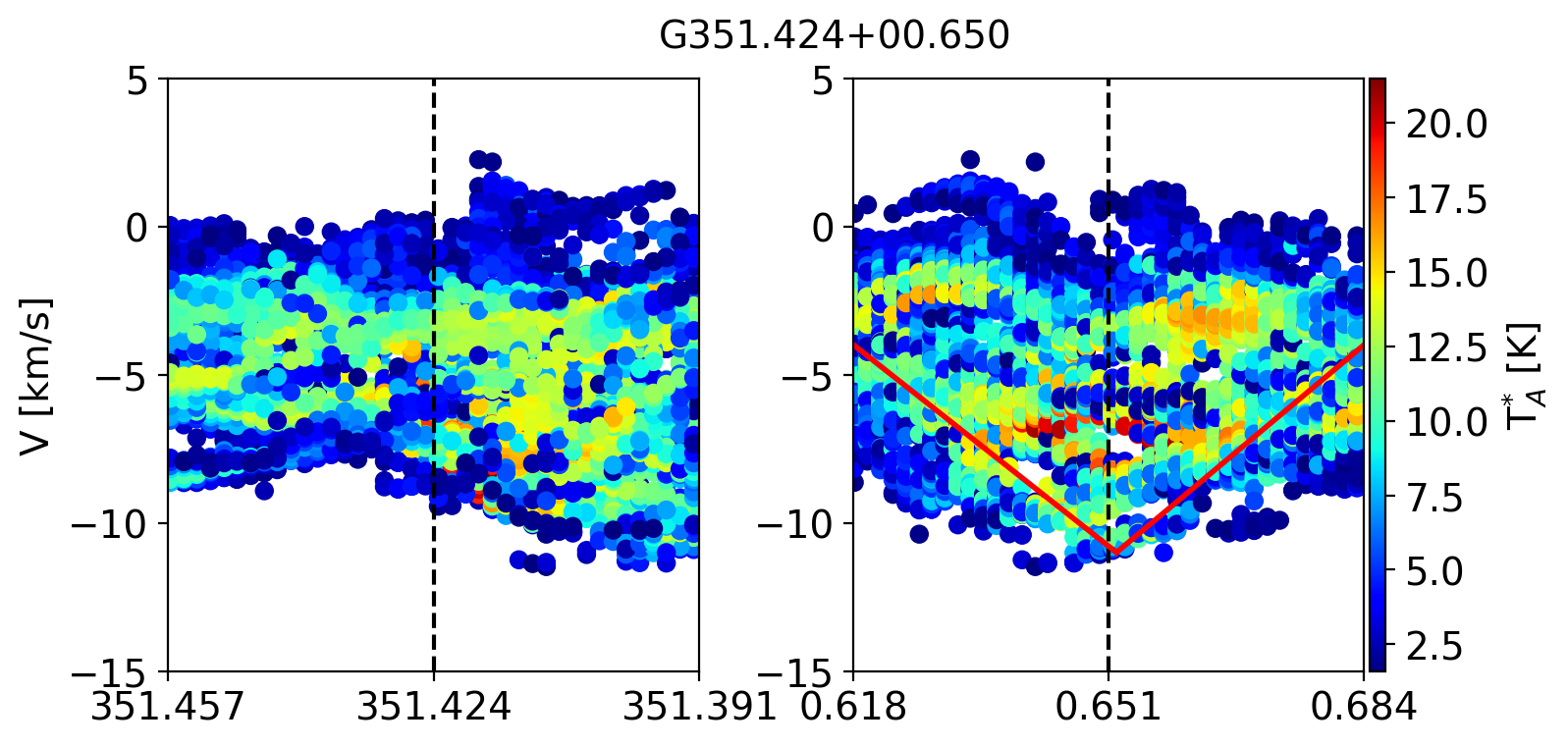}
    \includegraphics[width=0.45\linewidth]{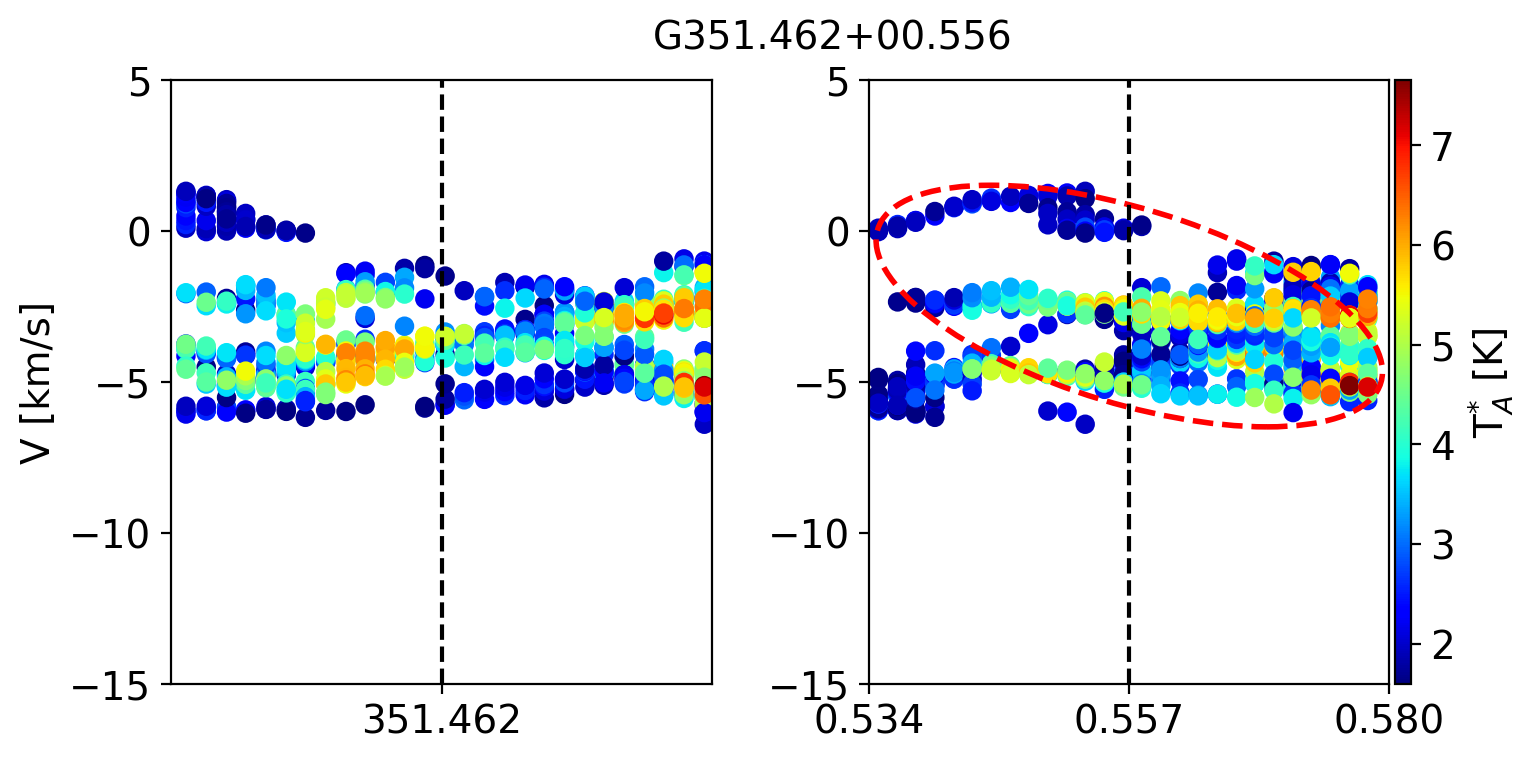}
    \includegraphics[width=0.45\linewidth]{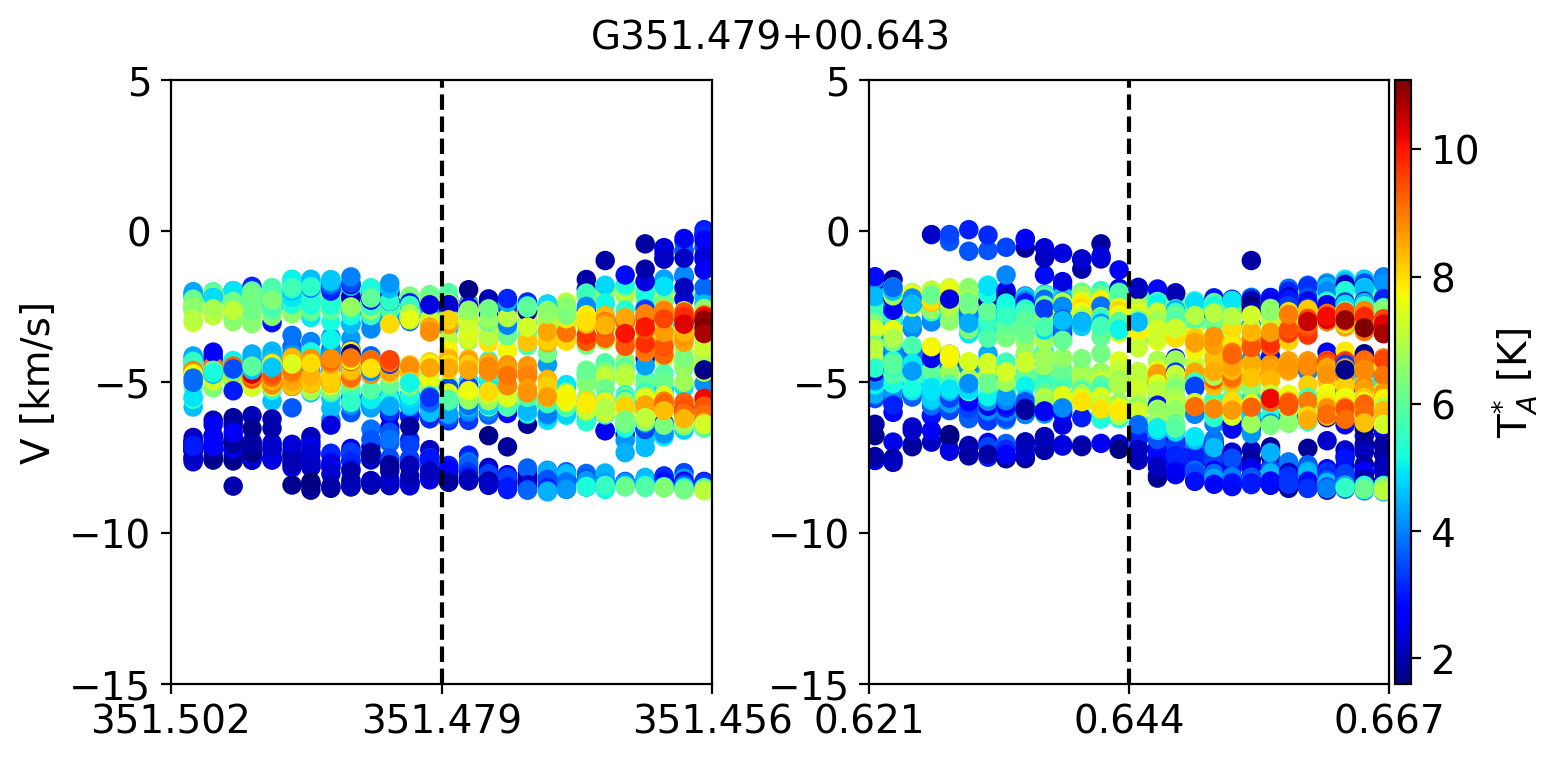}
    \caption{Same as Figure \ref{fig:lvplots_hiisources0}}
    \label{fig:lvplots_hiisources6}
\end{figure}

\begin{figure}[htbp!]
    \centering
    \includegraphics[width=0.45\linewidth]{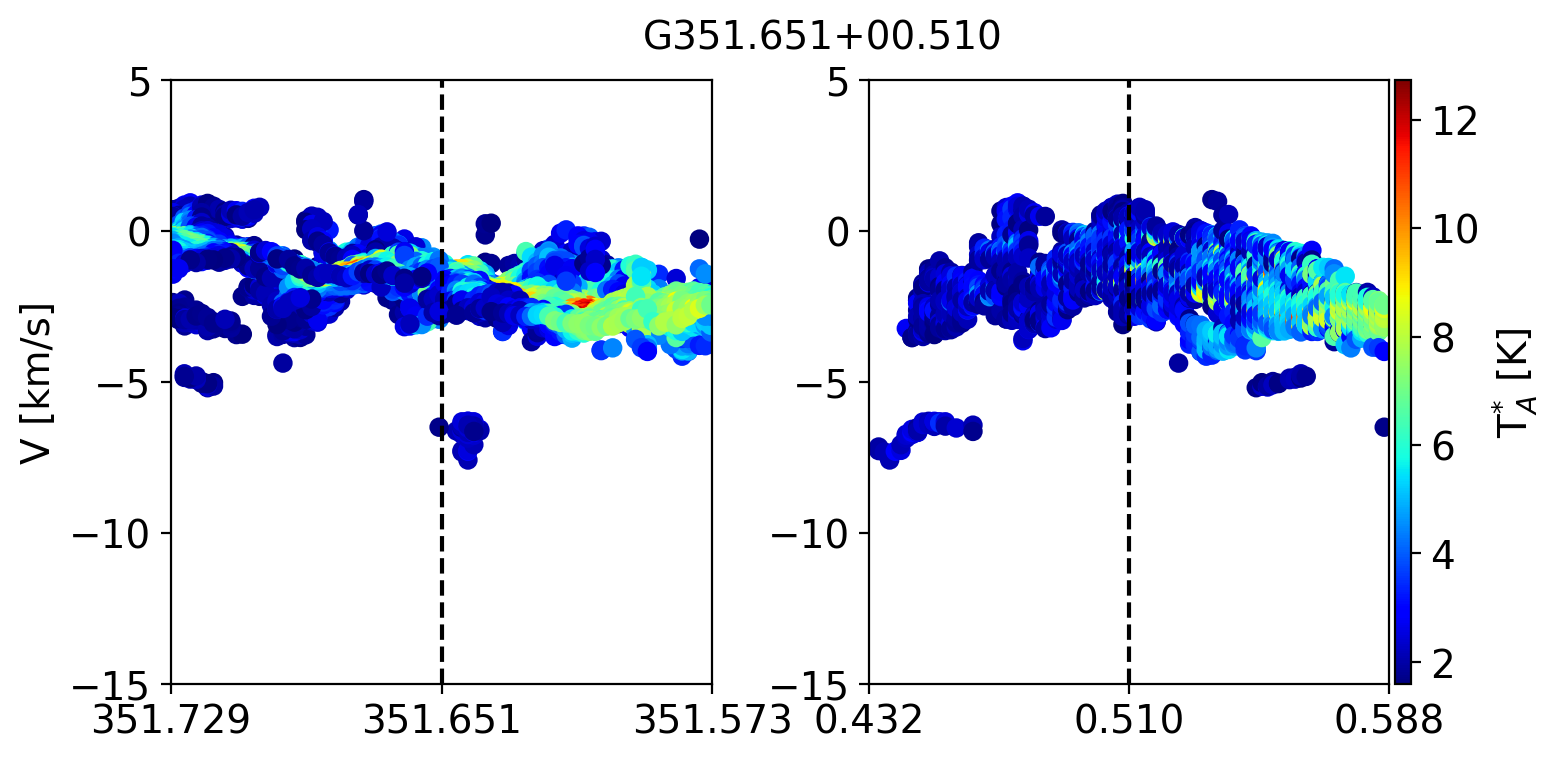}
    \includegraphics[width=0.45\linewidth]{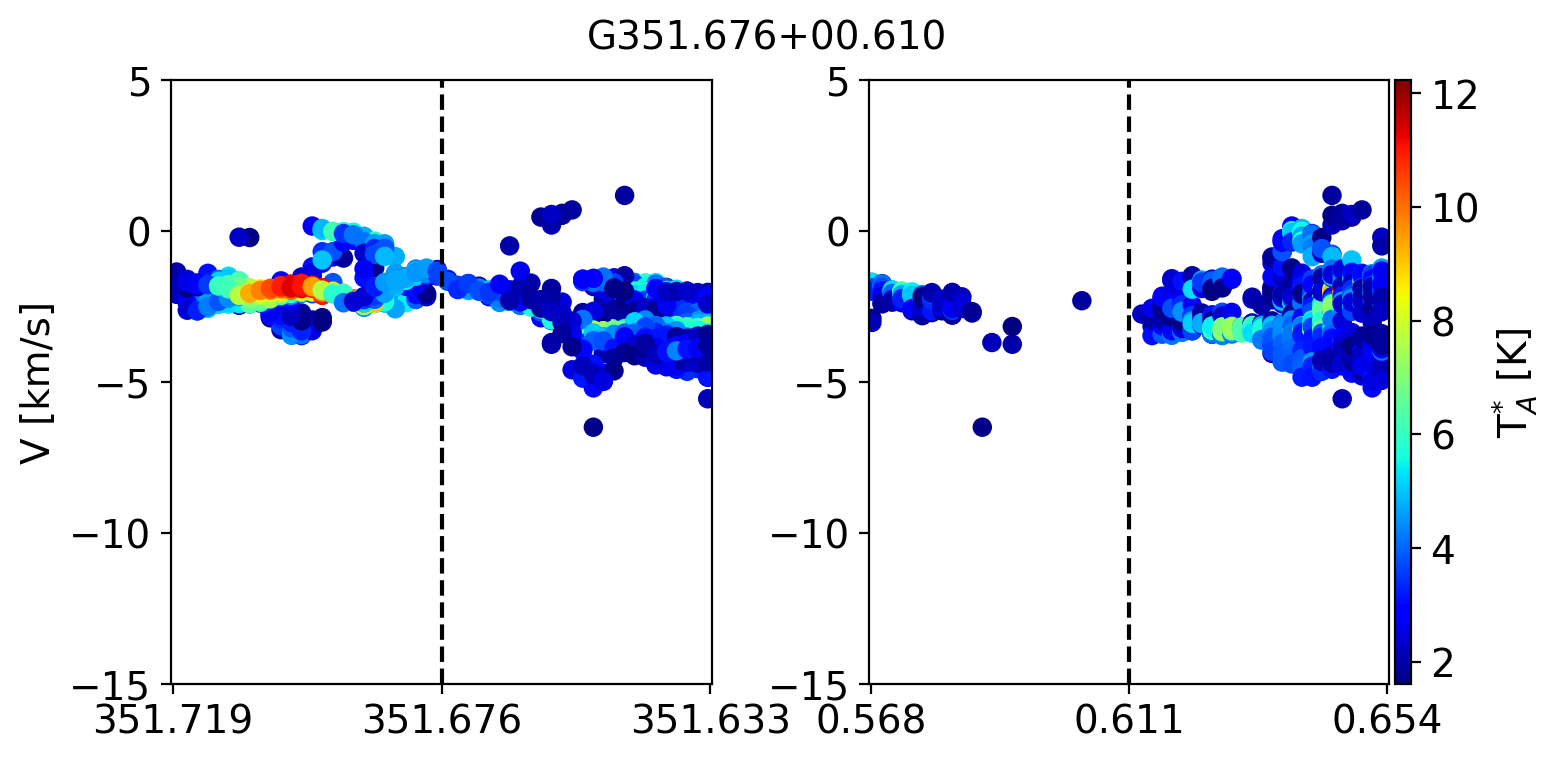}
    \includegraphics[width=0.45\linewidth]{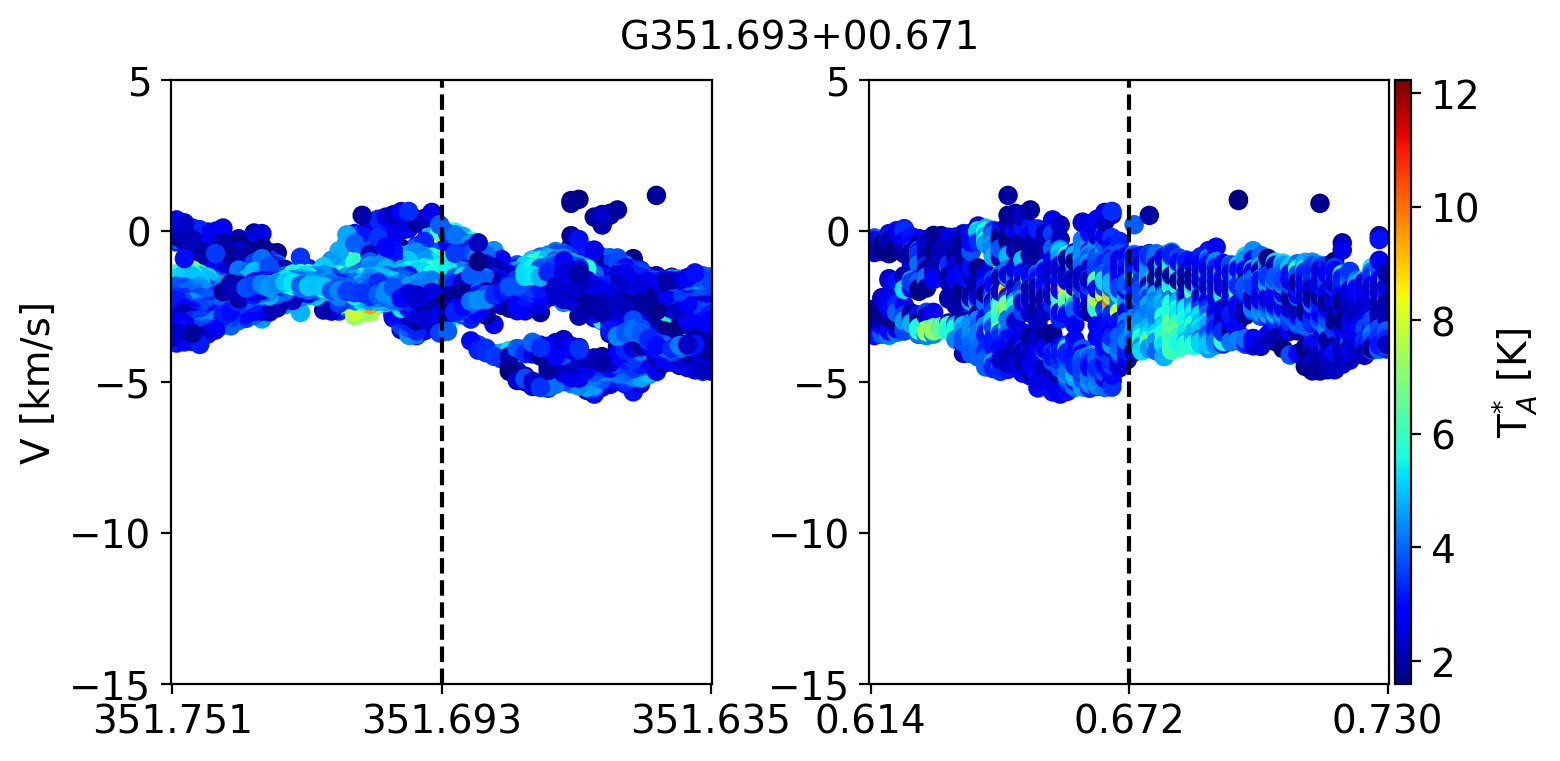}
    \includegraphics[width=0.45\linewidth]{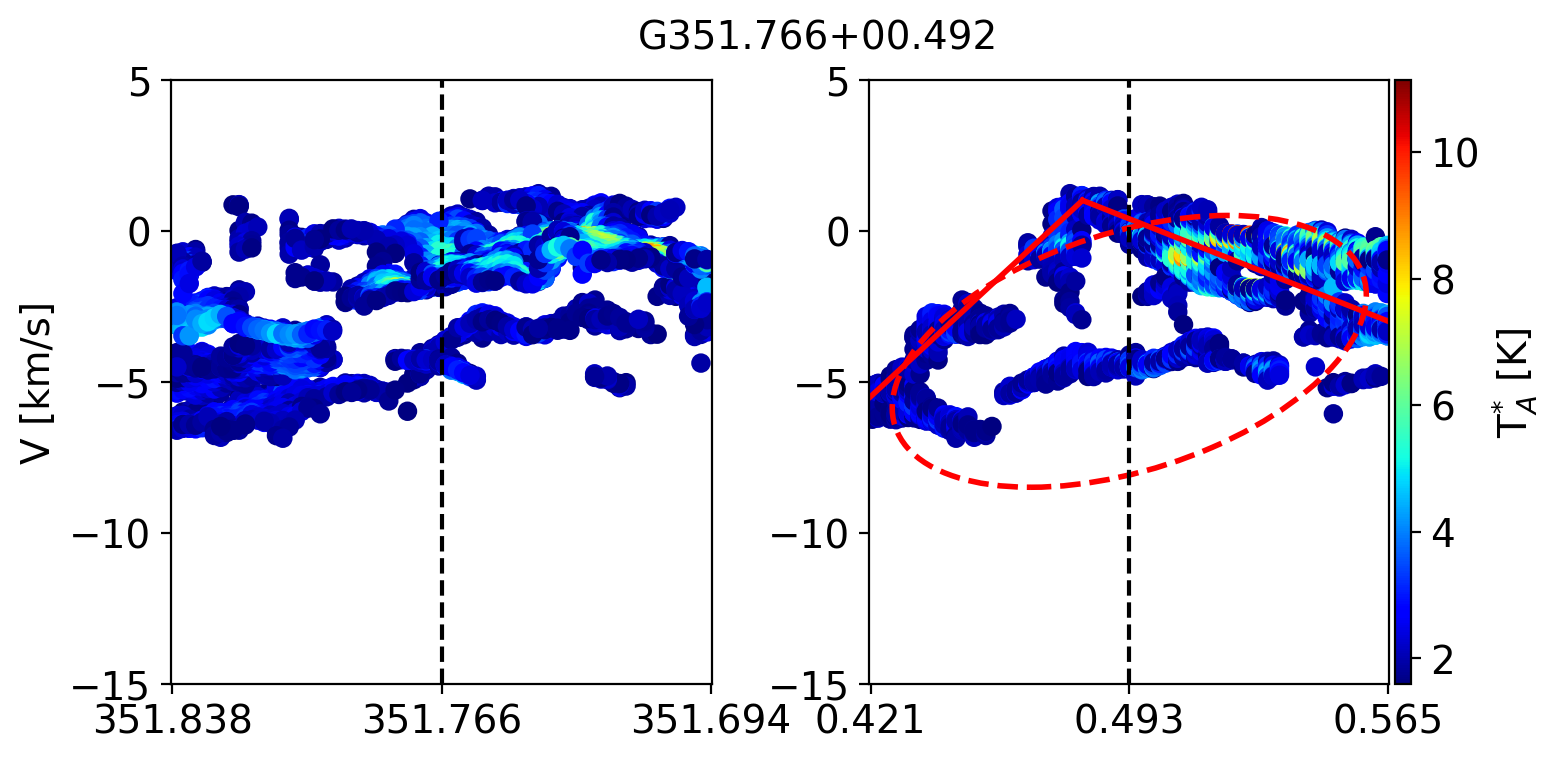}
    \includegraphics[width=0.45\linewidth]{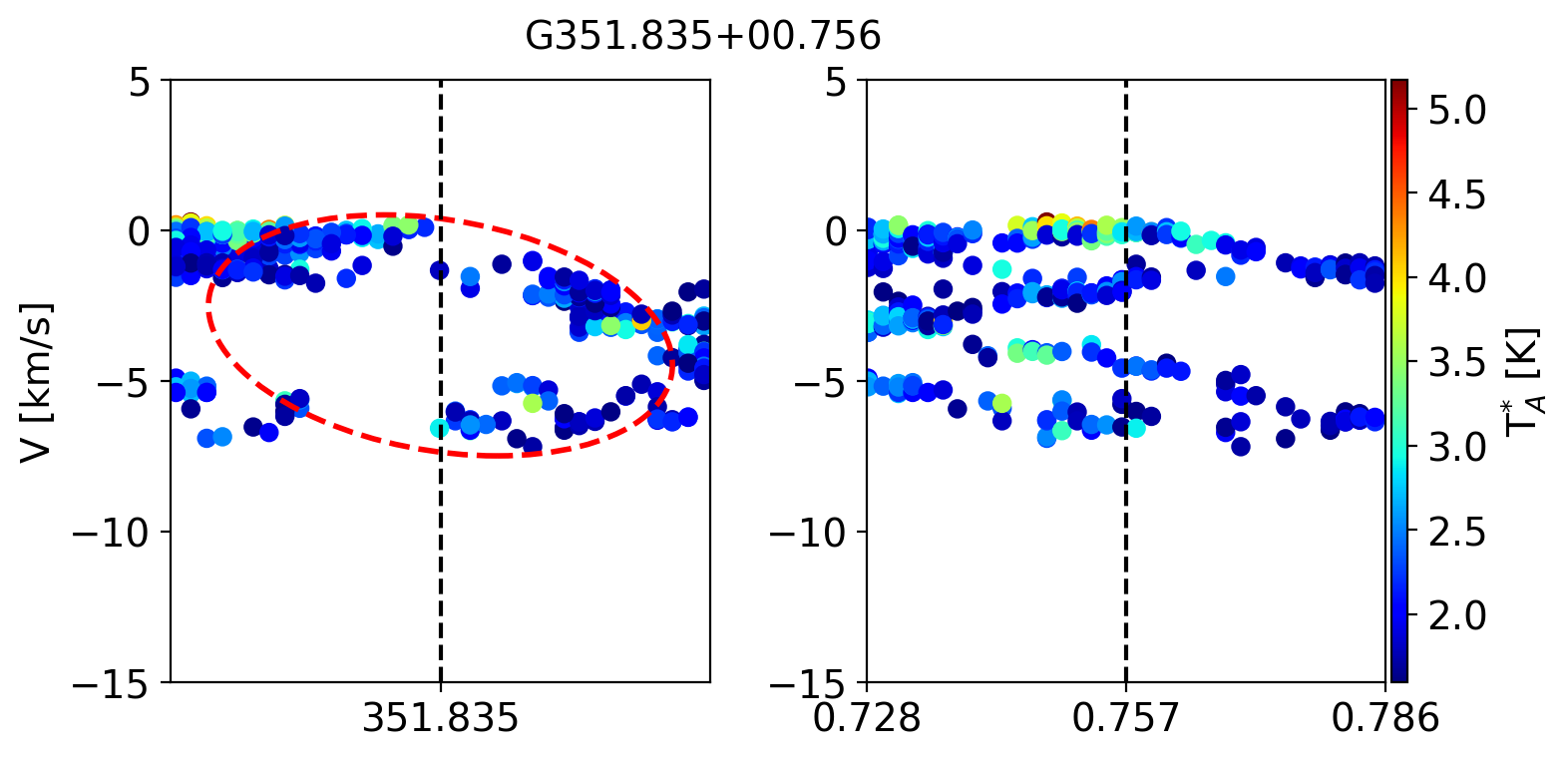}
    \includegraphics[width=0.45\linewidth]{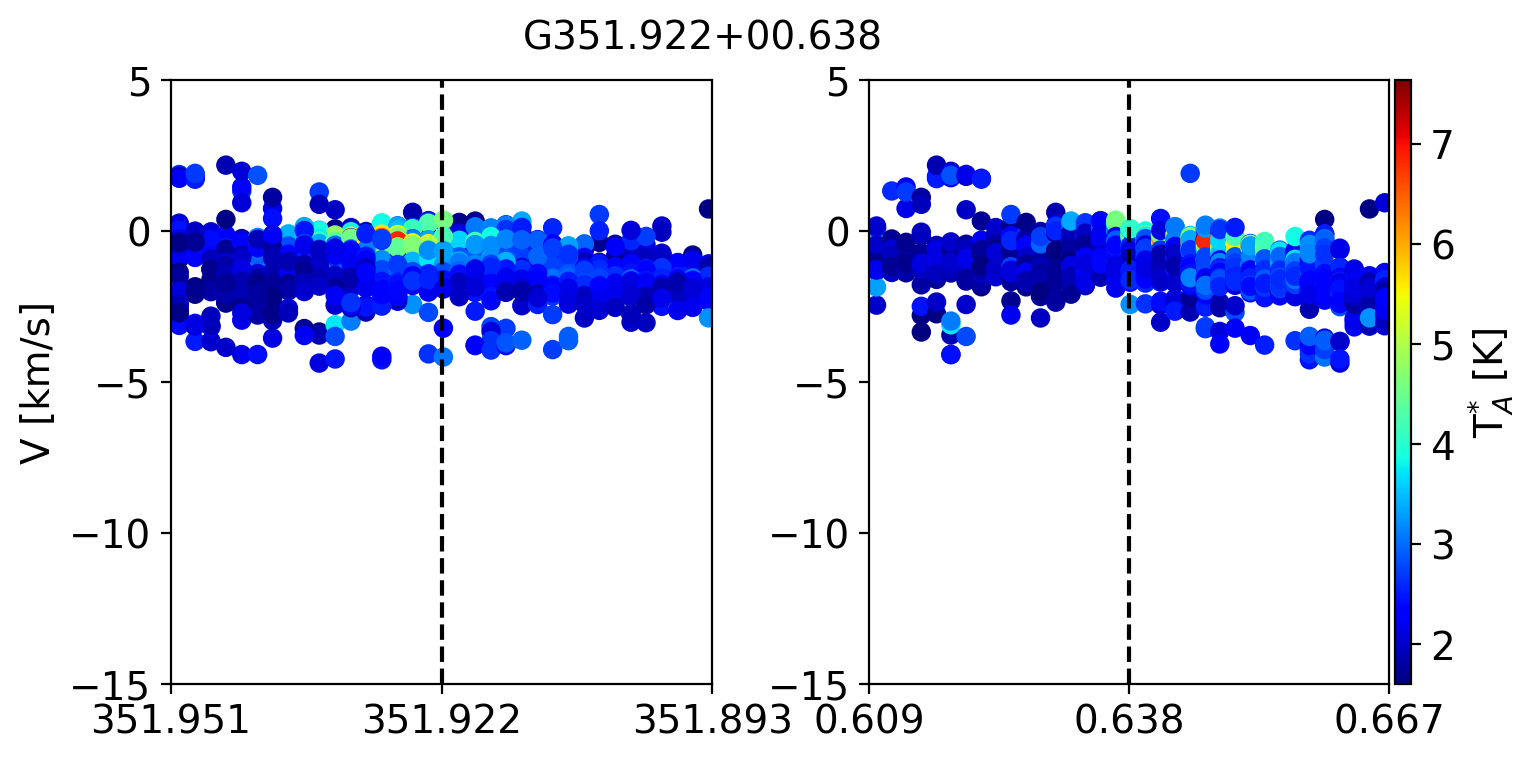}
    \includegraphics[width=0.45\linewidth]{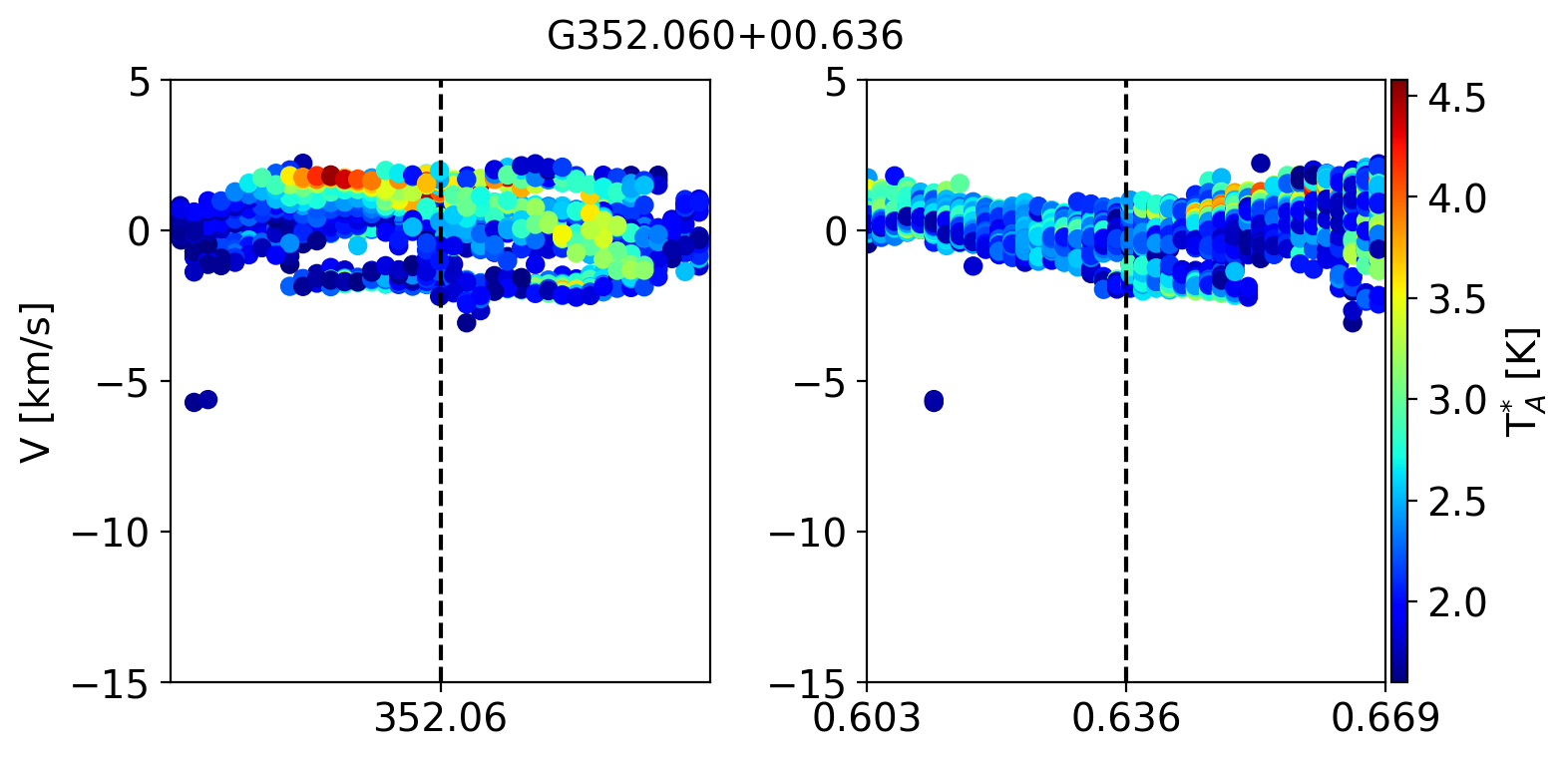}
    \includegraphics[width=0.45\linewidth]{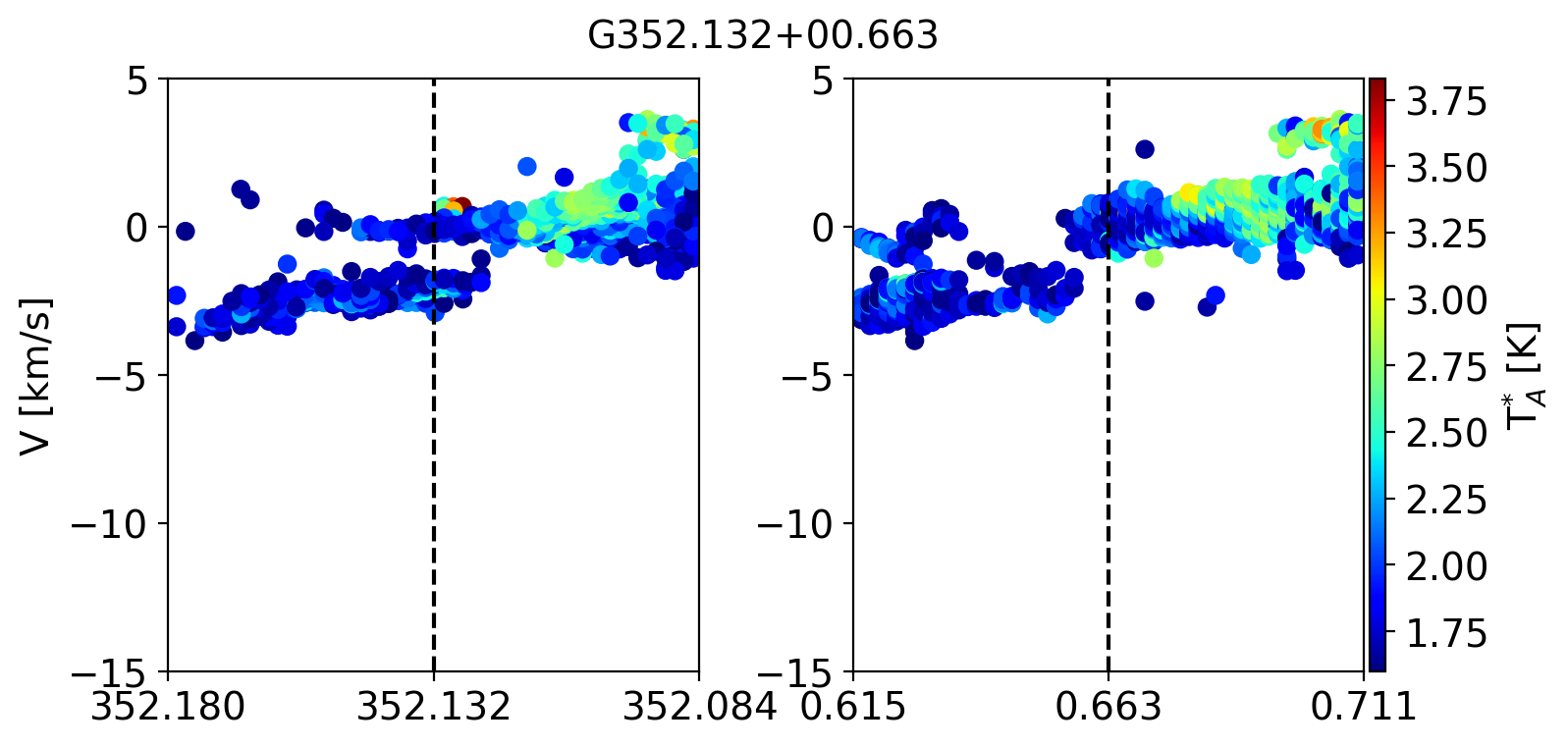}
    \includegraphics[width=0.45\linewidth]{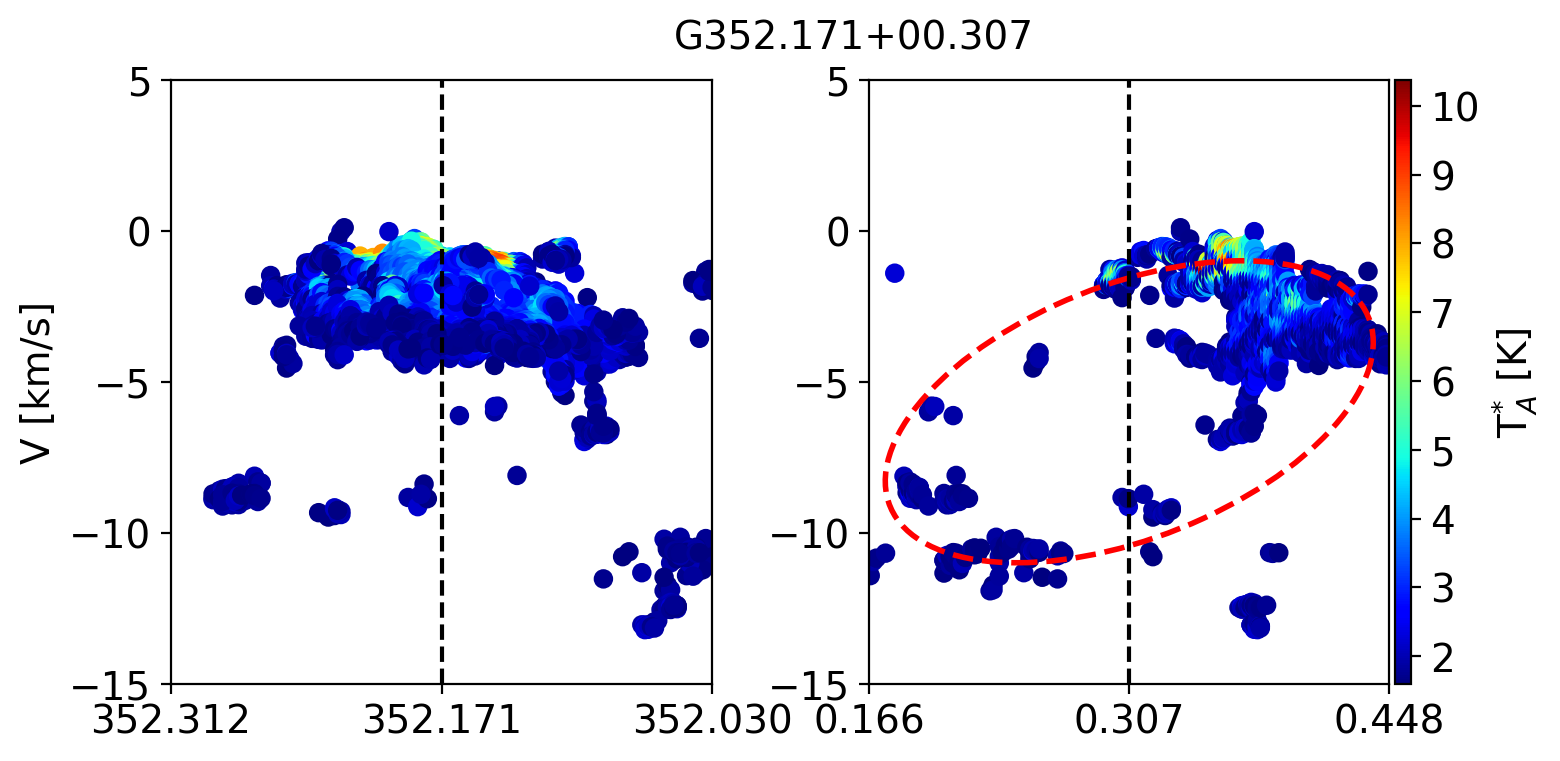}
    \includegraphics[width=0.45\linewidth]{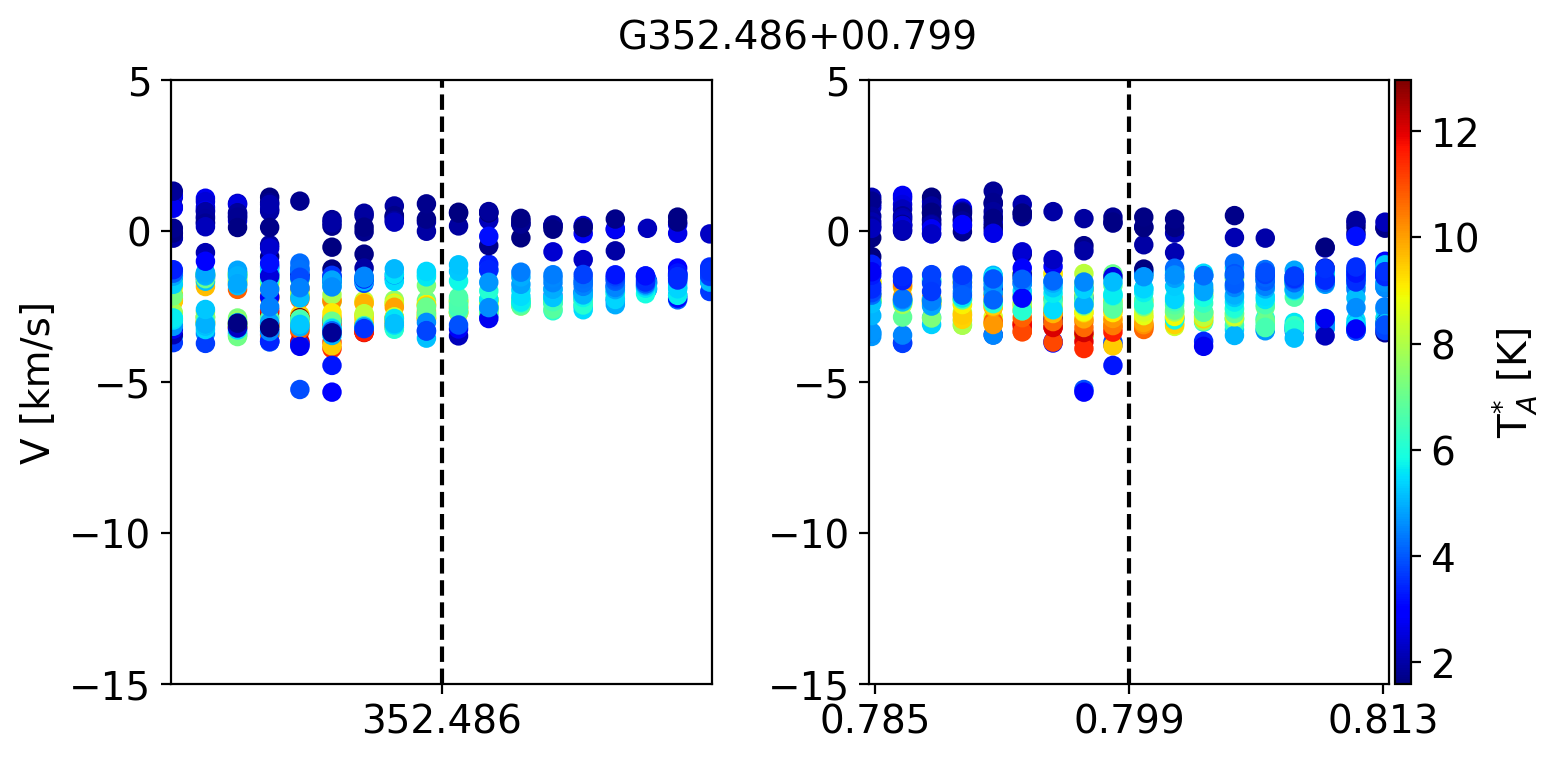}
    \caption{Same as Figure \ref{fig:lvplots_hiisources0}}
    \label{fig:lvplots_hiisources8}
\end{figure}

\begin{figure}[htbp!]
    \centering
    \includegraphics[width=0.45\linewidth]{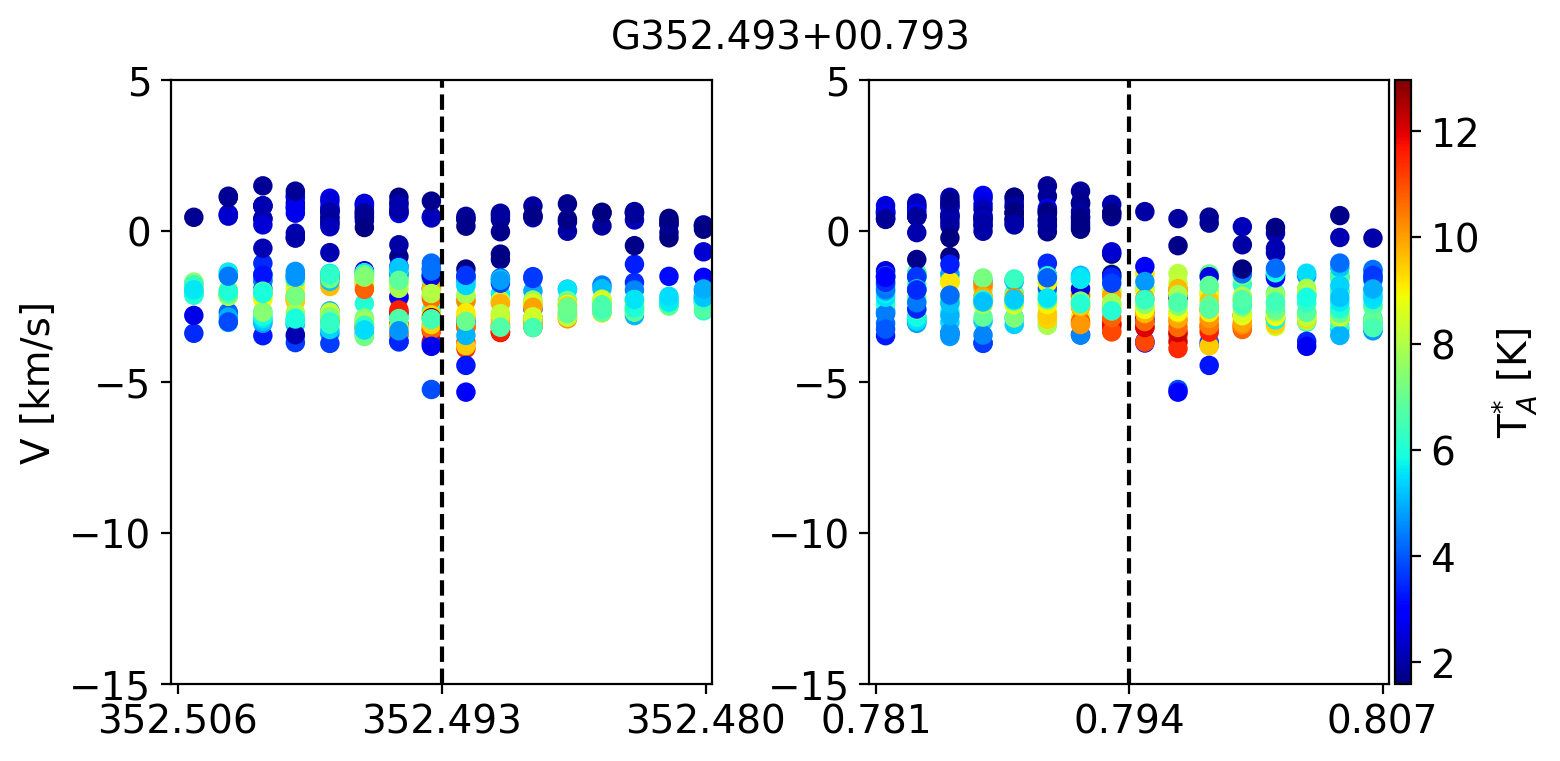}
    \includegraphics[width=0.45\linewidth]{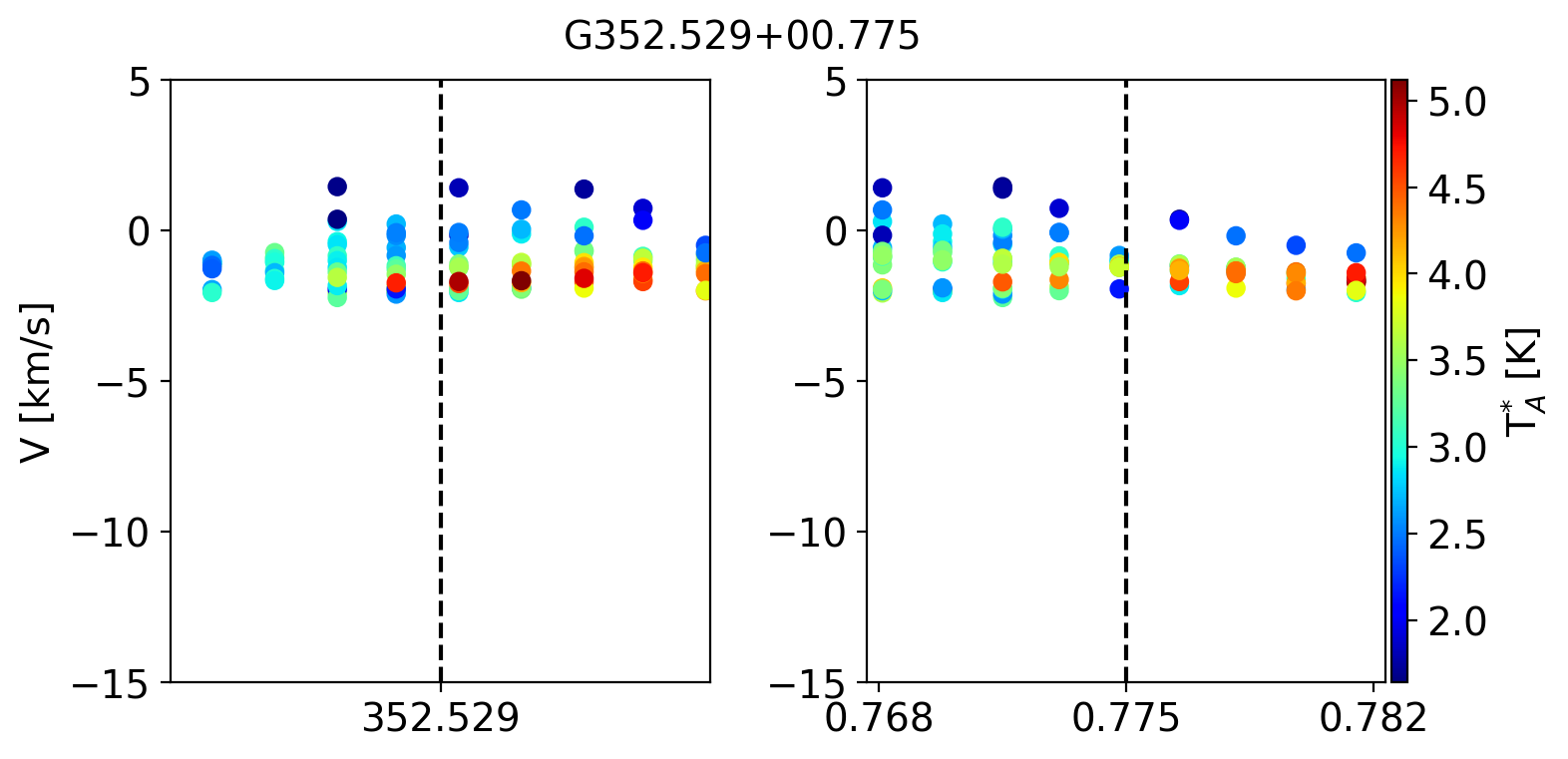}
    \caption{Same as Figure \ref{fig:lvplots_hiisources0}}
    \label{fig:lvplots_hiisources9}
\end{figure}

\begin{figure*}
    \centering
    \includegraphics[width=0.65\linewidth]{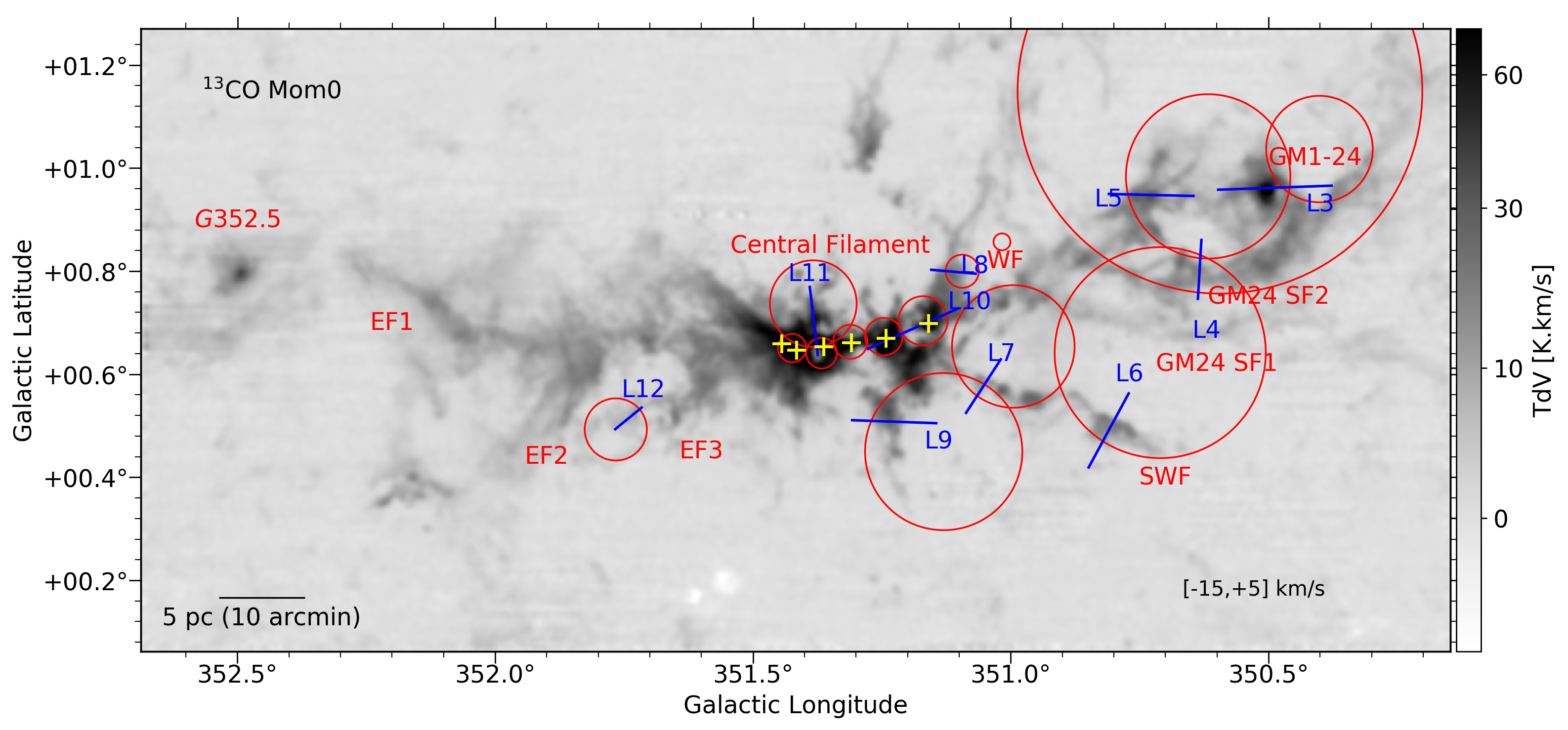}\\
    \includegraphics[width=5.5cm, height=4.5cm]{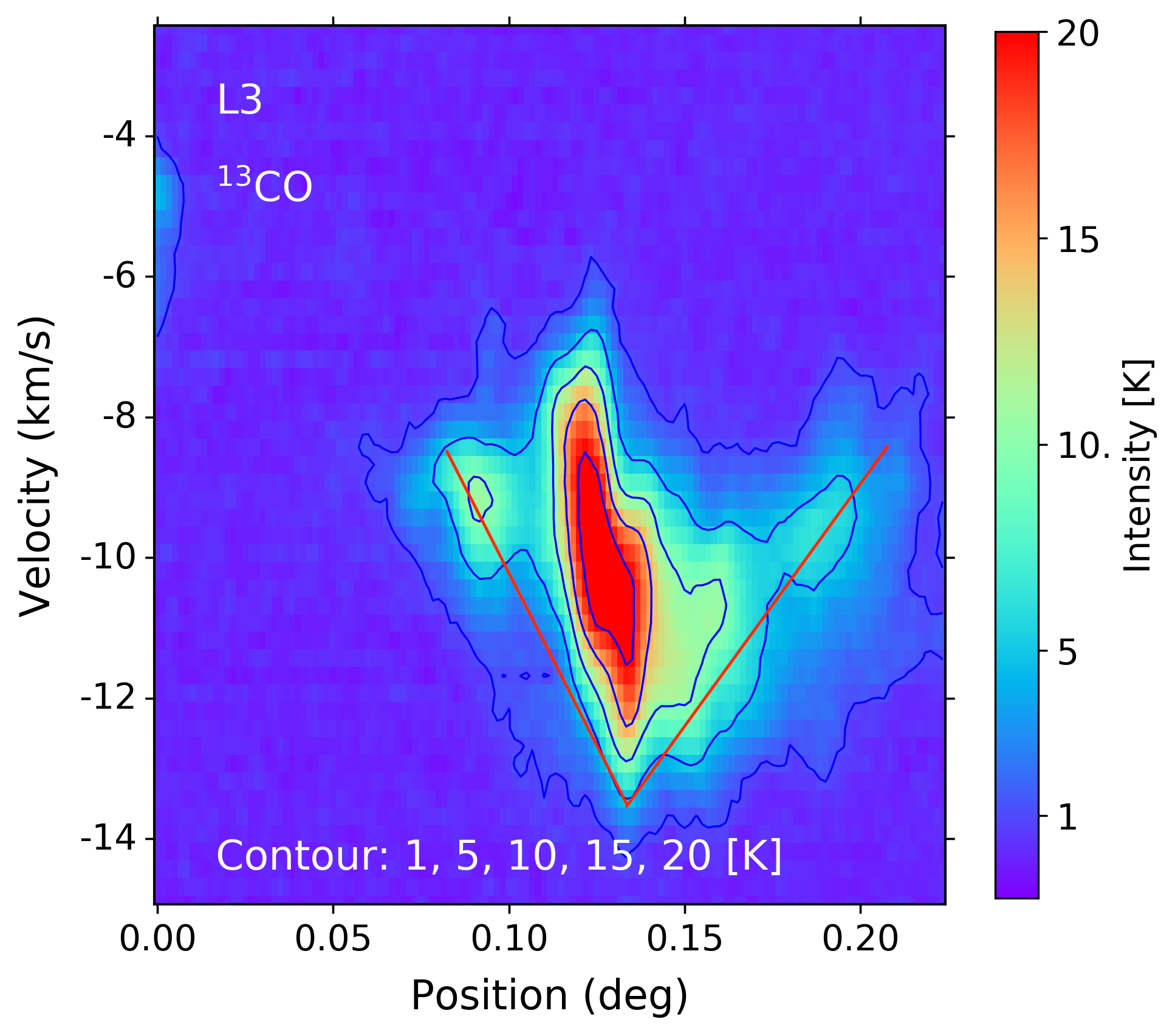}
    \includegraphics[width=5.5cm, height=4.5cm]{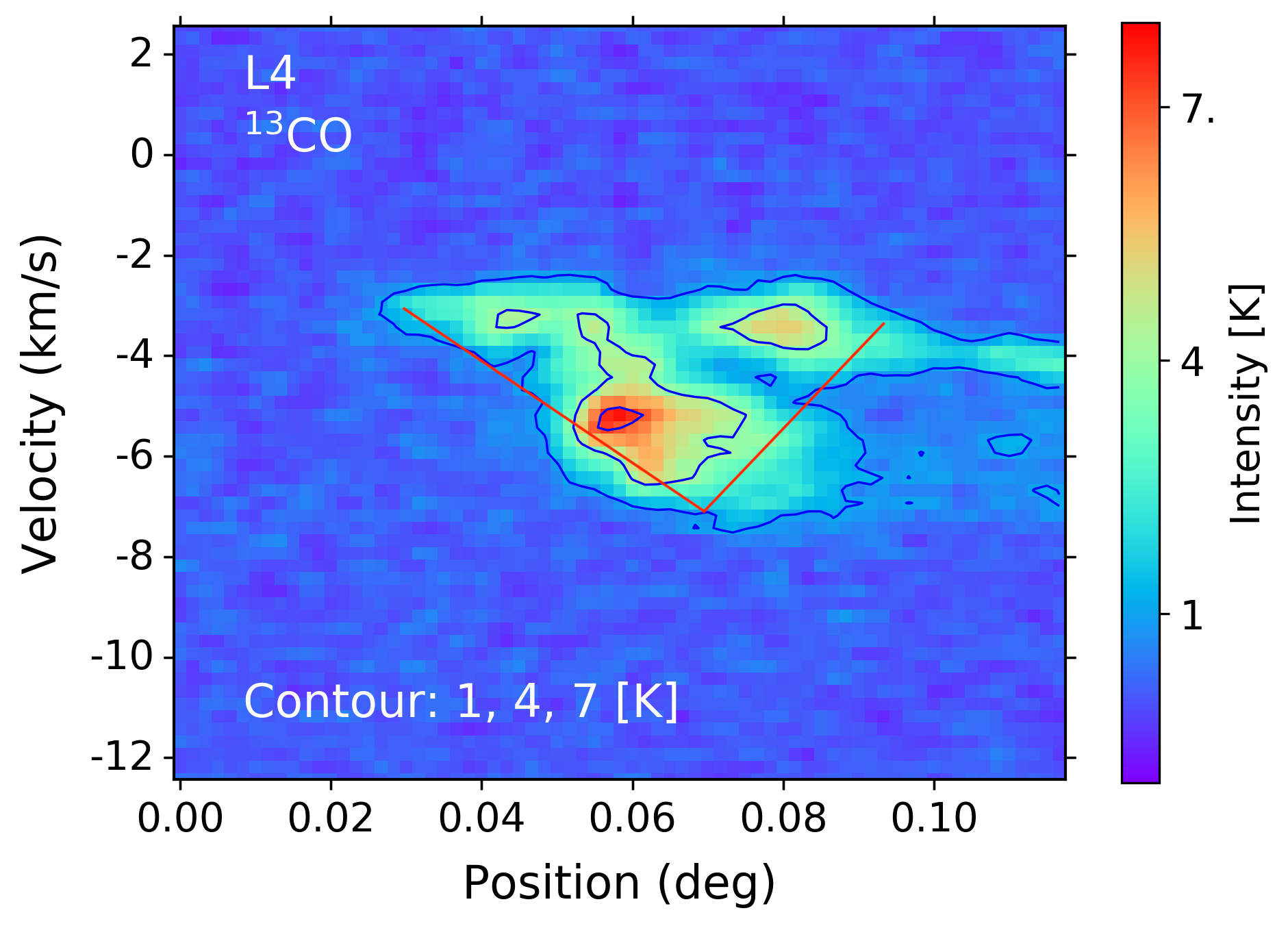}
    \includegraphics[width=5.5cm, height=4.5cm]{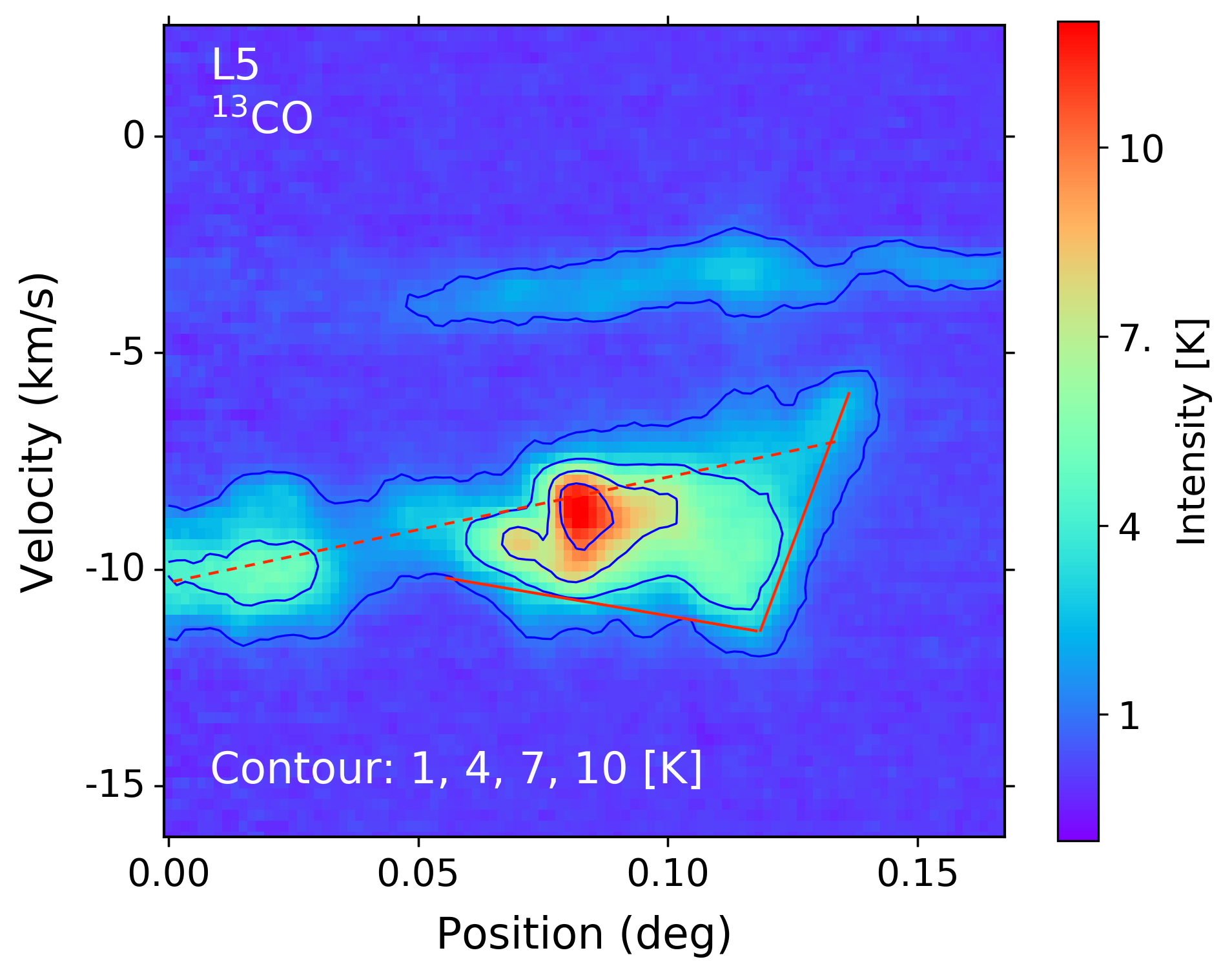}\\
    \includegraphics[width=5.5cm, height=4.5cm]{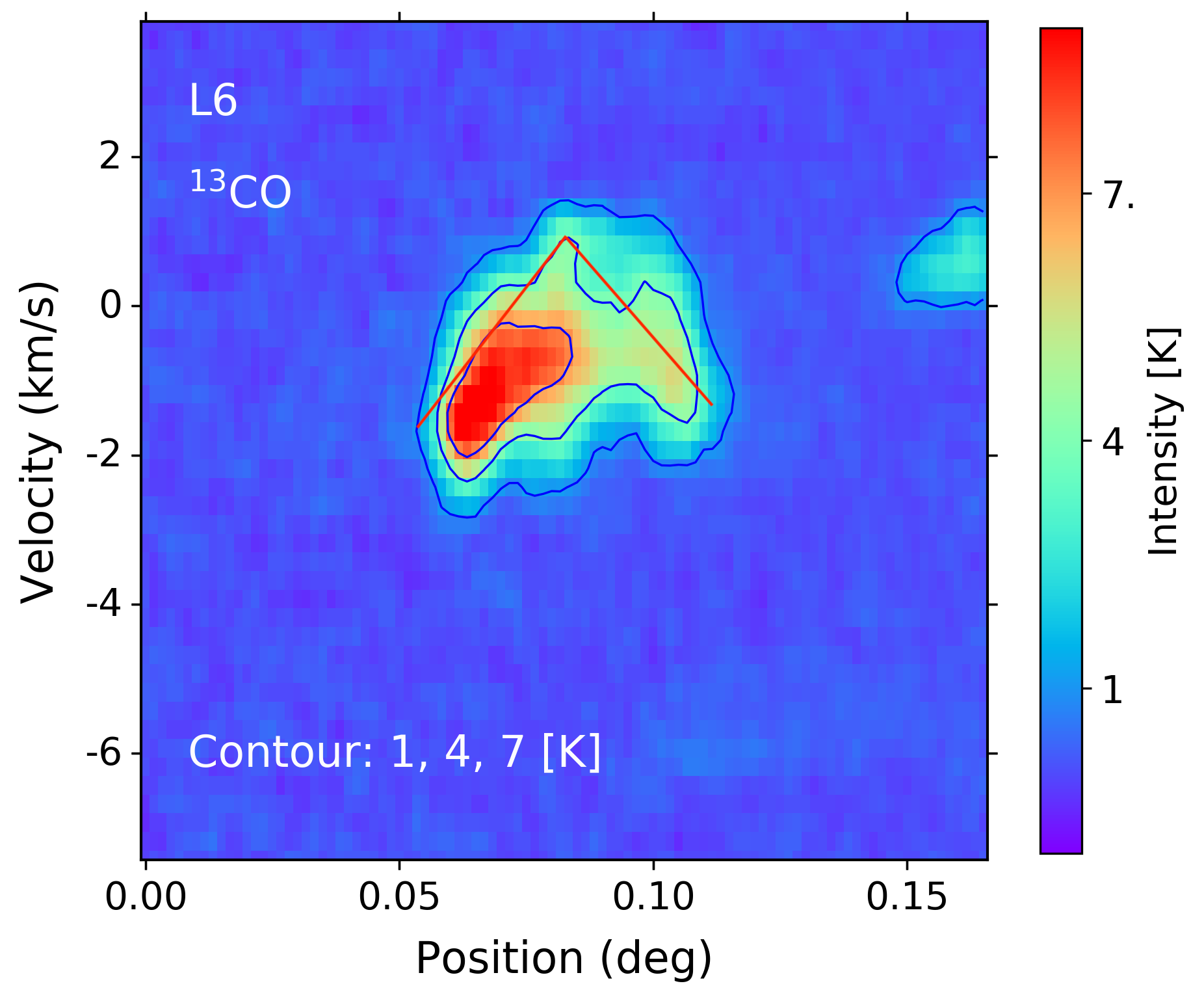}
    \includegraphics[width=5.5cm, height=4.5cm]{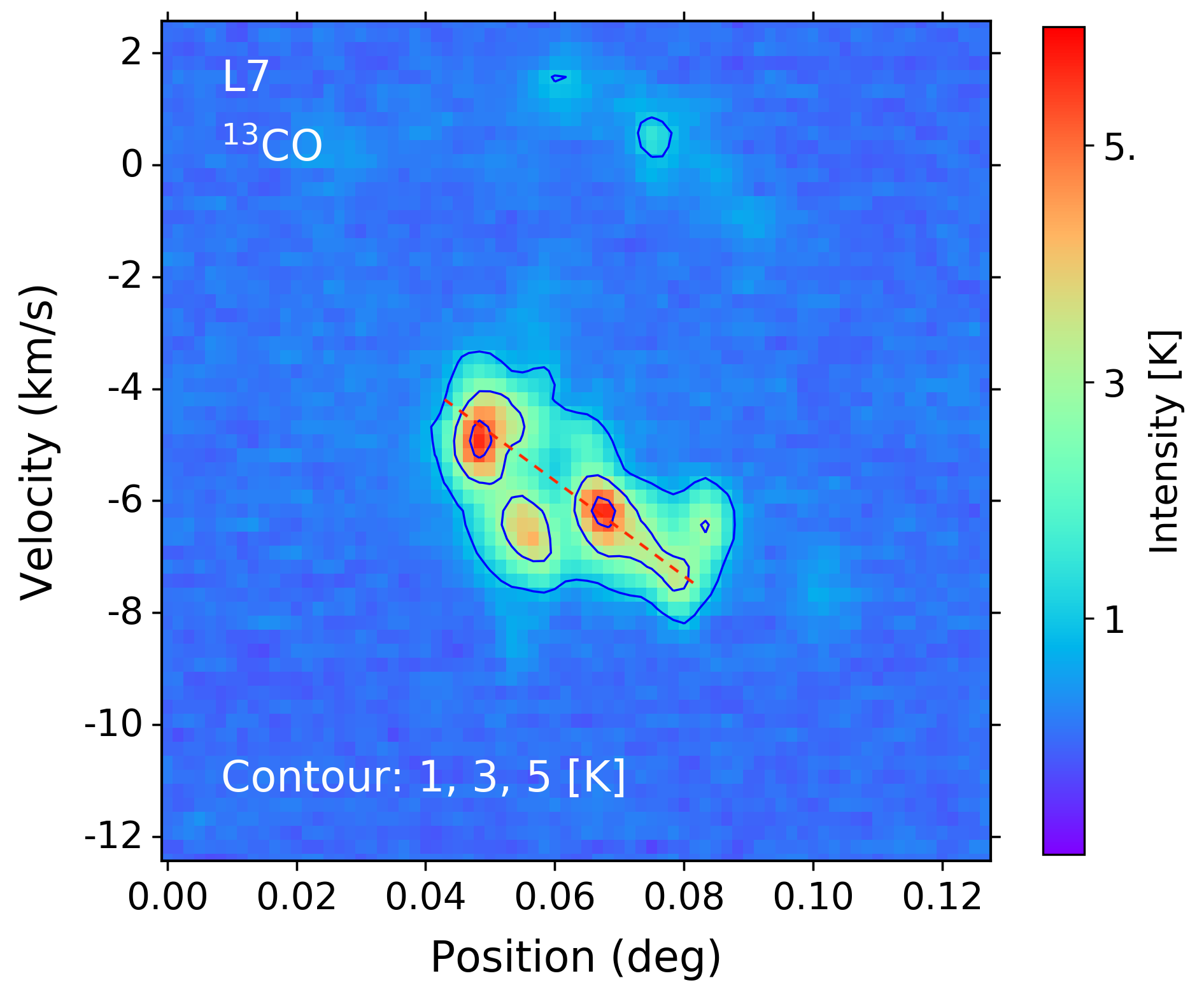}
    \includegraphics[width=5.5cm, height=4.5cm]{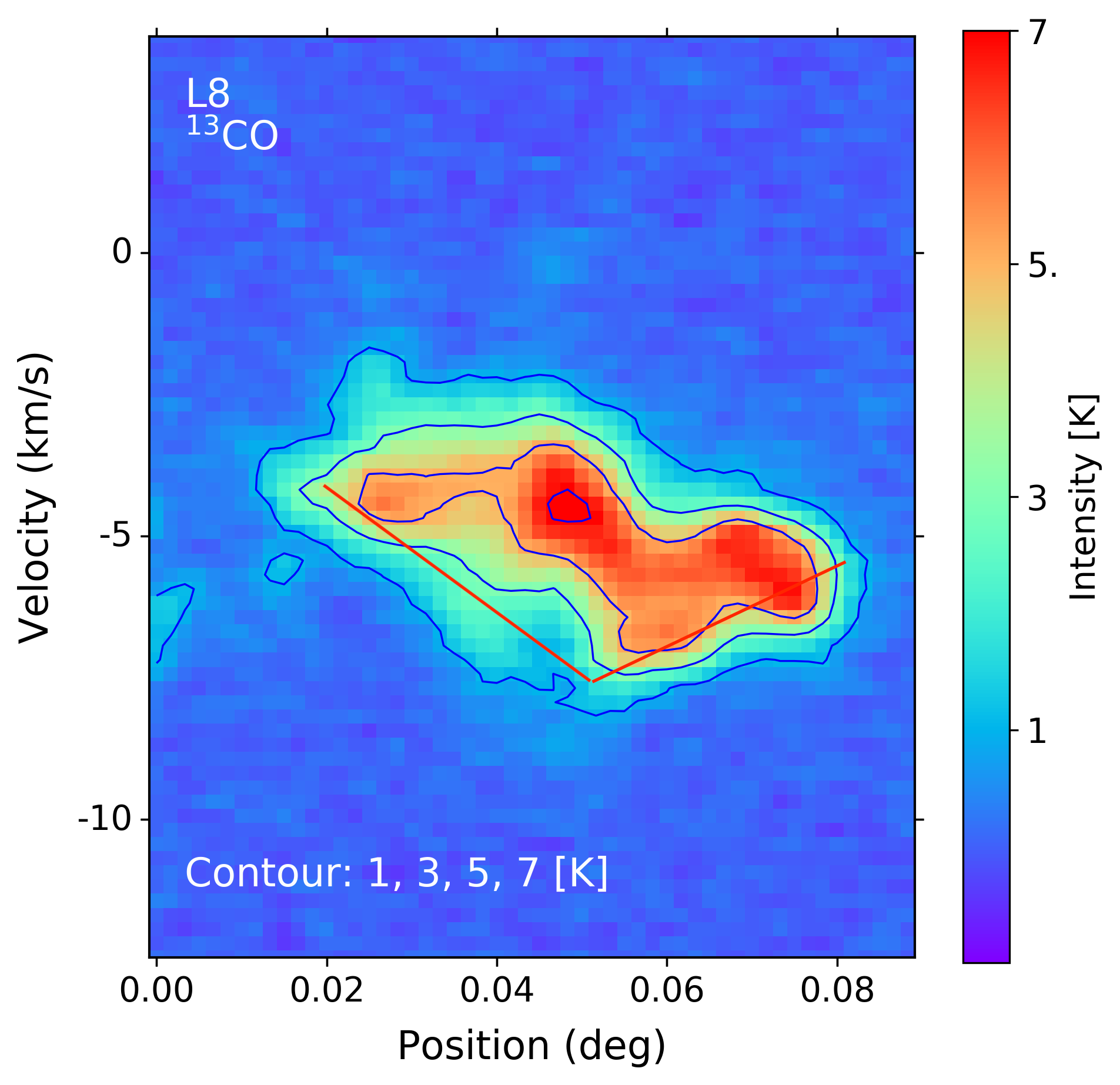}\\
    \includegraphics[width=5.5cm, height=4.5cm]{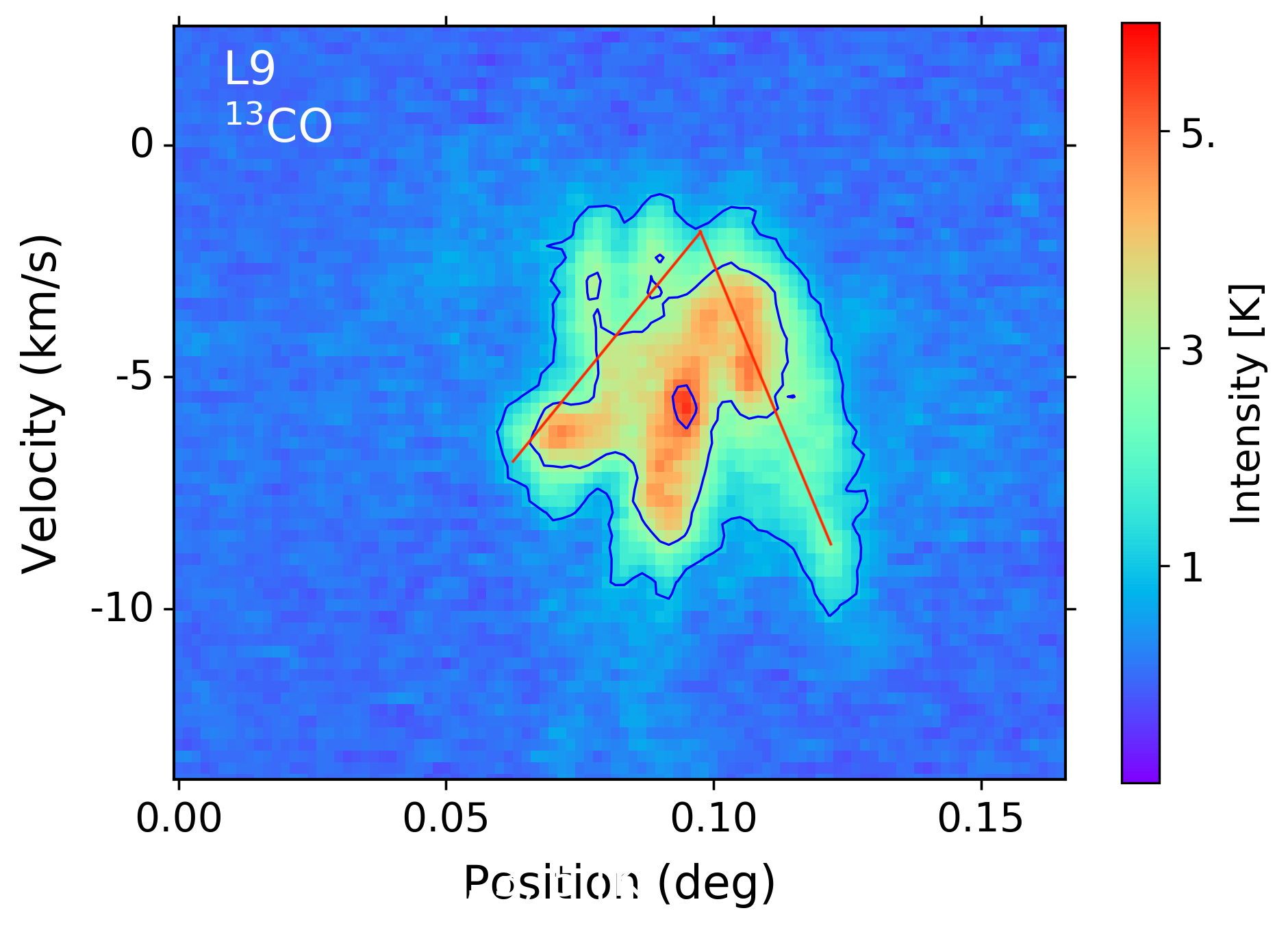}
    \includegraphics[width=5.5cm, height=4.5cm]{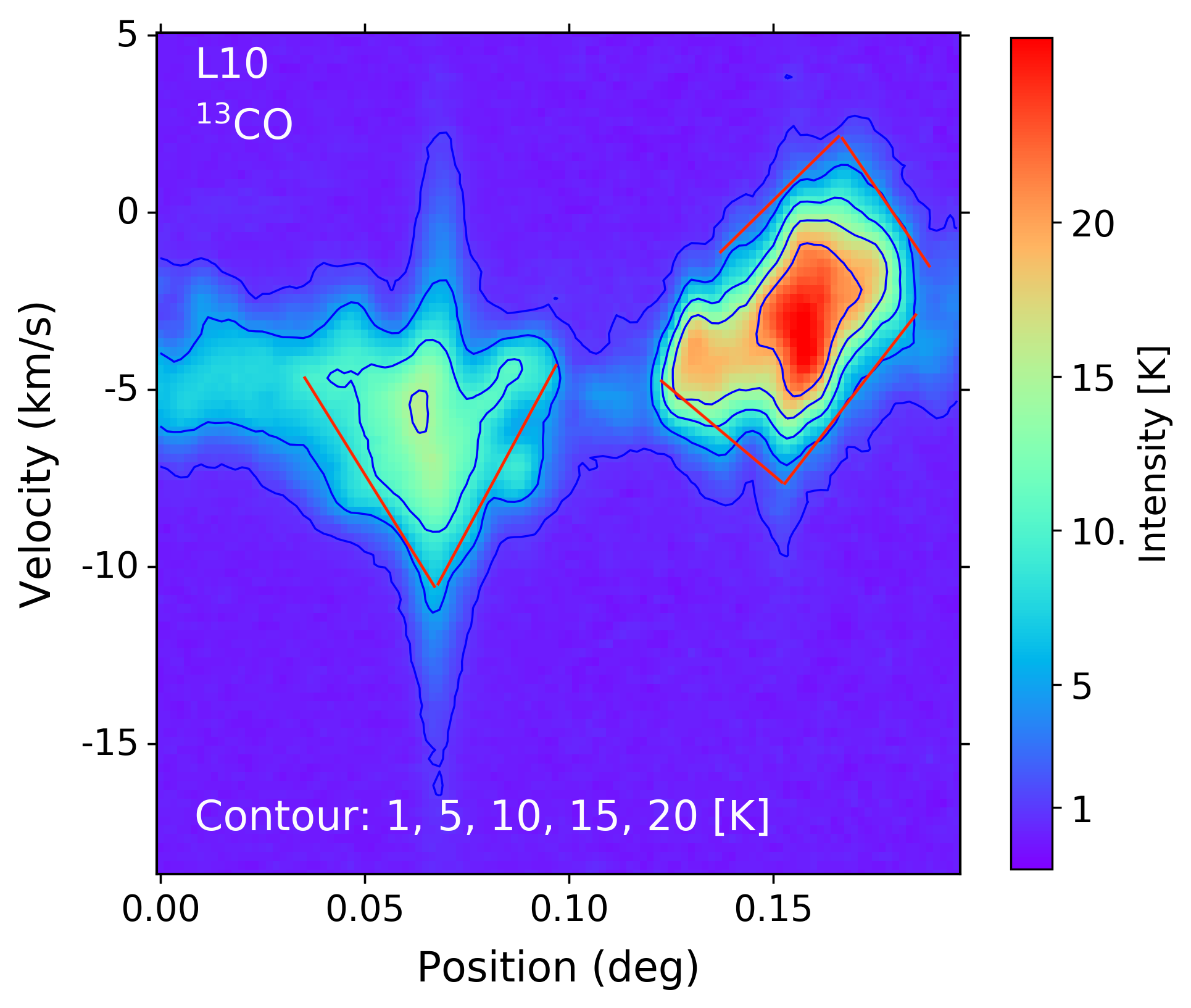}
    \includegraphics[width=5.5cm, height=4.5cm]{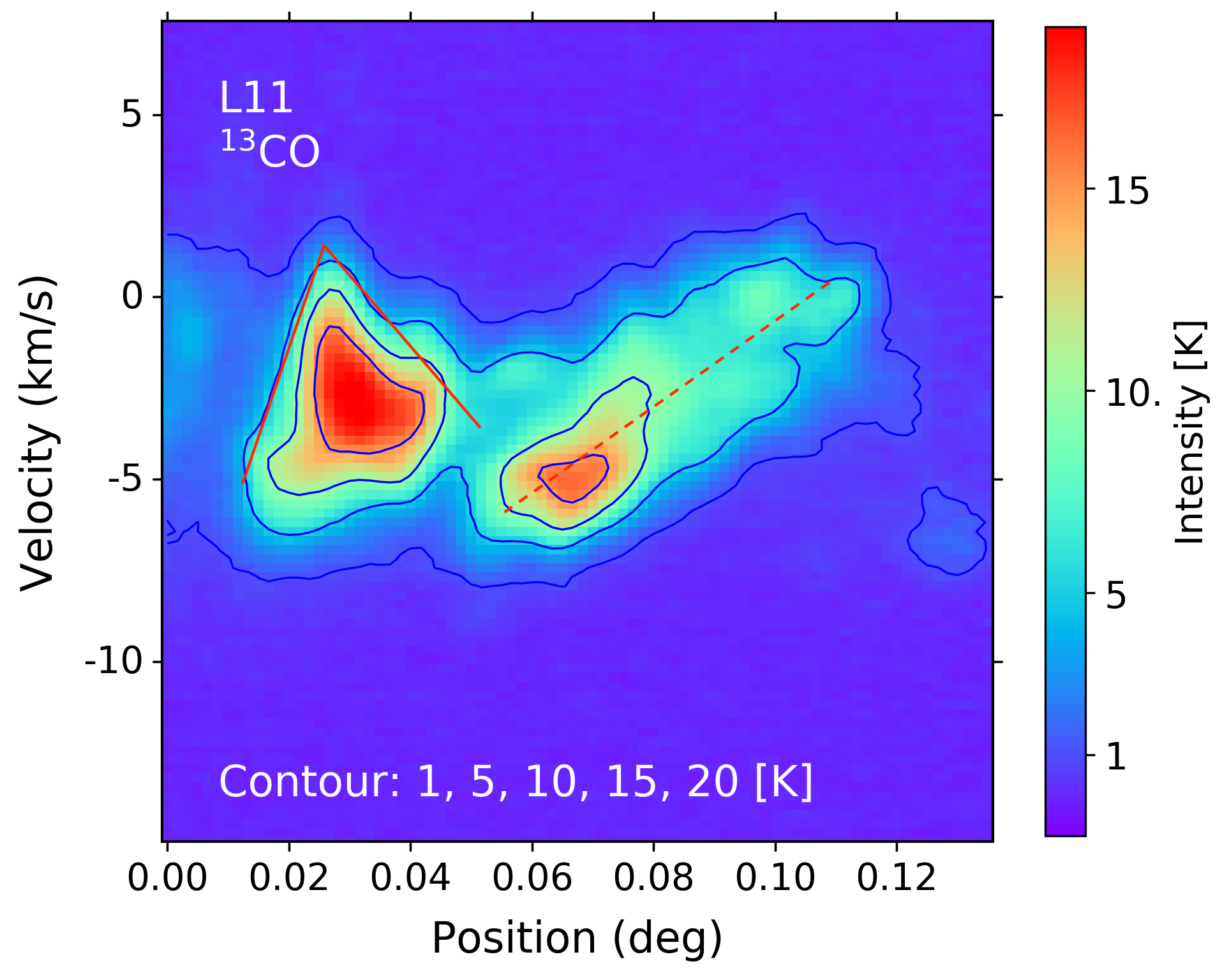}\\
    \includegraphics[width=5.5cm, height=4.5cm]{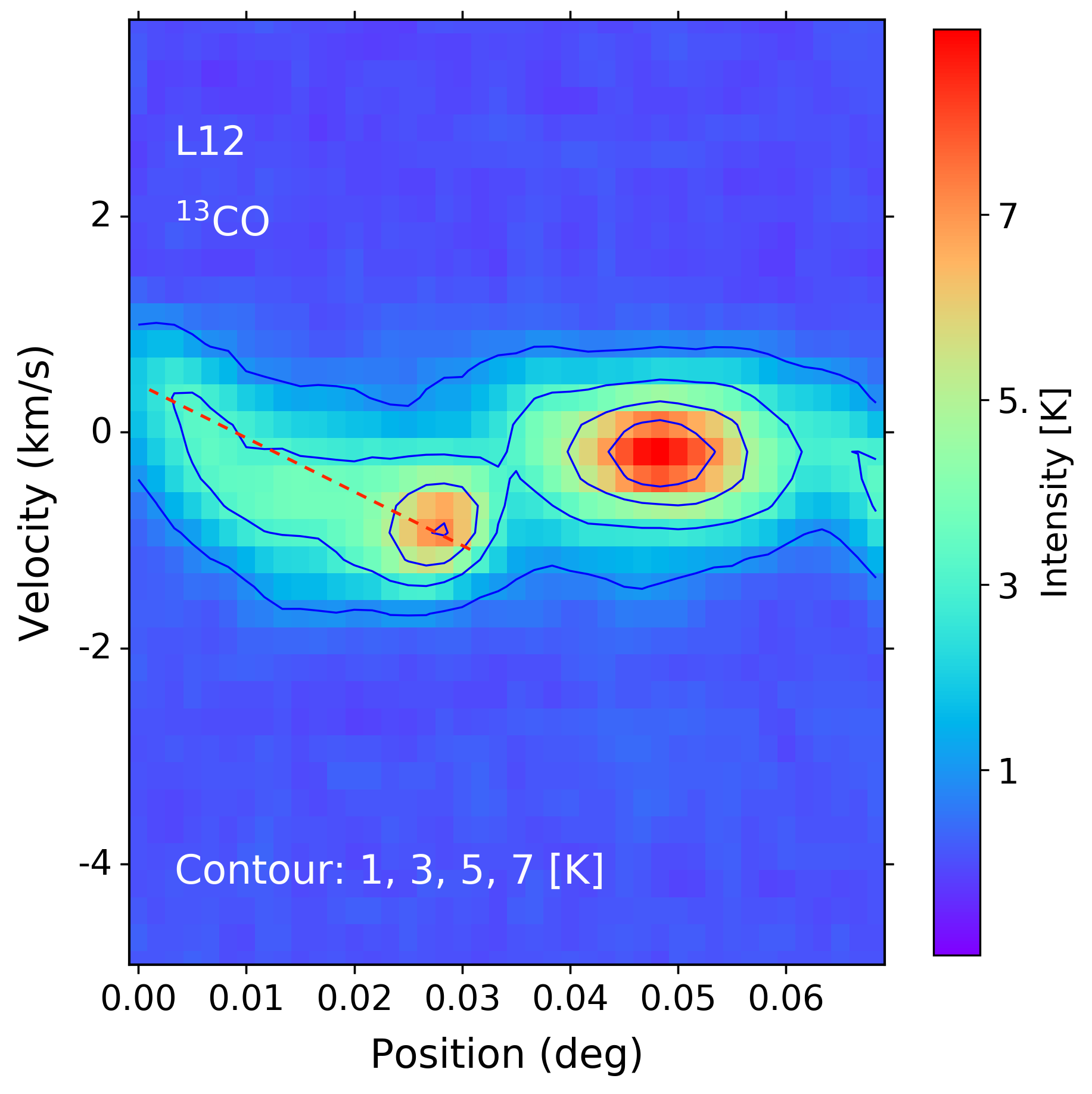}
    \caption{Top: $^{13}$CO moment0 map. H II regions that show signature of V-shaped velocity structure in Fig \ref{fig:lvplots_hiisources0} to \ref{fig:lvplots_hiisources9} are shown in red circles. Blue lines from L3 to L12 indicate the slices in which PV maps are constructed with widths of three beam size (0.016 deg). Bottom panels: PV maps along the blue lines indicated in the top panel image. Observed V-shaped velocity structure (red lines) and velocity gradient (dotted red lines) are indicated in each maps.}
\label{lvplots_added}
\end{figure*}

\begin{table*}[htbp!]
    \centering
 \caption{Expansion velocities (V$_{exp}$) of the H II regions derived from an elliptical fit to the $lv$ or $bv$ plots.}\smallskip
    \begin{tabular}{lcc}\hline\smallskip
      H II region   & Fit parameters (x, y, D1, D2, $\theta$) & V$_{exp}$ [km s$^{-1}$]  \\
                   & &  \\ \hline \hline
      G350.240+00.654  & 0.655, -7.5, 0.372, 10, -65  &  5.0 \\
      G350.401+01.037  & 1.037, -7.5, 0.207, 8, -40 &  4.0  \\
      G350.482+0.952   & 0.952, -10, 0.053, 8, 10 & 4.0 \\
      G350.505+0.957   & 0.957, -10, 0.053, 8, 10 & 4.0 \\
      G350.675+0.832   & 350.675, -5, 0.060, 7, 5 & 3.5 \\
      G350.71+0.642    & 0.642, -8, 0.408, 10., 40 & 5.0 \\
      G351.367+0.641   & 0.641, -5.5, 0.059, 8, 5  & 4.0 \\
      G351.42+0.638    & 0.638, -5, 0.029, 12, 5 & 4.0 \\
      G351.462+0.557   & 0.557, -2.5, 0.038, 8, 10 & 4.0 \\
      G351.766+0.493   & 0.493, -4, 0.120, 9, -20. & 4.5 \\
      G351.835+0.757   & 351.835, -3.5, 0.048, 8, -5 & 4.0\\ 
      G352.171+0.307   & 0.307, -6, 0.235, 10, -40 & 5.0  \\ \hline
    \end{tabular}

    \label{tab:vexp}
    \tablefoot{Fit parameters (central position in [l,v]) or [b,v], diameter in x-axis [deg], diameter in y-axis [km/s] and position angle [deg]) are given in Col.2. Derived expansion velocities V$_{exp}$ in km/s are given in Col. 3.}
\end{table*}

\begin{figure*}
    \centering
   \includegraphics[width=0.497\linewidth]{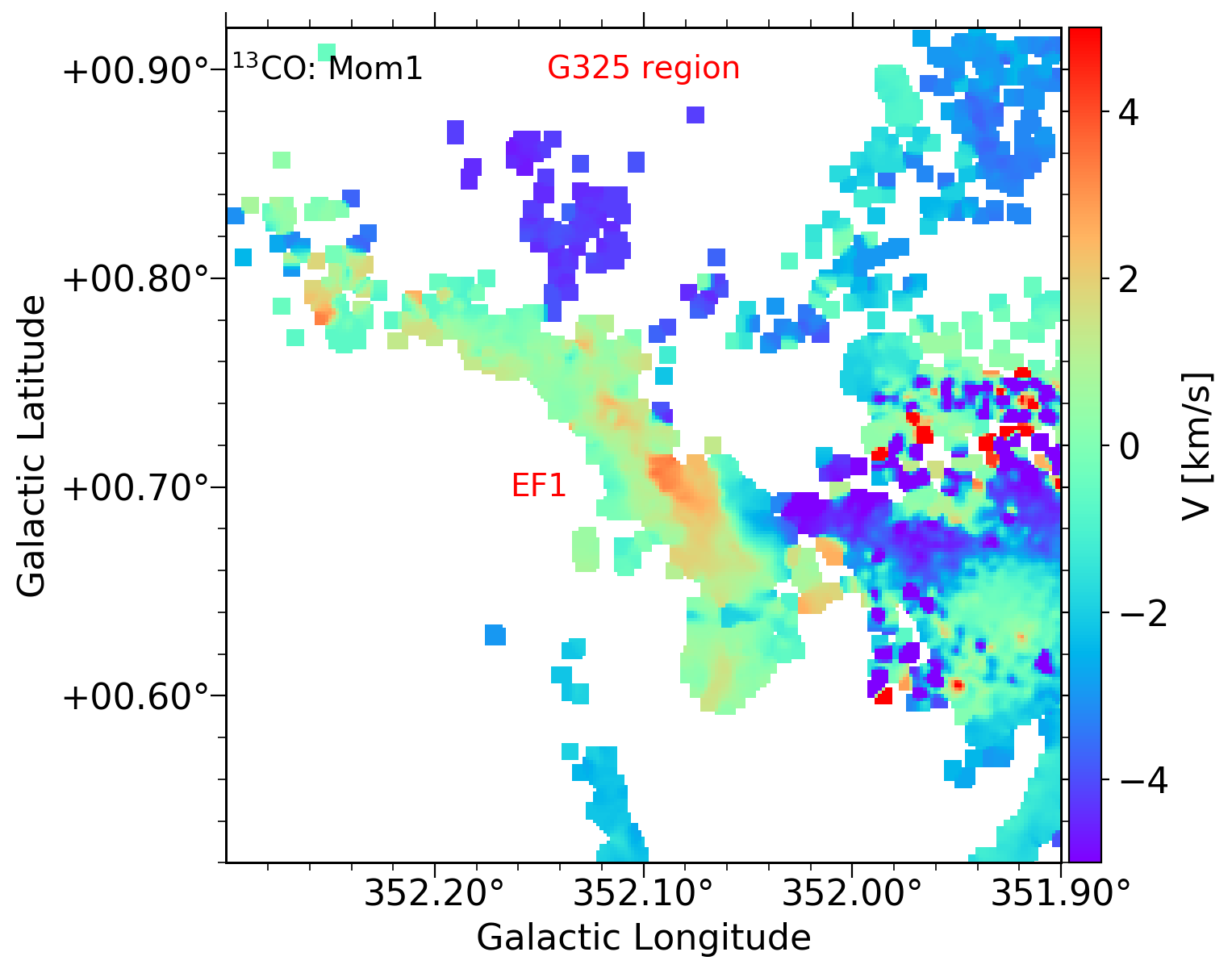}
   \caption{$^{13}$CO moment one map toward G352 region (EF1). The velocity gradient in hub-filament candidate EF1 filament is clearly observable. \label{mom1g352}}
\end{figure*}

\end{document}